%% file: thesis_v2.tex
\documentclass[11pt, a4paper, twoside]{Thesis} 
 \pdfoutput=1
 \usepackage{dcolumn}   
\usepackage{bm}        
\usepackage{amssymb}   
\usepackage{amsmath}
\usepackage{sidecap}
\usepackage{graphicx,psfrag,subfigure}
\usepackage{adjustbox}
\usepackage{color}
\usepackage{slashed}
\usepackage{xcolor}
\usepackage{wrapfig}
\usepackage{lscape}
\usepackage{rotating}
\usepackage{tikz}
\usepackage{adjustbox}
\usetikzlibrary{shapes.geometric, arrows}
\tikzstyle{startstop} = [rectangle, rounded corners, minimum width=2cm, minimum height=1cm,text centered, draw=black, fill=red!30,xshift=4cm]
\tikzstyle{io} = [trapezium, trapezium left angle=70, trapezium right angle=110, minimum width=1cm, minimum height=1cm, text centered, draw=black, fill=blue!30]
\tikzstyle{process} = [rectangle, minimum width=1cm, minimum height=1cm, text centered, draw=black, fill=orange!30]
\tikzstyle{decision} = [diamond, minimum width=3cm, minimum height=1cm, text centered, draw=black, fill=green!30]
\tikzstyle{arrow} = [thick,->,>=stealth]
\usepackage{multirow}

\newcommand{\bea}{\begin{eqnarray}}
\newcommand{\eea}{\end{eqnarray}}
\newcommand{\bee}{\begin{eqnarray*}}
\newcommand{\eee}{\end{eqnarray*}}
\newcommand{\be}{\begin{equation}}
\newcommand{\ee}{\end{equation}}

\usepackage{appendix}
\usepackage{lineno}
\modulolinenumbers[5]
\usepackage{array}
\usepackage{float}
\usepackage{placeins}
\usepackage{stackengine}
\usepackage{url}
\usepackage{numprint}
\usepackage{caption}
\usepackage{booktabs}  
\nprounddigits{3}
\newcolumntype{P}[1]{>{\centering\arraybackslash}p{#1}}
\newcolumntype{M}[1]{>{\centering\arraybackslash}m{#1}}
\setstackEOL{\#}
\setstackgap{L}{12pt}
\usepackage[utf8]{inputenc}
\usepackage{palatino}
\usepackage{mathtools}
\usepackage{layout}
\usepackage{afterpage}
\usepackage{lipsum}
\usepackage[Glenn]{fncychap}

\usepackage[colorlinks]{hyperref}
\hypersetup{
    colorlinks=true,
    linkcolor=blue,
    filecolor=magenta,      
    urlcolor=cyan,
}
\def\lsim{\mathrel{\raise.3ex\hbox{$<$\kern-.75em\lower1ex\hbox{$\sim$}}}}
\def\gsim{\mathrel{\raise.3ex\hbox{$>$\kern-.75em\lower1ex\hbox{$\sim$}}}}

\usepackage{fancyhdr}
\usepackage[avantgarde]{quotchap}
\usepackage{epsfig,epsf}
\usepackage{amsthm}
\usepackage{amsfonts}
\usepackage[all]{xy}
\usepackage{enumitem}
\usepackage{cite}
\usepackage{doi}
\usepackage{lipsum}
\usepackage{setspace}
\usepackage{comment} 
\usepackage{siunitx,bm}
\usepackage{amsmath,amssymb,bm}
\usepackage{bbold}
\usepackage{lipsum}
\usepackage{hyperref}
\usepackage{graphics,graphicx}
\usepackage{csquotes}
\usepackage{mathtools}
\usepackage{slashed}
\usepackage{color,colordvi}
\usepackage{soul}
\usepackage{xcolor}
\usepackage{array}
\usepackage{multirow}
\usepackage{silence}
\usepackage{booktabs}
\usepackage{tabularx,ragged2e,booktabs,caption}
\usepackage{comment}
\usepackage{slashed}
\usepackage{float}
\usepackage[normalem]{ulem}
\usepackage{bbold}
\usepackage{epsfig}
\usepackage{hyperref}
\usepackage{graphicx}
\usepackage{csquotes}
\usepackage{mathtools}
\usepackage{slashed}
\usepackage{color,colordvi}
\usepackage{soul}
\usepackage{xcolor}
\usepackage{booktabs,multirow,array}
\usepackage{multirow}
\usepackage{tabularx,ragged2e,booktabs,caption}
\usepackage{comment}
\usepackage{slashed}
\usepackage{float}
\usepackage[normalem]{ulem}
\usepackage{caption}
\newcommand{\lqcd}{\Lambda_{\text{QCD}}}
\newcommand{\msbar}{\overline{\text{MS}}}

\renewcommand{\(}{\left(}
\renewcommand{\)}{\right)}
\renewcommand{\{}{\left\lbrace}
\renewcommand{\}}{\right\rbrace}

\newcommand{\del}{\partial}
\newcommand{\nn}{\nonumber}

\newcommand{\order}[1]{\mathcal{O}\({#1}\)}

\def \im{\textrm{Im}}

\newcommand{\alphas}{\alpha_\mathrm{s}}

\newcommand{\gfermi}{G_\mathrm{F}}
\newcommand{\GeV}{\,\mathrm{GeV}}

\newcommand{\MeV}{\,\mathrm{MeV}}

\newcommand{\Mt}{M^2_{\tau}}
\newcommand{\hs}{\hspace{.4mm}}
\newcommand{\bs}{\hspace{1cm}}

\newcommand{\mtsq}{M_{\tau}^2}
\newcommand{\ordas}[1]{\mathcal{O}\left(\alpha_s^{#1}\right)}
\newcommand{\mh}{M_H}

\newcommand{\pow}[2]{#1^#2}
\newcommand{\as}{\alpha_\mathrm{s}}

\newcommand{\ord}[1]{\mathcal{O}\({#1}\)}

\AtBeginEnvironment{subappendices}{%
\chapter*{Appendix}
\addcontentsline{toc}{chapter}{Appendices}
\counterwithin{figure}{section}
\counterwithin{table}{section}
}
\graphicspath{{Pictures/}} 
\hypersetup{urlcolor=black, colorlinks=false} 
\thesistitle{ Renormalization Group Summation at High Orders and Implications to the Determination of Some Standard Model Parameters}
\supervisor{Prof. B. Ananthanarayan}
\degree{Doctor of Philosophy}
\degreemajor{The Faculty of Science}
\authors{Mohd Siddique Akbar Alam Khan}
\university{Indian Institute of Science}
\department{Centre for High Energy Physics}
\placelng{Bengaluru-560012, India}
\datesub{July 31, 2023}
\datesig{July 31, 2023}
\title{\ttitle} 

\begin{document}

\frontmatter 
\setstretch{1.6} 
\fancyhead{} 
\rhead{\thepage} 
\lhead{} 
\newcommand{\HRule}{\rule{\linewidth}{0.5mm}} 
\maketitle


\clearpage 

\clearpage
\newpage\null\thispagestyle{empty}\newpage

\Declaration

\setstretch{1.3} 
\dedication{\addtocontents{toc}{}
\begin{flushright}
 {\huge\calligra To My Family}  
\end{flushright}
	}
	\clearpage

\setstretch{1.3} 
\newpage\null\thispagestyle{empty}\newpage
\acknowledgements{\addtocontents{toc}{}
I extend my heartfelt gratitude to Prof. B. Ananthanarayan for his invaluable guidance, support, and encouragement to explore various topics. I am equally appreciative of my collaborators Prof. Daniel Wyler and Dr. Diganta Das, for their constant support and for clarifying many doubts while working on various projects. I am thankful to other collaborators, Prof. Nasrallah F. Nasrallah, Dr. Gauhar Abbas, and Ms. Vartika Singh, for the interesting projects that will soon be publicly available. \par
I am very thankful to former chairperson Prof. Justin David and current chairman Prof. Sudhir Vempati other CHEP faculties, especially Prof. Aninda Sinha, Dr. Prasad Hegde, and Dr. Chethan Krishnan, for creating a friendly environment and clarifying many doubts during my Ph.D. coursework. I am also very thankful to Prof. Apoorva Patel for the financial support. 

I express my gratitude to Mr. Keshava and Saravana for efficiently managing various official matters for me and to Mrs. Mallika for keeping my desk organized.

I am thankful to my seniors: Dr. Rahool, Dr. Aradhita, Dr. Shayan, Dr. Ratan, Dr. Shuili, Dr. Moid, Dr. Lata, Dr. Sipaz, Amandip, and others, for providing us with a friendly environment at CHEP. I am very grateful to Dr. Priyanka Lamba for the in-depth discussions that helped me to understand various aspects of physics and life.

I am also thankful to my batchmates: Dr. Ahamadullah, Dr. Pratik, Prabhat, Dr. Rhitaja, and Dr. Simran, as well as to my juniors: Adarsh, Pingal, Souvik, Sudeepan, Srijan, Dr. Sabarnya, Gulfisha, and Ranjini, for giving me a wonderful time at IISc.  I also cherish the enjoyable moments shared with my table tennis group  Satyajeet, Simran, Sagar, Rajat, Srashti, Mohit,  Tuhin, and Ujjwal, during the final stages of my Ph.D. I am also thankful to Suraj, Apoorva, Neha, Avinash, Saibal, and Apoorv for their friendship and for making my early days at IISc memorable. Special thanks to my friends, Prerana and Sahel, for their love and support, particularly during the writing phase of this thesis.

My special thanks to KG teacher, Ms. Kanti, for making those days wonderful and memorable. I also thank my high school teacher, D. K. Mishra, whose interesting way of teaching physics got me interested in physics. I am also grateful to Prem Singh, who believed in me even during my early struggles with coding, which later became one of my hobbies and greatly aided to my research. My sincere thanks to Prof. Sajjad Athar for his inspiring lectures on particle physics, which played a key role in pursuing particle physics as my career.
I thank my best friend, Dr. Monish, for his unwavering emotional support and all the help over the years. I also thank my batchmates from AMU, especially  Arshad, Mehroosh, and Rameez, whose presence enriched my life. I also thank Dr. Rehan for his guidance and support at various stages of my career. I thank my school friends: Ashish, Vikas, Amarendra, Vivek, Priyanka and Surabhi for giving me wonderful memories in Navodaya.

I am deeply grateful to my mother, Arfeen Begum, for her love, emotional support, and her spiritual teachings. To my father, Aftab Alam Khan, I owe immense thanks for getting me interested in physics and mathematics and for his support in pursuing education despite various challenges.\par
I thank my younger sister, Shafaque Naz, for adding joy and excitement to my life at home with her playful teasing. My special thanks to my grandfather, Mr. Iftikhar Khan, for his support during critical moments of my career. I also wish to thank my late maternal grandfather, Mr. Nabi Hasan, a remarkable man whose hospitality, assistance, and respect for others have impacted my life. He has been an inspiration to me, and  I am deeply grateful for the time I had with him.

I am indebted to my uncles: Parvez, Naukhej, Firdos, Shoeb, Aslam, Javed, Amir Hasan, Shareef, Kafeel, Haneef, Shakeel (late), Rafeeq, Aleem, and Zubair, and aunts: Nasreen, Parveen, Pappi, Anjum, and Aasiya for their love and support. I thank my aunts Sabina and Anjum, who hold a special place in my heart, for their care during my childhood. I am thankful to my cousins: Rizwan, Zuber, Kaifi, Sheebu, Tariq, Naushad, Khuldoon, Saad, Sahil, Haider, Zaman, Sahan, Ahsan, Nawaz, Arslan, Ashar, Afeefa, Lucky, Daraksha, Kausar, Nujhat, Zeenat, Shayda, Huma, Nazia, Tausiba, Goldie, Tarannum, Nusrat, Kashfa, and Huda as well as my adorable nieces: Fari and Pari for their love. I am also thankful to my brothers-in-law Arif Zaman and Aawesh, and sisters-in-laws Sabina, Shaista, Naureen, and Shabnam for their love and friendship. I am also thankful to my childhood friends from my village: Laik, Matin, Motees, Bablu, Maifooj, Shameem, Rizwan, Rehan and others, for giving me beautiful memories. I am very thankful to my late grandmother Anisun, late uncle Khalid Khan, and late aunt Shehnaaz, who provided me with a beautiful childhood but are no longer with us due to the COVID-19 pandemic.

To all others whose names may not be mentioned but who directly or indirectly supported and inspired me, I am genuinely thankful. This work would not have been possible without the collective efforts and contributions of countless individuals, and I am forever grateful for making this work accessible to everyone.

I acknowledge financial support from the Ministry of Human Resource Development (MHRD), Govt. of India. }
\clearpage 
\null\thispagestyle{empty}\newpage
\setstretch{1.3} 
\Synopsis{{ }
In perturbation theory, predictions from theories like Quantum Chromodynamics (QCD) are obtained by evaluating Feynman diagrams to high orders. Such calculations for results for various processes are already available in the literature, and their theoretical predictions depend on various parameters. With the availability of a large amount of data from experiments, it is possible to extract these parameters by comparing theoretical predictions with data. However, due to the finite order terms available from theory, any parameter determination depends on the perturbative scheme used and the choice of the renormalization scale. Once a renormalization scheme is fixed, the variation of the renormalization scale in a certain range can lead to large uncertainties, and optimizing perturbative series with respect to such free parameters is necessary. We have achieved such optimization using the renormalization group summed perturbation theory (RGSPT), and the resulting perturbative series is significantly less sensitive to the renormalization scale dependence.  It is a renormalization group (RG) improved version of the fixed order perturbation theory (FOPT), where the running RG-logarithms are summed to all orders using the RG equation. Once these running logarithms are summed, various operations such as analytic continuation, contour integrals, and Borel-Laplace transform are found to have enhanced convergence and scale variation improvement compared to a FOPT analysis. These operations are important in the precision determination of pQCD parameters using methods such as QCD sum rules.\par
In chapter~\eqref{Chapter1}, we give a brief introduction to the standard model of particle physics and its parameters. We also discuss the role of the renormalization group in QCD and various schemes based on it that are used in the literature for the precise determination of some of the standard model parameters.\par
In chapter~\eqref{Chapter2}, we discuss the RGSPT and the procedure of the running logarithm in this scheme. Derivation of the running of the strong coupling constant and quark masses in the $\msbar$ scheme as well as the relation between $\msbar$ and pole mass scheme, are also presented. These relations are very crucial for various other studies, such as ones based on QCD sum rules. \par
In chapter~\eqref{Chapter3}, we study the RG improvement of the QCD static energy between the two static heavy quarks sources. It is an important quantity that can be calculated using perturbative QCD as well as in lattice QCD simulations and has application in the precise determination of the strong coupling constant at low energies. The perturbative corrections to three-loop ($\as^4$) and some ultrasoft corrections to N$^4$LO ($\as^5$) were already calculated in the literature about a decade ago. We used asymptotic Pad\'e approximants to estimate the unknown perturbative contribution at four loops, which completes the static energy to $\ordas{5}$. The perturbative series is Fourier transformed (FT) to position space to compare our results with the lattice QCD.  FT results in additional pathological corrections from low-energy modes and is subtracted using a restricted version of the Fourier transform. We have provided the analytical expressions for the restricted and unrestricted Fourier-transformed version of static potential and static energy in the fixed-order perturbation theory (FOPT). These results of pQCD are compared with the Cornell-type potential from lattice simulations to obtain $\Lambda^{\msbar}_{\text{QCD}}$ for two-flavor QCD. \par 
In chapter~\eqref{Chapter4}, we discuss the determination of the strange quark mass $m_s$ and CKM matrix element $|V_{us}|$ from the moments of the Cabibbo suppressed spectral moments of hadronic $\tau$ decays using finite energy sum rule (FESR). Various determinations in the literature using FOPT and contour-improved perturbation theory (CIPT) have reported significant uncertainty from the renormalization scale variations. To reduce this uncertainty, we used RGSPT in our determinations and also performed updated analysis for FOPT and CIPT schemes to compare the results. We used the publicly available information on the spectral moment, evaluated at the fixed scale $s_0=\Mt$, which is an upper limit to the spectral moment integral, to calculate these parameters. We found that our results using RGSPT are significantly more stable with respect to the scale variations compared to the commonly used schemes such as FOPT and CIPT.\par
In chapter~\eqref{Chapter5}, we discuss the effects of the large kinematical $\pi^2-$corrections that arise when a perturbative series is analytically continued from spacelike to timelike regions. Analytic continuation is a key ingredient to compare theoretical higher-order calculations performed in deep Euclidean spacelike regions and experimental data obtained in the timelike regions. For FOPT, these corrections dominate the genuine perturbative corrections at higher orders and also result in the poor convergence of the perturbative series at low energies. These kinematical corrections have their origin in RG-logarithms and, therefore, can be naturally summed to all orders using RGSPT, leading to two-fold improvements, and the resulting series has improved convergence in addition to the reduced scale dependence. As an application, we have studied processes such $H\rightarrow b\hs b$,  $H\rightarrow g\hs g$, total hadronic Higgs decays, and electromagnetic R-ratio ($R_{em}$). $R_{em}$ is also used in the data-driven methods to calculate the continuum contribution to the hadronic vacuum polarization contribution to $\left(g-2\right)_\mu$. The formalism developed in the chapter has a significant role to play in studies related to the Borel-Laplace sum rule determinations using pQCD results. One of its applications is discussed in chapter~\eqref{Chapter6} for light quark mass determination.\par
In chapter~\eqref{Chapter6}, we discuss the determination of the light quark masses ($m_u$, $m_d$ and $m_s$) using the Borel-Laplace sum rule from the divergence of the axial vector current correlator. Theoretical inputs used in such analysis are the Borel transform of the polarization function of pseudoscalar current and the spectral density obtained from the polarization function. The spectral density gets extra kinematical $\pi^2-$terms due to analytic continuation, for which we used the formalism developed in chapter~\eqref{Chapter5}. A significant improvement in reducing the uncertainties coming from the scale variation and due to truncation of the perturbation series is obtained compared to FOPT studies. In fact, determinations from the Borel-Laplace rules using the FOPT series inherently suffer from large renormalization scale uncertainties, which are linear in nature. Using RGSPT, we have significantly reduced such uncertainties by nearly 10 times smaller in comparison to FOPT determination for $m_s$ case. \par
In chapter~\eqref{Chapter7}, we determine $\as$ and $\overline{m}_c$ and $\overline{m}_b$ from the moments of the vector and pseudoscalar heavy quark currents. The first four moments are already known to four loops ($\ordas{3}$) and have a significant renormalization scale dependence leading the large-scale uncertainties in determinations. In addition to these, determinations using $\msbar$ quark mass definition in the higher dimensional condensate terms result in unstable determinations from higher moments. On the experimental side, higher moments are relatively precisely known for which theoretical predictions from FOPT are unstable. We have used RGSPT to get improved determinations that do not suffer from the use of the quark mass definition in the condensate terms, and the results are much more stable compared to FOPT. \par
In chapter~\eqref{Chapter8}, we summarize the work discussed in this thesis and a brief discussion on possible future investigations.
	
}
\clearpage 

\setstretch{1.3} 
\listofpublications{\addtocontents{toc}{}
	Following are the work published during the Ph.D.:
\begin{enumerate}
\item  "QCD static energy using optimal renormalization and asymptotic Pad\'e-approximant methods,"~B.~Ananthanarayan, D.~Das and M.~S.~A.~Alam Khan.\newline Published in \href{https://doi.org/10.1103/PhysRevD.102.076008}{Phys. Rev. D \textbf{102} (2020) no.7, 076008}.
\item ``Renormalization group improved ms and |Vus| determination from hadronic \ensuremath{\tau} decays,'' B.~Ananthanarayan, D.~Das and M.~S.~A.~Alam Khan.\newline Published in \href{https://doi.org/10.1103/PhysRevD.106.114036}{Phys. Rev. D \textbf{106} (2022) no.11, 114036}.
\item ``Chiral Perturbation Theory Reflections on Effective Theories of the Standard Model," ~B.~Ananthanarayan, M.~S.~A.~Alam~Khan and D.~Wyler, \newline Published in Indian Journal of Physics, \href{https://doi.org/10.1007/s12648-023-02591-5}{DOI: 10.1007/s12648-023-02591-5 }.
\item  ``Renormalization group summation and analytic continuation from spacelike to timeline regions,''~M.~S.~A.~Alam Khan.\newline
Available as preprint: \href{https://link.aps.org/doi/10.1103/PhysRevD.108.014028}{Phys. Rev. D \textbf{108} (2023) no.1, 014028 }.
\end{enumerate}

The following articles are available as preprints and are part of the thesis:
\begin{enumerate}
		\item ``Renormalization group improved determination of light quark masses from Borel-Laplace sum rules,''~M.~S.~A.~Alam Khan.\newline
	Available as preprint: \href{https://arxiv.org/abs/2306.10266}{arXiv:2306.10266[hep-ph]} (Under review in Phys. Rev. D).
	\item  ``Renormalization group improved determination of $\alpha_s$, $m_c$, and $m_b$ from the low energy moments of heavy quark current correlators,''~M.~S.~A.~Alam Khan.\newline
	Available as preprint: \href{https://arxiv.org/abs/2306.10323}{arXiv:2306.10323 [hep-ph]} (Under review in Phys. Rev. D).
\end{enumerate}

}

\clearpage 

\pagestyle{fancy} 

\lhead{\bfseries\texttt{Contents}} 
\tableofcontents 

\lhead{\bfseries\texttt{List of Figures}} 
\listoffigures 

\lhead{\bfseries\texttt{List of Tables}} 
\listoftables 


\clearpage 

\setstretch{1.5} 

\lhead{\bfseries\texttt{Abbreviations}} 
\listofsymbols{ll} 
{

\textbf{CKM} &: Cabibbo-Kobayashi-Maskawa\\
\textbf{EW} &: Electro-Weak\\
\textbf{QCD} &:  Quantum Chromodynamics\\
\textbf{LQCD} &: Lattice Quantum Chromodynamics\\
\textbf{pQCD} &: Perturbative Quantum Chromodynamics\\
\textbf{QFT} &: Quantum Field Theory\\
\textbf{RG} &: Renormalization Group\\
\textbf{RGE} &: Renormalization Group Equation\\
\textbf{SM} &: Standard Model\\
\textbf{vev} &: Vacuum Expectation Value\\
\textbf{EFT} & : Effective Field Theory\\
\textbf{RGSPT}&: Renormalization Group Summed Perturbation Theory\\
\textbf{FOPT}&: Fixed Order Perturbation Theory\\
\textbf{$\msbar$}&: Modified Minimal Subtraction
}

%
%
%


\clearpage 




%
%
%
%


\mainmatter 

\pagestyle{fancy} 

\include{Chapters/Chap1}
\include{Chapters/Chap2}
\include{Chapters/Chap3}
\include{Chapters/Chap4}
\include{Chapters/Chap5}
\include{Chapters/Chap6}
\include{Chapters/Chap7}
\include{Chapters/Chap8}


\addtocontents{toc}{\vspace{2em}} 

\appendix 


\input{Appendices/AppendixA}

\addtocontents{toc}{} 

\backmatter

\label{Bibliography}

\lhead{\emph{Bibliography}} 


\bibliographystyle{plain}

\end{document}

%% file: Chapters/Chap1.tex

\chapter{Introduction} 

\label{Chapter1} 

\lhead{Chapter 1. \emph{Introduction}} 

 \onehalfspacing
The observable universe around us, to a good of accuracy, can be described by four fundamental forces: gravitational, electromagnetic, strong, and weak nuclear forces. These forces operate in certain ranges and play a vital role in sustaining life. The nuclear forces, for instance, are responsible for the burning of stars that acts as a constant source of energy, while the gravitational force keeps planets in their orbits around these stars, and the electromagnetic force governs the life processes within living organisms.

Human curiosity is naturally inclined to comprehend the inner workings of nature and describe them using a minimal set of rules. Newton's law of universal gravitation is one such example that provides an explanation that the force which binds various planets in orbit is the same as one responsible for falling objects on Earth. Another example is Maxwell's unification of electric and magnetic force into electromagnetic force. These are the early success in the history of physics in understanding a phenomenon from a unified perspective. \par 
Various sophisticated experiments in the last century have revealed that nature is not just the story of electrons, protons, and neutrons, but a plethora of other particles also exist. Some of them are even theoretically predicted and later confirmed in the experiments. A lot of scientific efforts have been dedicated to understanding various processes with better precision, and now we have a quantum field theoretical description of fundamental forces except gravity.\par

\section{Standard Model of Particle Physics}

\begin{figure}[ht]
	\centering
\includegraphics[width=.7\textwidth]{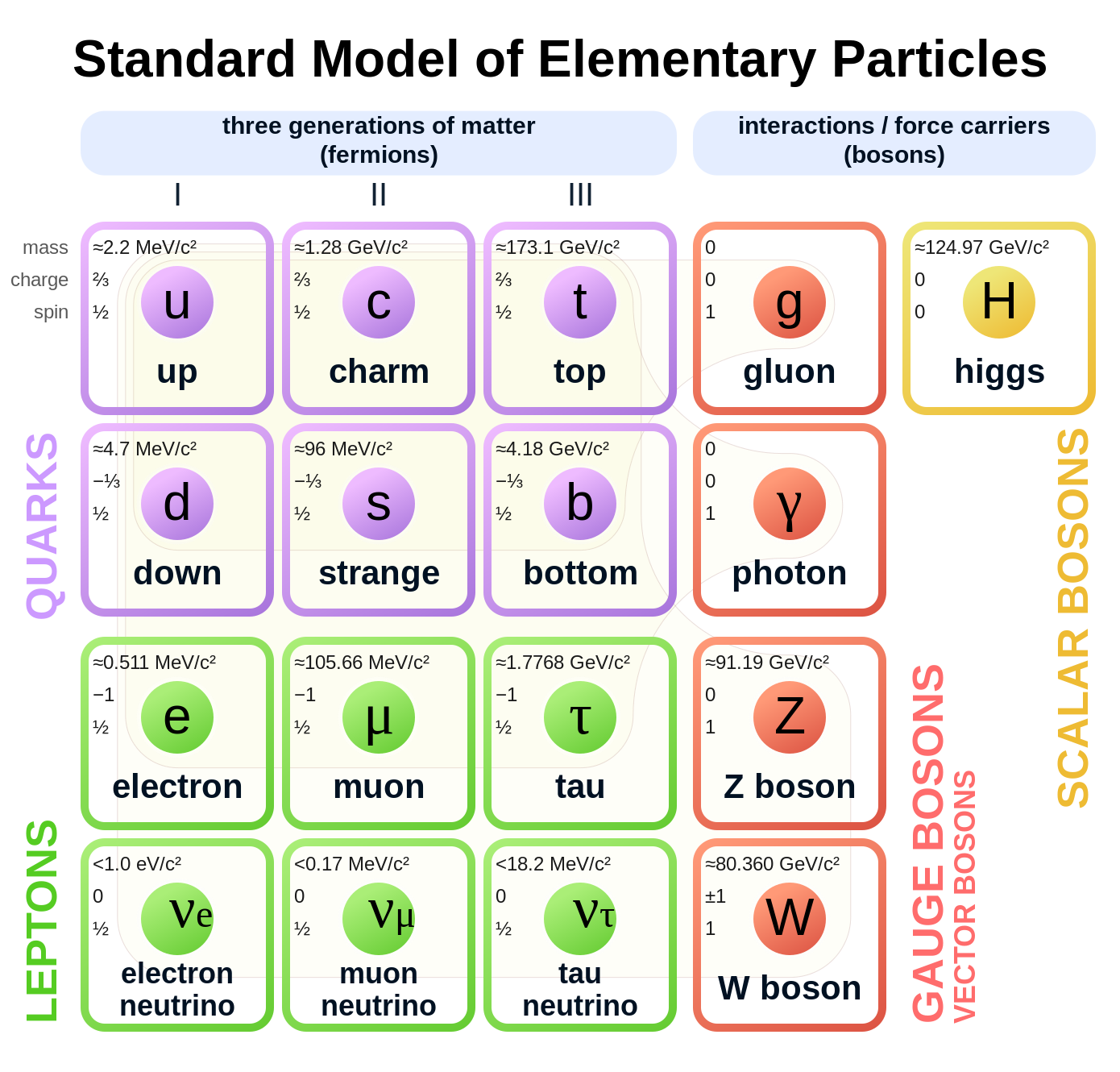}
\caption{Content of the SM\cite{wiki1} and some of the properties of the particles.}
	\label{fig:SM1}
\end{figure}

With such advancements, it is natural to ask if these particles and their interactions can be systematically understood in a systematic theoretical framework. One such framework is now known as the Standard Model of Particle Physics (SM), which describes three of the forces except gravity. It provides a theoretical formalism to understand the interactions of the quarks and leptons. These are spin-1/2, and the local gauge invariance demands the existence of the gauge bosons fields which have spin-1. To explain the masses of these particles, an extra spin-0 boson known as a Higgs is needed. The particle content of the SM is shown in Fig.~\eqref{fig:SM1}. SM has $SU(3)_C\times SU(2)_L\times U(1)_Y$ gauge symmetry where subscript $C$, $L$, and $Y$ refer to color, left-handed isospin, and hypercharge, respectively.\par
The strong interaction is responsible for the confining quark into hadrons. It is described by quantum chromodynamics (QCD), which describes the interaction among quarks and gluons and has $SU(3)_C$ color gauge symmetry. Due to confinement, quarks are never freely observed, but instead, colorless singlet states in the form of hadrons are observed in the experiments. \par
The weak interaction is needed to explain the radioactive decay of atoms and the decay of some hadrons produced by the high-energy cosmic ray particles in the Earth's atmosphere. In the Glashow-Weinberg-Salam (GWS) model~\cite{Glashow:1961tr,Weinberg:1967tq,Salam:1968rm}, the electromagnetic and weak interactions are unified, and the theory has $SU(2)_L\times U(1)_Y$ gauge symmetry. In this model, the left-handed doublet of fermions transforms as $SU(2)$ under weak isospin transformation, while right-handed fermions transform as singlets under isospin transformation.
\begin{align}
\text{Leptons :}\quad& \binom{\nu_e}{e}_L\quad e_R\nonumber\,,\\
\text{Quarks :}\quad& \binom{u}{d}_L\quad u_R,\hs  d_R.   
\label{eq:con_EW}
\end{align}
These particles are also assigned a weak hypercharge, and particles in a doublet have the same hypercharge, while all the right-handed ones have different values. Early models for the weak interactions only have only two charged gauge bosons $W^{\pm}$, and neutral charge currents are predicted in the GWS model, which led to the discovery of the $Z$ boson. Masses of these particles are generated via Higgs mechanism~\cite{Englert:1964et,Higgs:1964pj,Guralnik:1964eu}.\par
Despite many successful predictions, SM is still an incomplete theory as it does not explain baryon asymmetry in the universe, dark energy and dark matter, and non-zero masses of the neutrinos. These topics are also of special interest but not part of this thesis. In this thesis, various discussions and references are made in the context of the SM.  \par 
Given the rich content of the SM, it has 19 free parameters, which are as follows:
\begin{align*}
\text{Fermion masses:} & \quad m_u, \hs  m_d, \hs  m_s, \hs  m_c, \hs  m_b, \hs  m_t\,, \\
\text{Lepton masses:} & \quad m_e,\hs m_\mu, \hs m_\tau \\
\text{Higgs boson mass:} & \quad m_H \\
\text{Higgs vacuum expectation value:} & \quad v \\
\text{CKM mixing angles:} &\quad \theta_{12}, \theta_{23}, \theta_{13} \\
\text{CKM CP violation phase:} & \quad \delta \\
\text{Gauge couplings $\left\lbrace U(1), SU(2), SU(3)\right\rbrace$}: & \quad \left\lbrace g_1 , g_2 , g_3\right\rbrace \\
\text{QCD vacuum angle} : &\quad \theta_{\text{QCD}}
\end{align*}
where the relation of these parameters to the CKM matrix element $\vert V_{us}\vert=\cos\left(\theta_{13}\right)\hs \sin\left(\theta_{12}\right)$ and strong coupling $\as=\frac{g_3^2}{4\pi}$ is used in this thesis.

The values of these parameters are obtained by comparing theoretical predictions with experimental data. Their precise value depends on the amount of data available and how precisely a process can be calculated using theory. Theoretical calculations use both non-perturbative as well as perturbative tools depending on the interaction and energy involved in the processes. This thesis, however, focuses only on the perturbative aspects of the strong interactions and their applications in the determination of the precise value of some of the standard model parameters.\par
\section{Perturbation theory at higher orders and renormalization group improvement in QCD}
The development of the perturbation theory in the relativistic quantum field theory picture goes back to the issue of the divergences encountered in the higher-order calculation of the magnetic moment of an electron using Dirac's theory~\cite{Dirac:1927dy} of electrodynamics. In the late 1940s, the work of Tomonaga~\cite{Tomonaga:1946zz}, Schwinger~\cite{Schwinger:1948iu,Schwinger:1948yk}, and Feynman~\cite{Feynman:1949zx,Feynman:1949hz,Feynman:1950ir} resolved this problem of divergences and excellent agreement was obtained between theoretical predictions and their experimental value. The formalism developed in QED uses regularization and renormalization of the Feynman diagram for a physical process to systematically remove the divergences by absorbing them into parameters of the theory order by order. The resulting finite predictions can then be tested against the experimental predictions. This formalism has profound applications in the perturbative description of other interactions as well.\par
Stueckelberg and Petermann~\cite{StueckelbergdeBreidenbach:1952pwl}, Gell-Mann and Low~\cite{Gell-Mann:1954yli} used the idea of renormalization to relate coupling at different scales for QED. Wilson~\cite{Wilson:1973jj,Wilson:1971bg,Wilson:1971dh} generalized this concept to explain the physics of a system at different scales, which has enormous applications in effective field theories and other areas of physics, such as condensed matter, statistical physics, and cosmology, etc. There are various approaches to the renormalization group, and for details, we refer to Refs.~\cite{Schwartz:2014sze,Peskin:1995ev,Hollowood,Jakovac:2016zkg,Gies:2006wv,Polonyi:2001se,Burgess:2020tbq,Baldazzi:2021lym,Huang:2013zaa}. \par
QCD has a rich spectrum as it has various bound states of the quarks. At low energies, the interactions between pions and kaon can be explained using chiral perturbation theory (ChPT)~\cite{Weinberg:1978kz,Gasser:1983yg,Gasser:1984gg}. For heavy quark case, heavy quark effective theory (HQET)~\cite{Eichten:1989zv,Georgi:1990um,Mannel:1991mc} and for quarkonia non-relativistic QCD (NRQCD)~\cite{Caswell:1985ui,Bodwin:1994jh} can be used in the description of the heavy quark systems. These effective theories can be used to calculate various quantities using lattice QCD simulations, and their formulation also allows perturbative treatment as well, which results in the extraction of various parameters of the theory.\par
Due to asymptotic freedom~\cite{Gross:1973id,Politzer:1973fx}, quarks inside a hadron appear to be free when smashed with very highly energetic particles. In this case, perturbative treatment in terms of expansion in strong coupling constant ($\as$) is applicable. Various quantities can be calculated as an expansion in $\as$ and experimentally tested. At low energies ($\sim$ few $\GeV$), the value of $\as$ is large that requires higher-order calculations, and predictions are meaningful if the series is convergent. At each order of perturbation theory, the number of Feynman diagram significantly increase, and their computation is a challenge on its own. With the development of computational power, higher-order calculations are performed on computers that make use of codes such as FORM~\cite{Vermaseren:2000nd}, KIRA~\cite{Maierhofer:2017gsa}, and FIESTA~\cite{Smirnov:2015mct} etc.\par 
Only a few terms of the perturbation series are known for various processes. The current status of various higher-order series used in this thesis is as follows:
\begin{enumerate}
\item The QCD beta function coefficients are known to five-loop \cite{vanRitbergen:1997va,Gross:1973id,Caswell:1974gg, Jones:1974mm,Tarasov:1980au,Larin:1993tp,Czakon:2004bu,Baikov:2016tgj,Herzog:2017ohr}.
\item Quark mass anomalous dimension coefficients in the $\msbar$ scheme are known to five loops~\cite{Tarrach:1980up,Tarasov:1982plg,Larin:1993tq,Vermaseren:1997fq,Chetyrkin:1997dh,Baikov:2014qja,Luthe:2016ima,Luthe:2016xec}.
\item QCD static potential is known to three loops~\cite{Appelquist:1977es,Appelquist:1977tw,Susskind,Fischler:1977yf,Melles:1998dj,Schroder:1998vy, peter:1998ml, Peter:1997me,Anzai:2009tm,Smirnov:2009fh,Lee:2016cgz}
\item The Adler function for the vector current correlators for massless and with massive corrections is known to  $\order{\alpha_s^4}$ in Refs.~\cite{Appelquist:1973uz,Zee:1973sr,Chetyrkin:1979bj,Dine:1979qh,Gorishnii:1990vf,Surguladze:1990tg,Chetyrkin:1996ez,Baikov:2008jh,Baikov:2010je,Herzog:2017dtz} and Refs.~\cite{Chetyrkin:1990kr,Chetyrkin:1994ex,Chetyrkin:2000zk,Baikov:2004ku}, respectively.
    \item The Adler function of the scalar current correlator is known to $\ordas{4}$ \cite{Becchi:1980vz,Broadhurst:1981jk,Chetyrkin:1996sr,Baikov:2005rw,Gorishnii:1990zu,Gorishnii:1991zr,Herzog:2017dtz}.
    \item Higgs decay into pair of gluon is known to $\ordas{4}$~\cite{Kataev:1981gr,Chetyrkin:1997iv,Baikov:2006ch,Schreck:2007um,Herzog:2017dtz} 
\end{enumerate}
Theoretical predictions from such finite order information also introduce a scheme dependence that arises due to the definitions of the input parameters, such as quark masses and strong coupling constant. Various definitions of these parameters result in the rearrangement of the finite order perturbative series, which are also sensitive to the renormalization scale variations. A systematic approach to tackle scale and scheme dependence issues was first addressed by Stevenson~\cite{Stevenson:1981vj} using optimized perturbation theory and later improved in Ref.~\cite{Stevenson:1986cu}. The idea of this prescription is to make the predictions from finite order series less sensitive to the variation of the free parameters. Further development in this subject leads to formulations of other prescriptions, which include Grunberg's method of effective charge (MOC)~\cite{Grunberg:1982fw}, Brodsky-Lepage-Mackenzie (BLM) \cite{Brodsky:1982gc}, complete renormalization group improvement (CORGI)~\cite{Maxwell:1999dv,Maxwell:2000mm}, renormalization group summed perturbation theory (RGSPT) \cite{McKeon:1998tr,Ahmady:2002fd}, the principle of maximum conformality (PMC)~\cite{Brodsky:2013vpa}, etc. \par
In MOC, the scheme dependence of strong coupling constant and higher order terms cancel when the perturbation series is replaced by an effective charge that has a similar RG running as $\as$. The first two beta function coefficients ($\beta_0$ and $\beta_1$) are universal, but higher beta function coefficients are process dependent. In BLM, the momentum scale $Q$, in the strong coupling constant $\as(Q)$, is defined in such a way that the perturbative series is flavor independent. To remove scheme dependence from $\as$, its value is taken from some other process (such as $e^+e^-$). CORGI is closely based on the MOC but with the improvement that the renormalization scale dependence is eliminated by summing RG logarithms, the resulting series has correct UV behavior as predicted by asymptotic freedom. In PMC, the renormalization scale and scheme dependence are eliminated by choosing the momentum scale $Q^*$ such that all the beta function coefficients are absorbed in ($\as(Q^*$)). All these schemes have their own merits and are used for various physical processes for the determination of SM parameters. This is a brief overview of the various schemes used in the literature, and the rest of the discussion in this thesis is based on RGSPT. 

\section{Structure of the thesis}
This thesis details the properties of RGSPT and how it can be used to improve the predictions of perturbation series in QCD. The formalism of this prescription is rooted in the use of the renormalization group equation (RGE) to sum the running logarithms $\log(\mu^2/Q^2)$ arising from a given known order of the perturbation theory to all orders. The resulting series has an expansion in which running logarithms always appear with $\as$ such that $\as\hs \log(\mu^2/Q^2)\sim\order{1}$. A closed analytical form of the perturbative series can then be used in the extraction of various parameters of the theory. Various applications of this prescription can be found in Refs.~\cite{Ahmady:2002fd,Abbas:2012py,Ananthanarayan:2016kll,Ahmady:2002pa,Ahmed:2015sna,Abbas:2022wnz,Chishtie:2018ipg}.\par
In this thesis, we have discussed the extraction of 6 out of 19 parameters of the SM. They are:
\begin{enumerate}
     \item Landau pole $\Lambda^{\msbar}_{QCD}$ from 2-flavor QCD static energy~\cite{Ananthanarayan:2020umo}.
     \item Strange quark mass $m_s$ and CKM matrix element $|V_{us}|$ from hadronic $\tau-$decays~\cite{Ananthanarayan:2022ufx}.
     \item Light quark masses ($m_u$, $m_d$, $m_s$) using Borel-Laplace sum rules~\cite{AlamKhan:2023ili}. This work is an application of the formalism developed in my independent work on analytic continuation in pQCD using RGSPT~\cite{AlamKhan:2023dms}.
     \item Strong coupling constant $\as$, charm quark mass ($\overline{m}_c$) and bottom quark mass ($\overline{m}_b$) from relativistic sum rules~\cite{AlamKhan:2023kgs}.
 \end{enumerate}
 Some of these determinations use powerful tools such as QCD sum rules~\cite{Shifman:1978bx,Shifman:1978by}, which are based on the analytic properties of Green's function. These sum rules use both theoretical and experimental input on the spectral function and are based on the assumption of the quark-hadron duality~\cite{Poggio:1975af}. \par

In chapter~\eqref{Chapter2}, we give an introduction to RGSPT and explain the procedure of the RG summation and derive the RG summed expression for strong coupling constant and quark mass in the $\msbar$ scheme as well as quark mass relation between $\msbar$ on pole mass scheme. The procedure described in this section is used in the other sections as input in studying various processes.\par
In chapter~\eqref{Chapter3}, we study the RG improvement of the QCD static energy between the two static heavy quarks sources. It is an important quantity that can be calculated using perturbative QCD as well as in lattice QCD simulations and has application in the precise determination of the strong coupling constant at low energies. We use asymptotic Pad\'e approximants to estimate the unknown perturbative contribution at four loops, which completes the static energy to $\ordas{5}$. The perturbative series is Fourier transformed (FT) to position space to compare our results with the lattice QCD.  FT results in additional pathological corrections from low-energy modes and is subtracted using a restricted version of the Fourier transform. We have provided the analytical expressions for the restricted and unrestricted Fourier-transformed version of static potential and static energy in the fixed-order perturbation theory (FOPT). These results of pQCD are compared with the Cornell-type potential from lattice simulations~\cite{Karbstein:2018mzo} to obtain $\Lambda^{\msbar}_{QCD}$ for two flavors QCD. \par 
In chapter~\eqref{Chapter4}, the strange quark mass $m_s$ and CKM matrix element $|V_{us}|$ are determined from the moments of the Cabibbo suppressed spectral moments of hadronic $\tau$ decays using finite energy sum rule (FESR). Various determinations in the literature using FOPT and contour-improved perturbation theory (CIPT) have reported significant uncertainty from the renormalization scale variations. To reduce this uncertainty, we used RGSPT in our determinations and also performed updated analysis for FOPT and CIPT schemes to compare the results. We used the publicly available information on the spectral moment, evaluated at the fixed scale $s_0=\Mt$, which is an upper limit to the spectral moment integral, to calculate these parameters. We found that our results using RGSPT are significantly more stable with respect to the scale variations compared to the commonly used schemes such as FOPT and CIPT.\par
In chapter~\eqref{Chapter5}, we discuss the large kinematical $\pi^2-$corrections that arise when a perturbative series is analytically continued from spacelike to timelike regions. Analytic continuation is a key ingredient to compare theoretical higher-order calculations performed in deep Euclidean spacelike regions and experimental data obtained in the timelike regions~\cite{Pennington:1981cw}. For FOPT, these corrections dominate the genuine perturbative corrections at higher orders~\cite{Herzog:2017dtz} and also result in the poor convergence of the perturbative series at low energies~\cite{Aparisi:2021tym}. These kinematical corrections have their origin in RG-logarithms and, therefore, can be naturally summed to all orders using RGSPT, leading to two-fold improvements, and the resulting series has improved convergence in addition to the reduced scale dependence. As an application, we have studied processes such $H\rightarrow b\hs b$,  $H\rightarrow g\hs g$, total hadronic Higgs decay width, electromagnetic R-ratio ($R_{em}$). The $R_{em}$ is also used in the data-driven methods to calculate the continuum contribution to the hadronic vacuum polarization contribution to $\left(g-2\right)_\mu$. The formalism developed in the paper has a significant role to play in studies related to the Borel-Laplace sum rule determinations using pQCD results which are used in Ref.~\cite{AlamKhan:2023ili}.\par
In chapter~\eqref{Chapter6}, the masses of the light quarks ($m_u$, $m_d$ and $m_s$) are determined using the Borel-Laplace sum rule from the divergence of the axial vector current correlator. Theoretical inputs used in such analysis are the Borel transform of the polarization function of the pseudoscalar current and the spectral density obtained from the polarization function. The spectral density gets extra kinematical $\pi^2-$terms due to analytic continuation, for which we used the formalism developed in Ref.~\cite{AlamKhan:2023dms}. A significant improvement in reducing the uncertainties coming from the scale variation and due to truncation of the perturbation series is obtained compared to FOPT studies. In fact, determinations from the Borel-Laplace rules using the FOPT series inherently suffer from large renormalization scale uncertainties, which are, in fact, linear in nature~\cite {Chetyrkin:2005kn}. Using RGSPT, we have significantly reduced such uncertainties by nearly 10 times smaller in comparison to FOPT determination for $m_s$ case. \par
In chapter~\eqref{Chapter7}, we determine $\as$ and $\overline{m}_c$ and $\overline{m}_b$ from the low energy moments of the vector and pseudoscalar heavy quark currents. The first four moments are already known to four loops ($\ordas{3}$) and have a significant renormalization scale dependence leading the large-scale uncertainties in determinations. In addition to these, determinations using $\msbar$ quark mass definition in the higher dimensional condensate terms result in unstable determinations from higher moments. On the experimental side, higher moments are relatively precisely known for which theoretical predictions from FOPT are unstable~\cite{Kuhn:2007vp}. We have used RGSPT to get improved determinations that do not suffer from the use of the quark mass definition in the condensate terms, and the results are much more stable compared to FOPT. \par
In chapter~\eqref{Chapter8}, we give a summary and outlook of the thesis.

%% file: Chapters/Chap2.tex
\chapter{RGSPT}
\label{Chapter2}

\lhead{Chapter 2. \emph{RGSPT}}
\section{Motivation}
Predictions from the perturbation series in QCD depend on the scheme used and the free parameters present in it that can result in large uncertainties in these predictions. The renormalization group summed perturbation theory (RGSPT) is one of the prescriptions based on the renormalization group (RG) that sums the RG running logarithms to all orders in a closed form. In this chapter, we describe the procedure of the RG summation and expression for the running of the strong coupling constant, quark masses, and as well as quark mass relations in the pole and $\msbar$ scheme are presented.

 \onehalfspacing
    \section{Introduction\label{sec:intro_RGSPT}}
   \bs Various observables in the pQCD are calculated in the high energy limit, and the resulting perturbative series is expressed in terms of the $\as$ and masses of the quarks ($m_q$). The observables are independent of the scheme used and variations of the renormalization scale. This fact results in the cancellation renormalization scale dependence of the parameters from different orders of the perturbation series. Practically, all order coefficients of a perturbative series of a physical process are not available for QCD, and the effects of scale dependence can be seen in the fixed order results. These perturbative series also have issues of poor convergence behavior. These effects can lead to a substantial amount of theoretical uncertainty, and any various perturbative schemes used in the literature  bring them under control.\par
    RGSPT is a perturbative scheme where RGE is used to systematically sum the running RG logarithms accessible from a given order of the perturbation theory. This formalism was proposed in Ref.~\cite{McKeon:1998tr} and further developed in Ref.~\cite{Ahmady:2002fd} for the QCD process.  The closed-form results are found to be less sensitive to the renormalization scale, and hence we get a significant reduction in the theoretical uncertainties. These RG logarithms are of special importance when operations like Fourier transform, analytic continuation in the complex plane, and Borel-transformations, etc., are performed. Some of these operations are, in fact, key ingredients in the QCD sum rules~\cite{Shifman:1978bx,Shifman:1978by} which are widely used in the determination of the SM parameters. \par
    In section~\eqref{sec:rgspt_formalism}, we describe the procedure of summation of the running logarithm using RGSPT is presented. In section~\eqref{sec:asmq_rgspt}, we derive the expression for the running of the strong coupling constant, and quark masses in the $\msbar$ scheme. We also derive the relation of the quark masses in the $\msbar$ and pole scheme using RGSPT.
\section{Formalism}\label{sec:rgspt_formalism}
    In FOPT prescription, a perturbative series $S(Q^2)$ in pQCD can be written as:
    \begin{align}
    	\mathcal{S}(Q^2)\equiv \sum_{i,j} T_{i,j} x^i L^j \,,
    	\label{eq:Pseris}
    \end{align}
    where $x=\as(\mu)/\pi$, $L=\log(\mu^2/Q^2)$ is the RG logarithm, $\mu$ is the renormalization scale and $Q$ is the momentum scale of the process considered. The renormalization scale is introduced when a Feynman diagram is regularized in schemes such as dimensional regularization is used. After renormalization, this scale appears in the running logarithms and is usually summed by choosing $\mu^2=Q^2$. These logarithms, however, can be regenerated  using RG evolution of the perturbative series in Eq.~\eqref{eq:Pseris_uds} using its anomalous dimension, $\gamma_S(x)$, by solving:
    \begin{align}
    	\mu^2 \frac{d}{d\mu^2} \mathcal{S}(Q^2)&=\gamma_S(x) \hs\hs \mathcal{S}(Q^2)\,,
    	\label{eq:RGE}\\
    	\mu^2 \frac{d}{d\mu^2} x(\mu)&=\beta(x)\,,
    	\label{eq:alpha_S}
    \end{align}
    where anomalous dimension $\gamma_S(x)$ and $\beta(x)$ is given by:
    \begin{equation}\label{eq:anom_S}
    	\begin{aligned}
    		\gamma_S(x)&=\sum_{i=0}\gamma_i x^{i+1}\,,\\
    		\beta(x)&=-\sum_{i=0}\beta_i x^{i+2}\,,
    	\end{aligned}
    \end{equation}
   coefficients $\beta_i$ and $\gamma_i$ are presented in appendix~\eqref{app:mass_run}. It should be noted that expansion in Eq.~\eqref{eq:Pseris} has two issues related to convergence, which need to be addressed. First, $x$ has to be small enough so that the series is well convergent if we set $\mu^2\simeq Q^2$. Second is related to the problem of the large logarithms $L$ so that convergence is not spoiled by it. These running logarithms effects are enhanced once the RGE in Eq.~\eqref{eq:RGE} is solved. \par
   When perturbative series in Eq.~\eqref{eq:Pseris} is used for theoretical prediction for a process, the uncertainties from it should also have to be considered in the predictions. Such uncertainties include: 
   \begin{enumerate}
   	\item Uncertainties present in the experimental value of the $x$.
   	\item Truncation uncertainty due to finite numbers of terms known and usually estimated by the difference in the predictions when the known last term is ignored.
   	\item Scale variation usually calculated by setting $\mu^2=\xi Q^2$ and varying parameter $\xi \in \left[1/2,2\right]$. 
   \end{enumerate}
   Various prescriptions for the perturbation theory are designed to reorganize the finite order perturbation series such that the uncertainties from the above-mentioned problems can be minimized. It should be noted that the truncation uncertainty in the rest of the chapters of this thesis is calculated from the last term present in the perturbation series.\par
    In RGSPT, perturbative series in Eq.~\eqref{eq:Pseris} is organized as follows:
    \begin{equation}
    	\mathcal{S}^\Sigma(Q^2)= \sum_{i} x^i S_i (x\hs L)\,,
    	\label{eq:summed}
    \end{equation}
    where the goal is to obtain a closed-form expression for coefficients:
    \begin{equation}
    	S_i (u)=\sum_{j=0}^{\infty} T_{i+j,j} u^j\,,
    	\label{eq:summed_coefs}
    \end{equation}
    where $u\equiv x \hs L\hs$.   $S_i (u)$ are functions of one variable where $u\sim\mathcal{O}(1)$. The closed-form solution for them is obtained using RGE. It should be noted that coefficients $T_{i,0}$ are known as RG-inaccessible coefficients as they can only be obtained by direct Feynman diagram calculation and can not be obtained using RGE.  Coefficients $T_{i,j}$ for $j>0$ are known as RG-accessible coefficient and are obtained using RGE.\par 
    To explain the procedure for the RG summation, we start with a simple case where a perturbative series, in Eq.~\eqref{eq:Pseris} is known to $\order{x^5}$, have the form:
    \begin{align}
    		\mathcal{S}^{\left(5\right)}(Q^2)\equiv \sum_{i=0,j=0}^{5,i\ge j} T_{i,j} x^{i+1} L^j \,,\quad \text{ with }\gamma_{S}=0\,,
    \end{align}
     and we have a homogeneous differential RG equation from Eq.~\eqref{eq:RGE}.  The RGE in Eq.~\eqref{eq:RGE} results:
    \begin{align}
    	&0=x^2 \left(T_{1,1}-\beta _0 T_{0,0}\right)+L x^3 \left(2 T_{2,2}-2 \beta _0 T_{1,1}\right)+L^2 x^4 \left(3 T_{3,3}-3 \beta _0 T_{2,2}\right)\nonumber\\&+L^3 x^5 \left(4 T_{4,4}-4 \beta _0 T_{3,3}\right)+L^4 x^6 \left(5 T_{5,5}-5 \beta _0 T_{4,4}\right)+x^3 \left(-\beta _1 T_{0,0}-2 \beta _0 T_{1,0}+T_{2,1}\right)\nonumber\\&+L x^4 \left(-2 \beta _1 T_{1,1}-3 \beta _0 T_{2,1}+2 T_{3,2}\right)+L^2 x^5 \left(-3 \beta _1 T_{2,2}-4 \beta _0 T_{3,2}+3 T_{4,3}\right)\nonumber\\&+L^3 x^6 \left(-4 \beta _1 T_{3,3}-5 \beta _0 T_{4,3}+4 T_{5,4}\right)+x^4 \left(-\beta _2 T_{0,0}-2 \beta _1 T_{1,0}-3 \beta _0 T_{2,0}+T_{3,1}\right)\nonumber\\&+L x^5 \left(-2 \beta _2 T_{1,1}-3 \beta _1 T_{2,1}-4 \beta _0 T_{3,1}+2 T_{4,2}\right)+L^2 x^6 \big(-3 \beta _2 T_{2,2}-4 \beta _1 T_{3,2}\nonumber\\&\bs-5 \beta _0 T_{4,2}+3 T_{5,3}\big)+x^5 \left(-\beta _3 T_{0,0}-2 \beta _2 T_{1,0}-3 \beta _1 T_{2,0}-4 \beta _0 T_{3,0}+T_{4,1}\right)\nonumber\\&+L x^6 \left(-2 \beta _3 T_{1,1}-3 \beta _2 T_{2,1}-4 \beta _1 T_{3,1}-5 \beta _0 T_{4,1}+2 T_{5,2}\right)+\order{x^7}
    	\label{eq:RG_cancellation}
    \end{align}
     and we see the cancellation among the coefficients of different orders in the terms $x^n\hs L^m$. For FOPT prescription, the RG accessible terms can be trivially obtained by solving for the coefficients for $x^m L^n$.\par 
     Various terms in Eq.~\eqref{eq:RG_cancellation} are organized to show a particular pattern as $x^{n+2} L^n$-terms have the following recurrence relation:
     \begin{equation}
     	(n+1) T_{n+1,n+1}-\beta_0 (n+1) T_{n,n}=0\,,
     	\label{eq:rec1}
     \end{equation}
     and $x^{n+3} L^n$ have following recurrence relation:
     \begin{equation}
     	(n+1) T_{n+2,n+1}-\beta_0 (n+2) T_{n+1,n} -\beta_1 (n+1) T_{n,n}=0\,,
     	\label{eq:rec2}
     \end{equation}
     and so on.  Now, we can use these recurrence relations to find a closed-form solution for Eq.~\eqref{eq:summed_coefs}. To obtain that, we multiply $u^n+1$ in Eq.~\eqref{eq:rec1} and sum to all orders leading to:
     \begin{align}
     	0=\sum_{n=0}^{\infty} u^{n+1} \left( (n+1) T_{n+1,n+1}-\beta_0 (n+1) T_{n,n}\right)= \frac{d}{d\hs u} S_0[u] -u\hs\beta_0\hs \frac{d}{d\hs u} S_0[u]-\beta_0 S_0[u]\,.
     \end{align}
     Similarly, for Eq.~\eqref{eq:rec2}, we get:
     \begin{align}
     	\frac{d}{d\hs u}S_1[u]-u\beta_0\frac{d}{d\hs u}S_1[u]-2\beta_0\hs S_1[u]-\beta_1 u \frac{d}{d\hs u}S_0[u]-\beta_1 \hs S_0[u]=0 
     \end{align}
     where $u= x\hs L$. The solution to the above differential equations is given by:
     \begin{align}
     	S_0[x L ]&= \frac{T_{0,0}}{1-\beta _0 x\hs L}\label{eq:RGsol0}\,,\\
     	S_1[x L ]&= \frac{\beta _0 T_{1,0}-\beta _1 T_{0,0} \log \left(1-\beta _0 x\hs  L \right)}{\beta _0 \left(1-\beta _0 x\hs L\right)^2}\,.	\label{eq:RGsol1}
     \end{align}
     The form of Eq.~\eqref{eq:RGsol0} looks familiar and can be found in the textbooks when the summation of large logarithms for $\as$ at one-loop after renormalization is discussed. However, the appearance of $\log \left(1-\beta _0 x\hs  L \right)$ term in  Eq.~\eqref{eq:RGsol1} is something new that is obtained using RGSPT. This process can be repeated to get closed-form solutions for higher-order coefficients as well. \par
     The recurrence relations are very important if one wants to perform the Borel transformation of the series in renormalon-related studies. However, this method gets cumbersome when one goes to higher orders and especially when anomalous dimensions are also involved. A lot of hard work can be saved by applying RG to Eq.~\eqref{eq:summed}, and coefficients of $x^n$ are taken after replacing $L=u/x$. This procedure results in the same set of differential equations as one obtains following the recurrence relation method.\par 
    Now, having described the method above, we can find the RG summation for the perturbative series given in Eq.~\eqref{eq:Pseris} using RGE in Eq.~\eqref{eq:RGE}. After applying RGE, we get terms in which $L$ also appears when derivative w.r.t $x$ is taken and by substituting $L=u/x$, we  get a set of coupled differential equations for $S_i(u)$, which in compact form can be written as:
    \begin{equation}
    	\left(\sum_{i=0}^{n} \frac{\beta_{i}}{u^{n-i-1}}\frac{d}{d\hs u}\left(u^{n-i} S_{n-i}(u)\right)+\gamma_i S_{n-i}(u)\right)-S'_n(u)=0\,,
    	\label{eq:rec_all}
    \end{equation}
    The first three coefficients can be obtained by solving the above differential equation and are given by:
    \begin{align}
    	S_0(u)&=T_{0,0} w^{-\tilde{\gamma }_0}\,,\nonumber\\
    	 S_1(u)&=T_{1,0} w^{-\tilde{\gamma }_0-1}+T_{0,0} w^{-\tilde{\gamma }_0-1} \Big[(1-w) \tilde{\gamma }_1+\tilde{\beta }_1 \tilde{\gamma }_0 (w-\log(w)-1)\Big]\nonumber\\
    	S_2(u)&=T_{2,0} w^{-\tilde{\gamma }_0-2}-T_{1,0} w^{-\tilde{\gamma }_0-2} \Big[(w-1) \tilde{\gamma }_1+\tilde{\beta }_1 \left(\tilde{\gamma }_0 (-w+\log (w)+1)+\log (w)\right)\Big]\nonumber\\&+\frac{1}{2} T_{0,0} w^{-\tilde{\gamma }_0-2} \Big\lbrace-\tilde{\beta }_1 \tilde{\gamma }_1 \Big[1-w^2+2 \log (w)+2 (w-1) \tilde{\gamma }_0 (w-\log (w)-1)\Big]\nonumber\\&+(w-1) \Big[(w-1) \tilde{\beta }_2 \tilde{\gamma }_0+(w-1) \tilde{\gamma }_1^2-(w+1) \tilde{\gamma }_2\Big]+\tilde{\beta }_1^2 \tilde{\gamma }_0\big(\tilde{\gamma }_0-1\big) (w-\log (w)-1)^{2} \Big\rbrace\,,
    	\label{eq:RGsol3}
    \end{align}
    where $w\equiv 1-\beta_0 \hs u$, for anomalous dimension and higher order beta function coefficients, we have used $\tilde{X}\equiv X/\beta_0$. Now, we can check that solutions obtained in Eqs.~\eqref{eq:RGsol0}-\eqref{eq:RGsol1} can be obtained from solutions in Eq.~\eqref{eq:RGsol3} by setting $T_{0,0}=0$, $\tilde{\gamma}_i=0$ and $T_{i,0}\rightarrow T_{i-1,0}$. An important feature of the above procedure is that the most general term present in the summed coefficient using RGSPT is given by:
    \begin{equation}
    	\Omega_{n,a}\equiv\frac{\log^n(w)}{w^a}=\frac{\log^n(1-\beta_{0}\hs x(\mu)\log(\mu^2/Q^2))}{(1-\beta_{0}\hs x(\mu)\log(\mu^2/Q^2))^a}\,,
    	\label{eq:coef_rgspt}
    \end{equation}
    where $n$ is a positive integer and $a\propto \gamma_0/\beta_{0}$ appearing in Eq.~\eqref{eq:anom_S}. \par 
    It should be noted that for $\mu^2=Q^2$, both RG summed series in Eq.~\eqref{eq:summed} and Eq.~\eqref{eq:Pseris} agree with each other. However, the two approaches give different results when we compute the scale dependence by setting $\mu^2=\xi Q^2$ and vary $\xi\in [1/2,2]$. We have followed this strategy to compute the scale dependence in this thesis. For RGSPT, the presence of logarithm in the numerator and denominator in term~\eqref{eq:coef_rgspt} results in extra stability with respect to the variation in $\xi$ and the effects of large logarithms are also under control.\par  
    It is important to note that that the organization of perturbative series in Eq.~\eqref{eq:Pseris} is not unique, and a more general method can also be explored in future studies, which is described below~\footnote{We thank the examiner for suggesting this method for estimating scale dependence.}.\par Consider a perturbative series with vanishing anomalous dimension given by:
    \begin{equation}
        \mathcal{S}(Q^2)=\sum_{i=1}^\infty x^i\hs \sum_{j=0}^{i-1} T_{i,j} \log^j\left(\frac{\mu^2}{Q^2}\right)\,,
        \label{eq:demo1}
    \end{equation}
    which can also be rearranged as:
\begin{align}
\mathcal{S}\left(Q^2\right) & =\sum_{i=1}^{\infty} x^i \sum_{j=0}^{i-1} T_{i, j} \sum_{k=0}^j\left(\begin{array}{l}
j \\
k
\end{array}\right) \log ^{j-k}(\xi) \log^k\left(\frac{\mu^2}{\xi Q^2}\right) \\
& =\sum_{i=1}^{\infty} x^i \sum_{k=0}^{i-1} \log ^k\left(\frac{\mu^2}{\xi Q^2}\right) \sum_{j=0}^{i-k-1}\left(\begin{array}{c}
j+k \\
k
\end{array}\right) T_{i, j+k} \log ^j(\xi) \\&\equiv \sum_{i=1}^{\infty} x^i \sum_{j=0}^{i-1} T_{i, j}(\xi) L_{\xi}^j\label{eq1:temp1}\,,
\end{align}
where, $L_{\xi}\equiv\log\left(\frac{\mu^2}{\xi Q^2}\right)$ and $T_{i, j}(\xi=1)=T_{i, j}$ in Eq.~\eqref{eq1:temp1}. For $\xi\sim \ord{1}$, $\log(\xi)$ is not large and $\log\left(\frac{\mu^2}{Q^2}\right)\equiv L_1\sim L_\xi\sim\ord{x^{-1}}$. Another way of arriving at Eq.~\eqref{eq1:temp1} is to rescale the renormalization scale to $\mu^2\rightarrow \mu^2/\xi$ in Eq.~\eqref{eq:demo1} and expand $x(\mu/\sqrt{\xi})$ in terms of $x(\mu)$ and $\log(\xi)$ using \eqref{eq:as_fopt} discussed in section~\eqref{sec:asmq_rgspt}. \par
For the RGSPT series in Eq.~\eqref{eq:summed}, one can rearrange it in such a way that $x\hs L_\xi\sim \ord{1}$ and can be written in the following form:
\begin{align}
\mathcal{S}^{\Sigma_{\xi}}\left(Q^2\right) & =\sum_{i=1}^{\infty} x^i \sum_{j=0}^{i-1} T_{i, j}(\xi) L_{\xi}^j\label{eq:mid1}\,,\\&=\sum_{j=0}^{\infty} \sum_{i=j+1}^{\infty} x^i T_{i, j}(\xi) L_{\xi}^j\label{eq:mid2}\,,\\&=\sum_{i=1}^{\infty} x^i \hs \sum_{j=0}^{\infty}   x^j T_{i+j, j}(\xi) L_{\xi}^j\label{eq:mid3} \,,\\
& \equiv \sum_{i=1} x^i S_i\left[\xi, x \log \left(\frac{\mu^2}{\xi Q^2}\right)\right] \label{eq:mid4}.
\end{align}
It is clear from the above equations that the initial series in Eq.~\eqref{eq:mid1} is $\xi$-independent for each value of $i$, but the rearrangement leads to $\xi$ dependence in each $S_i(\xi, x\hs L_\xi)$ in Eq.~\eqref{eq:mid4}. If the series is truncated to some finite order $i$ then the $\xi$ dependence will be visible. However this dependence will get milder as more and more terms are added. This procedure can also be generalized for the perturbative series with non-vanishing anomalous dimensions and therefore left for future work.    

\section{Running of \texorpdfstring{$\mathbf{\as}$}{} and quark mass running in RGSPT}   \label{sec:asmq_rgspt}
    The closed-form solution for strong coupling constant and quark mass running in the $\msbar$ scheme can be found in Ref.~\cite{Chishtie:2018ipg}. In addition to this, an RG improved relation for the quark mass in pole and $\msbar$ can be found in Ref.~\cite{Ahmady:2004er} to N$^3$LO. These relations are relevant for this work, so we are briefly giving details of the calculations. \par 
   \subsection{RG summation of \texorpdfstring{$\as$}{} }
    The running of the strong coupling constant in the $\msbar$ scheme is obtained by solving the RG equation Eq.~\eqref{eq:alpha_S} numerically. Since there is no exact analytical solution is available, the series solution is obtained in the FOPT. Using five-loop beta function coefficients, an analytical expression is presented in the appendix in Eq.~\eqref{alphasmu} to $\order{x^5}$, it has the following form:
    \begin{equation}
	x(Q) = x \Bigg(1+ x \beta _0 L+x^2 \left(\beta _1 L+\beta _0^2 L^2\right)+x^3 \left(\beta _2 L+\frac{5}{2} \beta _1 \beta _0 L^2+\beta _0^3 L^3\right)\Bigg)+\order{x^5}
	\label{eq:as_fopt}
    \end{equation}
where $x\equiv x(\mu)$ and $L=\log(\mu^2/Q^2)$ and above fixed-order relation is valid for $\mu\sim\hs Q$. It should be noted that if no scale is specified in the argument of the $x$, then it is assumed to be evaluated at scale $\mu$, and this notation is followed in this thesis. \par
   Now, the summed series for Eq.~\eqref{eq:as_fopt} can be written as:
   \begin{equation}
   	x(Q)=\sum_{i=0}x^{i+1}S_i(x\hs L)\,.
   	\label{eq:xrun1}
   \end{equation}
  Inserting above equation in Eq.~\eqref{eq:alpha_S}, we get:
  \begin{align}
  	Q^2 \frac{d}{d\hs Q^2} x(Q)=-\sum_{i=0}\beta_{i} (x(Q))^{i+2}
  \end{align}
that results in the following set of coupled differential equations:
\begin{align}
0&=S_0'(u)-\beta _0 S_0(u)^2\,,\\
0&=\beta _1 S_0(u)^3+2 \beta _0 S_1(u) S_0(u)-S_1'(u)\\
0&=\beta _2 S_0(u)^4+3 \beta _1 S_1(u) S_0(u)^2+2 \beta _0 S_2(u) S_0(u)+\beta _0 S_1(u)^2-S_2'(u)\\
0&=	\beta _3 S_0(u)^5+4 \beta _2 S_1(u) S_0(u)^3+3 \beta _1 S_2(u) S_0(u)^2+3 \beta _1 S_1(u)^2 S_0(u)\nonumber\\&\bs+2 \beta _0 S_3(u) S_0(u)+2 \beta _0 S_1(u) S_2(u)-S_3'(u)\\
0&=\beta _4 S_0(u)^6+5 \beta _3 S_1(u) S_0(u)^4+4 \beta _2 S_2(u) S_0(u)^3+6 \beta _2 S_1(u)^2 S_0(u)^2\nonumber\\&\bs+3 \beta _1 S_3(u) S_0(u)^2+6 \beta _1 S_1(u) S_2(u) S_0(u)+2 \beta _0 S_4(u) S_0(u)\nonumber\\&\bs+\beta _1 S_1(u)^3+\beta _0 S_2(u)^2+2 \beta _0 S_1(u) S_3(u)-S_4'(u)
\end{align}
 The solution to the above equation results in the following expression for Eq.~\eqref{eq:xrun1}:
 \begin{align}
x(Q)=&\frac{x}{w}-\frac{x^2 \tilde{\beta }_1 L_w}{w^2}+x^3 \left(\frac{-\tilde{\beta }_1^2+\tilde{\beta }_2+\tilde{\beta }_1^2 L_w^2-\tilde{\beta }_1^2 L_w}{w^3}+\frac{\tilde{\beta }_1^2-\tilde{\beta }_2}{w^2}\right)\nonumber\\&+x^4 \Big\lbrace\frac{-\frac{\tilde{\beta }_1^3}{2}+\frac{\tilde{\beta }_3}{2}+\tilde{\beta }_1^3 \left(-L_w^3\right)+\frac{5}{2} \tilde{\beta }_1^3 L_w^2+\left(2 \tilde{\beta }_1^3-3 \tilde{\beta }_1 \tilde{\beta }_2\right) L_w}{w^4}\nonumber\\&\bs+\frac{\tilde{\beta }_1^3-\tilde{\beta }_2 \tilde{\beta }_1+\left(2 \tilde{\beta }_1 \tilde{\beta }_2-2 \tilde{\beta }_1^3\right) L_w}{w^3}+\frac{-\frac{\tilde{\beta }_1^3}{2}+\tilde{\beta }_2 \tilde{\beta }_1-\frac{\tilde{\beta }_3}{2}}{w^2}\Big\rbrace\nonumber\\&+x^5 \Big\lbrace\frac{1}{w^5}\bigg[\frac{7 \tilde{\beta }_1^4}{6}-3 \tilde{\beta }_2 \tilde{\beta }_1^2-\frac{1}{6} \tilde{\beta }_3 \tilde{\beta }_1+\frac{5 \tilde{\beta }_2^2}{3}+\frac{\tilde{\beta }_4}{3}+\tilde{\beta }_1^4 L_w^4-\frac{13}{3} \tilde{\beta }_1^4 L_w^3\nonumber\\&\bs+\left(6 \tilde{\beta }_1^2 \tilde{\beta }_2-\frac{3 \tilde{\beta }_1^4}{2}\right) L_w^2+\left(4 \tilde{\beta }_1^4-3 \tilde{\beta }_2 \tilde{\beta }_1^2-2 \tilde{\beta }_3 \tilde{\beta }_1\right) L_w\bigg]\nonumber\\&+\frac{1}{w^4}\bigg[-2 \tilde{\beta }_1^4+5 \tilde{\beta }_2 \tilde{\beta }_1^2-3 \tilde{\beta }_2^2+\left(5 \tilde{\beta }_1^2 \tilde{\beta }_2-5 \tilde{\beta }_1^4\right) L_w+\left(3 \tilde{\beta }_1^4-3 \tilde{\beta }_1^2 \tilde{\beta }_2\right) L_w^2\bigg]\nonumber\\&\bs+\frac{\frac{\tilde{\beta }_1^4}{2}-\tilde{\beta }_2 \tilde{\beta }_1^2-\frac{1}{2} \tilde{\beta }_3 \tilde{\beta }_1+\tilde{\beta }_2^2+\left(\tilde{\beta }_1^4-2 \tilde{\beta }_2 \tilde{\beta }_1^2+\tilde{\beta }_3 \tilde{\beta }_1\right) L_w}{w^3}\nonumber\\&\bs+\frac{\frac{\tilde{\beta }_1^4}{3}-\tilde{\beta }_2 \tilde{\beta }_1^2+\frac{2}{3} \tilde{\beta }_3 \tilde{\beta }_1+\frac{\tilde{\beta }_2^2}{3}\-\frac{\tilde{\beta }_4}{3}}{w^2}\Big\rbrace+\order{x^6}\label{eq:as_1}
 \end{align}
 where, $w=1-\beta_0 \hs x\hs L$, $L_w=\log(w)$ and  $\tilde{\beta}_i=\beta_i/\beta_{0}$. A similar inverse relation for Eq.~\eqref{eq:as_1} can be found in Eq.~(A29) of Ref.~\cite{Abbate:2010xh}.
\subsection{Quark mass running in \texorpdfstring{$\msbar$}{} scheme}
The quark mass ($m_q(Q)$) running in the RGSPT is obtained by using:
\begin{align}
Q^2 \frac{d}{d\hs Q^2} m_q(Q)=m_q(Q) \gamma_m(x(Q)) =-m_q(Q)  \sum_{i=0}   (x(Q))^{i+1} \hs\hs  \gamma_i 
\label{eq:RGE_mq}
\end{align}
    where $\gamma_m(x(Q))$ is quark mass anomalous dimension. In this case, the RG summed coefficients for $m(Q)$ follow coupled differential given in Eq.~\eqref{eq:rec_all}. The first three solutions are given by Eq.~\eqref{eq:RGsol3}. Since the final expression is very lengthy and can be obtained by directly solving Eq.~\eqref{eq:rec_all}, $\order{x^2}$ results are given by:
\begin{align}
m(q)=\frac{m(\mu)}{w^{\tilde{\gamma}_0}}\Bigg\lbrace&1+x \Big\lbrace\tilde{\beta }_1 \tilde{\gamma }_0-\tilde{\gamma }_1+\frac{-\tilde{\beta }_1 \tilde{\gamma }_0+\tilde{\gamma }_1-\tilde{\beta }_1 \tilde{\gamma }_0 L_w}{w}\Big\rbrace+x^2 \Big\lbrace\frac{1}{2} \tilde{\beta }_1^2 \tilde{\gamma }_0^2-\frac{1}{2} \tilde{\beta }_1^2 \tilde{\gamma }_0+\frac{1}{2} \tilde{\beta }_1 \tilde{\gamma }_1\nonumber\\&-\tilde{\beta }_1 \tilde{\gamma }_0 \tilde{\gamma }_1+\frac{1}{2} \tilde{\beta }_2 \tilde{\gamma }_0+\frac{\tilde{\gamma }_1^2}{2}-\frac{\tilde{\gamma }_2}{2}+\frac{1}{w^2}\Big[\frac{1}{2} \tilde{\beta }_1^2 \tilde{\gamma }_0^2-\frac{1}{2} \tilde{\beta }_1^2 \tilde{\gamma }_0+\frac{1}{2} \tilde{\beta }_2 \tilde{\gamma }_0-\frac{1}{2} \tilde{\beta }_1 \tilde{\gamma }_1\nonumber\\&-\tilde{\beta }_1 \tilde{\gamma }_0 \tilde{\gamma }_1+\frac{\tilde{\gamma }_1^2+\tilde{\gamma }_2}{2}+L_w^2 \left(\frac{1}{2} \tilde{\beta }_1^2 \tilde{\gamma }_0^2+\frac{1}{2} \tilde{\beta }_1^2 \tilde{\gamma }_0\right)+L_w \left(\tilde{\beta }_1^2 \tilde{\gamma }_0^2-\tilde{\beta }_1 \tilde{\gamma }_1 \tilde{\gamma }_0-\tilde{\beta }_1 \tilde{\gamma }_1\right)\Big]\nonumber\\&+\frac{\tilde{\beta }_1^2 \left(-\tilde{\gamma }_0^2\right)+\tilde{\beta }_1^2 \tilde{\gamma }_0+2 \tilde{\beta }_1 \tilde{\gamma }_0 \tilde{\gamma }_1-\tilde{\beta }_2 \tilde{\gamma }_0-\tilde{\gamma }_1^2+L_w \left(\tilde{\beta }_1 \tilde{\gamma }_0 \tilde{\gamma }_1-\tilde{\beta }_1^2 \tilde{\gamma }_0^2\right)}{w}\Big\rbrace+\order{x^3}\Bigg\rbrace
\end{align}

\subsection{Quark mass relation between \texorpdfstring{$\msbar$}{} and pole scheme}
Quark mass relations between pole and $\msbar$ scheme are known to four-loops~\cite{Chetyrkin:1999ys,Chetyrkin:1999qi,Melnikov:2000qh,Marquard:2015qpa,Marquard:2016dcn}. Since pole mass is RG invariant, we can use this fact to derive an RG summed expression for it. The summed series has the following form:
\begin{align}
	M_q=m_q(\mu)\sum_{i=0} x^i\hs S_i(x \hs \log\left(\frac{\mu^2}{(m(\mu))^2}\right)=m(\mu)	\mathcal{S}^\Sigma_m(Q^2)
\end{align}
where, $\mathcal{S}^\Sigma_m(Q^2)$ has anomalous dimension $\gamma_{S}(x)=\gamma_m(x)$. The coupled differential equations, in this case, are obtained as follows:
\begin{align}
0&=\left(1-\beta _0 u\right) S_0'(u)-\gamma _0 S_0(u)\,,\\0&=\left(1-\beta _0 u\right) S_1'(u)+S_0'(u) \left(2 \gamma _0-\beta _1 u\right)-\beta _0 S_1(u)-\gamma _1 S_0(u)-\gamma _0 S_1(u)\,,\\
0&=\left(1-\beta _0 u\right) S_2'(u)+S_0'(u) \left(2 \gamma _1-\beta _2 u\right)+S_1'(u) \left(2 \gamma _0-\beta _1 u\right)-\beta _1 S_1(u)-2 \beta _0 S_2(u)\nonumber\\&\bs-\gamma _2 S_0(u)-\gamma _1 S_1(u)-\gamma _0 S_2(u)\,,\\
0&=\left(1-\beta _0 u\right) S_3'(u)+S_0'(u) \left(2 \gamma _2-\beta _3 u\right)+S_1'(u) \left(2 \gamma _1-\beta _2 u\right)+S_2'(u) \left(2 \gamma _0-\beta _1 u\right)-\gamma _0 S_3(u)\nonumber\\&\bs-\beta _2 S_1(u)-2 \beta _1 S_2(u)-3 \beta _0 S_3(u)-\gamma _3 S_0(u)-\gamma _2 S_1(u)-\gamma _1 S_2(u)\,,\\
0&=\left(1-\beta _0 u\right) S_4'(u)+S_1'(u) \left(2 \gamma _2-\beta _3 u\right)+S_2'(u) \left(2 \gamma _1-\beta _2 u\right)+S_3'(u) \left(2 \gamma _0-\beta _1 u\right)\nonumber\\&\bs+S_0'(u) \left(2 \gamma _3-\beta _4 u\right)-\beta _3 S_1(u)-2 \beta _2 S_2(u)-3 \beta _1 S_3(u)-4 \beta _0 S_4(u)-\gamma _4 S_0(u)\nonumber\\&\bs-\gamma _3 S_1(u)-\gamma _2 S_2(u)-\gamma _1 S_3(u)-\gamma _0 S_4(u)\,,
\end{align}
where $u=x \hs \log(\mu^2/(m(\mu))^2)$. To $\order{x^2}$, we have following expression:
\begin{align}
M_q=&m_q(\mu) w^{-\tilde{\gamma }_0}\Bigg\lbrace1+x \Big\lbrace\tilde{\beta }_1 \tilde{\gamma }_0-\tilde{\gamma }_1+\frac{2 L_{\tilde{w}} \beta _0 \tilde{\gamma }_0^2-\tilde{\beta }_1 \tilde{\gamma }_0-L_{\tilde{w}} \tilde{\beta }_1 \tilde{\gamma }_0+\tilde{\gamma }_1+T_{1,0}}{\tilde{w}}\Big\rbrace\\&+  x^2\Big\lbrace \frac{1}{2} \tilde{\beta }_1^2 \tilde{\gamma }_0^2-\frac{1}{2} \tilde{\beta }_1^2 \tilde{\gamma }_0+\frac{1}{2} \tilde{\beta }_1 \tilde{\gamma }_1-\tilde{\beta }_1 \tilde{\gamma }_0 \tilde{\gamma }_1+\frac{1}{2} \tilde{\beta }_2 \tilde{\gamma }_0+\frac{\tilde{\gamma }_1^2}{2}-\frac{\tilde{\gamma }_2}{2}\nonumber\\& +\frac{1}{\tilde{w}} \bigg[\tilde{\beta }_1 \tilde{\gamma }_0 T_{1,0}-\tilde{\gamma }_1 T_{1,0}+\tilde{\beta }_1^2 \left(-\tilde{\gamma }_0^2\right)+\tilde{\beta }_1^2 \tilde{\gamma }_0-2 \beta _0 \tilde{\beta }_1 \tilde{\gamma }_0^2+2 \tilde{\beta }_1 \tilde{\gamma }_0 \tilde{\gamma }_1\nonumber\\& \bs-\tilde{\beta }_2 \tilde{\gamma }_0+2 \beta _0 \tilde{\gamma }_0 \tilde{\gamma }_1-\tilde{\gamma }_1^2+\left(2 \beta _0 \tilde{\beta }_1 \tilde{\gamma }_0^3-\tilde{\beta }_1^2 \tilde{\gamma }_0^2-2 \beta _0 \tilde{\gamma }_1 \tilde{\gamma }_0^2+\tilde{\beta }_1 \tilde{\gamma }_1 \tilde{\gamma }_0\right) L_{\tilde{w}}\bigg]\nonumber\\& +\frac{1}{\tilde{w}^2}\bigg[ L_{\tilde{w}} \Big(2 \beta _0 \tilde{\gamma }_0^2 T_{1,0}+2 \beta _0 \tilde{\gamma }_0 T_{1,0}-\tilde{\beta }_1 \tilde{\gamma }_0 T_{1,0}-\tilde{\beta }_1 T_{1,0}-4 \beta _0^2 \tilde{\gamma }_0^3-2 \beta _0 \tilde{\beta }_1 \tilde{\gamma }_0^3+\tilde{\beta }_1^2 \tilde{\gamma }_0^2\nonumber\\&\bs+2 \beta _0 \tilde{\beta }_1 \tilde{\gamma }_0^2+2 \beta _0 \tilde{\gamma }_1 \tilde{\gamma }_0^2+2 \beta _0 \tilde{\gamma }_1 \tilde{\gamma }_0-\tilde{\beta }_1 \tilde{\gamma }_1 \tilde{\gamma }_0-\tilde{\beta }_1 \tilde{\gamma }_1\Big)-\tilde{\beta }_1 \tilde{\gamma }_0 T_{1,0}+\tilde{\gamma }_1 T_{1,0}\nonumber\\&\bs+\frac{1}{2} \tilde{\beta }_1^2 \tilde{\gamma }_0^2+2 \beta _0 \tilde{\beta }_1 \tilde{\gamma }_0^2-\frac{1}{2} \tilde{\beta }_1^2 \tilde{\gamma }_0+\frac{1}{2} \tilde{\beta }_2 \tilde{\gamma }_0-\frac{1}{2} \tilde{\beta }_1 \tilde{\gamma }_1-2 \beta _0 \tilde{\gamma }_0 \tilde{\gamma }_1-\tilde{\beta }_1 \tilde{\gamma }_0 \tilde{\gamma }_1+\frac{\tilde{\gamma }_1^2}{2}+\frac{\tilde{\gamma }_2}{2}\nonumber\\&\bs+\left(2 \beta _0^2 \tilde{\gamma }_0^4+2 \beta _0^2 \tilde{\gamma }_0^3-2 \beta _0 \tilde{\beta }_1 \tilde{\gamma }_0^3+\frac{1}{2} \tilde{\beta }_1^2 \tilde{\gamma }_0^2-2 \beta _0 \tilde{\beta }_1 \tilde{\gamma }_0^2+\frac{1}{2} \tilde{\beta }_1^2 \tilde{\gamma }_0\right) L_{\tilde{w}}^2+T_{2,0}\bigg]   \Big\rbrace\nonumber\\&\bs+\order{x^3}\Bigg\rbrace\,,
\label{eq:Mpole2Ms}
\end{align}
where $\tilde{w}=1-\beta_0 \hs x \hs \log(\mu^2/(m(\mu))^2)$  and $L_{\tilde{w}}=\log(\tilde{w})$. and the coefficients can be obtained from 
\begin{align}
	T_{1,0}=4/3\,,\quad T_{2,0}=\left(-\frac{\zeta (3)}{6}+\frac{7 \pi ^2}{18}+\frac{2905}{288}+\frac{1}{9} \pi ^2 \log (2)+\left(-\frac{71}{144}-\frac{\pi ^2}{18}\right) n_f\right)\,.
\end{align}
We can also find the inverse relation for Eq.~\eqref{eq:Mpole2Ms}, and in this case, the summed series has the form:
\begin{align}
m_q(\mu)=M_q \sum_{i=0}x^i S_i(x \log(\mu^2/M_q^2))
\end{align}
The RGE for the quark mass in the $\msbar$ scheme is given by Eq.~\eqref{eq:RGE_mq}, and the summed expression to $\order{x^2}$ is obtained as:
\begin{align}
m_q(\mu)=M_q w^{\tilde{\gamma}_0}&\Bigg\lbrace  1+x \Big\lbrace \frac{\tilde{\beta }_1 \tilde{\gamma }_0-\tilde{\gamma }_1+\tilde{\beta }_1 \tilde{\gamma }_0 L_w+T_{1,0}}{w}-\tilde{\beta }_1 \tilde{\gamma }_0+\tilde{\gamma }_1\Big\rbrace\nonumber\\&+x^2\Big\lbrace \frac{1}{2} \tilde{\beta }_1^2 \tilde{\gamma }_0^2+\frac{1}{2} \tilde{\beta }_1^2 \tilde{\gamma }_0-\frac{1}{2} \tilde{\beta }_1 \tilde{\gamma }_1-\tilde{\beta }_1 \tilde{\gamma }_0 \tilde{\gamma }_1-\frac{1}{2} \tilde{\beta }_2 \tilde{\gamma }_0+\frac{\tilde{\gamma }_1^2}{2}+\frac{\tilde{\gamma }_2}{2}\nonumber\\&+\frac{1}{w}\bigg[-\tilde{\beta }_1 \tilde{\gamma }_0 T_{1,0}+\tilde{\gamma }_1 T_{1,0}+\tilde{\beta }_1^2 \left(-\tilde{\gamma }_0^2\right)+2 \tilde{\beta }_1 \tilde{\gamma }_0 \tilde{\gamma }_1-\tilde{\beta }_1^2 \tilde{\gamma }_0+\tilde{\beta }_2 \tilde{\gamma }_0-\tilde{\gamma }_1^2\nonumber\\&+L_w \left(\tilde{\beta }_1 \tilde{\gamma }_0 \tilde{\gamma }_1-\tilde{\beta }_1^2 \tilde{\gamma }_0^2\right)\bigg] +\frac{1}{w^2}\bigg[T_{2,0}+  \frac{1}{2} \tilde{\beta }_1^2 \tilde{\gamma }_0^2+\frac{1}{2} \tilde{\beta }_1^2 \tilde{\gamma }_0-\frac{1}{2} \tilde{\beta }_2 \tilde{\gamma }_0-\tilde{\gamma }_1 T_{1,0}\nonumber\\&+ \tilde{\beta }_1 \tilde{\gamma }_0 T_{1,0}+\frac{1}{2} \tilde{\beta }_1 \tilde{\gamma }_1-\tilde{\beta }_1 \tilde{\gamma }_0 \tilde{\gamma }_1+\frac{\tilde{\gamma }_1^2}{2}-\frac{\tilde{\gamma }_2}{2}+ L_w^2 \left(\frac{1}{2} \tilde{\beta }_1^2 \tilde{\gamma }_0^2-\frac{1}{2} \tilde{\beta }_1^2 \tilde{\gamma }_0\right)\nonumber\\&+L_w \left(\tilde{\beta }_1 \tilde{\gamma }_0 T_{1,0}-\tilde{\beta }_1 T_{1,0}+\tilde{\beta }_1^2 \tilde{\gamma }_0^2-\tilde{\beta }_1 \tilde{\gamma }_1 \tilde{\gamma }_0+\tilde{\beta }_1 \tilde{\gamma }_1\right) \bigg]\Big\rbrace+\order{x^3} \Bigg\rbrace\,,
\end{align}
and the coefficients $T_{i,0}$~\cite{Melnikov:2000qh,Marquard:2016dcn} are given by:
\begin{align}
	T_{1,0}=-4/3\,,\quad T_{2,0}=\left(\frac{\zeta (3)}{6}-\frac{7 \pi ^2}{18}-\frac{3161}{288}-\frac{1}{9} \pi ^2 \log (2)+\left(\frac{71}{144}+\frac{\pi ^2}{18}\right) n_f\right)
\end{align}
In addition to these relations, RG summation using RGSPT for various quantities is described in the coming chapters for various quantities.

%% file: Chapters/Chap3.tex
 \chapter{RG improvement of QCD Static Energy}
\label{Chapter3}

\lhead{Chapter 3. \emph{RG improvement of QCD Static Energy}}
 \onehalfspacing

\section{Motivation}
The perturbative QCD static potential and ultrasoft contributions, which together give the static energy, have been calculated to N$^3$LO and N$^4$LO in $\as$, respectively, by several authors. Using the renormalization group, and Pad\'e approximants, we estimate the four-loop (N$^4$LO) corrections to the static energy.
We also employ the renormalization group summed perturbation theory (RGSPT) and sum the logarithms of the perturbative series in order to reduce sensitivity to the renormalization scale in momentum space. The convergence behavior of the perturbative series is also improved in position space using the restricted Fourier transform scheme. Using RGSPT, we have extracted the value of $\Lambda^{\overline{\textrm{MS}}}_{\textrm{QCD}}$ at different scales for two active flavors by matching to the static energy from lattice QCD simulations.
\section{Introduction}
The static potential energy of quantum chromodynamics is the non-Abelian analog of the well-known Coulomb potential energy of quantum electrodynamics. The short distance part of this quantity is calculated in non-relativistic QCD (NRQCD)~\cite{Caswell:1985ui,Bodwin:1994jh} framework and involves the evaluation of Feynman diagrams. It has been studied extensively in recent years, and analytical results are known to three-loop. At four-loop order, contributions involving the ultrasoft gluons that start to contribute from three-loop order in perturbation series are known. Such contributions only appear for short distances when the system is weakly coupled. They are calculated using weakly coupled potential non-relativistic QCD (pNRQCD)~\cite{Pineda:1997bj,Brambilla:2004jw} formalism. Hence, we discuss only the weakly coupled regime of pNRQCD in this chapter. The static potential depends on the renormalization scale $\mu$, the ultrasoft factorization scale $\mu_{\rm us}$, and the magnitude of the three momentum transfer between the heavy sources $p\left(=|\mathbf{p}|\right)$. It should be noted that static potential obeys a homogeneous RGE and the $\mu$-dependence is due to finite order results available and such dependence disappears if in all order results are available. In this study, only $\overline{\text{MS}}$ renormalization scheme is used.\par
The first attempt to perform the full three-loop (numerical) calculations was by two independent groups \cite{Anzai:2009tm,Smirnov:2009fh}. 
The results were found to be in agreement. Analytical calculations were presented almost six years later in Ref.~\cite{Lee:2016cgz}. At three-loop order, ultrasoft gluons also appear, which are capable of changing singlet to octet state of the system and \textit{vice-versa}. Such contributions were first pointed in Ref.~\cite{Appelquist:1977es}, and were calculated in Ref.~\cite{Brambilla:1999qa} in pNRQCD. The renormalization group (RG) improvement of the ultrasoft terms at three-loop order was first discussed in Ref.~\cite{Pineda:2000gza}. The next order calculations for the ultrasoft terms can be found in Ref.~\cite{Brambilla:2006wp}, and their resummation is discussed in Ref.~\cite{Brambilla:2009bi}.

QCD is known to be non-perturbative at long distances, and this fact is manifest in lattice QCD (LQCD) simulations. The static energy, which includes the static potential and the ultrasoft contributions, can also be computed in LQCD simulations, and hence it becomes a topic of great interest to extract various parameters of the theory. Some recent LQCD simulation results for the static energy can be found in Refs.~\cite{Bazavov:2014soa,Karbstein:2018mzo,Karbstein:2014bsa,Takaura:2018vcy,Takaura:2018lpw,Bazavov:2019qoo}, and they support the Cornell potential type behavior for the heavy quark-antiquark system.

The potential energy between a heavy quark and anti-quark pair is an important ingredient to describe, among other things, non-relativistic bound states like quarkonia~\cite{Brambilla:1999xf,Brambilla:2004jw}, quark masses~\cite{Hoang:2000fm,Pineda:2001zq,Beneke:2014pta,Penin:2014zaa,Ayala:2014yxa,Kiyo:2015ufa,Kiyo:2015ooa,Beneke:2015zqa,Mateu:2017hlz,Peset:2018ria} and threshold production of top quarks~\cite{Hoang:2013uda,Beneke:2013jia,Beneke:2015kwa}, etc. These studies have pointed out the reliable determinations using QCD static potential requires subtraction of the leading $\ord{\lqcd}$ infrared renormalon present in it~\cite{Aglietti:1995tg}. This renormalon cancels with twice the magnitude of the renormalon contributions~\cite{Hoang:1998nz} present in the pole mass of the heavy quark~\cite{Beneke:1994sw,Bigi:1994em,Beneke:1994rs}, leaving static energy renormalon free. This has led to various definitions of the quark masses such as kinetic mass~\cite{Bigi:1994em}, potential subtracted (PS) mass~\cite{Beneke:1998rk}, 1S mass~\cite{Hoang:1998ng}, $\overline{\text{PS}}$~\cite{Yakovlev:2000pv}, renormalon subtracted mass~\cite{Pineda:2001zq} and MSR mass~\cite{Hoang:2008yj,Hoang:2017suc,Hoang:2017btd,Bris:2020uyb}.  It should be noted that the leading renormalon at $u=1/2$ in the Borel plane is absent for static potential in the momentum space, and renormalon at $u=3/2$ is suppressed~$\left(\lqcd^2/q^2\right)^{3/2}$~\cite{Beneke:1998rk,Takaura:2021yaj}. 

In this study, we have addressed the following issues:
\begin{itemize}
	\item At four-loop order, the constant term (RG inaccessible) contributing to the perturbative series for static potential is as yet unknown. It requires the calculation of four-loop Feynman diagrams. In the absence of such calculations, we use RGE and Pad\'e approximants~\cite{Chishtie:2001mf} to get the estimate for the four-loop coefficients. 
	\item The RG improvement of the static energy by summation of all RG-accessible running logarithms following the method advocated in Refs.~\cite{Maxwell:1999dv,Maxwell:2000mm,Maxwell:2001he,Ahmady:1999xg,Ahmady:2002fd,Ahmady:2002pa,Ananthanarayan:2016kll,Abbas:2012py}, and applied to the ultrasoft logarithms present in the static energy. All the leading and next-to-leading logarithms at each order in perturbation theory that can be accessed through the RG equation (RGE) are called RG-accessible logarithms. We call this renormalization group improved perturbative series as RG-summed or RGSPT series. It should be noted that our summation prescription discusses RG-improvement with respect to the renormalization scale, whereas Refs.~\cite{Pineda:2000gza,Brambilla:2009bi} deals with the RG-improvement with respect to the ultrasoft scale.
	\item The convergence of the static energy is improved to the four-loop order in position space using the Restricted Fourier Transform (RFT) proposed in Ref.~\cite{Karbstein:2013zxa}. The RFT is closely related to the potential subtracted scheme~\cite{Beneke:1998rk} that also removes leading long-distance contributions $\ord{\lqcd\hs r}$ from the pole mass of the heavy quark.
	\item We use the RG-summed and FOPT (unsummed) prescription of the static energy in momentum space to extract $\Lambda^{\overline{\textrm{MS}}}_{\textrm{QCD}}$ to four-loop from the LQCD inputs from Ref.~\cite{Karbstein:2018mzo}.
	\item In comparison to the unsummed series, the RG-summed static energy in momentum space gives a better fit to the static energy obtained from LQCD in Ref.~ \cite{Karbstein:2018mzo}.
	\item The four-loop RG-summed static energy is used as a trial case to show that these improvements (scale sensitivity and better fit) also persist at the higher order.
\end{itemize}

The scheme of this chapter is as follows: In section~\eqref{sec:pert_pot}, we discuss the perturbative treatment to the static energy and various pieces associated with this quantity in the weak coupling limit. In section~\eqref{sec:RG_ser}, we calculate some contributions to the static energy at the four-loop using the RGE. In section~\eqref{sec:Pade_est}, we use Pad\'e approximants to estimate all the four-loop coefficients, even the one that is not accessible with the RGE. Some of the estimates are found in agreement with RGE solutions. In section~\eqref{sec:RG_sum}, we use the formalism presented in chapter~\eqref{Chapter2} to perform all order RG-summation of certain running logarithms and show that this RG-improvement will bring down the sensitivity to the renormalization scale. In section~\eqref{sec:RFT}, we discuss the improvement of the static energy to the four-loop order in position space by removing the pathological uncontrolled contributions using RFT. The inputs from the previous sections are applied in section~\eqref{sec:lat_input} to fit the LQCD inputs for the static energy to extract the $\Lambda^{\overline{\textrm{MS}}}_{\textrm{QCD}}$ for two active flavors in momentum space. A discussion is presented in section \eqref{sec:conc} and we summarize our results in  section~\eqref{sec:summary4}. Appendix~\eqref{app:loop_coef} contains the known contribution to the static energy. Appendix~\eqref{app:Vr} contains the useful formula for calculating the restricted and unrestricted versions of position space static potential and the static energy. Appendix~\eqref{app:resVr} contains the final result of uncontrolled contribution to the static energy in position space to the four-loop order. 
\section{The Perturbative QCD-Static Energy \label{sec:pert_pot}}
A heavy quark and anti-quark system, with heavy quark mass $m_Q$ and relative velocity $v<1$, is non-relativistic in nature and various scales~\cite{Brambilla:2004jw,Beneke:1997zp} present in the system are hard scale $\sim \mathcal{O}(m_Q)$, soft scale $\sim\mathcal{O}(m_Q v)$, and ultrasoft scale $\sim\mathcal{O}(m_Q v^2)$. If these scales are well separated, we can integrate them one by one to study only relevant degrees of freedom. Integrating out the hard scale from QCD gives the non-relativistic QCD (NRQCD) in which the soft and the ultrasoft degrees of freedom are dynamical. Contributions to the static energy at different orders were calculated using this framework. The one-loop perturbative calculations for massless quarks were first performed in the late 1970s and can be found in Refs.~\cite{Appelquist:1977es,Appelquist:1977tw,Susskind,Fischler:1977yf} and in the massive case in Ref.~\cite{Billoire:1979ih}. Two-loop massless calculations appeared in Refs.~\cite{Melles:1998dj,Schroder:1998vy, peter:1998ml, Peter:1997me} while the massive case were considered in Refs.~\cite{Melles:2000ml,Melles:2000ey, sumino:2002ms, Hoang:2000fm}.\par
The pNRQCD is obtained from NRQCD by integrating out the soft scale, and this formalism is best suitable for studying the threshold systems. The heavy quark and anti-quark systems in this formalism are described by color singlet fields $S$ and color octet fields $O$. The gauge fields are multipole expanded about the inter-quark separation, and the gauge invariant Lagrangian for pNRQCD\cite{Brambilla:1999xf,Brambilla:2004jw}, at leading order in $1/m_Q$ is given by
\begin{align}
	\mathcal{L}_{\text {pNRQCD }}=& \mathcal{L}_{\text {light}}-\frac{1}{4} F_{\mu \nu}^{a} F^{\mu \nu a}+\operatorname{Tr}\left(\mathrm{S}^{\dagger}\left(i \partial_{0}-V_{s}(r,\mu_{us})+\ldots\right) \mathrm{S}\right)\nonumber\\&+\operatorname{Tr}\left(\mathrm{O}^{\dagger}\left(i D_{0}-V_{o}(r,\mu_{us})+\ldots\right) \mathrm{O}\right)+g V_{A}(r,\mu_{us}) \operatorname{Tr} \left(\mathrm{O}^{\dagger} \boldsymbol{r} \cdot \mathbf{E} \mathrm{S}+\mathrm{S}^{\dagger} \boldsymbol{r} \cdot \mathbf{E} \mathrm{O}\right)\nonumber \\ & +g\hs \frac{V_{B}(r,\mu_{us})}{2} \operatorname{Tr}\left(\mathrm{O}^{\dagger} \boldsymbol{r} \cdot \mathbf{E} \mathrm{O}+\mathrm{O}^{\dagger} \mathbf{O} \boldsymbol{r} \cdot \mathbf{E}\right)\, ,
	\label{eq:lag_pnrqcd}
\end{align}
where $\mathcal{L}_{\text {light}}$ is Lagrangian for light quarks, $iD_0 O \equiv i \partial_0 O-g \left[A_0(\mathbf{R},t),O\right]$ and $\mathbf{E}$ are chromoelectric field strength. $V_s(r,\mu_{us})$ and $V_o(r,\mu_{us})$ are singlet and octet potential at leading order in $r$ while $V_A$ and $V_B$ appear at sub-leading order in $r$. It should be noted that these potentials will also depend on the renormalization scale $\mu$ for any finite order calculation. The potentials at higher order in $1/m_Q$ are hidden in ellipses that disappear in the static limit $m_Q\rightarrow\infty$. The gauge fields are already multipole expanded about inter-quark separation ($\mathbf{R}$) in Eq.~\eqref{eq:lag_pnrqcd}) and therefore, $F^{\mu \nu a}\equiv F^{\mu \nu a}(\mathbf{R},t)$. 

In the weak coupling regime, the system has a hierarchy of scales:	
\begin{equation}
	m_Q\gg m_Q~v\gg m_Q~v^2\gg\Lambda^{\overline{\textrm{MS}}}_{\textrm{QCD}},
	\label{wc_limit}
\end{equation}
and in this limit, the static energy can be written as a perturbative series in the strong coupling constant $\alpha_{s}$. The singlet static energy to known orders can be written as:
\begin{equation}
	E_{\mathrm{0}}(p,\mu)=V_s(p,\mu,\mu_{\mathrm{us}})+\delta^{\rm us}(p,\mu,\mu_{\mathrm{us}})\, ,
	\label{E0}
\end{equation}
where $V_s(p,\mu,\mu_{\mathrm{us}})$ is the perturbative static potential  and $\delta^{\rm us}(p,\mu,\mu_{\mathrm{us}})$ are contribution from ultrasoft gluons. The static potential encodes the interaction of the quarks and gluons degrees of freedom in the singlet state and is known to three loop. It takes the following form:
\begin{equation}
	V_s(p,\mu,\mu_{us})=\frac{-4 \pi^2 C_F}{p^2}\sum _{i=0}^n \sum _{j=0}^i x^{i+1} L^j \left(T_{i,j,0}+\sum _{k=1}^{i-2} \theta (i-3) \widetilde{T}_{i,j,k}^{\rm us} \log ^k\left(\frac{\mu _{us}^2}{p^2}\right)\right)+\order{x^{n+2}}\,.
	\label{Vpert}
\end{equation}
In the above, $L\equiv \log\left(\frac{\mu^2}{p^2}\right)$ is running logarithm, and $C_F=4/3$ is color factor of the $SU(3)$ representation. The expansion parameter in the above equation is defined as $x\equiv(\alpha_s(\mu)/\pi)$, and any argument of $x$ indicates the scale at which it is evaluated. The coefficients $T_{i,0,0}$ are the $ i^{th}-$loop perturbative contributions. The one-loop coefficient $T_{1,0,0}$ can be found in Refs.~\cite{Billoire:1979ih,Fischler:1977yf}, the two-loop coefficient $T_{2,0,0}$ in Refs.~\cite{Schroder:1998vy,peter:1998ml,Peter:1997me} and the three-loop coefficient $T_{3,0,0}$ in Refs.~\cite{Smirnov:2009fh,Anzai:2009tm,Lee:2016cgz}. All these coefficients are collected in appendix~\eqref{app:loop_coef}. The coefficients of infrared logarithms, $\widetilde{T}_{i,j,k}^{\rm us}$, in the static potential and can be found in Refs.~ \cite{Brambilla:1999qa,Brambilla:2006wp}.\par
In effective field theory language, the static potential is a matching coefficient that depends upon the factorization scale $\mu_{\rm us}$. The presence of the logarithmic terms in the static potential at three-loop were first pointed out in the Ref.~\cite{Appelquist:1977es} and they act as a source of infrared divergences. The ultrasoft part $\delta^{\rm us}$ is now known to next to the leading order (NLO)~\cite{Brambilla:1999qa,Brambilla:2006wp} but it contributes to the static energy from three-loop order (briefly discussed in section~\eqref{sec:RFT}). They carry the information of the dynamical ultrasoft gluon degrees of freedom with the ultraviolet cut-off $\mu_{\rm us}$. This scale acts as a source of ultra-violet divergences for these gluons. Both divergences, however, cancel with each other for the static energy, which results in non-analytic dependence $\left(\sim \alpha_{s}^n \log^m\left(\alpha_{s}\right)\right)$ in terms of the expansion parameter and the total energy in Eq.~\eqref{E0}) takes the form:
\begin{align}
	E_0(p,\mu)&=-\frac{4 \pi^2 C_F }{p^2} \sum _{i=0}^n \sum _{k=0}^{\tiny{\substack{(i-2)\\ \times\theta (i-3)}}} x(p)^{i+1} T_{i,0,k}\log ^k(x(p))+\order{x(p)^{n+2}}\nonumber\\& \equiv -\frac{4 \pi^2 C_F }{p^2} W(x(p),L=0)\, ,
	\label{E01}
\end{align}
where $T_{i,0,0}\overset{i>2}{=} T^{\rm pert}_{i,0,0}+\delta T^{\rm us}_{i,0,0}$ and constant terms ($\delta T^{\rm us}_{i,0,0}$) are the contributions from the ultrasoft gluons and can be found in Ref.~\cite{Brambilla:2006wp}.

The ultrasoft contributions to the static energy are now known to the four-loop order, but the perturbative correction to static potential is yet to be calculated at this order. Some of the contributions to the static energy at this order can be accessed using RGE, and they are discussed in the next section~\eqref{sec:RG_ser}.

\section{RG Solutions of the four-loop contributions\label{sec:RG_ser}}
We can rewrite Eq.~\eqref{E01} in terms of the coupling at the renormalization scale $\mu$ using Eq.~\eqref{alphasmu}) as:
\begin{align}
	E_0(p,\mu)&=-\frac{\left(4 \pi^2 C_F \right) }{p^2} \sum _{i=0}^n \sum _{j=0}^i \sum _{k=0}^{\tiny{\substack{(i-j-2)\\ \times \theta (i-j-3)}}} x^{i+1} L^j \log ^k(x) T_{i,j,k}\nonumber\\&=-\frac{4 \pi^2 C_F }{p^2} W(x,L)\, .
	\label{E0exp}
\end{align}
Despite the explicit dependence, the all-order series should be independent of the renormalization scale $\mu$. Mathematically, this implies that any perturbative series $W(x,L)$ must satisfy the RGE:
\begin{equation}
	\mu^2 \frac{d^2}{d\mu^2} W(x, L)=\left(\frac{\partial }{\partial L} + \beta(x) \frac{\partial }{\partial x}\right)W(x, L) = 0\,.
	\label{RG_W}
\end{equation} 
The QCD beta function $\beta$ is defined in terms of $x$ as:
\begin{equation}
	\beta(x) = \mu^2 \frac{d^2}{d\mu^2} x = -\sum_{i=0}^{\infty} \beta_i x^{i+2}\, ,
	\label{beta}
\end{equation}	
where the coefficients $\beta_i$ up to five-loop order~\cite{Gross:1973id, Caswell:1974gg,Jones:1974mm,Tarasov:1980au,Larin:1993tp,vanRitbergen:1997va,Czakon:2004bu,Baikov:2016tgj,Herzog:2017ohr} are given in appendix~\eqref{app:mass_run}. The RGE can be used to solve iteratively for the RG-accessible terms in terms of the QCD $\beta_i$ coefficients and the lower order RG-inaccessible coefficients $T_{i,0,k}$. 

The RGE for $W(x,L)$ along with known results to three-loop and QCD beta functions, allow us to extract the RG-accessible four-loop coefficients $T_{4,1,0}$,\hs $T_{4,2,0}$,\hs $T_{4,3,0},\hs T_{4,1,1}$ and $T_{4,4,0}$. To obtain them, we note that 
\begin{align}
	&\left(\frac{\partial }{\partial L}+\beta(x) \frac{\partial}{\partial x}\right) W(x,L)\nonumber \\ & = \left( T_{1,1,0}-\beta_0\right)x^2+(T_{2,1,0}-\beta_1 - 2 \beta_0 T_{1,0,0}x^3+ \left(2T_{2,2,0}-2\beta_0 T_{1,1,0}\right)x^3 L\nonumber \\ &+(T_{3,1,0}-\beta_2-2\beta_1 T_{1,0,0}-3\beta_0 T_{2,0,0}x^4+\big(-3\beta_{1} T_{2,1,0}-2\beta_1 T_{1,1,0}+2T_{3,2,0}\big)x^4 L\nonumber \\ &+ \left(3T_{3,3,0}-3\beta_{1} T_{2,2,0}\right)x^4 L^2+\big(T_{4,1,0}-\beta_3-2\beta_2 T_{1,0,0}-3\beta_1 T_{2,0,0}-4\beta_{1} T_{3,0,0}-\beta _0 T_{3,0,1}\nonumber \\ &+\log (x) T_{4,1,1}-4 \beta _0 \log (x) T_{3,0,1}\big)x^5 +(2 T_{4,2,0}-2 T_{1,1,0} \beta _2-3 T_{2,1,0} \beta _1-4 \beta _0 T_{3,1,0}x^5 L\nonumber\\&+(3T_{4,3,0}-3\beta_1 T_{2,2,0}-4\beta_{1} T_{3,2,0}x^5 L^2+\left(4T_{4,4,0}-4\beta_{0} T_{3,3,0}\right)x^5 L^3+ \order{x^6} =0\, .
	\label{eq:rg_W}
\end{align} 
Coefficients of $x^i L^j \log^k(x)$ of the above equation gives the RG-accessible coefficients
\begin{align}
	T_{4,1,0}&= 4 \beta_0 T_{3,0,0}\beta_0 +\beta _0 T_{3,0,1}+3 \beta_1 T_{2,0,0}+2 \beta_2 T_{1,0,0}+\beta_3 \, ,\nonumber\\
	T_{4,2,0}&=6 \beta_0^2 T_{2,0,0}+7 \beta_1 \beta_0 T_{1,0,0}+3 \beta_2 \beta_0+\frac{3 \beta_1^2}{2} \, ,\nn \\
	T_{4,3,0}&=4 \beta_0^3 T_{1,0,0}+\frac{13}{3} \beta_1 \beta_0^2\, ,,\quad
	T_{4,4,0}=\beta_0^4\, ,\quad 	T_{4,1,1}=4 \beta _0 T_{3,0,1}\,
	\label{RG_sol}
\end{align}
The coefficients $T_{i,0,k}$ can not be obtained using RGE and are known as the RG-inaccessible terms. The calculation of such terms involves the evaluation of all the Feynman diagrams relevant to that order. Such calculations are yet to be performed, so we use asymptotic Pad\'e approximant (APAP) to estimate the unknown coefficient, $T_{4,0,0}$, which is discussed in the next section. 

\section{Pad\'e estimate of the four-loop contribution \label{sec:Pade_est}}
Pad\'e approximants are rational functions that can be used to estimate the higher-order terms of a series from its lower-order coefficients. Both the original series and the Pad\'e approximants have the same Taylor expansion to a given order, and the next term is taken as its prediction. They are used in Refs.~\cite{Samuel:1995jc,Ellis:1996zn,Ellis:1997sb,Elias:1998bi, Chishtie:1998rz,Ahmady:1999xg, Elias:2000iw,Chishtie:2000ex}  in the past to improve the higher-order perturbative results for QCD.  It is worth mentioning that they were first used for the static potential in Ref.~\cite{Chishtie:2001mf},  and we are extending the results using the same method for static energy to the four-loop order. The procedure used is explained in this section.

Note that the ultrasoft corrections to the static energy are already known to the $\ord{\as^4}$ order, so we do not need to predict them using Pad\'e approximants. They can be added to the predicted values at the end of calculations. We rewrite the perturbative series in Eq.~\eqref{E0exp}, without the ultrasoft corrections as $\widetilde{W}(x,L)$, in the following form:
\begin{equation}\label{eq:Wxl2}
	\widetilde{W}(x,L)= 1 + R_1 x + R_2 x^2 + R_3 x^3 + R_4 x^4 +\cdots + R_N x^N + \cdots\, ,
\end{equation}
where \begin{equation}\label{eq:Rs}
	R_i\equiv\sum _{j=0}^i L^j T_{i,j,0}\, .
\end{equation}
In general, if the series coefficients $\{R_1, R_2, R_3, \cdots, R_N \}$ are known then the Pad\'e approximant for this series is denoted by $\widetilde{W}^{[N-M|M]}$, and given as:
\begin{equation}
	\widetilde{W}^{[N-M|M]} = \frac{1+A_1 x+A_2 x^2+\cdots+A_{N-M} x^{N-M}}{1+B_1 x+\cdots+B_M x^{M}}\, ,
\end{equation}
can be used to estimate $R_{N+1}$. For example, if only the NLO term of Eq.~\eqref{eq:Wxl2}, $R_1$, is known then the next coefficient, $R_2$, is estimated from $\widetilde{W}^{[0|1]}$ by Taylor expanding it for small $x$ as
\begin{equation}
	\widetilde{W}^{[0|1]} = \frac{1}{1-R_1 x} = 1 + R_1 x + R_1^2 x^2 + \order{x^3}\, ,
\end{equation}
{\emph i.e.,} $R_1^2$ is the Pad\'e approximant prediction for $R_2$.\par
For the static energy, the coefficients $R_1, R_2, R_3$ are known and can be read off by comparing the $x^i$-th terms in Eq.~\eqref{E0exp} with Eq.~\eqref{eq:Wxl2}. To predict the unknown four-loop coefficient $R_4$ in Eq.~\eqref{eq:Wxl2}, we use the approximant $\widetilde{W}^{[2|1]}$ and its series expansion, for small $x$, is given by:
\begin{align}\label{eq:W21}
	\widetilde{W}^{[2|1]} &= \frac{1 + A_1x + A_2 x^2}{1 - B_1 x} \nonumber\\&\simeq1+x (A_1+B_1)+x^2 \left(A_1 B_1+A_2+B_1^2\right)+x^3 \left( A_1 B_1^2+A_2 B_1+B_1^3\right)\nonumber \\ &\bs+x^4 (A_1 B_1^3+A_2 B_1^2+B_1^4)+O\left(x^5\right)\, .
\end{align}
~Now, first we solve for $A_1$, $A_2$ and $B_1$ in terms of known $R_1,R_2$ and $R_3$ of Eq.~\eqref{eq:Wxl2} and the solutions are:
\begin{align}
	A_1&= \frac{R_1 R_2-R_3}{R_2}\, ,\\
	A_2&= \frac{R_2^2-R_1 R_3}{R_2}\, ,\\
	B_1&= \frac{R_3}{R_2}\, .
\end{align}
In the next step, the coefficient of $x^4$ in Eq.~\eqref{eq:W21} is taken as the prediction for $R_4$ which, in terms of the lower order $R_i$'s, can be written as:
\begin{equation}
	R_4^{\text{pred}}=\frac{R_3^2}{R_2}\, .
\end{equation}
In large $L$ limit, we get the the following form for $R_4^{\text{pred}}$ compatible with Eq.~\eqref{eq:Rs} as:
\begin{equation}
	R_4^{\text{pred}} = T_{4,0,0}^{\text{Pad\'e}} + T_{4,1,0}^{\text{Pad\'e}} L + T_{4,2,0}^{\text{Pad\'e}} L^2 + T_{4,3,0}^{\text{Pad\'e}} L^3 + T_{4,4,0}^{\text{Pad\'e}} L^4 \, .
\end{equation}
These predictions can be further improved by knowing the asymptotic behavior of the series, and this extra piece of information is taken as input to get a more precise approximation. This improvement is termed APAP and can be found in Ref.~\cite{Ellis:1996zn}.

The error associated with such approximation in asymptotic limit~\cite{Ellis:1996zn,Chishtie:2001mf} is given by:
\begin{equation}
	\delta_{[N/M]}\approx - \frac{N^M A^{M}}{D^{M}}\, ,
\end{equation}
where $D= N+M(1+a_p)+b_p$ and $A, a_p, b_p$ are fitting parameters. We get APAP results in terms of $R_4^{\text{pred}}$ by:
\begin{equation}
	R_4^{\rm APAP} =\frac{R_4^{\text{pred}}}{(1+\delta_{[N/M]})}\, .
\end{equation}
Repeating the procedure for known lower order $R_i's$, we can fix the constants $A, a_p$, and $b_p$ for a fixed $M$.  It is worth mentioning that among the different choices of Pad\'e approximant for APAP, for a given order, $M=1$ and $a_p=0,~b_p=0$ gives the best result compatible with RG for the static energy, and hence this particular choice is used in this chapter.

Following the procedure explained above, we get the four-loop Pad\'e prediction in the large $L$ limit  as:
\begin{align}\label{ps0}
	T_{4,0,0}^{\text{Pad\'e}}\nonumber  =& \frac{13 \beta_1 T_{1,0,0}^3}{\beta_0}+\frac{169}{9}\frac{ \beta_1^2 T_{1,0,0}^2}{ \beta_0^2}-\frac{62 }{9 }\frac{\beta_1 \beta_2 T_{1,0,0}}{\beta_0^2} +\frac{43}{9}\frac{\beta_1 \beta_2}{\beta_0}+\frac{313}{27}\frac{\beta_1^3 T_{1,0,0}}{ \beta_0^3}-\frac{55}{9}\frac{ \beta_1^2 \beta_2}{ \beta_0^3}\nonumber \\ &-\frac{160}{9}\frac{ \beta_1 T_{2,0,0} T_{1,0,0}}{ \beta_0}-\frac{62}{9}\frac{ \beta_2 T_{1,0,0}^{2}}{ \beta_0 }+\frac{43}{9}\frac{\beta_1 T_{3,0,0}}{\beta_0} + \frac{62}{9}\frac{ \beta_2 T_{2,0,0}}{ \beta_0} +\frac{19}{9}\frac{\beta_2^2}{\beta_0^2} +\frac{46 }{9} T_{2,0,0}^2\nonumber \\ &- \frac{571}{36}\frac{ \beta_1^2 T_{2,0,0}}{ \beta_0^2}+\frac{52}{9} T_{1,0,0}^4 -\frac{113 }{9}T_{2,0,0} T_{1,0,0}^2+ \frac{1429}{324}\frac{\beta_1^4}{\beta_0^4}+\frac{8 }{3} T_{3,0,0} T_{1,0,0}\, ,\\
	T_{4,1,0}^{\text{Pad\'e}} =& \frac{8}{3} \beta_0 T_{3,0,0}+\frac{61}{9} \beta_1 T_{2,0,0}-2 \beta_0 T_{1,0,0}^3+\frac{8}{3} \beta_2 T_{1,0,0}-\frac{461}{108 }\frac{ \beta_1^3}{\beta_0^2} -\frac{34}{9} \beta_1 T_{1,0,0}^2 +\frac{43}{9}\frac{\beta_1 \beta_2}{\beta_0}\nonumber \\ &+\frac{10}{3} \beta_0 T_{2,0,0} T_{1,0,0}-\frac{43}{18}\frac{\beta_1^2 T_{1,0,0}}{ \beta_0}\, ,\\
	T_{4,2,0}^{\text{Pad\'e}} =& \frac{17}{3} \beta_0^2 T_{2,0,0}+\frac{22}{3} \beta_1 \beta_0 T_{1,0,0}+\frac{8}{3} \beta_2 \beta_0+\frac{79}{36} \beta_1^2+\frac{1}{3} \beta_0^2 T_{1,0,0}^2 \, ,\\
	T_{4,3,0}^{\text{Pad\'e}}= &4 T_{1,0,0} \beta_0^3+\frac{13}{3} \beta_0^2 \beta_1, \\ T_{4,4,0}^{\text{Pad\'e}} =&\beta_0^4\, .
\end{align}

We can see that for $T_{4,4,0}$ and $T_{4,4,3}$, the predictions from Pad\'e and the renormalization group are in perfect agreement. For the other RG-accessible coefficient $T_{4,1,0}$ and $T_{4,2,0}$, the predictions are different. However, the numerical difference for $T_{4,2,0}$ is not more than 2.2\% for active quark flavors $n_f\le 6$. However, $T_{4,1,0}$ has larger deviations $\ge2\%$ for $n_f>2$ from RGE prediction. For this reason, we will restrict our discussion in the next sections only to two active flavors for the four-loop. The Pad\'e prediction for the unknown constant term at the four-loop order for the static potential can be obtained by setting $T_{i,j,k}\xrightarrow{k>0}0$ and $\delta T^{\rm us}_{i,0,0}\rightarrow 0$. \par	
Interestingly in the large $n_f$ limits both the Pad\'e approximant and solutions of RGE for RG-accessible coefficients $T_{4,i,0}$ give the same values
\begin{align}
	T_{4,1,0}^{\text{Pad\'e}} &\xrightarrow{n_f\rightarrow\infty}\frac{125}{8748}n_f ^4\, , \hfill\quad &T_{4,1,0}^{\text{RG}}&\xrightarrow{n_f\rightarrow\infty}\frac{125}{8748}n_f^4\, ,\nonumber\\
	T_{4,2,0}^{\text{Pad\'e}}&\xrightarrow{n_f\rightarrow\infty}\frac{25 }{1944}n_f^4\, , \quad &T_{4,2,0}^{\text{RG}}&\xrightarrow{n_f\rightarrow\infty}\frac{25}{1944}n_f^4\, ,\nonumber\\
	T_{4,3,0}^{\text{Pad\'e}}&\xrightarrow{n_f\rightarrow\infty}\frac{5}{972} n_f^4\, , \quad &T_{4,3,0}^{\text{RG}}&\xrightarrow{n_f\rightarrow\infty}\frac{5}{972}n_f^4\, ,\nonumber\\
	T_{4,4,0}^{\text{Pad\'e}}&\xrightarrow{n_f\rightarrow\infty}\frac{1}{1296}n_f^4\, ,\quad &T_{4,4,0}^{\text{RG}}&\xrightarrow{n_f\rightarrow\infty}\frac{1}{1296}n_f^4\nonumber\, .
\end{align}
For RG-inaccessible coefficient $$T_{4,0,0}^{\text{Pad\'e}}\xrightarrow{n_f\rightarrow\infty}\left(\frac{-5 T_f n_f}{9}\right)^4.$$
In fact, a similar pattern has been observed for known lower orders: $$T_{i,0,0}^{\text{Pad\'e}}\xrightarrow{n_f\rightarrow\infty}\left(\frac{-5 T_f n_f}{9}\right)^i.$$ ~Now we have an estimate for $T_{4,0,0}$, but the truncated perturbation series also suffers from scale dependence. The scale sensitivity of the perturbative series can be minimized using RG-summation of running logarithms, and the procedure is discussed in the next section. 
\section{RG Improvement in the momentum space}\label{sec:RG_sum}
The issue with the perturbative series in the QCD is to account for the RG running of all the parameters. The RGSPT, discussed in chapter~\eqref{Chapter2}, accounts for the RG running by summation of all the RG-accessible logarithms. The RG-accessible logarithms at each order in the perturbation theory are defined as the leading and the next-to-leading logarithms that can be accessed through the processes-dependent RGE. The RG summation becomes interesting from the three-loop order due to the presence of the ultrasoft terms, and this issue is discussed in detail for static energy in this chapter. 

The perturbative series has the following form: 
\begin{align}
	&W(x,L)=\sum _{i=0}^n \sum _{j=0}^i \sum _{k=0}^{\tiny{\substack{(i-j-2)\\ \times \theta (i-j-3)}}} x^{i+1} L^j \log ^k(x) T_{i,j,k}
	\label{eq:eq1T2}\, ,
\end{align}
where the series coefficients are $T_{i,j,k}$. To obtain RG-summed perturbation series which we call $W^{(n)}_{\rm RG\Sigma}$, we rewrite $W(x,L)$ as
\begin{align}
	&W^{(n)}_{\rm RG\Sigma} = \sum _{i=0}^n \sum _{k=0}^{(i-2) \theta (i-3)} x^{i+1} \log ^k(x) S_{i,k}(x L)\, ,
	\label{eq:RGsummed}
\end{align}
where intermediate quantities $S_{i,k}(x L)$ are the resummed series obtained by summing terms:
\begin{equation}\label{eq:S}
	\sum _{n=i}^{\infty} (x~L )^{n-i} T_{n,n-i,k}\, .
\end{equation}
We substitute Eq.~\eqref{eq:eq1T2} in Eq.~\eqref{RG_W}, which leads to a recursion relation between the series coefficients. We multiply the recursion relation with $(x L)^{k-1}$ with appropriate $k$ and sum it from $n=k$ to infinity, which following Eq.~\eqref{eq:S}, give differential equations for $S_{i,k}(x L)$. The solution to these differential equations results in the closed form expression for $S_{i,k}(x L)$.\par
The RG-summed solutions $S_{i,0}(x L)$ are calculated to the two-loop order in Ref.~\cite{Ahmady:2002fd} and are given below:
\begin{align}
	S_0=&\frac{1}{w}\,,\quad S_1=w^{-2}\left(T_{1,0,0}-\tilde{B}_1 L_w\right)\,,\label{eq:S0}\\ 
	S_2=&w^{-2}\left(\tilde{B}_1^2-\tilde{B}_2\right)+w^{-3}\bigg[T_{2,0,0}-\tilde{B}_1^2+\tilde{B}_2-\tilde{B}_1 L_w \left(\tilde{B}_1+2 T_{1,0,0}\right)+\tilde{B}_1^2 L_w^2\bigg]\,,\label{eq:S2}
\end{align} 
where $w=(1-\beta_{0} u)$ and $\tilde{B}_i=\beta_i/\beta_0$ and $L_w\equiv\log(w)$. The RG-summation for $e^+ e^-$ process to three-loop order is also discussed in Ref.~\cite{Ahmady:2002pa}, which is a special case of the results of this chapter in the limit where ultrasoft coefficients are taken zero. \par
The static energy requires a new series representation, compatible with the RGE, to incorporate the ultrasoft logarithms. These logarithms form a separate recurrence relation among the coefficients. The RG-summation of ultrasoft terms at the three-loop order is obtained by collecting the coefficients of $x^n L^{n-4}\log(x)$ in Eq.~\eqref{eq:rg_W}, which results in the following recurrence relation among the coefficients:
\begin{align}
	(n-3) T_{n,n-3,1}-n \beta_0 T_{n-1,n-3,1}=0\, .
\end{align}
Collecting $x^n L^{n-4}$ terms, we get the following recurrence relation:
\begin{align}
	&(n-3)T_{n,n-3,0}-n\beta_{0} T_{n-1,n-4,0} -\beta_0T_{n-1,n-4,1}\nonumber\\&-(n-1)\beta_1T_{n-2,n-4,0} -(n-2)\beta_2 T_{n-3,n-4,0}\nonumber\\&-(n-3) \beta_3T_{n-4,n-4,0}=0\, .
\end{align}
Notice that presence of the $\beta_0T_{n-1,n-4,1}$ terms in the above equation is new and differs from $e^+ e^-$ case in Ref.~\cite{Ahmady:2002pa}.

For the four-loop order, coefficients of $x^n L^{n-5}\log^2(x)$ terms give the following recurrence relation for ultrasoft terms:
\begin{equation}
	(n-4)T_{n,n-4,2}-n \beta _0 T_{n-1,n-5,2}=0\, .
\end{equation}
Collecting $x^n L^{n-5}\log(x)$ terms, we get the following recurrence relation:
\begin{align}
	(n-4)&T_{n,n-4,1}-(n-1) \beta _1 T_{n-2,n-5,1}-n \beta _0 T_{n-1,n-5,1}-2 \beta _0 T_{n-1,n-5,2}=0\, .
\end{align}
Collecting $x^n L^{n-5}$ terms, we get the following recurrence relation:
\begin{align}
	&(n-4)T_{n,n-4,0}-(n-4)\beta _4 T_{n-5,n-5,0}-(n-3)\times \beta _3 T_{n-4,n-5,0}-(n-2) \beta _2 T_{n-3,n-5,0} \nonumber\\&-(n-1) \beta _1 T_{n-2,n-5,0}-\beta _1 T_{n-2,n-5,1}-n \beta _0 T_{n-1,n-5,0}\nonumber\\&-\beta _0 T_{n-1,n-5,1}=0\, ,
\end{align}
multiplying $u^{k-1}$, where $k=n-3$ for the three-loop and $k=n-4$ for the four-loop, to the recurrence relations and then summing $n$ from $k$ to $\infty$ gives separate differential equations for the above recurrence relations. All these recursion relations can be written in a general-differential equation for the static energy as:
\begin{align}
	\sum _{i=0}^n &\Big(\theta (i-k,-k)-\theta (i-k-2,k-1)\Big)\times \Bigg( (\delta _{i,n}-u \beta (n-i))\frac{d}{d u}S_{i,k}(u)\nonumber \\&-(i+1) \beta (n-i) S_{i,k}(u)-(k+1) \theta (i-k-3) \beta (n-i) S_{i,k+1}(u)\Bigg)=0\, .
\end{align}
The solution to different RG-summed series, to the order we are interested in, are given by:
\begin{align}
	&S_{3,1}(w)=\frac{T_{3,0,1}}{w^4}, \quad S_{4,2}(w)=\frac{T_{4,0,2}}{w^5}\, ,\label{eq:S42}\\
	S_{4,1}(w)&=\frac{1}{w^5}\left(-4 \tilde{B}_1 T_{3,0,1} L_w-2 T_{4,0,2} L_w+T_{4,0,1}\right)\,.
\end{align}
Similarly, other solutions are:
\begin{align}
	S_{3,0}(w)&=\frac{1}{w^2}\left(-\frac{\tilde{B}_1^3}{2}+\tilde{B}_2 \tilde{B}_1-\frac{\tilde{B}_3}{2}\right)+\frac{1}{w^4}\left(T_{3,0,0}-2 \tilde{B}_1^2 T_{1,0,0}+2 \tilde{B}_2 T_{1,0,0}-\frac{\tilde{B}_1^3}{2}+\frac{\tilde{B}_3}{2}\right)\nonumber\\&+\frac{L_w}{w^3}\left(2 \tilde{B}_1 \tilde{B}_2-2 \tilde{B}_1^3\right) +\frac{L_w}{w^4} \left(-2 \tilde{B}_1^2 T_{1,0,0}-3 \tilde{B}_1 T_{2,0,0}+2 \tilde{B}_1^3-3 \tilde{B}_2 \tilde{B}_1-T_{3,0,1}\right)\nonumber\\&+\frac{L_w^2}{w^4} \left(3 \tilde{B}_1^2 T_{1,0,0}+\frac{5 \tilde{B}_1^3}{2}\right)-\frac{\tilde{B}_1^3 L_w^3}{w^4} +\frac{1}{w^3}\bigg(2 \tilde{B}_1^2 T_{1,0,0}-2 \tilde{B}_2 T_{1,0,0}+\tilde{B}_1^3-\tilde{B}_2 \tilde{B}_1\bigg) \, ,
\end{align}

	\begin{align}
		S_{4,0}(w)&=\frac{1}{w^5}\bigg[T_{4,0,0}-\tilde{B}_1^3 T_{1,0,0}-3 \tilde{B}_1^2 T_{2,0,0}+\tilde{B}_3 T_{1,0,0}+3 \tilde{B}_2 T_{2,0,0}+\frac{7 \tilde{B}_1^4}{6}-3 \tilde{B}_2 \tilde{B}_1^2\nonumber\\&-\frac{1}{6} \tilde{B}_3 \tilde{B}_1+\frac{5 \tilde{B}_2^2}{3}+\frac{\tilde{B}_4}{3}\bigg]+\frac{L_w}{w^5}\Big[6 \tilde{B}_1^3 T_{1,0,0}-3 \tilde{B}_1^2 T_{2,0,0}-8 \tilde{B}_2 \tilde{B}_1 T_{1,0,0}-4 \tilde{B}_1 T_{3,0,0}\nonumber\\&-\tilde{B}_1 T_{3,0,1}+4 \tilde{B}_1^4-3 \tilde{B}_2 \tilde{B}_1^2-2 \tilde{B}_3 \tilde{B}_1-T_{4,0,1}\Big]+ \frac{L_w^2}{w^5}\Big[7 \tilde{B}_1^3 T_{1,0,0}+6 \tilde{B}_1^2 T_{2,0,0}\nonumber\\&+4 \tilde{B}_1 T_{3,0,1}-\frac{1}{2} 3 \tilde{B}_1^4+6 \tilde{B}_2 \tilde{B}_1^2+T_{4,0,2}\Big]+\frac{L_w^3}{w^5}\left(-4 \tilde{B}_1^3 T_{1,0,0}-\frac{1}{3} 13 \tilde{B}_1^4\right)+\frac{\tilde{B}_1^4 L_w^4}{w^5}\nonumber\\&+\frac{1}{w^4}\Big[2 \tilde{B}_1^3 T_{1,0,0}+3 \tilde{B}_1^2 T_{2,0,0}-2 \tilde{B}_2 \tilde{B}_1 T_{1,0,0}-3 \tilde{B}_2 T_{2,0,0}-2 \tilde{B}_1^4+5 \tilde{B}_2 \tilde{B}_1^2-3 \tilde{B}_2^2\Big]\nonumber\\&+\frac{L_w^2}{w^4}\left(3 \tilde{B}_1^4-3 \tilde{B}_1^2 \tilde{B}_2\right)+\frac{L_w}{w^4}\left(-6 \tilde{B}_1^3 T_{1,0,0}+6 \tilde{B}_2 \tilde{B}_1 T_{1,0,0}-5 \tilde{B}_1^4+5 \tilde{B}_2 \tilde{B}_1^2\right)\nonumber\\&+\frac{1}{w^2}\left(\frac{\tilde{B}_1^4}{3}-\tilde{B}_2 \tilde{B}_1^2+\frac{2}{3} \tilde{B}_3 \tilde{B}_1+\frac{\tilde{B}_2^2}{3}-\frac{\tilde{B}_4}{3}\right)+\frac{L_w}{w^3}\left(\tilde{B}_1^4-2 \tilde{B}_2 \tilde{B}_1^2+\tilde{B}_3 \tilde{B}_1\right)\nonumber\\&+\frac{1}{w^3}\left(-\tilde{B}_1^3 T_{1,0,0}+2 \tilde{B}_2 \tilde{B}_1 T_{1,0,0}-\tilde{B}_3 T_{1,0,0}+\frac{\tilde{B}_1^4}{2}-\tilde{B}_2 \tilde{B}_1^2-\frac{1}{2} \tilde{B}_3 \tilde{B}_1+\tilde{B}_2^2\right)\, .\label{eq:S40}
	\end{align}

\begin{figure}[ht]
	\centering
	\includegraphics[width=0.45\textwidth]{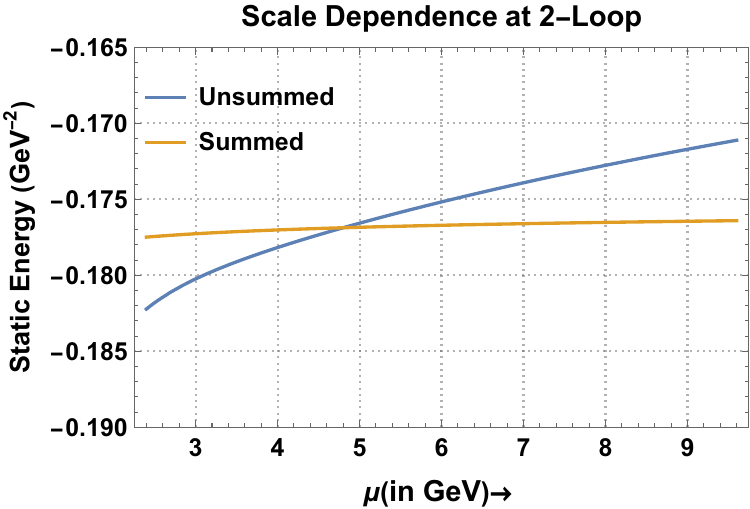}
	\includegraphics[width=0.45\textwidth]{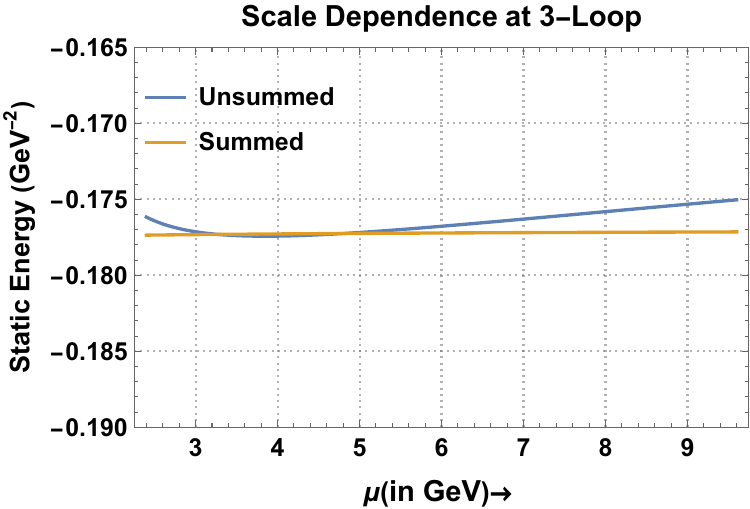}
	\includegraphics[width=0.45\textwidth]{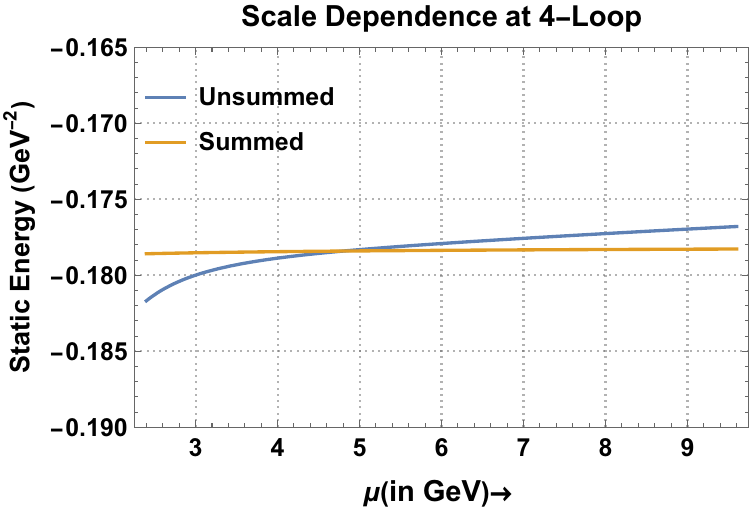}
	\caption{Renormalization scale dependence of the resummed and the unsummed static energy at different loops.}
	\label{fig:234_summed}
\end{figure}

The importance of the summation of all accessible logarithms can be seen in Fig.~\eqref{fig:234_summed}. The scale dependence of the RG-summed series Eq.~\eqref{eq:RGsummed} around momentum $p =\space m^{\overline{\rm MS}}_b=4.17\GeV$ is almost negligible in the range $m^{\overline{\rm MS}}_b/2 \le \mu \le 2 m^{\overline{\rm MS}}_b $ while the unsummed series in Eq.~\eqref{E0exp} has significant $\mu$ dependence. The scale sensitivity of unsummed series decreases order by order, but the advantage of RG-summed series provides results less sensitive to scale with the same available information. This theoretical improvement provides us an opportunity to extract various parameters from available experimental data with less scale sensitivity.

\section{Restricted Fourier transform \label{sec:RFT}}	
Let us begin by noting that the ultrasoft part of the static energy is calculated in position space, whereas the perturbative part is carried out in momentum space. To make any phenomenological study, it is necessary to bring all the contributions of static energy into the same space. Bringing the momentum space results in position space via a Fourier transform may seem natural, but such a transformation induces pathological contributions to the static energy. These undesirable contributions originate from the non-perturbative, small momentum modes, which must be removed explicitly. After removing such contributions, the convergence behavior of the static energy improves drastically. This scheme was first addressed in Ref.~\cite{Karbstein:2013zxa} and termed the Restricted Fourier transform (RFT) and it is related to the potential subtracted scheme~\cite{Beneke:1998rk}. This scheme has already been discussed in detail for the static energy to the three-loop. Here we discuss it very briefly and provide corresponding results for the four-loop order.

Note that the ultrasoft terms are already known to the four-loop order. Therefore, what we are providing is the complete calculation of the static energy in the position and momentum space to the four-loop order. The final result for the uncontrolled contribution to the position space static energy in the RFT scheme is given in appendix~\eqref{app:resVr}, but an overview of the calculation is given here.

The position space version of the static potential, $V(r,\mu,\mu_{us})$, from momentum space potential, say $\tilde{V}(p,\mu,\mu_{us})$ to distinguish between the two basis, in RFT scheme is defined by:
\begin{equation} 
	V\left(r,\mu,\mu_{us};\mu_{f}\right) =\int_{\left[\mathbf{p} |>\mu_{f}\right.} \frac{\mathrm{d}^{3} p}{(2 \pi)^{3}} \mathrm{e}^{i \mathbf{p} \cdot \mathbf{r}} \tilde{V}(p,\mu,\mu_{us}=V(r,\mu,\mu_{us}-\delta V\left(r, \mu_{us},\mu_{f}\right)\, ,
\end{equation}
where $\mu_{f}$ is a perturbative scale chosen such that $\mu>\mu_f\gg \Lambda^{\overline{\textrm{MS}}}_{\textrm{QCD}}$ and uncontrolled terms, $\delta V\left(r,\mu,\mu_{us};\mu_{f}\right)$, in the potential is
\begin{equation}
	\delta V\left(r, \mu,\mu_{us};\mu_{f}\right)=\int_{\left[\mathbf{p} |<\mu_{f}\right.} \frac{\mathrm{d}^{3} p}{(2 \pi)^{3}} \mathrm{e}^{i \mathbf{p} \cdot \mathbf{r}} \tilde{V}(p,\mu,\mu_{us};\mu_f)\, .
\end{equation}
The static energy, given by Eq.~\eqref{E0}, in RFT scheme is:
\begin{equation}
	E_0\left(r, \mu;\mu_{f}\right)=V_s\left(r,\mu, \mu_{us};\mu_{f}\right)+\delta^{\rm us}\left(r,\mu, \mu_{us};\mu_{f}\right)\,.
\end{equation}
The ultrasoft gluonic contribution to the static energy, at order $r^2$ in multipole expansion, is given by: 
\begin{align}
	&\delta^{\rm us}(r,\mu,\mu_{us})=-i\frac{g^{2}}{N_{c}} T_{F} V_{A}^{2} \frac{r^{2}}{d-1} \int_{0}^{\infty} d t \text{ } e^{-i t\left(V_{o}-V_{s}\right)} \times\left\langle 0\left|\mathbf{E}^{a}(t) \phi(t, 0)_{a b}^{\operatorname{adj}} \mathbf{E}^{b}(0)\right| 0\right\rangle\, ,
\end{align}
where $T_F=1/2$, $N_c$ is the number of colors, $\phi(t, 0)_{a b}^{\operatorname{adj}} $ is Wilson line in the adjoint representation connecting two points at temporal separation $t$, $ \mathbf{E}^{a/b}$ is choromoelectric field strength, and $V_A$ is matching coefficient which appears at order $r^2$ in multipole expansion. This quantity has been calculated to NLO in Refs.~\cite{Brambilla:1999qa,Brambilla:2006wp} and final result, sub-leading in $r$, is given by:
\begin{align}
	\delta^{\mathrm{\rm us}}\left(r,\mu_{us}\right)=&-C_{F} \frac{\alpha_{s}(\mu_{us})}{\pi} \frac{r^{2}}{3} V_{A}^{2}(V_{o}(r)-V_{s}(r))^{3}\times\left(\delta^{\rm us}_{3-loop}+ \frac{\alphas(\mu_{us})}{\pi}\delta^{\rm us}_{4-loop}\right)\, ,
	\label{Vus4}
\end{align}
where, 
\begin{align}
	&\delta^{\rm us}_{3-loop}= 2 \log \left(\frac{V_{o}(r)-V_{s}(r)}{\mu_{\mathrm{us}}}\right)-\frac{5}{3}+2 \log 2\, ,
	\shortintertext{and}
	&\delta^{\rm us}_{4-loop}=  C_{1} \log ^{2} \left(\frac{V_{o}(r)-V_{s}(r)}{\mu_{us}}\right)+C_{2} \log \left(\frac{V_{o}(r)-V_{s}(r)}{\mu_{us}}\right)+D\, .
	\label{Vus2}
\end{align}
Coefficients $C_1$, $C_2$ and $D$ are given in Ref.~\cite{Brambilla:2006wp} and we have presented them in appendix~\eqref{app:loop_coef}.
Next to the leading order expression for $V_{o}(r)-V_{s}(r)$ is given by:
\begin{equation}
	\left(V_{o}-V_{s}\right)(r) =\frac{C_A \alpha _s(\mu )}{2 r}\bigg(1+\frac{\alpha _s(\mu )}{\pi } \left(T_{1,0,0}+(2 \gamma_E + \log(\mu ^2 r^2))\beta_0\right)\bigg)+\order{\alpha_s^3(\mu)}\, ,
	\label{dVNLO}
\end{equation}
and we can see that if we substitute Eq.~\eqref{dVNLO} to Eq.~\eqref{Vus4} then ultrasoft contributions to the three- and four-loop order can be obtained from NLO results.
\par Due to the presence of the extra scale in the problem $\mu_{\mathrm{us}}$, it is desirable to expand all the quantities in terms of $\alpha_{s}(\mu)$. This expansion induces mixed logarithms at the four-loop of form $\frac{1}{r}\log(\mu^2_{us} r^2)\log(\mu^2 r^2)$ in the ultrasoft contributions. To simplify the calculation we can write:
\begin{equation}
	\alpha_s(\mu_{us})=\alpha_s(\mu)\left(1+\frac{\alpha_s(\mu)}{\pi} \beta_0 \log \left(\frac{\mu^2}{\mu^2_{us}}\right)\right)+\order{\alpha_s^3(\mu)}\, ,
\end{equation}
and,
\begin{equation}
	\log \left(r^2 \mu _{us}^2\right)=\log \left(\mu ^2 r^2\right)-\log \left(\frac{\mu ^2}{\mu _{us}^2}\right)\, ,
\end{equation}
so that the ultrasoft scale will appear only in running logarithms.\par 
The uncontrolled term for $\left(V_{o}-V_{s}\right)(r)$ in RFT scheme, in next to the leading order is given by:
\begin{equation}
	\delta\left(V_{o}-V_{s}\right)(r,\mu_f)=C_A \mu_f x \Bigg[H_1+x \Bigg(H_1 T_{1,0,0}+\beta _0 \big(2 H_2+H_1 L_{\mu_f}\big)\Bigg)\Bigg]+\order{x^3}, 
\end{equation}
where $L_{\mu_f}=\log(\frac{\mu^2}{\mu^2_f})$ and $H_i's$ functions  defined by Eq.~\eqref{hyp} and are proportional to generalized hypergeometric function. The only matching coefficient left is $V_A$ and it is taken as $$V_A=\frac{C_A \alpha_{s}}{2 r}\frac{1}{V_o-V_s}+\order{\alpha_{s}},$$ same as calculated in Ref.~\cite{Karbstein:2013zxa}. These contributions will enter in Eq.~\eqref{Vus4} and the final expression is given in appendix~\eqref{app:resVr}.\par
Pad\'e estimates and the ultrasoft contributions give us numerical estimate for $T_{4,0,0}=T_{4,0,0}^{\text{Pad\'e}}+\delta T_{4,0,0}^{\text{\rm us}}$. Now, the restricted version for the static potential, the ultrasoft contributions, and the static energy in position space can be constructed using Eq.~\eqref{FFT} and Eq.~\eqref{UnFT}.\par
\begin{figure}[ht]
	\centering
	\includegraphics[width=.7\linewidth]{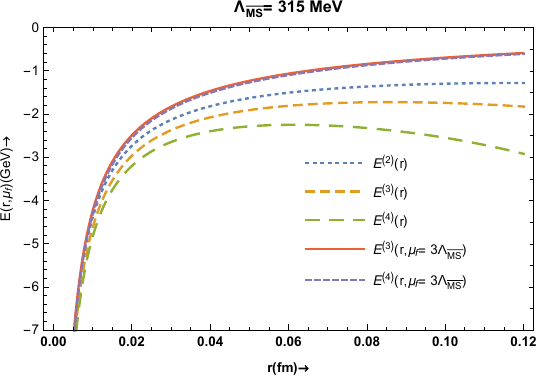}
	\caption{Restricted and unrestricted static energy to a given order (in superscript). $\Lambda^{\overline{\textrm{MS}}}_{\textrm{QCD}}=315\MeV$ is used as input.}
	\label{fig:restE}
\end{figure}

 
Using $\Lambda^{\overline{\textrm{MS}}}_{\textrm{QCD}}=315 \MeV$, we can see in Fig.~\eqref{fig:restE} that the four-loop corrections to the static energy in RFT scheme makes a very small contribution but, without removing pathological contributions, behavior of the same quantity in unrestricted scheme has very bad for $r>0.05~{\rm fm}$ at the four-loop.
\par
In the next section, we have performed the extraction of $\Lambda^{\overline{\textrm{MS}}}_{\textrm{QCD}}$ in momentum space using LQCD inputs from Ref.~\cite{Karbstein:2018mzo}.

\section{Fitting Perturbative results to lattice data \label{sec:lat_input}}
In this section, we fit the perturbative static energy to the lattice data and extract the value of the $\Lambda^{\overline{\textrm{MS}}}_{\textrm{QCD}}$ from RG-summed and unsummed case. Due to the absence of all order results, the extracted quantity $\Lambda^{\overline{\textrm{MS}}}_{\textrm{QCD}}$ depends on the choice of renormalization scale. To reduce the scale dependence, we exploit the RGSPT for static energy to extract $\Lambda^{\overline{\textrm{MS}}}_{\textrm{QCD}}$.
The parameterization of lattice data to the Cornell potential is given in position space in Ref.~\cite{Karbstein:2018mzo} for two-flavor QCD 
\begin{equation}
	E_{\rm lat}(r)=V_0-\frac{\alpha}{r}+\sigma r,
	\label{lp}
\end{equation}
where $\alpha =0.326\pm0.005$ and $\sigma= 7.52\pm0.55 {\rm fm}^{-2}$ are constants for $n_f=2$ flavor. The parameter $V_0$ is a potential offset needed to match the perturbative static energy. However, $V_0$ is not needed in the case of momentum space analysis\cite{Karbstein:2018mzo}. It is important to note that these parameters are correlated and have $cor(\alpha,\sigma )=-0.17$, which has to be taken into account in sampling from normal distribution.\par 
The Fourier transform of lattice static energy, in Eq.~\eqref{lp}, to momentum space is given by:
\begin{equation}
	E_{\rm lat}(p)=-\frac{4 \pi \alpha}{p^2}-\frac{8\pi \sigma}{p^4}\, .
	\label{lpm}
\end{equation}
To extract $\Lambda^{\overline{\textrm{MS}}}_{\textrm{QCD}}$, we minimize the square-deviation~\cite{Karbstein:2018mzo}:
\begin{equation}
	\Delta(\Lambda^{\overline{\textrm{MS}}}_{\textrm{QCD}})=	\int_{p_{min}}^{p_{max}}dp \left(E_{0}\left(p,\mu\right)- E_{\rm lat}\left(p\right) \right)^2 .
\end{equation}
It should be noted that the strategy for sampling is analogous to the one discussed in Ref.~\cite{Karbstein:2018mzo}. The matching region is chosen $1500\MeV\le p\le3000 \MeV$ and momentum values are randomly sampled from a uniform distribution with $p_{min}\in[1500,2250]\MeV$ and $p_{max}\in[2250,3000]\MeV$ such that $p_{max}-p_{min}\ge375\MeV$. It is important to note that five-loop running of $\alpha_{s}$ is used to extract $\Lambda^{\overline{\textrm{MS}}}_{\textrm{QCD}}$.\par

\begin{figure}[ht]
	\centering
	\includegraphics[width=.7\linewidth]{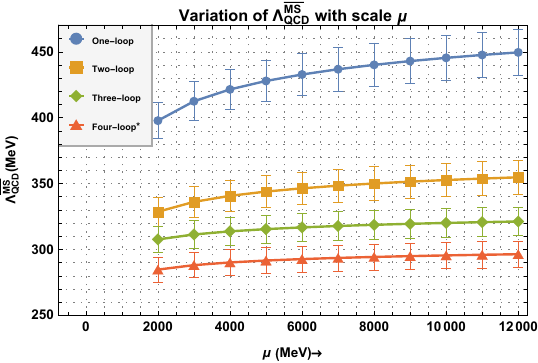}
	\caption{Renormalization scale dependence of $\Lambda^{\overline{\textrm{MS}}}_{\textrm{QCD}}$ at different loop order with error bars. The full Pad\'e estimated value for $T_{4,0,0}$ is used from Eq.~\eqref{pade4}.}
	\label{sc_dep_lam}
\end{figure}

Following the procedure explained above, the scale dependence of the extracted (for 500 samples for $\{p,\alpha,\sigma\}$) $\Lambda^{\overline{\textrm{MS}}}_{\textrm{QCD}}$ can be seen from Fig.~\eqref{sc_dep_lam} at different loop order and at different renormalization scales.
 \par
In Table~\eqref{lambda_3}, we present our determinations of $\Lambda^{\overline{\textrm{MS}}}_{\textrm{QCD}}$ for different orders of the perturbative static energy and for different choices of the renormalization scale. In addition to the reduction in the errors, the central values are closer to each other for different choices of $\mu$ at higher order in the perturbation series.
\begin{table}[ht]
	\begin{center}
		\begin{tabular}{|c|c|c|c|c|}
			\hline
			\multirow{3}{*}{Loop} & \multicolumn{4}{c|}{$\Lambda^{\overline{\textrm{MS}}}_{\textrm{QCD}}(\text{in }\MeV)$} \\
			\cline{2-5}
			&\multicolumn{1}{c|}{unsummed}&\multicolumn{3}{c|}{RG-Summed}\\
			\cline{2-5}
			& $\mu=p$& $\mu=2.25\GeV$& $\mu=4.17\GeV$&$\mu=6.5 \GeV$\\
			\hline			
			1&$398.3\pm13.7$&$402.3\pm13.0$& $422.7\pm15.0$&$434.9\pm15.8$\\
			2&$328.8\pm11.2$&$330.9\pm11.0$&$341.3\pm11.6$&$347.4\pm12.0$\\
			3&$307.3\pm10.9$&$308.4\pm10.6$&$313.7\pm10.9$&$317.0\pm11.1$\\
			\hline
		\end{tabular}
	\end{center}
	\caption{\label{lambda_3}$\Lambda^{\overline{\textrm{MS}}}_{\textrm{QCD}}$ at different loop-orders}
\end{table}\par
At the four-loop order (for 2000 samples of $\alpha,\sigma,p$), we make the following three choices for the unknown coefficient $T_{4,0,0}$: 
\begin{equation}
	T_{4,0,0}=\{T_{4,0,0}^{\text{Pad{\' e}}}+\delta T_{4,0,0}^{us},\frac{T_{4,0,0}^{\text{Pad{\' e}}}}{2}+\delta T_{4,0,0}^{us},\frac{T_{4,0,0}^{\text{Pad{\' e}}}}{4}+\delta T_{4,0,0}^{us}\}\,,
	\label{pade4}
\end{equation}
and the corresponding results are presented in Table~\eqref{lambda_4}.
\begin{table}[h]
	\begin{center}
		\begin{tabular}{|c|c|c|c|c|}
			\hline
			\multirow{3}{*}{$T_{4,0,0}$} & \multicolumn{4}{c|}{$\Lambda^{\overline{\textrm{MS}}}_{\textrm{QCD}}(\text{in }\MeV)$} \\
			\cline{2-5}
			&\multicolumn{1}{c|}{unsummed}&\multicolumn{3}{c|}{RG-Summed}\\
			\cline{2-5}
			& $\mu=p$& $\mu=2.25\GeV$&$\mu=4.17\GeV$&$\mu=6.5 \GeV$\\			
			\hline			
			I&$284.6\pm10.0$&$285.6\pm9.6$&$290.1\pm9.7$&$292.7\pm9.8$\\ 
			II&$296.3\pm10.3$&$297.0\pm10.0$&$300.0\pm10.1$&$302.0\pm10.2$\\
			III&$302.9\pm10.4$&$303.4\pm10.3$&$305.5\pm10.3$&$307.0\pm10.4$\\ 
			\hline
		\end{tabular}
	\end{center}
	\caption{\label{lambda_4}$\Lambda^{\overline{\textrm{MS}}}_{\textrm{QCD}}$ at different loop-order for choices of $T_{4,0,0}$ according to Eq.~\eqref{pade4}).}
\end{table}

The RG-summed and unsummed series give very similar values of $\Lambda^{\overline{\textrm{MS}}}_{\textrm{QCD}}$ in case of the renormalization scale is chosen in the middle of the matching region. However, the RG-summed static energy provides a better fit to the lattice static energy than the unsummed one, which can be seen in Table~\eqref{lambda5} below:
\begin{table}[ht]
	\begin{center}
		\begin{tabular}{|c|c|c|c|}
			\hline
			\multirow{3}{*}{Loop} & \multicolumn{3}{c|}{Percent of cases (in \%) with $\left(\Delta_S(\mu)-\Delta_U\right) <0$} \\
			\cline{2-4}
			& $\mu=2.25\GeV$& $\mu=4.17\GeV$&$\mu=6.5 \GeV$\\
			\hline
			1&85.45& 81.60& 79.70\\
			2&97.10& 95.50& 94.75\\
			3&99.94&99.92& 99.84\\	
			\hline
		\end{tabular}
	\end{center} 
	\caption{\label{lambda5} Percentage of cases where the RG-summed series gives a small square deviation compared to the unsummed one at different loop order.}
\end{table}

The RG-summed static energy gives a better fit, and this improvement persists even at the next higher order. For example, at the four-loop order, we have found that the RG-summed static energy produces a significantly better fit to lattice static energy compared to unsummed perturbative static energy for different cases when we choose $T_{4,0,0}$ according to Eq.~\eqref{pade4}.
\section{Discussions \label{sec:conc}}
The static energy between a heavy quark and an anti-quark is an important quantity in QCD, as it can be calculated in both perturbation theory and in LQCD simulations. The existence of a matching region, where both perturbation theory and lattice agree, provides an opportunity to extract the parameters of the theory. If we assume the light quarks are massless and the heavier one decouples, then the only parameter left in the theory is the strong coupling constant. Precise calculation of the static energy thus becomes a very important goal for precision physics involving strong interactions.

\subsection{Pad\'e estimate}
We have extended the results for static energy to the four-loop order using the renormalization group and Pad\'e approximant. Among the seven coefficients at the four-loop, the estimates for $T_{4,3,0}$ and $T_{4,4,0}$, using APAP, are in exact agreement with the solutions of the renormalization group. Deviation for coefficient $T_{4,2,0}$ is within 2.2\% for $n_f\le6$ but $T_{4,1,0}$ deviates more than 2\% for $n_f>2$. Since the deviation is less than 2\% for RG-accessible coefficient for light flavor, we have used APAP estimate for $T_{4,0,0}$ to extract the $\Lambda^{\overline{\textrm{MS}}}_{\textrm{QCD}}$ to the four-loop order with different choices given in Eq.~\eqref{pade4}. The large-$n_f$ limits for RG-accessible terms from both methods are in perfect agreement. There are also some contribution to $T_{4,0,0}$ from ultrasoft terms which demands this quantity must be Fourier transformed in order to get complete the RG-inaccessible terms in momentum space. 

\subsection{Position space improvement}
The static energy from LQCD simulations is mostly parameterized in position space, and hence the perturbative static energy is Fourier transformed to the position space. This quantity in position space suffers from pathological contributions stemming from the non-perturbative regions and has to be removed explicitly. This is achieved using the Restricted Fourier transform advocated in Ref.~\cite{Karbstein:2013zxa}, which improves the convergence behavior for $r\sim0.12$ fm. The ultrasoft and the static potential have an explicit dependence on another scale $\mu_{\mathrm{\rm us}}$, which is absent in total energy. This scale dependence should also be canceled for the static energy in the RFT scheme. The final expression for uncontrolled contributions to the static energy is provided in appendix~\eqref{app:resVr}. The four-loop contribution to static energy in the RFT scheme has very little effect and can be seen in Fig.~\eqref{fig:restE}. The static energy in the RFT scheme is provided in the section~\eqref{sec:RFT} for the four-loop order and can be used in future studies.
\subsection{RG-improvement of the static energy and \texorpdfstring{$\Lambda^{\overline{\textrm{MS}}}_{\textrm{QCD}}$}{} extraction}
The RG-summed static energy in momentum space is used in this chapter to extract $\Lambda^{\overline{\textrm{MS}}}_{\textrm{QCD}}$ by fitting the static energy from perturbative to the static energy from LQCD from Ref.~\cite{Karbstein:2018mzo} for two active flavors. Its value for the three-loop from the RG-summed static energy is found to be $308.4\pm10.6 (2.25\GeV) \MeV$, $313.7\pm10.9 (4.17\GeV))\MeV$ and $317.0\pm11.1(6.5\GeV)\MeV$ where the quantity in the parenthesis is the renormalization scale. For the unsummed static energy, this parameter is found to be $307.3\pm10.9\MeV$. The RG-summed version of static energy has been observed to provide not only a better fit to the lattice energy but also give less standard deviation if the renormalization scale is chosen in the middle of the matching region. A similar trend also persists for the next order, but we have used less sample size since the exact calculations are not available. Our finding of $\Lambda^{\overline{\textrm{MS}}}_{\textrm{QCD}}$ from RG-summed and unsummed series agrees within error bars to the findings of Ref.~\cite{Karbstein:2018mzo}.

\section{Summary} \label{sec:summary4}
To summarize, the QCD static potential is known to the three-loop order, and the ultrasoft terms which first appear at the three-loop order are known to the four-loop order. In section~\eqref{sec:pert_pot}, we describe the perturbative and the ultrasoft part of the static static energy. The main results of this chapter are the following
\begin{itemize}
	\item In section~\eqref{sec:RG_ser}, using the RGE, we determine the RG-accessible coefficients at four-loop order, which is shown in Eq.~\eqref{RG_sol}.
	\item The constant term of the four-loop coefficient can not be determined using RGE. In  section~\eqref{sec:Pade_est}, we use the Pad\'e approximant method to obtain this term which is given in equation Eq.~\eqref{ps0}. 
	\item In section \eqref{sec:RG_sum}, we apply for the RGSPT to QCD static energy beyond two-loop order to sum up the RG-accessible running logarithms to all orders in perturbation theory. The RG-summed series is defined in equation Eq.~\eqref{eq:RGsummed} and the subsequent quantities are given in Eq.~\eqref{eq:S0}-\eqref{eq:S2}, Eq.~\eqref{eq:S42}-\eqref{eq:S40}. The RG-summed series ensures the expected reduction in sensitivity to the renormalization scale, as shown in Fig.~\eqref{fig:234_summed}.
	\item In section \eqref{sec:RFT}, we use the Restricted Fourier Transform scheme to improve the convergence behavior of the static energy in the position space to four-loop order. 
	\item Using the RG-summed series Eq.~\eqref{eq:RGsummed} (see definition in Eq.~\eqref{E0exp}) and the lattice QCD parametrization Eq.~\eqref{lpm}, we fit the the QCD scale $\Lambda^{\overline{\textrm{MS}}}_{\textrm{QCD}}$ to the lattice data. Our fit results at different loop orders can be found Table~\eqref{lambda_3}-\eqref{lambda_4} and scale dependence at these orders in Fig.~\eqref{sc_dep_lam}.
	\item The uncertainties associated with our extraction of the $\Lambda^{\overline{\textrm{MS}}}_{\textrm{QCD}}$ is discussed in section \eqref{sec:conc}.
\end{itemize}\par
In summary, we have used a variety of techniques, theoretical and numerical, and have rendered the picture of the QCD static energy as a very useful tool to obtain a clear handle on $\Lambda^{\overline{\textrm{MS}}}_{\textrm{QCD}}$ which is one of the fundamental parameters of the QCD, thereby confirming the results in a large number of other studies. We have also studied the consistency of the picture by invoking Pad\'e approximants as well as renormalization group summation in order to achieve these ends. We also provide improvement in position space using RFT- scheme. Our findings in this chapter provide better control over variation of the renormalization scale for finite order results available for the static energy. It also discusses scale dependence of the extracted $\Lambda^{\overline{\textrm{MS}}}_{\textrm{QCD}}$ at different orders of perturbation theory as an application of the method.

\begin{subappendices}

 \section{The QCD-Static Coefficients at Different Loop Order \label{app:loop_coef}}
	The known results for the static energy are presented here. The coefficients of the perturbative part $V^{\rm pert}$ at different loop orders are listed below:
	\begin{align}
		\textbf{The one-loop terms:}\quad&T_{1,0,0}=\frac{31}{12}-\frac{5 n_f}{18} \,, \quad T_{1,1,0}=\beta _0 \,,\\
		\textbf{The two-loop terms:}\quad &T_{2,0,0}=28.5468-4.14714 n_f+\frac{25 n_f^2}{324}\, ,\nonumber\\& T_{2,1,0}=2 T_{1,0,0} \beta _0+\beta _1\, ,\quad T_{2,2,0}=\beta _0^2\, ,\\
		\textbf{The three-loop terms:}\quad &T_{3,0,0}=209.884-51.4048 n_f+2.90609 n_f^2 -0.0214335 n_f^3\, , \nonumber\\& T_{3,3,0}=\beta _0^3\,,T_{3,1,0}=2 T_{1,0,0} \beta _1+3 T_{2,0,0} \beta _0+\beta _2\,,\nonumber\\&  T_{3,2,0}=3 T_{1,0,0} \beta _0^2+\frac{5 \beta _1 \beta _0}{2}\,.
	\end{align}
	The ultrasoft contribution to the three-loop RG-inaccessible coefficient is given by: 
	\begin{align}
		\delta \widetilde{T}^{us} _{3,0,0}=\frac{1}{72} \pi ^2 C_A^3 (6 \ell_1-5)\,,
	\end{align}
	and contribution to the four-loop is given by:
	\begin{align}
		\delta \widetilde{T}^{us}_{4,0,0}=&\frac{C_A^4\pi^2}{2592}\Big[18 \pi ^2 \gamma_E +141 \gamma_E-6 L_2 \left(66 \ell_1+6 \pi ^2+47\right)+198 L_2^2-3 \left(47+6 \pi ^2\right) L_{\pi }\nonumber\\& +72 \pi ^2 \ell_1+894 \ell_1+432 \zeta (3)-81 \pi ^2-1241\Big]+\frac{C_A^3\pi^2}{1728}\Big[ 432 \ell_1T_{1,0,0}-216 T_{1,0,0}\nonumber\\& +60 \pi ^2 \beta _0+60 \gamma \beta _0-536 \beta _0+\left(165-60 \beta _0\right) L_{\pi }+6 L_2 \left(-20 \beta _0+132 \ell_1+55\right)\nonumber\\&-396 L_2^2-144 \beta _0 \ell_1^2+480 \beta _0 \ell_1-1320 \ell_1+66 \pi ^2-165 \gamma +1474\Big]
	\end{align}
	where $\ell_1= \log(C_A \pi )+\gamma_E $, $L_{\pi}=\log(\pi)$ and $L_{2}=\log(2)$. The constant terms appearing in Eq.~\eqref{Vus2} can be found in the Ref.~\cite{Brambilla:2006wp} and are given by:
	\begin{align}
		C_1=& \frac{2}{3} \beta_0\,,\quad\quad
		C_2= \frac{1}{54} \bigg(C_A \left(-12 \pi ^2-149+66 \log (2)\right)+4 n_f T_f (10-6 \log (2))\bigg)\, ,\nn\\
		D=& \frac{C_A}{9}\Bigg(\bigg[\frac{9 \pi ^2}{4}+\frac{1241}{36}+\frac{11 \log ^2(2)}{2}-\frac{\gamma_E}{2}\left(\pi ^2 +\frac{47 }{6}\right)-12 \zeta (3)- \left(\pi ^2+17\right) \log (2)\nonumber\\&+\frac{1}{2} \pi ^2 \log (\pi )\nonumber+\frac{47 }{12}\log (\pi )\bigg]+ n_f T_f \left[\frac{5}{6} \left(\gamma_E +\log\left(64/\pi \right)\right)-\frac{\pi ^2}{3}-\frac{67}{9}-2 \log ^2(2)\right] \Bigg)\, .
	\end{align}
	
	\section{Position Space Potential }\label{app:Vr}
	The unrestricted Fourier integrals of logarithms to position space is given by:
	\begin{align}
		\int \frac{d^3 \mathbf{p} }{(2 \pi)^3} &e^{-\text{i} \mathbf{p}.\mathbf{r}}\frac{ 4 \pi }{\mathbf{q}^2} \log^{m} \left(\frac{\mu^2}{\mathbf{q}^2}\right) = \frac{1}{r} \sum_{j=0}^m \binom{m}{j} \log^{j}\left( \mu^2 r^2\right) \partial_{\eta}^{m-j}y\left(\eta\right)\vert_{\eta=0}
		\label{FFT}
	\end{align} 
	and the RFT of these logarithms are given by:
	\begin{align}
		\int \frac{d^3 \mathbf{p}}{(2 \pi)^3} e^{-i \mathbf{p}\mathbf{r}}\frac{4\pi}{\mathbf{q}^2}& \log^m\left(\frac{\mu^2}{\mathbf{q}^2}\right)
		\Theta\left(\mu_{f}-|\mathbf{p}|\right)\nonumber\\&= -\frac{\mu_{f}}{\pi} \sum_{j=0}^{m}  \binom{m}{j} \log ^{j}\left(\frac{\mu^{2}}{\mu_{f}^2}\right)(-2)^{m-j}\left[\partial_{\eta}^{m-j}f(\eta,r \mu_f) \right]_{\eta=0}
		\label{UnFT}
	\end{align}
	Here $y(\eta)$ and $f(\eta,\beta)$ are given by:
	\begin{align}
		&y(\eta) \equiv e^{\tiny{ \left(2 \gamma_{E} \eta+\sum_{l=2}^{\infty} \eta^{l} \frac{(2^{\mathrm{l}}-1-(-1)^{l} )\zeta(l)}{l}\right)}}=\frac{\Gamma (1-2 \eta )}{\Gamma (1-\eta ) \Gamma (\eta +1)}\,,\nn \\
		&f(\eta,\beta) \equiv \frac{\Gamma(\eta)-\Gamma\left(\eta,\mathrm{i} \beta\right)}{\left(\mathrm{i}\beta\right)^{1+\eta}}+\mathrm{c.c.}
	\end{align}
	and c.c. stands for complex conjugate. Writing $L_{\gamma}=2\gamma_E -\log (\mu^2 r^2)$, the unrestricted Fourier transforms for static potential without the ultrasoft terms can be written as:
	\begin{align}
		V(r)=\sum_{i=0}^{4} \sum_{j=0}^{i}x^{i+1} V_j(r)+\order{x^6}
	\end{align}
	\begin{align}
		V_0(r)&=\frac{1}{r} \,, \quad V_1(r)=\frac{L_{\gamma}}{r} \,, \quad V_2(r)=\frac{1 }{ r}(L_{\gamma}^2+\frac{\pi ^2}{3}) \,, \quad 
		V_3(r)=\frac{L_{\gamma}}{r}(L_{\gamma}+\pi ^2) +\frac{16}{r} \zeta (3)\,,\nonumber\\
		V_4(r)&=\frac{L_{\gamma}}{r} \left(L_{\gamma}^3+2 \pi ^2 L_{\gamma}+64 \zeta (3)\right)+\frac{19 \pi ^4}{15 r}
	\end{align}
	\par
	Restricted Fourier transform contains hypergeometric functions with array of $\frac{1}{2}$ in first argument and $\frac{3}{2}$ in the second argument and if we define:
	\begin{align}
		\text{Si}(r \mu_f)\equiv \left(\mu_f r\right)\times H_1 , \quad \quad
		_nF_{n+1}\big(\frac{1}{2},\frac{1}{2},\dots;\frac{3}{2},\frac{3}{2},\frac{3}{2},\dots;-\frac{1}{4} r^2 \mu _f^2\big)\equiv H_{n}
		\label{hyp}
	\end{align}
	then the uncontrolled contribution to static potential without the ultrasoft term is given by:
	\begin{align}
		\delta V(r,\mu_f)=\frac{2 \mu_f}{\pi}\sum_{i=0}^{4} \sum_{j=0}^{i}x^{i+1} \delta V_j(r,\mu_f)+\order{x^6}
	\end{align}
	where, 
	\begin{align}
		\delta V_0(r,\mu_f)=&H_1\,,\quad 
		\delta V_1(r,\mu_f)=2 H_2-H_1 L_{\mu_f} \,,\nonumber\\ \delta V_2(r,\mu_f)=&-4 H_2 L_{\mu_f}+\frac{1}{3} H_1 \left(3 L_{\mu_f}^2-\pi ^2\right)+8 H_3\,,\nn\\
		\delta V_3(r,\mu_f)=&-24 H_3 L_{\mu_f}+H_2 \left(6 L_{\mu_f}^2-2 \pi ^2\right)+H_1 \left(-L_{\mu_f}^3+\pi ^2 L_{\mu_f}-16 \zeta (3)\right)+48 H_4\,,\nn \\
		\delta V_4(r,\mu_f)=&-192 H_4 L_{\mu_f}+384 H_5+\frac{1}{5} H_1 \left(5 L_{\mu_f}^4-10 \pi ^2 L_{\mu_f}^2+320 \zeta (3) L_{\mu_f}-3 \pi ^4\right)\nonumber\\&+\frac{1}{5} H_2 \left(-40 L_{\mu_f}^3+40 \pi ^2 L_{\mu_f}-640 \zeta (3)\right)+\frac{1}{5} H_3 \left(240 L_{\mu_f}^2-80 \pi ^2\right)
	\end{align}
	\section{Restricted Version of the Static Energy in Position Space\label{app:resVr}} 
	Uncontrolled contribution to the static energy in RFT scheme is given by:
	\begin{align}
		\delta E&(r,\mu,\mu_f)=-\left(2 C_F H_1 \mu_f x \right)\Bigg\lbrace1+x \big(2 \tilde{H}_2 T_{1,1,0}+T_{1,1,0} L_{\mu_f}+T_{1,0,0}\big)+x^2 \big(4 \tilde{H}_2 T_{2,2,0} L_{\mu_f}\nonumber\\&+2 \tilde{H}_2 T_{2,1,0}+8 \tilde{H}_3 T_{2,2,0}+T_{2,0,0}+T_{2,2,0} L_{\mu_f}^2+T_{2,1,0} L_{\mu_f}\big) +x^3 \Big(T_{3,0,0}+2 \tilde{H}_2 T_{3,1,0}\nonumber\\&+8 \tilde{H}_3 T_{3,2,0}+L_{\mu_f} \big(4 \tilde{H}_2 T_{3,2,0}+24 \tilde{H}_3 T_{3,3,0}+T_{3,1,0}\big) +48 \tilde{H}_4 T_{3,3,0}+L_{\mu_f}^2 \big(6 \tilde{H}_2 T_{3,3,0}\nonumber\\&+T_{3,2,0}\big)+T_{3,3,0} L_{\mu_f}^3+ \frac{1}{144} \pi ^2 C_A^3 \big(12 \tilde{H}_2+12 \log \left(H_1\right)+12 L_{us}-12 \gamma _E -10+24 L_2\big) \Big)\nonumber\\&+x^4\Big\lbrace T_{4,0,0}+2 \tilde{H}_2 T_{4,1,0}+8 \tilde{H}_3 T_{4,2,0}+48 \tilde{H}_4 T_{4,3,0}+384 \tilde{H}_5 T_{4,4,0}+\frac{\pi ^4 C_A^3 }{144} \Big[12 \tilde{H}_2-9 \beta _0\nonumber\\&+12 \log \left(H_1\right)+12 L_{us}-9 \gamma _E -8+18 L_2-3 L_{\pi}\Big]+L_{\mu_f}^2 \big(6 \tilde{H}_2 T_{4,3,0}+48 \tilde{H}_3 T_{4,4,0}+T_{4,2,0}\big)\nonumber\\&+T_{4,4,0} L_{\mu_f}^4+L_{\mu_f} \Big[\frac{1}{144} \pi ^2 C_A^3 \big(48 \beta _0 \tilde{H}_2+12 T_{1,1,0}-48 \gamma _E \beta _0-40 \beta _0 +96 \beta _0 L_2+48 \beta _0 L_{us}\nonumber\\&+48 \beta _0 \log \left(H_1\right)\big)+T_{4,1,0}+24 \tilde{H}_3 T_{4,3,0}+192 \tilde{H}_4 T_{4,4,0}+4 \tilde{H}_2 T_{4,2,0}\Big]+L_{\mu_f}^3 \big(8 \tilde{H}_2 T_{4,4,0}\nonumber\\&+T_{4,3,0}\big)+\frac{\pi ^2 C_A^3 }{144} \Big[36 \tilde{H}_2 T_{1,0,0}+24 \tilde{H}_2 T_{1,1,0}-96 \gamma _E \beta _0 \tilde{H}_2-40 \beta _0 \tilde{H}_2+168 \beta _0 \tilde{H}_3\nonumber\\&+72 \beta _0 \tilde{H}_2 \log \left(H_1\right)+39 \tilde{H}_2+72 \beta _0 L_{us}\tilde{H}_2+36 T_{1,0,0} \log \left(H_1\right)+36 L_{us}T_{1,0,0}-36 \gamma _E T_{1,0,0}\nonumber\\&-18 T_{1,0,0}+72 L_2 T_{1,0,0}-84+12 \gamma _E ^2 \beta _0-15 \gamma _E \beta _0-\frac{134 \beta _0}{3}-48 \beta _0 L_2 ^2+70 \beta _0 L_2-5 \beta _0 L_{\pi}\nonumber\\&-12 \beta _0 \log ^2\left(H_1\right)-48 \beta _0 L_2 \log \left(H_1\right)+40 \beta _0 \log \left(H_1\right)-24 \beta _0 L_{us}\log \left(H_1\right)+39 \log \left(H_1\right)\nonumber\\&-12 \beta _0 (L_{us})^2+40 \beta _0 L_{us}-48 \beta _0 L_{us}L_2+39 L_{us}+72 \zeta (3)-\frac{117 \gamma _E }{4}+\frac{117 L_2}{2}-\frac{39 L_{\pi}}{4}\Big]\Bigg\rbrace \Big\rbrace
	\end{align}
	where $L_{us}=\log \left(\frac{ C_A x}{2}\right)$, $L_{\mu_f}=\log \left(\frac{\mu ^2}{\mu_f^2}\right)$, $L_{\pi}=\log(\pi)$, $L_{2}=\log(2)$ and $\tilde{H_i}\equiv H_i / H_1$. Note that $H_i's$ are defined by Eq.~\eqref{hyp} in appendix~\eqref{app:Vr}.
	
\end{subappendices}

%% file: Chapters/Chap4.tex
\chapter{Renormalization group improved \texorpdfstring{$m_s$}{} and \texorpdfstring{$\vert V_{us}\vert$}{} determination from hadronic \texorpdfstring{$\tau$ }{}decays.}
\label{Chapter4}

\lhead{Chapter 4. \emph{Renormalization group improved \texorpdfstring{$m_s$}{} and \texorpdfstring{$\vert V_{us}\vert$}{} determination from hadronic \texorpdfstring{$\tau$ }{}decays.}}
\section{Motivation}
			We determine the strange quark mass (\texorpdfstring{$m_s$}{}) and quark mixing element \texorpdfstring{$\vert V_{us}\vert $}{}, and their joint determination from the Cabibbo suppressed hadronic $\tau$ decays in various perturbative schemes. We have improved this analysis compared to the previous analysis based on the renormalization group summed perturbation theory (RGSPT) prescription by replacing the theoretical longitudinal contributions with phenomenological parametrization and the RGSPT coefficients are used for the dimension-4 Adler functions. The improved analysis results in the extraction of \texorpdfstring{$m_s(2\GeV)=98\pm19\MeV$}{} and \texorpdfstring{$\vert V_{us}\vert=0.2191\pm0.0043$}{} from the RGSPT scheme.
		\section{Introduction}
	The hadronic decays of the $\tau$ leptons have been of constant interest for determining various parameters of the Standard Model (SM) of particle physics. The availability of experimental data on the strange and non-strange decay modes for the hadronic $\tau$ decays has opened the window for the determination of various parameters relevant for the quantum chromodynamics (QCD), namely the strong coupling constant $\alpha_s$, the strange quark mass $m_s$, the vacuum condensates, the low-energy chiral couplings, and the quark mixing element $\vert V_{us}\vert$ of the Cabibbo–Kobayashi–Maskawa (CKM) matrix (see Refs.~\cite{Davier:2005xq,Pich:2013lsa} for details). \par
	On the theoretical side, the QCD contributions to the hadronic $\tau$ decays are studied by evaluating the current correlator using the Operator Product Expansion (OPE)\cite{Wilson:1969zs}. The OPE factorizes the long- and short-distance contributions. The long-distance information is encoded into the vacuum condensates. The short distance part is written as the perturbative series in the strong coupling constant and quark masses. The vacuum condensates can also be evaluated using chiral perturbation theory (ChPT) \cite{Gasser:1984gg}, lattice QCD \cite{FlavourLatticeAveragingGroup:2019iem}, and Renormalization Group (RG) optimized perturbation theory \cite{Kneur:2015dda,Kneur:2020bph}. The short-distance contributions require the evaluation of the Feynman diagrams. It is also known that some of the contributions to the hadronic vacuum polarization function are not captured by the OPE, and it is a quark-hadron duality violation. These duality-violating terms are parameterized in a model-dependent way and fitted to experimental data and should also be added to the OPE contributions \cite{Boito:2017cnp}.\par
	The longitudinal component of the QCD Adler function, corresponding to the zero angular momentum state, has been calculated to $\order{\alpha_s^4}$~\cite{Becchi:1980vz,Broadhurst:1981jk,Chetyrkin:1996sr,Baikov:2005rw,Gorishnii:1990zu,Gorishnii:1991zr}. It has poor convergence behavior and raises the question of the method's applicability in the extraction of strange quark mass. This problem can be cured by replacing these contributions with their phenomenological input and has been used in Refs.~\cite{Gamiz:2002nu,Jamin:2001zq,Jamin:2001zr,Maltman:2001gc,Maltman:2001sv} for $m_s$ and $\vert V_{us}\vert$ extractions from the experimental moment data. These improvements have resulted in much better control over the theoretical uncertainties in the $m_s$ and $\vert V_{us}\vert$ determinations.\par
	The hadronic $\tau$ decays have been extensively studied using various perturbative schemes. These schemes differ in how strong coupling constant and quark masses are evaluated along the contour in the complex plane using their renormalization-group (RG) properties. The most commonly used schemes in the extraction of strange quark mass and CKM matrix element from the Cabibbo suppressed hadronic $\tau$ decay are fixed-order perturbation theory (FOPT) and contour-improved perturbation theory (CIPT). For the hadronic $\tau$ decays, the FOPT suffers from the problem of large logarithms along the contour in the complex energy plane, and the higher-order spectral moments are very sensitive to scale variations. In the CIPT scheme, direct numerical evaluation of coupling constant and masses along a contour in the complex energy plane using their RGE does not suffer from the problem of the large logarithms. However, scale dependence is still the major source of theoretical uncertainties for higher moments.\par Recently, RG-summed perturbation theory (RGSPT) has been used in Ref.~\cite{Ananthanarayan:2016kll} in the strange quark mass determination. The behavior of polarization and Alder functions in the complex contour was also studied for RGSPT, CIPT, FOPT, and the method of effective charges (MEC) in great detail. However, the numerical impact of the theoretical uncertainties from perturbation series truncation and scale dependence were excluded. We improve the previous analysis by:
	\begin{itemize}
		\item Including the RGSPT coefficients for dimension-4 Adler functions.
		\item Replacing the divergent longitudinal perturbative QCD expressions for the Adler function with the phenomenological parametrization used in Refs.~\cite{Gamiz:2002nu,Maltman:2001gc}. This replacement significantly reduces the theoretical uncertainties.
		\item Performing $\vert V_{us}\vert$ as well as the joint $m_s$ and $\vert V_{us}\vert$ determinations for the first time using RGSPT.
		\item The effects of the variation of the $m_s$ and $|V_{us}|$ with the variation of the moments calculated at different energies ($s_0<\mtsq$) is also included and found to be constituting an important source of uncertainty. 
		\item Using the five-loop QCD $\beta-$function and anomalous dimensions for the running of the strong coupling constant and quark masses.
	\end{itemize}
	This chapter is organized as follows: Section~\eqref{sec:formalism_ms_vus} provides a brief overview of the various quantities that are needed for the extraction of $m_s$ and $\vert V_{us}\vert$. A short introduction to RGSPT is given in section~\eqref{sec:RGSPT}. Section~\eqref{sec:OPE_contributions} explains the OPE contributions to the Adler function. The behavior of leading-order mass corrections to the Adler functions in different schemes used in this chapter is studied in section~\eqref{sec:dim_2_behaviour}. The higher-order term of the perturbation series becomes very important for the higher moments, and two prescriptions for the truncation of the perturbation series are also defined in this section. In section~\eqref{sec:rev_pheno}, the phenomenological parametrization of longitudinal contributions is briefly discussed. Then, we move to section~\eqref{sec:ms_pert}, where strange quark mass is extracted using only the perturbative QCD (pQCD) contributions calculated from OPE. The weighted average results for the $m_s(\mtsq)$ extraction  using this method in CIPT, FOPT, and RGSPT schemes are presented in Table~\eqref{tab:mspertweighted}. The details of uncertainties can be found in the appendix~\eqref{app:pQCD_mass}. In section ~\eqref{sec:pheno_ms}, $m_s(\mtsq)$ determination using the phenomenological parametrization for the longitudinal component is performed, and results are presented in Table~\eqref{tab:msphenoweighted2}. Details of the strange quark mass determinations from moments are presented in the appendix~\eqref{app:pheno_mass}. The determination of $\vert V_{us}\vert$ using external input for $m_s$ is performed in section~\eqref{sec:Vusextraction}. The weighted average results using the OPAL and ALEPH data are presented in Table~\eqref{tab:Vus_weightedmean12} and Table~\eqref{tab:Vus_Aleph1}, respectively. The details of determinations from the moments as well as the uncertainties coming from various sources are presented in appendix~\eqref{app:vus}. In section~\eqref{sec:joint_msVus} the joint extraction of $m_s$ and $\vert V_{us}\vert$ is performed. We provide a summary and conclusion in section~\eqref{sec:summary_ms_vus}. We also provide supplementary inputs needed for this chapter in the appendices~\eqref{app:mass_run}, \ref{app:summed_sol}, \ref{app:adlercoef}, \ref{app:dim4corrections}. Details of $m_s$ and $\vert V_{us}\vert$ determinations from moments can be found in appendix~\eqref{app:determination_details}.
	\section{Formalism}\label{sec:formalism_ms_vus}
	An important quantity for the study of hadronic $\tau$ decay width~\cite{Braaten:1991qm,Pich:1999hc,Narison:1988ni} is the two-point current correlator:
	\begin{equation}
		\Pi^{V/A}_{\mu \nu,ij} (p^2) \equiv i \int dy \text{ }e^{i p y}\text{ }\langle\Omega\left|T\{J^{V/A}_{\mu,ij}\left(y\right)J^{V/A}_{\nu,ij}\left(0\right)^{\dagger}\}\right|\Omega\rangle
		\label{cor}
	\end{equation}
	where $|\Omega\rangle$ denotes the physical vacuum, $J^{V/A}_{\mu,ij}\left(y\right)=\left(\bar{q}_j\gamma_{\mu}/\left(\gamma_{\mu}\gamma_{5}\right)q_i\right)\left(x\right)$ is the hadronic vector/axial current, and the indices $i$ and $j$ denote the flavors of light quarks. The current correlator can be calculated perturbatively using OPE \cite{Wilson:1969zs} as a power expansion in $1/p$, and the corresponding coefficients are the operator of that dimension. Purely perturbative corrections appear up to dimension-2 in the OPE expansion, and the long-distance corrections corresponding to the vacuum condensates start from dimension-4.\par
	Using Lorentz decomposition, the current correlator in Eq.~\eqref{cor} can be decomposed into the longitudinal (with angular momentum $J=0$) and transverse ($J=1$) components as:
	\begin{align}
		\Pi^{V/A}_{\mu \nu,ij} (p^2) &=\left(p_{\mu}p_{\nu}-g_{\mu,\nu}\right) \Pi^{V/A,T}_{ij} (p^2)+ p_{\mu} p_{\nu} \Pi^{V/A,L}_{ij}(p^2)\nonumber\\
  &=\left(p_{\mu}p_{\nu}-g_{\mu,\nu}\right) \Pi^{V/A,L+T}_{ij} (p^2)+g_{\mu,\nu}\hs p^2 \hs\Pi^{V/A,L}_{ij}(p^2)\,.
		\label{eq:cor_decompose}
	\end{align}
	The $L/T$ correlators are related to experimentally measurable semi-hadronic $\tau$ decay rate ($R_{\tau}$), defined by:
	\begin{equation}
		R_{\tau}\equiv\frac{\Gamma\left(\tau^-\rightarrow\text{hadrons } \nu_{\tau}\left(\gamma\right)\right)}{\Gamma\left(\tau^-\rightarrow e^- \nu_{\tau}\left(\gamma\right)\right)}=R_{\tau,V}+R_{\tau,A}+R_{\tau,S}\,,
	\end{equation}
	where, $R_{\tau,S}$ is contributions from the strange channels containing odd numbers of kaons. The non-strange channel is resolved into vector $R_{\tau, V}$ and axial vector channel $R_{\tau,A}$ components containing. $R_{\tau}$ is related to the imaginary part of the current correlators in Eq.~\eqref{eq:cor_decompose} by:
	\begin{align}
		R_{\tau}\left(s_0\right)=12 \pi &\int_{0}^{s_0} \frac{ds}{s_0}\left(1-\frac{s}{s_0}\right)^2 \times \left[\left(1+\frac{2 s}{s_0}\right) \text{Im}\left(\Pi^T\right)(s)+ \text{Im}\left(\Pi^L\right)(s)\right]\,.
		\label{eq:Rratio_ms_vus}
	\end{align}
	It should be noted that these current correlators also carry the information about mixing among the quark flavors and can be written as:
	\begin{equation}
		\Pi^J\equiv \sum_{i=d,s}\left|V_{ui}\right|^2 \left[\Pi^{V,J}_{ui}\left(s\right)+\Pi^{A,J}_{ui}\left(s\right)\right]\,,\label{R_tau1}
	\end{equation}
	and $\vert V_{ij}\vert$ are the elements of the CKM matrix. \par
	To study the invariant mass distribution of final-state hadrons, we need moments from the hadronic $\tau$ decay rate, defined by\cite{LeDiberder:1992zhd}:
	\begin{align}
		R^{kl}_{\tau}\left(s_0\right)&\equiv \int_{0}^{s_0} ds \left(1-\frac{s}{s_0}\right)^k \left(\frac{s}{s_0}\right)^l \frac{d R_{\tau}}{ds}\,,
		\label{eq:rklmomentdef}
	\end{align}
 Using integration by parts, we can convert Eq.~\eqref{eq:rklmomentdef} into the following form:
	\begin{align}
		R^{kl}_{\tau}\left(s_0\right)=- i \pi  &\oint\limits_{\left|x_c\right|=1} \frac{dx_c}{x_c}\times\big\lbrace 3 \mathcal{F}^{L+T}_{kl}(x_c) \mathcal{D}^{L+T}\left(s_0 x_c\right)+4 \mathcal{F}^{L}_{kl}(x_c)\mathcal{D}^{L}\left(s_0 x_c\right)\big\rbrace\,,
		\label{Rtau_def}
	\end{align}
	where $x_c=s/s_0$ and $\mathcal{D}^{L+T/L}(s)$ are known as the Adler functions.

 \par Usually, the experimental value of the moments defined above are provided for  $s_0=\mtsq$ in the literature. However, their values at different energy can be calculated using the experimental data on the spectral functions provided in the Refs.~\cite{OPAL:1998rrm,OPAL:2004icu,Davier:2013sfa,Boito:2020xli}. \par The Adler function satisfies the homogeneous renormalization group equation (RGE) and is related to the current correlators by the relation:
	\begin{align}
		\mathcal{D}^{L+T}\left(s\right)&\equiv -s\frac{d}{ds}\left(\Pi^{L+T}\left(s\right)\right)\,,\\ \mathcal{D}^{L}\left(s\right)&\equiv \frac{s}{\Mt}\frac{d}{ds}\left(s\hspace{1 mm} \Pi^{L}\left(s\right)\right).
		\label{eq:Ds}
	\end{align}
	The resulting quantity in Eq.~\eqref{Rtau_def} is an expansion in the strong coupling constant, quark masses, and condensates of higher dimension operators. It explicitly depends on the CKM matrix element and the electroweak corrections. These terms are not shown in Eq.~\eqref{Rtau_def} but factored out in Eq.~\eqref{eq:Rtauexpanded} and Eq.~\eqref{eq:delRkl}.
	The kinematic kernels $ \mathcal{F}^{kl}_{L+T}\left(x_c\right)$ and $ \mathcal{F}^{kl}_{L}\left(x_c\right)$ appearing in the Eq.~\eqref{Rtau_def} are given by:
	\begin{align}
		\mathcal{F}^{kl}_{L+T}(x_c)\equiv &2\left(1-x_c\right)^{3+k} \sum_{n=0}^{l} \frac{l!}{\left(l-n\right)! n!}\left(x_c-1\right)^n\times \frac{(6+k+n)+2(3+k+n) x_c}{(3+k+n)(4+k+n)}\,,\\
		\mathcal{F}^{kl}_{L}(x_c)\equiv &3\left(1-x_c\right)^{3+k} \sum_{n=0}^{l} \frac{l!}{\left(l-n\right)! n!}\frac{\left(x_c-1\right)^n} {(3+k+n)}\,,
	\end{align}
	and their explicit form used in this chapter is presented in the Table~\eqref{tab:klmoments}.
	\begin{table}
		\begin{center}
			\begin{tabular}{|c|c|c|}	\hline
				$(k, l)$ & $\mathcal{F}_{L+T}^{k l}(x)$ & $\mathcal{F}_{L}^{k l}(x)$ \\	\hline
				(0,0) & $(1-x)^{3}(1+x)$ & $(1-x)^{3}$ \\
				(1,0) & $\frac{1}{10}(1-x)^{4}(7+8 x)$ & $\frac{3}{4}(1-x)^{4}$ \\
				(2,0) & $\frac{2}{15}(1-x)^{5}(4+5 x)$ & $\frac{3}{5}(1-x)^{5}$ \\
				(3,0) & $\frac{1}{7}(1-x)^{6}(3+4 x)$ & $\frac{1}{2}(1-x)^{6}$ \\
				(4,0) & $\frac{1}{14}(1-x)^{7}(5+7 x)$ & $\frac{3}{7}(1-x)^{7}$\\ \hline
			\end{tabular}
			\caption{Kinematic kernels used in this study.}
			\label{tab:klmoments}
		\end{center}
	\end{table}
	\par
	Performing the contour integral defined in Eq.~\eqref{Rtau_def}, we can write $R^{kl}_{\tau}$\cite{Braaten:1991qm} as:
	\begin{align}  \label{eq:Rtauexpanded}
		R^{kl}_{\tau}\left(s_0\right)=3 \left(\vert V_{ud}\vert^2+\vert V_{us}\vert ^2\right)& S_{EW} \bigg\lbrace1+\delta_{EW}^{kl}+\delta^{\left(0\right),kl}\nonumber\\&+\sum_{n=2,4...}\left(\cos^2\left(\theta_C\right) \delta^{\left(n\right),kl}_{ud}+ \sin^2 \left(\theta_C\right) \delta^{\left(n\right),kl}_{us}\right)\bigg\rbrace\,,
	\end{align}
	where $\theta_C=\sin^{-1}\left(\vert V_{us}\vert/\sqrt{\left(|V_{us}|^2+|V_{ud}|^2\right)}\right)$ is the Cabibbo angle, $\delta^{00}_{EW}=0.0010$ and $S_{EW}=1.0201\pm0.0003$ are one-loop RG-improved electroweak corrections \cite{Braaten:1990ef,Erler:2002mv}. The $\delta^{(n),kl}_{ud/us}$~\cite{Braaten:1991qm,Baikov:2004tk} are contributions from the  $n^{th}$-dimension OPE to the polarization function to the quantity:
\begin{align}
    \delta^{kl}_{ij}&=2 \pi i \oint\limits_{\left|s\right|=s_0}\frac{ds}{s_0}\left(1-\frac{s}{s_0}\right)^{k+2}\Bigg\lbrace\left(1+2\frac{s}{s_0}\right) \left(\frac{s}{s_0}\right)^l \Pi^{L+T}_{ij}(s)-2  \left(\frac{s}{s_0}\right)^{l+1} \Pi^{L}_{ij}(s)\Bigg\rbrace\nonumber\,.
\end{align}
 The most important quantity of interest in the determination of the strange quark mass is the $SU(3)$ breaking terms $\delta R^{kl}_{\tau}\left(s_0\right)$~\cite{ALEPH:1999uux} defined as
	\begin{align}  \label{eq:delRkl}
		\delta R^{kl}_{\tau}\left(s_0\right)&\equiv\frac{R^{kl}_{\tau,V+A}\left(s_0\right)}{\left\vert V_{ud}\right\vert^2} -\frac{R^{kl}_{\tau,S}\left(s_0\right)}{\left\vert V_{us}\right\vert^2} =3 S_{EW}\sum_{n\ge2}\left(\delta^{\left(n\right),kl}_{ud}-\delta^{\left(n\right),kl}_{us}\right)\,,
	\end{align}
	which is free from instanton and renormalon contributions and vanishes in the chiral limit. It is an experimentally measurable quantity that is used as input from Table~\eqref{tab:rkl_exp} along with the theoretical quantities appearing in Eq.~\eqref{eq:delRkl} in the strange quark mass determination. \par
	The value of strong coupling constant $\alpha_s(M^2_Z)=0.1179\pm 0.0010$ has been taken from Ref.~\cite{Zyla:2020zbs} and evolved to $\tau$ lepton mass scale using five-loop $\beta-$function using package REvolver \cite{Hoang:2021fhn}.  Its value at $\tau$ lepton mass is $\alpha_s(M^2_{\tau})=0.3187\pm 0.0083$ and has been used in this chapter.
	\section{Inputs from RGSPT}\label{sec:RGSPT}
	 In the case of hadronic $\tau$ decays, where weighted integrals along the complex contour are involved, the RG plays a key role as the running logarithms become very important. Their summation is necessary to perform the perturbative analysis properly. The formalism developed in chapter~\eqref{Chapter2} for RGSPT can be used to sum these logarithms, and the resulting fixed order truncated series has less sensitivity to scale variations even for higher moments than FOPT and CIPT. We give a brief overview of the RGSPT in the rest of the section explicitly relevant to this chapter.\par
	The perturbative series describing a QCD process is given by:
	\begin{equation}
		\mathcal{S}(x,m)=x^{n_1} m^{n_2} \sum_{i=0}x^i L^j \hspace{.4mm}T_{i,j}\,,	\label{eq:series_expanded}
	\end{equation}
	where $x\equiv x(\mu^2)$, $L\equiv \log(\mu^2/q^2)$ and $m=m(\mu^2)$. We can rewrite the series as follows:
	\begin{equation}
		\mathcal{S}^{RG\Sigma}=x^{n_1} m^{n_2}\sum_{i=0}x^i\hspace{.4mm}S_{i}[x \hspace{.4 mm} L]\,,
		\label{ser_summed}
	\end{equation}
	where the $S_{i}[x L]$-coefficients are given by:
	\begin{equation}
		S_{i}[x\hspace{.4mm}L]=\sum_{n=i}^{\infty} T_{n,n-i}  (x\hspace{.4mm}L)^{n-i}\,.
		\label{eq:summed_coefs_ms_vus}
	\end{equation}
	The RGE for Eq.~\eqref{eq:series_expanded} is given by:
	\begin{equation}
		\left(\mu^2 \frac{d}{d\mu^2} -\gamma_{a}(x) \right)\mathcal{S}(x,m)=\left(\beta(x) \partial_x+ \gamma_m(x) \partial_m+\partial_L-\gamma_{a}(x)\right)\mathcal{S}(x,m)=0\,,
	\end{equation}
	where $\gamma_{a}(x)=\sum_{i=0}x^{i+1}\gamma_{a}^{\left(i\right)}$ is the anomalous dimension associated with  $\mathcal{S}(x,m)$. We can collect the terms corresponding to summed coefficients defined in Eq.~\eqref{eq:summed_coefs_ms_vus}. This process results in a set of coupled differential equations for $S_{i}[w]$, which can be summarized as:
	\begin{align}
		\sum _{i=0}^n \bigg[\beta _i &(\delta_{i,0}+w-1)  S_{n-i}'(w)+S_{n-i}(w) \left(n_2 \gamma_i+\beta _i (n_1-i+n)+\gamma_a^{(i)}\right)\bigg]=0\,.
		\label{summed_de}
	\end{align}
	Here we have substituted $w=1-\beta_0 x L$, which simplifies the solutions of differential equations.\par
	The solution to the first three summed coefficients, relevant for dimension-0 and dimension-2 Adler functions, appearing in the Eq.~\eqref{summed_de} can be found in the appendix~\eqref{app:summed_sol}. It should be noted that the RGE for dimension-4 operators mix perturbative coefficients with condensates; hence they do not obey Eq.~\eqref{ser_summed}. The RG-summed perturbative coefficients are relevant for Eq.~\eqref{eq:dim4LTAdler} and Eq.~\eqref{eq:dim4LAdler} in the appendix~\eqref{app:dim4corrections}.
	\section{OPE Contributions to the QCD Adler Function}\label{sec:OPE_contributions}
	\subsection{Leading order contribution}
	Dimension-zero is the leading perturbative contribution to the current correlator in the massless limit and has been calculated to $\order{\alpha_s^4}$ \cite{Appelquist:1973uz,Zee:1973sr,Chetyrkin:1979bj,Dine:1979qh,Gorishnii:1990vf,Surguladze:1990tg,Chetyrkin:1996ez,Baikov:2008jh,Baikov:2010je,Herzog:2017dtz}. It receives a contribution only from the transverse piece of the current correlator, which is identical for both vector and axial-vector channels and thus cancels in the Eq.~\eqref{eq:delRkl}. The Adler functions obtained using the OPE can be organized as follows:
	\begin{align}
		\mathcal{D}^{L+T}(s)&=\sum_{n=0,2,4,\cdots}\frac{1}{s^\frac{n}{2}}\mathcal{D}_{n}^{L+T}(s)\,,\\
		\mathcal{D}^{L}(s)&=\frac{1}{M_{\tau}^2}\sum_{n=2,4,\cdots} \frac{1}{s^{\frac{n}{2}-1}}\mathcal{D}_{n}^{L}(s)\,,
	\end{align}
	where Adler functions in the RHS of the above equations are expansion in $\alpha_s$, $m_q$, and the quark and gluon condensate terms. Their definition gets clearer if we take a contour integration along $s=\mtsq e^{i \phi}$ and the coefficient of $(\mtsq)^{-n}$ are called operators of dimension $2n$.
	The massless Adler functions are given by
	\begin{align}
		\mathcal{D}_{0}^{L+T,V/A}(s)&=\frac{1}{4\pi}\sum_{i}x(-s)^i \tilde{K}^{L+T}_{i}\,,\\
		\mathcal{D}_{0}^{L,V/A}(s)&=0\,,
	\end{align}
	where $x(-s)=x(q^2)=\alpha_s(q^2)/\pi$ and $\tilde{K}^{L+T}_{i}$ are the coefficient of Adler function at $i^{th}-$loop which can be found in appendix~\eqref{app:adlercoef}. The RG running of dimension-zero $``L+T"-$component of the Adler function is given by:
	\begin{align}
		\mu^2 \frac{d}{d\hspace{.4 mm} \mu^2}\mathcal{D}^{ L+T,V/A}_0(s)=\Bigg(\beta(x) \frac{\partial}{\partial x}+\frac{\partial}{\partial L}\Bigg) \mathcal{D}^{L+T,V/A}_0(s)=0\,,
	\end{align}
	where $L=\log(\frac{\mu^2}{-s})$, and the QCD beta function ($\beta(x)$) is defined as:
	\begin{align}
		\mu^2\frac{d}{d\mu^2}x(\mu^2)=\beta(x(\mu^2))=-\sum_i \beta_i x(\mu^2)^{i+2} \,. \label{beta_function}
	\end{align}
	The coefficients of the beta function $\beta_i$'s are known up to the five-loops and are presented in the appendix~\eqref{app:mass_run}.
	
	\subsection{The Dimension-2 contributions to the Adler Function}
	The leading  order mass corrections to hadronic $\tau$ decay rate come from the dimension-2 Adler function. The $\mathcal{D}^{L+T}(s)$ Adler function is known to $\order{\alpha_s^3}$ \cite{Baikov:2004ku,Baikov:2004tk,Chetyrkin:1993hi,Gorishnii:1986pz,Generalis:1989hf,Bernreuther:1981sp} while $\mathcal{D}^{L}$ is known to $\order{\alpha_s^4}$ \cite{Becchi:1980vz,Broadhurst:1981jk,Chetyrkin:1996sr,Baikov:2005rw,Gorishnii:1990zu,Gorishnii:1991zr} and their analytic expression can be found in the appendix~\eqref{app:dim2adler_ms_vus}.
	The RG running of dimension-2 operators is given by:
	\begin{align}
		\mu^2 \frac{d}{d\hspace{.4 mm} \mu^2}\mathcal{D}^{J}_2(s)=&\Bigg\lbrace \frac{\partial}{\partial L}+\beta(x) \frac{\partial}{\partial x}+2 \gamma_{m}(x) \frac{\partial}{\partial m_i}\Bigg\rbrace \mathcal{D}^{J}_2(s)=0\,,
	\end{align}
	where the QCD beta function and the quark mass anomalous dimension ($\gamma_{m}$) are known to the five-loop and can be found in the appendix~\eqref{app:mass_run}.\par
	The $SU(3)$ breaking contributions from the Adler function in the determination of quark masses is the difference:
	\begin{equation}
		\delta \mathcal{D}^{J,V+A}_{2}(s)\equiv\mathcal{D}^{J,V+A}_{2,ud}(s)-\mathcal{D}^{J,V+A}_{2,us}(s)
	\end{equation}
	where $J=(L+T)/L$ and the analytic expressions can be found in the appendix~\eqref{app:dim2adler_ms_vus}. These contributions are used in Eq.~\eqref{eq:delRkl} to evaluate the leading order mass correction term $\delta^{(2),kl}_{ud}-\delta^{(2),kl}_{us}$.
	
	The absence of a coefficient $\order{\alpha_s^4}$ for the $``L+T"-$ Adler function induces an additional theoretical uncertainty in the predictions from the perturbation theory. This missing piece can be estimated by $\tilde{d}^{L+T}_{4}\sim (\tilde{d}^{L+T}_{3})^2/\tilde{d}^{L+T}_{2}\approx4067$ and this value is used in the strange quark mass determinations in this chapter.\par
	The renormalization group running of different coefficients for CIPT and FOPT coefficients can be found in the Refs.~\cite{Pich:1998yn,Pich:1999hc}. The RG-summed coefficients can be obtained from appendix~\eqref{app:summed_sol} by setting $\left\lbrace n_1,n_2\right\rbrace=\left\lbrace0,2\right\rbrace$.
	
	\subsection{The Dimension-4 contributions to the Adler Function}\label{sec:dim_4_Adler}
	The OPE expansion at dimension-4 involves contributions from perturbative and quark, and gluon condensates \cite{Pich:1999hc,Chetyrkin:1985kn}. However, these contributions  are suppressed by a factor of $(\frac{1}{M_{\tau}^2})^2$, and they have the following form:
	\begin{align}
		\mathcal{D}^{L+T,V/A}_{4,ij}(s)&=\frac{1}{s^2}\sum_{n=0}\tilde{\Omega}^{L+T}_n(s) x(-s)^n\,,\\
		\mathcal{D}^{L,V/A}_{4,ij}(s)&=\frac{1}{M^2_{\tau}\hs s}\bigg\lbrace\frac{3}{2\pi^2}\sum_{n=0}\tilde{\Omega}_n^L(s) x(-s)^n-\langle\left(m_i \mp m_j\right)\left(\bar{q}_i q_i \mp \bar{q}_j q_j\right)\rangle\bigg\rbrace\,,
	\end{align}
	where the upper/lower sign corresponds to the V/A component. The $\tilde{\Omega}^{L+T/L}$ coefficients are given by:
	\begin{align}
		\tilde{\Omega}^{L+T}(s)=&\hspace{1mm}\frac{1}{6}\langle G^2\rangle+2\langle m_i \bar{q}_i q_i+m_j\bar{q}_j q_j\rangle \tilde{q}^{L+T}_n\pm \frac{8}{3}\langle m_j \bar{q}_i q_i+m_i \bar{q}_j q_j\rangle\tilde{t}^{L+T}_n+\sum_{k}\left\langle m_k \bar{q}_k q_k\right\rangle\nonumber\\&-\frac{3}{\pi^2}\bigg\lbrace\left(m_i^4+m_j^4\right) \tilde{h}^{L+T}_n-m^2_i m^2_j \tilde{g}_n^{L+T}\pm\frac{5}{3} m_i m_j\left(m^2_i+m^2_j\right) \tilde{k}_n^{L+T}\nonumber\\&\bs+\sum_{k}m^4_k \tilde{j}^{L+T}_n+2\sum_{k\neq l}m^2_k m^2_l \tilde{u}^{L+T}_n\bigg\rbrace \,, \\
		\tilde{\Omega}^L(s)=&\left(m^2_i+m^2_j\right)\tilde{h}_n^{L}\pm \frac{3}{2}m_i m_j \tilde{k}^L+\sum_{k} m_k^2 \tilde{j}^L_n\,,
	\end{align}
	where $\langle m_i \bar{q}_jq_j\rangle\equiv\langle0\vert m_i \bar{q}_j q_j\vert0\rangle(-\xi^2 s)$, $m_i=m_j(-\xi^2s)$, $\langle G^2\rangle\equiv\langle0\vert G^2\vert0\rangle(-\xi^2 s)$ and $\xi$ is the scale parameter to keep track of the dependence of the renormalization scale.  The RG-evolution of the perturbative coefficients and the condensates can be found in the Ref.~\cite{Pich:1999hc}.
	
	The relevant OPE corrections to strange quark mass determination are as follows:
	\begin{align}
		\delta &\mathcal{D}^{L+T}_4(s)\equiv\mathcal{D}^{L+T,V+A}_{4,ud}(s)-\mathcal{D}^{L+T,V+A}_{4,us}(s)\nonumber\\&=\frac{-4\delta O_4}{s^2} \sum_{n=0}\tilde{q}^{L+T}_n x(-s)^n+\frac{6}{\pi^2s^2}m_s(-s)^4(1-\epsilon_d^2)\sum_{n=0}\lbrace\left(1+\epsilon_d^2\right)\tilde{h}^{L+T}_n -\epsilon_u^2\tilde{g}^{L+T}_n\rbrace x(-s)^n\,,
		\label{eq:dim4LTAdler}
	\end{align}
	\begin{align}
		\delta& \mathcal{D}_4^{L}(s)\equiv\mathcal{D}^{L,V+A}_{4,ud}(s)-\mathcal{D}^{L,V+A}_{4,us}(s)\nonumber\\&=\frac{2\delta O_4}{s\mtsq}-\frac{3}{\pi^2 s \mtsq}m_s^4(1-\epsilon_d^2)\sum_{n=0}\lbrace(1+\epsilon_d^2)(\tilde{h}^{L}_n+\tilde{j}^{L}_n)+\epsilon_u^2(2 \tilde{h}^{L}_n-3\tilde{k}^{L}_n+\tilde{j}_n)\rbrace x(-s)^n\,,
		\label{eq:dim4LAdler}
	\end{align}
	where $\delta O_4=\langle0|m_s \overline{s}s-m_d \overline{d}d|\rangle(-\xi^2 s)$ with $\epsilon_u=m_u/m_s$ and  $\epsilon_u=m_u/m_s$. Using the numerical values
	\begin{align}
		&v_s=0.738\pm0.029   \text{\cite{Albuquerque:2009pr}}\,, \quad f_{\pi}=92.1\pm0.8\MeV\,,\quad m_{\pi}=139.6\MeV \text{\cite{Zyla:2020zbs}}\nonumber\\
		& \epsilon_d=0.053\pm0.002\,, \quad \epsilon_u=0.029\pm0.003\text{\cite{Leutwyler:1996eq}}\,,
	\end{align}
	$\delta O_4$ can be estimated similar to Ref.~\cite{Pich:1999hc} as:
	\begin{align}
		\delta O_4&=\left(v_s m_s-m_d\right)\langle0|\overline{d}d|0\rangle\simeq-\frac{m_s}{2 \hat{m}}\left(v_s-\epsilon_d\right)f_{\pi}^2 m_{\pi}^2\nonumber\\&=-\left(1.54\pm.08\right)\times10^3 \GeV^{-4}\,.
	\end{align}
	
	\section{The behavior of Leading Order Perturbative Mass Corrections in Different Renormalization Schemes}\label{sec:dim_2_behaviour}
	The FOPT and CIPT are the two versions of perturbative theory for the QCD analysis of the $\tau$ decay frequently used in the literature, and it has been further extended by including an RGSPT version of perturbation theory~\cite{Abbas:2012py,Ananthanarayan:2016kll}. It has been shown in Ref.~\cite{Abbas:2012py} that the $\delta^{(0)}$ contributions from the RGSPT scheme approach CIPT at higher orders of the perturbation theory, and the corresponding numerical value of the strong coupling constant lies closer to the CIPT value. A similar behavior is also observed in this study for the higher dimensional operators, but with the advantage that the scale dependence for higher moments is under control in the case of RGSPT compared to the FOPT and CIPT. This behavior is due to the fact that both schemes sum the same logarithms. Before moving on to the strange quark mass determination, the convergence behavior of the leading order mass corrections must be analyzed carefully for different schemes. This exercise is performed in the rest of the section. \par
	The leading order mass corrections to moment $\delta R^{kl,2}_{\tau}$ in Eq.~\eqref{eq:delRkl} are given by:
	\begin{equation}
		\delta R^{kl,D=2}=24\frac{ m_s(\xi^2\mtsq)^2 }{ M_{\tau}^2}S_{EW} \left(1-\epsilon _d^2\right)\Delta_{kl}(x,\xi)\,,
		\label{eq:rkl2}
	\end{equation}
	where,
	\begin{equation}
		\Delta_{kl}(x,\xi)\equiv\frac{3}{4}\Delta_{kl}^{L+T}(x,\xi)+\frac{1}{4}\Delta_{kl}^{L}(x,\xi)\,,
		\label{eq:Delta}
	\end{equation}
	and the $\Delta_{kl}^{J}(x,\xi)$ are the contributions from the Adler functions $\delta\mathcal{D}^{J}$ involving Eqs.~\eqref{eq:su3adlerLT}, \eqref{eq:su3adlerL}, \eqref{eq:summed_D2} evaluated along a contour in the complex plane with the kernels presented in Table~\eqref{tab:klmoments}. These functions are calculated differently in various schemes explained in the later subsections. \par
	It should be noted that the leading-order mass corrections are presented to remember where the perturbative series is truncated in the prescription I.
	\subsection{CIPT scheme}\label{sec:CIPT_intro}
	In CIPT, the masses and the strong coupling evolved along the contour in the complex plane by solving the RGE numerically. By construction, it does not suffer from the problem of large logarithm along the contour.
	Following the Refs.~\cite{Pich:1998yn,Pich:1999hc,Pich:2020gzz}, dimension-2 contribution to $``L+T"-$component moments can be organized in terms of contour integrals:
	\begin{equation}
		\Delta^{L+T}_{kl}(x,\xi)=-\frac{1}{4\pi i}\sum_{n=0}\tilde{d}^{L+T}_n(\xi)\oint_{|x_c|=1}\frac{dx_c}{x_c^2} \mathcal{F}_{kl}^{L+T}(x_c)\hs\frac{m_s^2(-\xi^2 M^2_{\tau} x_c)}{m_s^2(M^2_{\tau})} x^n(-\xi^2 M^2_{\tau} x_c)\,,
		\label{eq:delta_lt}
	\end{equation}
	and for the longitudinal component:
	\begin{equation}
		\Delta^{L}_{kl}(x,\xi)=\frac{1}{2\pi i}\sum_{n=0}\tilde{d}^{L}_n(\xi)\oint_{|x_c|=1}\frac{dx}{x_c} \mathcal{F}_{kl}^{L}(x_c)\frac{m_s^2(-\xi^2 M^2_{\tau} x)}{m_s^2(\mtsq)} x^n(-\xi^2 M^2_{\tau} x_c)\,,
		\label{eq:delta_l}
	\end{equation}
	The dimension-2 contributions to $\Delta_{kl}^{L+T}$ for $x(\mtsq)=0.3187/\pi$ contributions of different orders are given by:
	\begin{align}     \label{eq:dim2_CI}
		\Delta ^{L+T}_{0,0}&=\{0.7717,0.2198,0.0777,-0.0326,-0.135\}\nonumber\,,\\
		\Delta ^{L+T}_{1,0}&=\{0.9247,0.3324,0.1951,0.0866,-0.0375\}\nonumber\,,\\
		\Delta ^{L+T}_{2,0}&=\{1.0605,0.4410,0.3202,0.2302,0.1019\}\nonumber\,,\\
		\Delta ^{L+T}_{3,0}&=\{1.1883,0.5504,0.4567,0.4021,0.2897\}\nonumber\,,\\
		\Delta ^{L+T}_{4,0}&=\{1.3130,0.6634,0.6073,0.6065,0.5337\}\,.
	\end{align}
	which shows good convergence up to $(2,0)-$moment. The longitudinal contributions are:
	\begin{align}
		\Delta ^L_{0,0}&=\{1.6031,1.1990,1.1583,1.3023,1.6245\}\nonumber\,,\\
		\Delta ^L_{1,0}&=\{1.3832,1.1358,1.1970,1.4642,1.9856\}\nonumber\,,\\
		\Delta ^L_{2,0}&=\{1.2563, 1.1158, 1.2635, 1.6553, 2.4004\}\nonumber\,,\\
		\Delta ^L_{3,0}&=\{1.1783, 1.1204, 1.3494, 1.8740, 2.8757\}\nonumber\,,\\
		\Delta ^L_{4,0}&=\{1.1301, 1.1418, 1.4517, 2.1216, 3.4196\}\,,
		\label{eq:CIPT_L}
	\end{align}
	and we can see that longitudinal contributions show divergent behavior. The total perturbative contributions of dimension-2 is obtained using Eq.~\eqref{eq:Delta} are:
	\begin{align}
		\Delta_{0,0}&=\{0.9795, 0.4646, 0.3478, 0.3011, 0.3050\}\nonumber\,,\\
		\Delta_{1,0}&=\{1.0393, 0.5333, 0.4456, 0.4310, 0.4682\}\nonumber\,,\\
		\Delta_{2,0}&=\{1.1094, 0.6097, 0.5560, 0.5865, 0.6765\}\nonumber\,,\\
		\Delta_{3,0}&=\{1.1858, 0.6929, 0.6799, 0.7701, 0.9362\}\nonumber\,,\\
		\Delta_{4,0}&=\{1.2673, 0.7830, 0.8184, 0.9853, 1.2552\}\,.
		\label{eq:Delta_CIPT}
	\end{align}
	It is clear from Eq.~\eqref{eq:Delta_CIPT} that the pathological longitudinal contributions are a restricting factor in getting any reliable determination from CIPT unless we truncate the perturbative series to the minimum term.
	\subsection{FOPT scheme}
	In FOPT, the perturbative series for the Adler function is truncated to a given order in $\alpha_s(\mu)$, and running logarithms are integrated analytically along the contour in the complex energy plane\cite{Pich:1998yn,Beneke:2008ad}. The $\Delta_{kl}^{J}$ for FOPT is evaluated by inserting Eq.~\eqref{eq:su3adlerLT1},\eqref{eq:su3adlerL1} in Eq.~\eqref{eq:rkl2} and can be written as:
	\begin{equation}
		\Delta_{kl}^{J}(x,\xi)=\sum_{i=0}^{4} \sum_{j=0}^{i} x^i(\xi^2 \mtsq)\hs\tilde{d}^{J}_{i,j}H^{kl,J}_{j}(x,\xi) \,,
	\end{equation}
	where $H_i^{kl,J} (\xi)$ are evaluated analytically:
	\begin{align}
		H^{kl,L+T}_n(x,\xi)\equiv&\frac{-1}{4\pi i}\oint_{|x_c|=1} \frac{dx_c}{x_c^2}\mathcal{F}^{L+T}_{kl}(x_c)\log^n\left(\frac{-\xi^2}{ x_c}\right)\,,\\
		H^{kl,L}_n(x,\xi)\equiv&	\frac{1}{2\pi i}\oint_{|x_c|=1} \frac{dx_c}{x_c}\mathcal{F}^{L}_{kl}(x_c)\log^n\left(\frac{-\xi^2}{x_c}\right)\,.
	\end{align}
	\par Evaluating the above integrals, the $\Delta_{kl}^{L+T}$ contribution, using FOPT, at different orders of perturbative series is given by:
	\begin{align}
		\Delta ^{L+T}_{0,0}&=\{1.0000,0.4058,0.2575,0.1544,0.0163\}\nonumber\,,\\
		\Delta ^{L+T}_{1,0}&=\{1.0000,0.5072,0.4168,0.3679,0.2971\}\nonumber\,,\\
		\Delta ^{L+T}_{2,0}&=\{1.0000,0.5782,0.5366,0.5414,0.5429\}\nonumber\,,\\
		\Delta ^{L+T}_{3,0}&=\{1.0000,0.6323,0.6330,0.6892,0.7636\}\nonumber\,,\\
		\Delta ^{L+T}_{4,0}&=\{1.0000,0.6758,0.7140,0.8189,0.9654\}\,.
		\label{eq:dim2_FO}
	\end{align}
	$\Delta_{kl}^{L}$ are given by:
	\begin{align}
		\Delta ^L_{0,0}&=\{1.0000,0.9468,1.1319,1.3807,1.7855\}\nonumber\,,\\
		\Delta ^L_{1,0}&=\{0.7500,0.7482,0.9442,1.2183,1.6559\}\nonumber\,,\\
		\Delta ^L_{2,0}&=\{0.6000,0.6229,0.8184,1.1006,1.5520\}\nonumber\,,\\
		\Delta ^L_{3,0}&=\{0.5000,0.5360,0.7271,1.0098,1.4662\}\nonumber\,,\\
		\Delta ^L_{4,0}&=\{0.4286,0.4718,0.6570,0.9371,1.3937\}\,.
		\label{eq:FOPT_L}
	\end{align}
	We can see that the longitudinal piece has a bad convergence in the FOPT scheme. The total contribution $\Delta_{kl}$ is:
	\begin{align}
		\Delta_{0,0}&=\{1.0000, 0.5410, 0.4761, 0.4610, 0.4586\}\nonumber\,,\\
		\Delta_{1,0}&=\{0.9375, 0.5675, 0.5486, 0.5805, 0.6368\}\nonumber\,,\\
		\Delta_{2,0}&=\{0.9000, 0.5894, 0.6071, 0.6812, 0.7952\}\nonumber\,,\\
		\Delta_{3,0}&=\{0.8750, 0.6082, 0.6565, 0.7694, 0.9393\}\nonumber\,,\\
		\Delta_{4,0}&=\{0.8571, 0.6248, 0.6997, 0.8484, 1.0725\}\,.
		\label{eq:Delta_FOPT}
	\end{align}
	We can see that the convergence behavior of dimension-2 contribution in Eq.~\eqref{eq:Delta_FOPT} is not very different from the CIPT scheme in Eq.~\eqref{eq:Delta_CIPT}.
	\subsection{RGSPT Scheme}
	In RGSPT, masses and coupling are fixed at some renormalization scale, but the RG-summed running logarithms are evolved around the contour. Interestingly, contour integration can be done analytically, similar to FOPT. However, due to the summation of the running logarithms, the resulting perturbative contributions are very much closer to the CIPT numbers.\par
	The perturbative series in RGSPT scheme for the dimension-2 Adler function has the form:
	\begin{equation}
		\Delta_{kl}^{J}(x,\xi)=\sum_{i=0}^4 \sum_{n=0}^i \sum_{m=0}^n x^i\left(\xi^2\mtsq\right) \hs\tilde{T}^{J}_{i,n,m} K^{kl,J}_{n,m}(x,\xi)\,,
	\end{equation}
	which is obtained by inserting Eq.~\eqref{eq:summed_D2} into Eq.~\eqref{eq:rkl2} and the corresponding contour integrals $K^{kl,J}_{n,m}(x,\xi)$ have the following form:
	\begin{align}
		K^{kl,L+T}_{n,m}(x,\xi)\equiv&\frac{-1}{4\pi i}\oint_{|x_c|=1} \frac{dx_c}{x_c^2}\mathcal{F}^{L+T}_{kl}(x_c)\frac{\log^n(1-\beta_0x(\xi^2\mtsq)\log(-\xi^2/ x_c))}{(1-\beta_0 x(\xi^2\mtsq)\log(-\xi^2/ x_c))^m}\,,\\
		K^{kl,L}_{n,m}(x,\xi)\equiv&	\frac{1}{2\pi i}\oint_{|x_c|=1} \frac{dx_c}{x_c}\mathcal{F}^{L}_{kl}(x_c)\frac{\log^n(1-\beta_0x(\xi^2\mtsq)\log(-\xi^2 /x_c))}{(1-\beta_0 x(\xi^2\mtsq)\log(-\xi^2 /x_c))^m}\,.
	\end{align}
	 The $\Delta^{L+T}_{i,0}$ contributions for different moments are given by:
	\begin{align}
		\Delta ^{L+T}_{0,0}&=\{0.8878, 0.2307, 0.0799, -0.0328, -0.1561\}\nonumber\,,\\
		\Delta ^{L+T}_{1,0}&=\{0.9990, 0.3690, 0.2263, 0.1220, -0.0176\}\nonumber\,,\\
		\Delta ^{L+T}_{2,0}&=\{1.0885, 0.4931, 0.3736, 0.2980, 0.1691\}\nonumber\,,\\
		\Delta ^{L+T}_{3,0}&=\{1.1652, 0.6095, 0.5246, 0.4957, 0.4043\}\nonumber\,,\\
		\Delta ^{L+T}_{4,0}&=\{1.2336, 0.7212, 0.6806, 0.7153, 0.6891\}\,.
	\end{align}
	and the $\Delta^{L}_{i,0}$ have the form:
	\begin{align}
		\Delta ^L_{0,0}&=\{1.4048, 1.2210, 1.2280, 1.3899, 1.7560\}\nonumber\,,\\
		\Delta ^L_{1,0}&=\{1.1360, 1.1034, 1.2194, 1.5005, 2.0514\}\nonumber\,,\\
		\Delta ^L_{2,0}&=\{0.9687, 1.0302, 1.2287, 1.6169, 2.3536\}\nonumber\,,\\
		\Delta ^L_{3,0}&=\{0.8538, 0.9808, 1.2477, 1.7375, 2.6655\}\nonumber\,,\\
		\Delta ^L_{4,0}&=\{0.7697, 0.9459, 1.2727, 1.8617, 2.9888\}\,,
		\label{eq:RGSPT_L}
	\end{align}
	and the $\Delta(x)$ behave as:
	\begin{align}
		\Delta_{0,0}&=\{1.0171, 0.4783, 0.3669, 0.3229, 0.3219\}\nonumber\,,\\
		\Delta_{1,0}&=\{1.0332, 0.5526, 0.4746, 0.4666, 0.4996\}\nonumber\,,\\
		\Delta_{2,0}&=\{1.0585, 0.6274, 0.5874, 0.6277, 0.7153\}\nonumber\,,\\
		\Delta_{3,0}&=\{1.0874, 0.7023, 0.7054, 0.8061, 0.9696\}\nonumber\,,\\
		\Delta_{4,0}&=\{1.1176, 0.7774, 0.8286, 1.0019, 1.2640\}\,,
		\label{eq:Delta_RGSPT}
	\end{align}
	We can see from the numerical values provided in Eq.~\eqref{eq:CIPT_L}, Eq.~\eqref{eq:FOPT_L} and Eq.~\eqref{eq:RGSPT_L} that the longitudinal contributions have a convergence issue, and it is difficult to get the reliable determinations using them as input. However, the important ingredient in the mass determination is $\Delta_{i,j}$, defined in Eq.~\eqref{eq:Delta}. We can see from the numerical values presented in Eq.~\eqref{eq:Delta_CIPT}, Eq.~\eqref{eq:Delta_FOPT}, and Eq.~\eqref{eq:Delta_RGSPT} that these inputs can be taken in the mass determination if we truncate the perturbation series to the term which gives minimum contribution to it. This minimum term of the perturbative series is taken as the truncation uncertainty. This prescription has already been advocated in Ref.~\cite{Chen:2001qf}, and we have termed this procedure of truncation as \textbf{prescription I}. Another choice is to use all available terms of the perturbation series coefficients of the Adler function, including the estimate for the unknown $\order{\alpha_s^4}$ term of the $``L+T"-$component of the dimension-2 Adler function and termed as \textbf{prescription II}. These prescriptions have some advantages and disadvantages, which will be discussed later.
	\begin{table}
		\begin{center}
			\begin{tabular}{|c|c|c|}
				\hline
				\multirow{1}{*}{\shortstack{Moments\\(k,l)}}&\multicolumn{2}{c|}{$\delta R^{kl}_{\tau}$}\\\cline{2-3}
				&ALEPH &OPAL\\	\hline
				\cline{2-3}
				(0,0)&$0.374 \pm 0.133$&$0.332 \pm 0.10 $\\ \hline
				(1,0)&$0.398 \pm0.077$&$0.326 \pm 0.078 $\\ \hline
				(2,0)&$0.399 \pm 0.053$&$0.340 \pm 0.058$\\ \hline
				(3,0)&$0.396 \pm 0.042$&$0.353 \pm 0.046$\\ \hline
				(4,0)&$0.395 \pm 0.034$&$0.367 \pm 0.037$\\ \hline
			\end{tabular}
		\end{center}
		\caption{Spectral moments from ALEPH \cite{ALEPH:1999uux,Chen:2001qf} and OPAL\cite{OPAL:2004icu}. OPAL moments are calculated using current value of $\vert V_{us}\vert =0.2243\pm0.0008$ quoted in the PDG\cite{Zyla:2020zbs}.}
		\label{tab:rkl_exp}
	\end{table}
	\section{Phenomenological contribution to the longitudinal sector}\label{sec:rev_pheno}
	We can see from the section~\eqref{sec:dim_2_behaviour} that although the contributions from the $``L+T"$ part of the dimension-2 has a better convergence for CIPT and RGSPT relative to the FOPT, the longitudinal contributions are forcing us to truncate the higher-order terms. These pathological contributions get enhanced for higher moments and restrict one to use only the leading-order term of the perturbation series. This problem is cured by replacing the longitudinal perturbative series contributions with the phenomenological contributions from the chiral perturbation theory \cite{Gamiz:2002nu,Jamin:2001zq,Jamin:2001zr,Maltman:2001gc,Maltman:2001sv}. These contributions carry significantly less theoretical uncertainty and agree well with the corresponding pQCD results, as shown in Ref.~\cite{Gamiz:2002nu}. With these advantages at hand, the strange quark mass determination using the pQCD contribution from the $``L+T"-$component of the Adler function combined with phenomenological contributions for longitudinal contributions in section~\eqref{sec:pheno_ms} can be performed. \par
	The relevant quantities of interest for phenomenological contributions to $R^{kl,L}_{ij,V/A}$, the longitudinal component of Eq.~\eqref{Rtau_def}, are vector/axial-vector spectral functions $\rho^{V/A}_{ij}(s)$. They are related by:
	\begin{equation}
		R^{kl,L}_{ij,V/A}=-24\pi^2\int_{0}^{1}dx_c\left(1-x_c\right)^{2+k}x_c^{l+1} \rho^{V/A,L}_{ij}(\mtsq x_c)\,.
		\label{eq:Rkl_pheno}
	\end{equation}
	The pseudoscalar spectral function receives contributions from pion and kaon mass poles and higher resonances in the strange and non-strange channels. We are using the Maltman and Kambor \cite{Maltman:2001sv} parametrization of the pseudoscalar spectral function for the $us$ and $ud$ channels in our analysis, which is given by:
	\begin{equation}
		s^2 \rho^{A,L}_{us}(s)=2 f_K^2 m_K^2\delta(s-m_K^2)+\sum_{i=1,2}2 f_i^2 M_i^2 B_i(s)\,.
		\label{eq:rho_pseudoscalar}
	\end{equation}
	Here $f_i$ and $M_i$ are the decay constants and masses of resonances, and $B_i(s)$ is the Breit-Wigner resonance function taking the form:
	\begin{equation}
		B_i(s)=\frac{1}{\pi} \frac{\Gamma_i M_i}{(s-M_i^2)^2+\Gamma_i^2 M_i^2}\,,
	\end{equation}
	where $\Gamma_i $ is the decay width of the resonances. The spectral function for the $ud$ channel is obtained by replacing the kaon terms with the pion in Eq.~\eqref{eq:rho_pseudoscalar}. For the resonance contributions to pseudoscalar $ud$ and $us$ channels appearing Eq.~\eqref{eq:rho_pseudoscalar}, we have used the following data:
	\begin{table}[H]
		\centering
		\begin{tabular}{|c|c|c|c|c|}
			\hline
			\text{ }&$\pi(1300)$&$\pi(1800)$&$K(1460)$&$K(1800)$\\\hline
			$M_i(\MeV)$& 1300&1810&1482&1830\\\hline
			$\Gamma_i(\MeV)$&400&215&335&250\\\hline
			$f_i(\MeV)$&$2.2\pm0.46$&$0.19\pm0.19$&$21.4\pm2.8$&$4.5\pm 4.5$\\\hline
		\end{tabular}
		\caption{Masses and decay width are taken from PDG \cite{Zyla:2020zbs} and decay constants from Ref.~ \cite{Maltman:2001gc}.}
	\end{table}
	The vector component of the spectral function receives dominant contributions from the scalar channels $K\pi$,$K\eta$, and $K \eta'$ and the spectral function has the following form~\cite{Jamin:2001zr}:
	\begin{equation}
		\rho^{V,L}_{uj}(s)=\frac{3\Delta^2_{K\pi}}{32\pi^2}\sum_{i=\{\pi,\eta,\eta'\}}\sigma_{Ki}|F_{Ki}(s)|^2\,,
		\label{eq:rho_vector}
	\end{equation}
	where $\Delta_{K\pi}\equiv M_K^2-M_\pi^2$. The phase space factor $\sigma_{Ki}(s)$ are given by:
	\begin{equation}
		\sigma_{Ki}(s)=\theta(s-(M_K+M_i)^2)\sqrt{\left(1-\frac{\left(M_K+M_i\right)^2}{s}\right)\left(1+\frac{\left(M_K-M_i\right)^2}{s}\right)}\,.
	\end{equation}
	The strangeness changing scalar form factors $F_{Ki}(s)$ are defined by:
	\begin{equation}
		\langle\Omega|\partial^{\mu}\left(\overline{s}\gamma_{\mu}u |\Omega\rangle \right)\equiv-i \sqrt{\frac{3}{2}} \Delta_{K\pi} F_{K \pi}(s)\,,
	\end{equation}
	and can be found in Ref.~\cite{Jamin:2001zq}. Detailed discussion on the application of these form factors in the extraction of strange quark mass can be found in the Refs.~\cite{Jamin:2001zr,Gamiz:2002nu}.
	\section{Strange quark mass determination from pQCD}\label{sec:ms_pert}
	The strange quark mass determination in this section is based on the method used in Refs.~\cite{Pich:1999hc,Chen:2001qf,Ananthanarayan:2016kll}. In addition, we have employed different schemes to perform the comparative study. The strange quark mass determination from hadronic $\tau$ decays using RGSPT has been performed Ref.~\cite{Ananthanarayan:2016kll}. However, the uncertainties coming from the truncation of perturbative series and the scale dependence of strange quark masses were neglected. We have improved the previous determination using pQCD inputs by including these uncertainties and the determinations made in the two prescriptions mentioned in section~\eqref{sec:dim_2_behaviour}.\par
	It should be noted that the higher dimensional OPE contributions ($d>4$) to the Adler functions, which are numerically small \cite{Pich:1999hc} and not considered in this analysis. The strange quark mass is determined by supplying experimental and theoretical inputs to Eq.~\eqref{eq:delRkl}. The RHS of the equation is provided with theory inputs from dimension-2 contributions Eqs.~\eqref{eq:su3adlerLT}, \eqref{eq:su3adlerL} and dimension-4 with Eqs.~\eqref{eq:dim4LTAdler},\eqref{eq:dim4LAdler}. These quantities are evaluated along the complex contour in different schemes, as explained in section~\eqref{sec:dim_2_behaviour}. We present our weighted averaged determinations for $m_s(\mtsq)$ from different moments in Table~\eqref{tab:mspertweighted} for different schemes. The details of various sources of uncertainty in the two prescriptions are presented in appendix~\eqref{app:pQCD_mass}.\par
\begin{figure}[H]
			\centering
			\begin{subfigure}
				\centering
				\includegraphics[width=.48\linewidth]{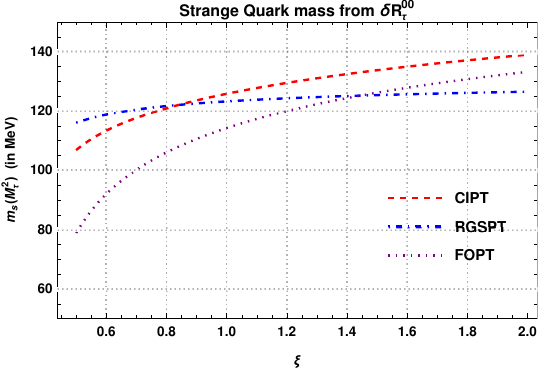}
			\end{subfigure}
			\begin{subfigure}
				\centering
				\includegraphics[width=.48\linewidth]{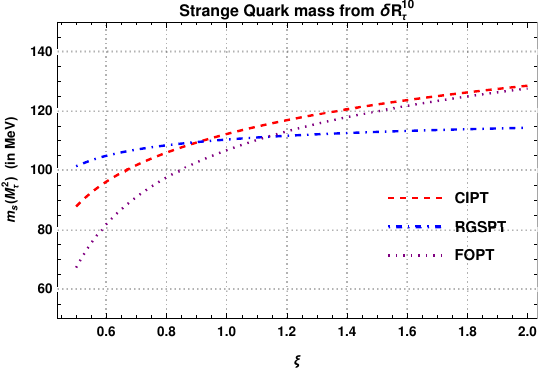}
			\end{subfigure}
		\newline
			\begin{subfigure}
				\centering
				\includegraphics[width=.48\linewidth]{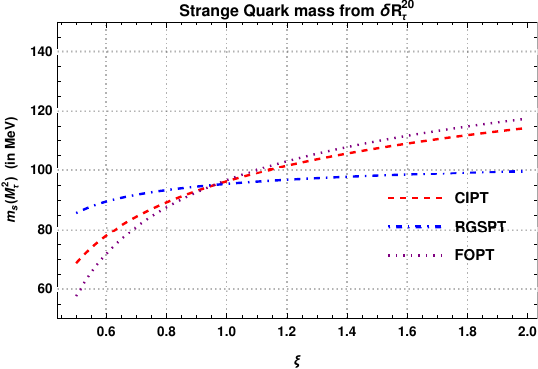}
			\end{subfigure}
			\begin{subfigure}
				\centering
				\includegraphics[width=.48\linewidth]{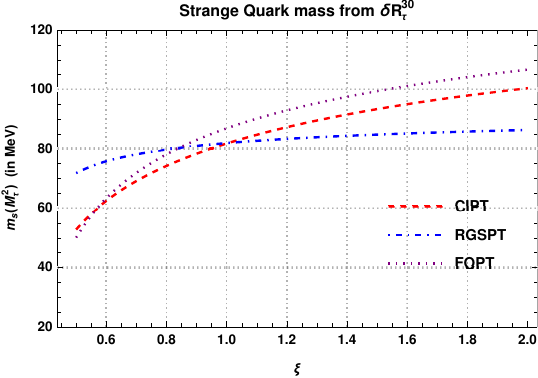}
			\end{subfigure}
			\caption{ The scale variation of the strange quark mass in different perturbative schemes using pQCD inputs.}
			\label{fig:ms_scdep}
		\end{figure}

	We can see from Table~\eqref{tab:mspertweighted} that the strange quark mass determination from different schemes agrees with each other within uncertainty. It is also evident from the tables presented in appendix~\eqref{app:pQCD_mass} that the uncertainties in the final strange quark mass are higher in prescription I than in prescription II mainly due to the truncation of the perturbative series. We also emphasize that the systematic comparison of the behavior of perturbative series in different schemes can only be made in prescription II, where the same order information is used. The RGSPT provides better control over the theoretical uncertainty by minimizing the renormalization scale dependence. The scale dependence of the strange quark mass for various moments is shown in Fig.~\eqref{fig:ms_scdep}, plotted using prescription II. These plots indicate that the strange mass from the RGSPT scheme is stable for a wider range of scale variations for the moments under consideration. It should be noted that the uncertainties associated with renormalization scale dependence are included only in the range $\xi\in\left[0.75,2.0\right]$ in the strange mass determination in Table~\eqref{tab:mspertweighted}. \par It should be noted that the poor convergence of the longitudinal contributions restricts this method to be applicable in the lower energies $s_0<\mtsq$. Additional uncertainties in the determinations of $m_s$ arise due to the variations of the upper limit of the moment in the integral $s_0$ defined in Eq.~\eqref{eq:Rratio_ms_vus}. These are estimated using the phenomenological determination, discussed in the next section, and are also included in Table~\eqref{tab:mspertweighted}. Further details on the numerical uncertainties using pQCD inputs can be found in the appendix~\eqref{app:pQCD_mass}.

		%
		%
		
		\begin{table}[H]
			\centering
			\begin{tabular}{|c|c|c|c|c|}
				\hline
				 Scheme&\multicolumn{2}{c|}{$m_s(\mtsq)$ in prescription I (in $\MeV$)}&\multicolumn{2}{c|}{$m_s(\mtsq)$ in prescription II (in $\MeV$)} \\
				\cline{2-3}\cline{4-5}
				\text{}&ALEPH&OPAL&ALEPH&OPAL\\ \hline
				CIPT& $117.7\pm28.5$&$105.9\pm27.5$&$93.2\pm24.1$&$84.1\pm22.7$\\ \hline
				FOPT&$129.3\pm33.5$&$116.2\pm29.2$&$94.2\pm25.4$&$85.2\pm23.9$\\ \hline
				RGSPT&$120.2\pm23.4$&$107.7\pm25.1$&$89.4\pm16.4$&$80.1\pm15.8$\\ \hline
			\end{tabular}
			\caption{ Weighted average of the strange quark mass in different perturbative schemes. }
			\label{tab:mspertweighted}
		\end{table}

	\section{Strange quark mass determination using phenomenological inputs}\label{sec:pheno_ms}
	The determination of the strange quark mass in this section is similar to the one used in section~\eqref{sec:ms_pert}, but now the longitudinal Adler function is replaced with the phenomenologically parameterized contributions, as discussed in section~\eqref{sec:rev_pheno}. It should be noted that the $``L+T"$ component of the Adler function at dimension-2 is known to $\order{\alpha_s^3}$ and we do not use its estimate for the $\order{\alpha_s^4}$ coefficient in the determination of $m_s(\mtsq)$ using phenomenological inputs. Contributions from the last known term of the perturbation series of the Adler function are taken as the total truncation uncertainty, similar to that of the previous section. \par
\begin{figure}[H]
			\centering
			\begin{subfigure}
				\centering
				\includegraphics[width=.48\linewidth]{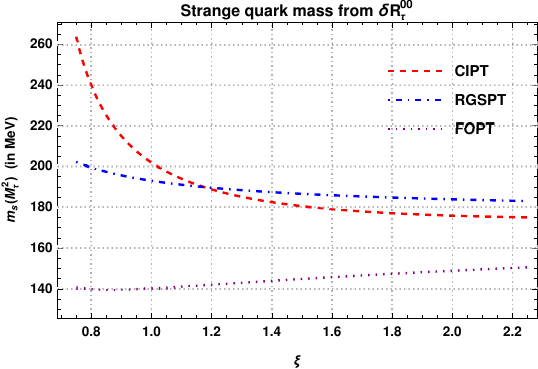}
			\end{subfigure}
			\begin{subfigure}
				\centering
				\includegraphics[width=.48\linewidth]{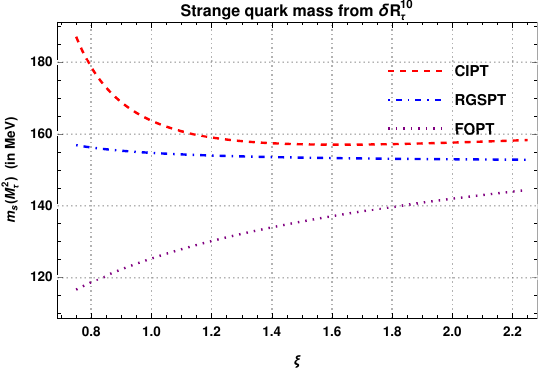}
			\end{subfigure}
		\newline
			\begin{subfigure}
				\centering
				\includegraphics[width=.48\linewidth]{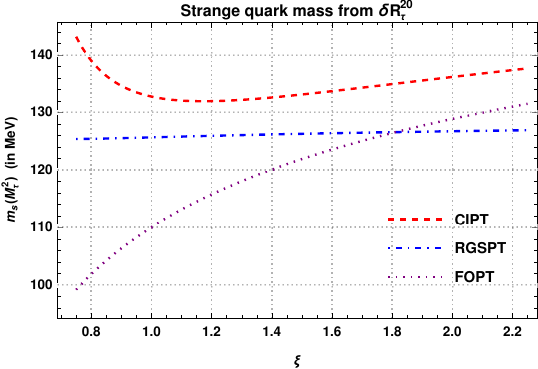}
			\end{subfigure}
			\begin{subfigure}
				\centering
				\includegraphics[width=.48\linewidth]{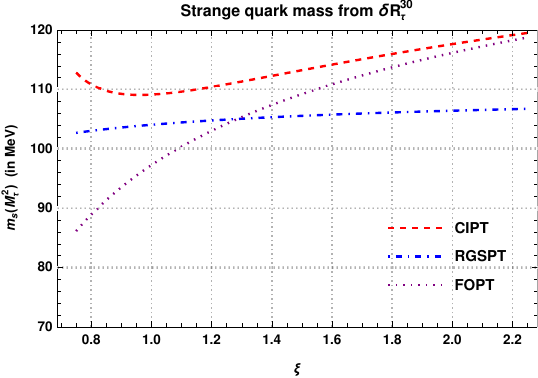}
			\end{subfigure}
			\caption{The scale variation of the strange quark mass obtained from phenomenological inputs in different perturbative schemes.}
			\label{fig:ms_scdeppheno}
		\end{figure}

	Following the discussions of section~\eqref{sec:rev_pheno}, we now have all the necessary ingredients for the strange quark mass determination. Using the transverse contributions used in section~\eqref{sec:ms_pert} and combining them with the input from section~\eqref{sec:rev_pheno}, we determine the strange quark in different schemes. We present our result for the weighted average in Table~\eqref{tab:msphenoweighted2} and further details of the determinations from moments in the appendix~\eqref{app:pheno_mass}. The scale dependence in the $m_s(\mtsq)$ is presented in Fig.~\eqref{fig:ms_scdeppheno} using prescription II. As observed in the previous section, the strange quark mass determinations from the RGSPT scheme are stable over the wider range of scale variation for moments under consideration. The determination of the $m_s$ from the traditional spectral moments is sensitive to the variation of the $s_0$. A typical 5\% variation of the $s_0$ from $\mtsq$ in the range $s_0\in\left[3,\mtsq\right]$ induces variations of $\sim 6-13\%$ in the $m_s$ determinations from moments using the OPAL data. Unfortunately, such variations can not be calculated for the ALEPH moments as the strange spectral function is not publicly available. These uncertainties are estimated from the determinations using the OPAL data.

		\begin{table}[H]
			\centering
			\begin{tabular}{|c|c|c|c|c|}
				\hline
				Scheme&\multicolumn{2}{c|}{$m_s(\mtsq)$ in prescription I (in $\MeV$)}&\multicolumn{2}{c|}{$m_s(\mtsq)$ in prescription II (in $\MeV$)} \\
				\cline{2-3}\cline{4-5}
				\text{}&ALEPH&OPAL&ALEPH&OPAL\\ \hline
				CIPT& $123.3\pm22.3$&$106.3\pm21.5$&$125.1\pm25.1$&$107.5\pm23.9$\\ \hline
				FOPT&$136.6\pm35.0$&$119.5\pm35.4$&$115.8\pm30.1$&$101.6\pm28.3$\\ \hline
				RGSPT&$123.1\pm21.1$&$107.0\pm21.2$&$117.7\pm20.1$&$102.0\pm19.5$\\ \hline
			\end{tabular}
			\caption{The weighted average of strange quark mass in the different perturbative schemes. Phenomenological inputs for the longitudinal contributions are used.}
			\label{tab:msphenoweighted2}
		\end{table}

	\section{Determination of \texorpdfstring{$\vert V_{us}\vert$}{}}\label{sec:Vusextraction}
	The data on strange and non-strange spectral moments for the hadronic $\tau$ decay provided by ALEPH \cite{ALEPH:2005qgp,Davier:2008sk,Davier:2013sfa}, HFLAV\cite{HFLAV:2019otj}, and OPAL \cite{OPAL:2004icu} collaborations can be used to determine the CKM matrix element $\vert V_{us}\vert$. These experimental moments along with the theoretical moments calculated with strange quark mass as input from other sources, can be used to determine $\vert V_{us}\vert$ using the following relation:
	\begin{equation}\label{eq:vus_calc}
		\vert V_{us}\vert=\sqrt{\frac{R_{\tau,S}^{kl}}{R^{kl}_{\tau,V+A}/|V_{ud}|^2-\delta R^{kl}_{\tau,th}}}\,,
	\end{equation}
	where $R^{kl}_{\tau,S}$ and $R^{kl}_{\tau,V+A}$ are experimental inputs and $\delta R^{kl}_{\tau,th}$ is the theory input, in which $m_s(2\GeV)$ is taken as an external input. This method has already been used previously in Refs.~\cite{Maltman:2007ic,Gamiz:2004ar,Gamiz:2006xx,Gamiz:2004gh,Gamiz:2005gh,Gamiz:2002nu}, and it has been observed that the uncertainties are dominated by the experimental data available for the strange component. An additional source of uncertainties is pointed out in Ref.~\cite{Maltman:2007ic} due to $s_0$ variations which can be cured using a different analysis based on the non-spectral weight functions. However, we have restricted this analysis only to the traditional weight functions.\par
	Using $m_s(2\GeV)=93\pm11\MeV$\cite{Zyla:2020zbs} as an external input and ALEPH data \cite{ALEPH:2005qgp,Davier:2008sk,Davier:2013sfa}, we have presented our determination for $\vert V_{us}\vert$ in Table~\eqref{tab:Vus_Aleph1}.

	\begin{table}
		\centering
		\begin{tabular}{|c|c|c|}
			\hline
			Scheme&\multicolumn{2}{c|}{$\vert V_{us}\vert$}\\
			\cline{2-3}
			\text{}&pQCD inputs&Phenomenological inputs\\
			\hline
			CIPT&$0.2174\pm0.0045$&$0.2168\pm0.0044$\\\hline
			FOPT&$0.2183\pm 0.0055$&$0.2179\pm0.0055$\\\hline
			RGSPT&$0.2178\pm0.0046$&$0.2170\pm0.0045$\\
			\hline
		\end{tabular}
		\caption{$\vert V_{us}\vert$ from ALEPH data in different perturbative scheme using $\delta R^{kl}_{\tau,th} $ from section~\eqref{sec:ms_pert} and section~\eqref{sec:pheno_ms}.}
		\label{tab:Vus_Aleph1}
	\end{table}
	The latest branching fraction of hadronic $\tau$ decays into the non-strange channel and slightly more precise strange component from HFLAV\cite{HFLAV:2019otj} can be used to get a more precise determination of $\vert V_{us}\vert$ from this method. The results for different schemes are presented in table \ref{tab:Vus_HFLAV1}.\par   The uncertainties shown in these tables are dominated by those coming from the variation of $s_0\in\left[2.5,\mtsq\right]$ and experimental uncertainty in strange $R_{\tau,S}$ contributions. It should be noted that uncertainties coming from the variation of the $s_0$ in Table~\eqref{tab:Vus_Aleph1} and Table~\eqref{tab:Vus_HFLAV1} are calculated using the experimental data on the spectral function from Refs.~\cite{OPAL:2004icu,Davier:2013sfa}.

	\begin{table}
		\centering
		\begin{tabular}{|c|c|c|}
			\hline
			Scheme&\multicolumn{2}{c|}{$\vert V_{us}\vert$}\\
			\cline{2-3}
			\text{}&pQCD inputs&Phenomenological inputs\\
			\hline
			CIPT&$0.2195\pm0.0047$&$0.2189\pm0.0044$\\\hline
			FOPT&$0.2205\pm 0.0043$&$0.2200\pm0.0037$\\\hline
			RGSPT&$0.2199\pm0.0046$&$0.2191\pm0.0043$\\
			\hline
		\end{tabular}
		\caption{$\vert V_{us}\vert$ from HFLAV data in different perturbative scheme using $\delta R^{kl}_{\tau,th} $ from section~\eqref{sec:ms_pert} and section~\eqref{sec:pheno_ms}.}
		\label{tab:Vus_HFLAV1}
	\end{table}

	The $\vert V_{us}\vert$ determinations from ALEPH \cite{ALEPH:2005qgp,Davier:2008sk,Davier:2013sfa} and  HFLAV\cite{HFLAV:2019otj} are based on $(0,0)-$moment. A detailed analysis for higher moments can be performed using the OPAL\cite{OPAL:2004icu} data, where $(k,0)$ moments for $k=0,\cdots,4$ are also available. These moments are correlated and their correlation should also be considered in the full analysis. Given the large uncertainties in their strange components and unknown precise higher dimensional OPE corrections, we have neglected these correlations among various moments in our determinations from this data.\par
	Using the strange and non-strange moments from OPAL and the theoretical inputs for $\delta R^{kl}_{\tau,th} $ from Section~\eqref{sec:ms_pert}, we present the weighted average of the determination $\vert V_{us}\vert$ from prescription I and prescription II in the Table~\eqref{tab:Vus_weightedmean12}. Details of the $\vert V_{us}\vert$ from moments along with the sources of uncertainties can be found in Table~\eqref{tab:VusAllpert1_Opal} and Table~\eqref{tab:VusAll2pert_Opal}, respectively. We can observe from these tables that the RGSPT is slightly more sensitive to the strange quark mass taken as input, but the overall theory uncertainty coming in this scheme is lesser than CIPT and FOPT in prescription II. We can also see that the divergent nature of the longitudinal component is still an issue causing a large theoretical uncertainty dominating in the higher moments in prescription II in Table~\eqref{tab:VusAll2pert_Opal}.\par
	These shortcomings are slightly improved in the phenomenological determination and can be seen in Table~\eqref{tab:VusAll1pheno_Opal} and Table~\eqref{tab:VusAll2pheno_Opal}. Again, these determinations suffer from large $s_0$ and  theoretical uncertainties, especially those coming from the strange quark mass in the higher moments. It is worth emphasizing that prescription I reduces the dependence on the spectral moments in the  $\vert V_{us}\vert$ determination. The weighted average of these results for $\vert V_{us}\vert$ are presented in Table~\eqref{tab:Vus_weightedmean12}. The RGSPT scheme is slightly more sensitive to the variation of the $s_0$, which, compared to CIPT, dominates in the final average presented in Table~\eqref{tab:VusAll1pheno_Opal} and can be seen in the table presented in appendix~\eqref{app:vus}. 

		\begin{table}[H]
			\centering
			\begin{tabular}{||c||c|c||c|c||}
				\hline\hline
				Scheme&\multicolumn{2}{c|}{$\vert V_{us}\vert$ in prescription I}&\multicolumn{2}{c|}{$\vert V_{us}\vert$ in prescription II}\\
				\cline{2-3}\cline{4-5}
				\text{}&pQCD inputs&Pheno. inputs&pQCD inputs&Pheno. inputs\\
				\hline\hline
				CIPT&$0.2220\pm0.0050$&$0.2212\pm0.0047$&$0.2232\pm0.0048$&$0.2212\pm0.0045$\\ \hline
				FOPT&$0.2212\pm 0.0059$&$0.2220\pm0.0054$&$0.2240\pm 0.0059$&$0.2224\pm0.0054$\\ \hline
				RGSPT&$0.2222\pm0.0051$&$0.2215\pm0.0048$&$0.2238\pm0.0049$&$0.2215\pm0.0047$\\ \hline\hline
			\end{tabular}
			\caption{The weighted average of the determination of $\vert V_{us}\vert$ from the OPAL data in the different perturbative schemes.}
			\label{tab:Vus_weightedmean12}
		\end{table}

	\section{Joint \texorpdfstring{$m_s$}{} and \texorpdfstring{$\vert V_{us}\vert$}{} determination}\label{sec:joint_msVus}
	The experimental moments provided by the OPAL collaboration in Ref.~\cite{OPAL:2004icu} can be used for the joint extraction of $m_s$ and $\vert V_{us}\vert$. It should be noted that the moments provided in Ref.~\cite{OPAL:2004icu} are correlated, and a proper analysis will require their correlations to be taken into account. Given the uncertainties present in the data, we are disregarding these correlations and restrict ourselves to simplified analysis.\par Using the phenomenological parametrization for the longitudinal contributions and perturbative $``L+T"-$component from section~\eqref{sec:rev_pheno} and section~\eqref{sec:OPE_contributions}, we fit the $m_s(2\GeV)$ and $\vert V_{us}\vert$ to Eq.~\eqref{eq:vus_calc} for moments $(k,0)$ with $k=0,1,\cdots,4$. The central values of the joint fit are presented in Table~\eqref{tab:msVusAll_Opal1}. It should be noted that the joint fit gives very close numerical value in both prescription and the difference arise beyond the quoted values in Table~\eqref{tab:msVusAll_Opal1}. These joint fits give smaller values for $m_s(2\GeV)$ and $\vert V_{us}\vert$ compared to the PDG average \cite{Zyla:2020zbs}, but very close to the findings of Gamiz et al. \cite{Gamiz:2004ar} for CIPT and RGSPT.
	\begin{table}
		\centering
		\begin{tabular}{||c||c|c||c|c||}
			\hline\hline
			Scheme&\multicolumn{4}{c|}{Phenomenological inputs}\\
			\cline{2-5}
			\text{}&\multicolumn{2}{c|}{prescription I}&\multicolumn{2}{c|}{prescription II}\\
			\cline{2-3}\cline{4-5}
			\text{}&$m_s(2\MeV)$&$\vert V_{us}\vert$&$m_s(2\MeV)$&$\vert V_{us}\vert$\\
			\hline
			CIPT&75&0.2199&75&0.2199\\\hline
			FOPT&46&0.2227&46&0.2227\\\hline
			RGSPT&73&0.2199&73&0.2199\\
			\hline\hline
		\end{tabular}
		\caption{Joint determination of $m_s(2\GeV)$ and $\vert V_{us}\vert$ from OPAL data with moment~(k,0) with k={0,1,2,3,4} in different perturbative scheme from pQCD inputs as well as using phenomenological inputs.}
		\label{tab:msVusAll_Opal1}
	\end{table}
	
	\section{Summary and conclusion}\label{sec:summary_ms_vus}
	The hadronic $\tau$ decays are important ingredients for extracting various QCD parameters. We have used perturbative schemes CIPT, FOPT, and RGSPT in the extraction of $m_s$, $\vert V_{us}\vert$, and their joint determinations from the experimental inputs available from ALEPH \cite{ALEPH:2005qgp,Davier:2008sk,Davier:2013sfa}, HFLAV\cite{HFLAV:2019otj}, and OPAL \cite{OPAL:2004icu} moments of hadronic $\tau$ decays. To reach the goal, we first calculate the RGSPT coefficients for the dimension-4 operator and use them for determinations of $m_s$ and $\vert V_{us}\vert$. Dimension-6 OPE corrections are known to NLO, and their RG-improvement is discussed in Refs.~\cite{Lanin:1986zs,Adam:1993uu,Boito:2015joa}. Four quark condensates present in these contributions are estimated using the vacuum saturation approximation~\cite{Shifman:1978bw} and found to be numerically very small and are not considered in this study. Higher dimensional OPE corrections are not fully known and are neglected in this study. \par
	The moments calculated using the perturbation theory suffer from convergence issues, so we have employed two prescriptions. The central values of the strange quark mass determinations from prescription I are less spread for different moments than in prescription II. The theoretical uncertainties arising from truncation and scale dependence dominate for higher moments in prescription I for CIPT and FOPT. However, RGSPT has better control over the scale dependence for a wider range of scale variation, even for higher moments as shown in Fig.~\eqref{fig:ms_scdep} and Fig.~\eqref{fig:ms_scdeppheno}. This improvement results in a more precise determination in RGSPT compared to FOPT and CIPT schemes.\par
	The important results of this study for the $m_s(\mtsq)$ determination are presented in Table~\eqref{tab:mspertweighted},\eqref{tab:msphenoweighted2} and for $\vert V_{us}\vert$ determinations in Tables~\eqref{tab:Vus_weightedmean12},~\eqref{tab:Vus_Aleph1}~and \eqref{tab:Vus_HFLAV1}. The joint $m_s$ and $\vert V_{us}\vert$ determination results are presented in Table~\eqref{tab:msVusAll_Opal1}. It should be noted that the ALEPH moments used in the $m_s$ determinations in this study are based on the old $\vert V_{us}\vert$ calculated in Ref.~\cite{Chen:2001qf}. The strange quark mass determinations from the moments are very sensitive to the value of $\vert V_{us}\vert$, and hence we do not consider them in the final average. However, the experimental data for the strange and non-strange moments provided by the OPAL collaboration in Ref.~\cite{OPAL:2004icu} with the current value of $\vert V_{us}\vert=0.2245\pm0.0008$\cite{Zyla:2020zbs} as an input can be used to provide the most updated determination of the strange quark mass.\par We give our final determination for $m_s(\mtsq)$, which comes from the weighted average of strange quark mass determination using RGSPT scheme from Table~\eqref{tab:msphenoweighted2}:
	\begin{equation}
		m_s(\mtsq)=102.0\pm19.5\MeV\bs\text{(OPAL,RGSPT)}\,,
	\end{equation}
	which corresponds to the strange quark mass at $2\GeV$ :
	\begin{equation}
		m_s(2\GeV)=98\pm19\MeV\,
	\end{equation}
	Using ALEPH\cite{Davier:2008sk,Davier:2013sfa} moment, the $\vert V_{us}\vert$ determinations along with their deviation from PDG\cite{Zyla:2020zbs} ( $\vert V_{us}\vert =0.2243\pm0.0008$) and CKM unitarity fit value ($\vert V_{us}\vert=0.2277\pm0.0013$) are:
	\begin{align}
		\vert V_{us}\vert&=0.2168\pm0.0044\quad(1.7\hspace{.5mm}\sigma,2.6\hspace{.5mm}\sigma)&\hs(\text{for CIPT}) \,,\\
		\vert V_{us}\vert&=0.2170\pm0.0045\quad(1.6\hspace{.5mm}\sigma,2.3\hspace{.5mm}\sigma)&\hs(\text{for RGSPT})\,,
	\end{align}
	and from HFLAV\cite{HFLAV:2019otj}:
	\begin{align}
		\vert V_{us}\vert&=0.2189\pm0.0044\quad(1.2\hspace{.5mm}\sigma,1.9\hspace{.5mm}\sigma)&\hs(\text{for CIPT}) \,,\\
		\vert V_{us}\vert&=0.2191\pm0.0043\quad(1.2\hspace{.5mm}\sigma,1.9\hspace{.5mm}\sigma)&\hs(\text{for RGSPT})\,.
	\end{align}
	The weighted average of the $\vert V_{us}\vert$ determinations from OPAL\cite{OPAL:2004icu} using phenomenological inputs is presented in Table~\eqref{tab:Vus_weightedmean12}. The most precise determinations for $\vert V_{us}\vert$ from this table come from CIPT and RGSPT:
	\begin{align}
		\vert V_{us}\vert&=0.2212\pm0.0047, 0.2212\pm0.0045&\hs(\text{for CIPT}) \,,\\
		\vert V_{us}\vert&=0.2215\pm0.0048,0.2215\pm0.0047&\hs(\text{for RGSPT})\,.
	\end{align}%
	The mean values of determinations in these schemes are:
	\begin{align}
		\vert V_{us}\vert&=0.2212\pm 0.0045\quad(0.7\hspace{.5mm}\sigma,1.4\hspace{.5mm}\sigma)&\hs(\text{for CIPT}) \,,\\
		\vert V_{us}\vert&=0.2215\pm 0.0047\quad(0.6\hspace{.5mm}\sigma,1.3\hspace{.5mm}\sigma)&\hs(\text{for RGSPT}) \,.
	\end{align}
	We give our final determinations by the weighted average of these results as:
	\begin{align}
		\vert V_{us}\vert&=0.2189\pm0.0044\quad(1.2\hspace{.5mm}\sigma,1.9\hspace{.5mm}\sigma)&\hs(\text{for CIPT}) \,,\\
		\vert V_{us}\vert&=0.2191\pm0.0043\quad(1.2\hspace{.5mm}\sigma,1.9\hspace{.5mm}\sigma)&\hs(\text{for RGSPT})\,.
	\end{align}%
	The values obtained for $\vert V_{us}\vert$ using OPAL data agree with the PDG average within uncertainties. However, the $\vert V_
	{us}\vert$ determination from ALEPH \cite{ALEPH:2005qgp,Davier:2008sk,Davier:2013sfa} and  HFLAV\cite{HFLAV:2019otj} are more than $1.2\sigma$ and $1.9\sigma$ away from the PDG\cite{Zyla:2020zbs} average and CKM unitarity fit value. It should be noted that the PDG average is already in tension with $2.2\hspace{.5mm}\sigma$ with the CKM unitarity.\par
	The dependence of our determinations on the choice of the moments and their correlation is not considered in this study, and we expect that these can be further improved using non-spectral weights used by Maltman et al. in Refs.~\cite{Kambor:2000dj,Maltman:2006st,Hudspith:2017vew}.

\begin{subappendices}
	
	\section{Anomalous dimension for vacuum energy }
		
	Vacuum anomalous dimension for dimension-4 operators, which has been recently computed to five-loop\cite{Baikov:2018nzi}. Their analytic expression of the diagonal component relevant to this study is given by :
		\begin{align}
			\hat{\gamma}_0^{di}&\equiv  \frac{3}{16\pi^2}\gamma^{ii}_n x^n\nonumber\\&=\frac{3}{16\pi^2}\Bigg\lbrace-1-\frac{4 x}{3}+x^2 \left(\frac{2 \zeta (3)}{3}-\frac{223}{72}\right)+x^3 \left(\frac{346 \zeta (3)}{9}-\frac{1975 \zeta (5)}{54}+\frac{13 \pi ^4}{540}-\frac{3305}{1296}\right)\nn\\&\bs+x^4 \Bigg(\frac{6121 \zeta (3)^2}{864}-\frac{11881 \pi ^4}{8640}+\frac{1680599 \zeta (3)}{2592}+\frac{36001 \zeta (7)}{96}+\frac{93925 \pi ^6}{326592}\nonumber\\&\bs\bs-\frac{59711 \zeta (5)}{48}+\frac{16141627}{248832}\Bigg) \Bigg\rbrace\,.
		\end{align}
		\section{The RGSPT Coefficients Relevant The Dimension-0 and The Dimension-2 Adler Functions} \label{app:summed_sol}
		The first three summed series coefficients are presented as
		\begin{equation}
		\resizebox{\textwidth}{!}{$
			\begin{aligned}
			S_0[w]=& w^{-n_2 \tilde{\gamma }_0-n_1}\\S_1[w]=&w^{-n_2 \tilde{\gamma }_0-n_1-1} \Bigg(T_{1,0}-n_1 \tilde{\beta }_1 L_w+n_2 \times \big(-\tilde{\beta }_1 \tilde{\gamma }_0+\tilde{\gamma }_1+w \tilde{\beta }_1 \tilde{\gamma }_0-\tilde{\beta }_1 \tilde{\gamma }_0 L_w-w \tilde{\gamma }_1\big)\Bigg)\\
			S_2[w]=&w^{-n_2 \tilde{\gamma }_0-n_1-2}\Bigg\lbrace T_{2,0}+T_{1,0} \left(n_2 \left(\tilde{\gamma }_1-\tilde{\beta }_1 \tilde{\gamma }_0\right)+n_2 w \left(\tilde{\beta }_1 \tilde{\gamma }_0-\tilde{\gamma }_1\right)+L_w \left(-n_2 \tilde{\beta }_1 \tilde{\gamma }_0-n_1 \tilde{\beta }_1-\tilde{\beta }_1\right)\right)\nonumber\\&+\bigg\lbrace n_1 \left(\tilde{\beta }_2-\tilde{\beta }_1^2\right)+n_2 \left(-\frac{1}{2} \tilde{\beta }_1^2 \tilde{\gamma }_0-\frac{1}{2} \tilde{\beta }_1 \tilde{\gamma }_1+\frac{1}{2} \tilde{\beta }_2 \tilde{\gamma }_0+\frac{\tilde{\gamma }_2}{2}\right)+w \bigg[n_1 \left(\tilde{\beta }_1^2-\tilde{\beta }_2\right)+n_2 \left(\tilde{\beta }_1^2 \tilde{\gamma }_0-\tilde{\beta }_2 \tilde{\gamma }_0\right)\nonumber\\&+n_2^2 \left(-\tilde{\beta }_1^2 \tilde{\gamma }_0^2+2 \tilde{\beta }_1 \tilde{\gamma }_1 \tilde{\gamma }_0-\tilde{\gamma }_1^2\right)+L_w \big(n_2^2 \left(\tilde{\beta }_1 \tilde{\gamma }_0 \tilde{\gamma }_1-\tilde{\beta }_1^2 \tilde{\gamma }_0^2\right)+n_1 n_2 \left(\tilde{\beta }_1 \tilde{\gamma }_1-\tilde{\beta }_1^2 \tilde{\gamma }_0\right)\big)\bigg]\nonumber\\&+L_w^2 \bigg[\frac{1}{2} n_2^2 \tilde{\beta }_1^2 \tilde{\gamma }_0^2+\frac{1}{2} n_1^2 \tilde{\beta }_1^2+\frac{1}{2} n_2 \tilde{\beta }_1^2 \tilde{\gamma }_0+n_1 \left(n_2 \tilde{\beta }_1^2 \tilde{\gamma }_0+\frac{\tilde{\beta }_1^2}{2}\right)\bigg]+w^2 \bigg[n_2^2 \left(\frac{1}{2} \tilde{\beta }_1^2 \tilde{\gamma }_0^2-\tilde{\beta }_1 \tilde{\gamma }_1 \tilde{\gamma }_0+\frac{\tilde{\gamma }_1^2}{2}\right)\nonumber\\&+n_2 \left(-\frac{1}{2} \tilde{\beta }_1^2 \tilde{\gamma }_0+\frac{1}{2} \tilde{\beta }_1 \tilde{\gamma }_1+\frac{1}{2} \tilde{\beta }_2 \tilde{\gamma }_0-\frac{\tilde{\gamma }_2}{2}\right)\bigg]\bigg\rbrace\Bigg\rbrace
		\end{aligned}$}
	\end{equation}
		where $\tilde{X}_i\equiv X_i/\beta_0$, and the rest of them can be found by solving Eq.~\eqref{summed_de} with the boundary conditions that $S_i[1]=T_{i,0}$ and for simplification of the expressions, we have taken $T_{0,0}=1$.
		\section{Perturbative coefficients Relevant for the Dimension-4 Corrections and their RGSPT coefficients}\label{app:dim4corrections}
		The RG-inaccessible coefficients needed for the dimension-4 operators are calculated in\ Refs.~\cite{Chetyrkin:1985kn,Generalis:1989hf,Broadhurst:1985js,Generalis:1990iy,Loladze:1985qk,Becchi:1980vz,Surguladze:1990sp,Bagan:1985zp,Pascual:1981jr,Jamin:1992se,Generalis:1990id} and their values are:
		\begin{equation}
		\resizebox{\textwidth}{!}{$
			\begin{aligned}
			&p0_0=0,\quad p0_1=1,\quad p0_2=7/6,\quad q0_0=1,\quad q0_1=-1,\quad q0_2=-131/24,\quad hl0_0=1,\quad kl0_0=1\nonumber\\
			&t0_0=0,\quad t0_1=1,\quad t0_2=17/2,\quad h0_0=1,\quad g0_0=1,\quad g0_1=94/9-4/3\zeta_3,\quad k0_0=0,\quad k0_1=1\,.
		\end{aligned}$}
		\end{equation}
		\par Perturbative coefficients involving the condensates terms described in section~\eqref{sec:dim_4_Adler} are:
		\begin{align}
			&p^{L+T}_0(w)=0,\quad p^{L+T}_1(w)=p0_1,\quad p^{L+T}_2(w)=\frac{\beta _1 p0_1}{\beta _0}+\frac{p0_2-\frac{\beta _1 p0_1}{\beta _0}}{w}\,,\nonumber\\&r^{L+T}_0(w)=0,\quad r^{L+T}_1(w)=\frac{\gamma _0 p0_1}{6 \beta _0 w}-\frac{\gamma _0 p0_1}{6 \beta _0},\quad r^{L+T}_2(w)=\frac{\frac{\beta _1 \gamma _0 p0_1}{6 \beta _0^2}-\frac{\gamma _0 p0_2}{6 \beta _0}}{w}\,,\nonumber\\&
			q^{L+T}_0(w)=q0_0,\quad q^{L+T}_1(w)=\frac{q0_1}{w},\quad q^{L+T}_2(w)=\frac{q0_2-\frac{\beta _1 q0_1 \log (w)}{\beta _0}}{w^2}\,,\nonumber\\&
			t^{L+T}_0(w)=0,\quad t^{L+T}_1(w)=\frac{t0_1}{w},\quad t^{L+T}_2(w)=\frac{t0_2-\frac{\beta _1 t0_1 \log (w)}{\beta _0}}{w^2}\,.
		\end{align}
		\par
		The RGSPT coefficients for the coefficients of $m^4$ to $\order{\alpha_s}$ are:
			\begin{equation}
			\resizebox{\textwidth}{!}{$
				\begin{aligned}
			k^{L+T}_0(w)=&\frac{\gamma^{ii}_0 t0_1 \left(2 (1-w) w^{-\frac{4\gamma _0}{\beta _0}}-2 w^{-\frac{4\gamma _0}{\beta _0}}+2\right)}{10 w \left(2 4\gamma _0-2 \beta _0\right)}\,,\quad g^{L+T}_0(w)=g0_0 w^{-\frac{4\gamma _0}{\beta _0}},\quad
			j^{L}_0(w)=0,\quad j^{L}_1(w)=0\,,\\
			k^{L+T}_1(w)=&\frac{\gamma^{ii}_0 w^{-\frac{4\gamma _0}{\beta _0}-2}}{10 \beta _0 \left(\beta _0-4\gamma _0\right){}^2} \Bigg(\beta _1 t0_1 \left(w^{\frac{4\gamma _0}{\beta _0}} \left(\beta _0 (\log (w)+1)-4\gamma _0 \log (w)\right)-\beta _0 w\right)-\beta _0 t0_2 \left(\beta _0-4\gamma _0\right) \left(w^{\frac{4\gamma _0}{\beta _0}}-w\right)\Bigg)\nonumber\\&+k0_1 w^{-\frac{4\gamma _0}{\beta _0}-1}+\frac{\frac{\gamma^{ii}_1 t0_1}{w}-\gamma^{ii}_1 t0_1 w^{-\frac{4\gamma _0}{\beta _0}-1}}{10 4\gamma _0}\,\\
			h^{L+T}_0(w)=&h0_0 w^{-\frac{4 \gamma _0}{\beta _0}}+\frac{q0_1 \left(\frac{1}{w}-w^{-\frac{4 \gamma _0}{\beta _0}}\right)}{2 \left(\beta _0-4 \gamma _0\right)}-\frac{q0_0 \gamma_1 w^{-\frac{4 \gamma _0}{\beta _0}} \left(\beta _0 \left(w^{\frac{4 \gamma _0}{\beta _0}}-1\right)-4 \gamma _0 (w-1)\right)}{2 \beta _0 \gamma _0 \left(\beta _0-4 \gamma _0\right)}\nonumber\\&\frac{q0_0 w^{-\frac{4 \gamma _0}{\beta _0}} \left(\beta _0^2 \left(-4 \beta _0+3 \beta _1+16 \gamma _0\right) \left(w^{\frac{4 \gamma _0}{\beta _0}}-1\right)+12 \beta _1 \gamma _0 \left(4 \gamma _0 (-w+\log (w)+1)-\beta _0 \log (w)\right)\right)}{24 \beta _0^2 \gamma _0 \left(\beta _0-4 \gamma _0\right)}	\,,\\
			h^{L+T}_1(w)=&w^{-\frac{4 \gamma _0}{\beta _0}-1} \left(\frac{h0_0 \left(4 \beta _0 \gamma _1-4 \beta _1 \gamma _0 (\log (w)+1)\right)}{\beta _0^2}+h0_1+\frac{\beta _1 q0_0 \log (w) \left(3 \beta _1 \gamma _0 (\log (w)+2)-2 \beta _0 \left(\beta _0+3 \gamma _1\right)\right)}{3 \beta _0^4}\right)\nonumber\\&+\frac{q0_0 \left(\beta _0^3 \left(-4 \beta _1+16 \gamma _1+6 \gamma _2\right)+2 \beta _0^2 \left(3 \left(\beta _2 \gamma _0+4 \gamma _1^2\right)-\beta _1 \left(8 \gamma _0+9 \gamma _1\right)\right)+6 \beta _1 \beta _0 \gamma _0 \left(\beta _1-8 \gamma _1\right)\right) w^{-\frac{4 \gamma _0}{\beta _0}-1}}{6 \beta _0^4\left(\beta _0+4 \gamma _0\right)}\nonumber\\&+\frac{q0_0 \left(3 \beta _2 \gamma _0+\gamma _1 \left(4 \beta _0-3 \beta _1-16 \gamma _0+12 \gamma _1\right)-12 \gamma _2 \gamma _0\right)}{6 \gamma _0 \left(\beta _0^2-16 \gamma _0^2\right)}+\frac{\gamma^{ii}_2 q0_0}{2\left(\beta _0-4 \gamma _0\right)}-\frac{ q0_0 \left(\gamma^{ii}_2-8 \frac{\beta _1^2 \gamma _0^2}{\beta_0^4}\right) w^{-\frac{4 \gamma _0}{\beta _0}-1}}{2\left(\beta _0+4 \gamma _0\right)}\nonumber\\
			&+\frac{1}{\left(\beta _0-4 \gamma _0\right)}\bigg\lbrace\frac{q0_0 \left(\beta _0^3 \gamma _2-\beta _0^2 \left(\beta _2 \gamma _0+\gamma _1 \left(\beta _1+4 \gamma _1\right)\right)+\beta _1 \beta _0 \gamma _0 \left(\beta _1+8 \gamma _1\right)-4 \beta _1^2 \gamma _0^2\right) w^{1-\frac{4 \gamma _0}{\beta _0}}}{\beta _0^4}\nonumber\\&+\frac{\frac{q0_1 \left(-4 \beta _0+3 \beta _1+16 \gamma _0-12 \gamma _1\right)}{24 \gamma _0}+w^{-\frac{4 \gamma _0}{\beta _0}} \left(q0_1 \left(\frac{\beta _0}{6 \gamma _0}-\frac{\beta _1-4 \gamma _1}{8 \gamma _0}-\frac{2 \gamma _1}{\beta _0}+\frac{2 \beta _1 \gamma _0 (\log (w)+1)}{\beta _0^2}-\frac{2}{3}\right)-\frac{q0_2}{2}\right)}{w}\nonumber\\&+\frac{\frac{q0_2}{2}-\frac{\beta _1 q0_1 \log (w)}{2 \beta _0}}{w^2}+\frac{2 q0_1 \left(\beta _0 \gamma _1-\beta _1 \gamma _0\right) w^{-\frac{4 \gamma _0}{\beta _0}}}{\beta _0^2}\bigg\rbrace\\
			g^{L+T}_1(w)&=-\frac{\beta _1 4\gamma _0 g0_0 \log (w) w^{-\frac{4\gamma _0}{\beta _0}-1}}{\beta _0^2}+\frac{w^{-\frac{4\gamma _0}{\beta _0}-1} \left(4 \beta _0^2 g0_1+4 \beta _0 \gamma _1 g0_0 (1-w)-4 \beta _1 4\gamma _0 g0_0 (1-w)\right)}{4 \beta _0^2}\,\\
			h^{L}_0(w)=&-\frac{\gamma^{ii}_0 w^{-\frac{4\gamma _0}{\beta _0}} \left(\beta _0^2 \left(\beta _1-\gamma _1\right) \left(w^{\frac{4\gamma _0}{\beta _0}}-1\right)-\beta _0 4\gamma _0 \left(\beta _1 \log (w)-\gamma _1 (w-1)\right)+\beta _1 4\gamma _0^2 (-w+\log (w)+1)\right)}{8 \beta _0^2 4\gamma _0 \left(\beta _0-4\gamma _0\right)}\nonumber\\&+hl0_0 w^{-\frac{4\gamma _0}{\beta _0}}+\frac{\gamma^{ii}_1 \left(1-w^{-\frac{4\gamma _0}{\beta _0}}\right)}{8 4\gamma _0}-\frac{\gamma^{ii}_0(1- w^{1-\frac{4\gamma _0}{\beta _0}})}{8 x \left(\beta _0-4\gamma _0\right)}\,,
		\end{aligned}$}
\end{equation}
	\begin{equation}
				\resizebox{\textwidth}{!}{$
					\begin{aligned}		h^{L}_1(w)=&w^{-\frac{4 \gamma _0}{\beta _0}} \bigg(-\frac{2 \beta _1^2 \gamma _0}{\beta _0^4}+\frac{4 \beta _1 \gamma _1}{\beta _0^3}+\frac{\beta _1 \gamma _1}{2 \beta _0^2 \gamma _0}-\frac{2 \gamma _1}{3 \beta _0 \gamma _0}-\frac{2 \gamma _1^2}{\beta _0^2 \gamma _0}+\frac{2 \beta _1}{3 \beta _0^2}-\frac{\beta _2}{2 \beta _0^2}+hl0_0 \left(\frac{4 \beta _1 \gamma _0}{\beta _0^2}-\frac{4 \gamma _1}{\beta _0}\right)+\frac{\frac{2 \gamma _1}{3 \gamma _0}+\frac{\gamma^{ii}_2}{2}}{\left(\beta _0+4 \gamma _0\right)}\nonumber\\&+\left(\frac{2 \beta _1 \gamma _1}{\beta _0^3}-\frac{2 \beta _1^2 \gamma _0}{\beta _0^4}\right) \log (w)\bigg)+\frac{\left(\beta _0^3 \gamma _2-\beta _0^2 \left(\beta _2 \gamma _0+\gamma _1 \left(\beta _1+4 \gamma _1\right)\right)+\beta _1 \beta _0 \gamma _0 \left(\beta _1+8 \gamma _1\right)-4 \beta _1^2 \gamma _0^2\right) w^{1-\frac{4 \gamma _0}{\beta _0}}}{\beta _0^4 \left(\beta _0-4 \gamma _0\right)}\nonumber\\&+\frac{-\frac{\gamma _1 \left(\beta _1-4 \gamma _1\right)}{\gamma _0}+\beta _2-4 \gamma _2}{2 \left(\beta _0^2-16 \gamma _0^2\right)}+w^{-\frac{4 \gamma _0}{\beta _0}-1} \Bigg\lbrace hl0_0 \left(-\frac{4 \beta _1 \gamma _0}{\beta _0^2}+\frac{4 \gamma _1}{\beta _0}-\frac{4 \beta _1 \gamma _0 \log (w)}{\beta _0^2}\right)+hl0_1+\frac{\beta _1^2 \gamma _0 \log ^2(w)}{\beta _0^4}\nonumber\\&+\left(\frac{2 \beta _1^2 \gamma _0}{\beta _0^4}-\frac{2 \beta _1 \gamma _1}{\beta _0^3}-\frac{2 \beta _1}{3 \beta _0^2}\right) \log (w)+\frac{1}{6 \beta _0^4 \left(\beta _0+4 \gamma _0\right)}\bigg\lbrace2 \beta _0^2 \left(3 \left(\beta _2 \gamma _0+4 \gamma _1^2\right)-\beta _1 \left(8 \gamma _0+9 \gamma _1\right)\right)\nonumber\\&\bs+\beta _0^3 \left(-4 \beta _1+16 \gamma _1+6 \gamma _2\right)+6 \beta _1 \beta _0 \gamma _0 \left(\beta _1-8 \gamma _1\right)+24 \beta _1^2 \gamma _0^2-3 \beta _0^4 \gamma^{ii}_2\bigg\rbrace\Bigg\rbrace\,,\\
			k^{L}_0(w)=&-\frac{w^{1-\frac{4 \gamma _0}{\beta _0}}-1}{3 \left(\beta _0-4 \gamma _0\right) x}-\frac{1}{9 \gamma _0}+\frac{\beta _1-4 \gamma _1}{12 \gamma _0 \left(\beta _0-4 \gamma _0\right)}+\frac{4 \left(\beta _0 \gamma _1-\beta _1 \gamma _0\right) w^{1-\frac{4 \gamma _0}{\beta _0}}}{3 \beta _0^2 \left(\beta _0-4 \gamma _0\right)}\nonumber\\&\bs\bs+w^{-\frac{4 \gamma _0}{\beta _0}}\left(kl0_0+\frac{-3 \beta _0 \left(\beta _1-4 \gamma _1\right)+4 \beta _0^2-12 \beta _1 \gamma _0 (\log (w)+1)}{36 \beta _0^2 \gamma _0}\right)\,,\\
			k^{L}_1(w)=&kl0_1 w^{-\frac{4 \gamma _0}{\beta _0}-1}+\frac{2 \beta _1 \log (w) w^{-\frac{4 \gamma _0}{\beta _0}-1} \left(2 \beta _0 \gamma _1 w+\beta _1 \gamma _0 (\log (w)-2 w)\right)}{3 \beta _0^4}+\frac{\gamma _1 \left(4 \gamma _1-\beta _1\right)+\gamma _0 \left(\beta _2-4 \gamma _2\right)}{3 \gamma _0 \left(\beta _0^2-16 \gamma _0^2\right)}\nonumber\\&+\frac{3 \gamma _0 \gamma^{ii}_2+4 \gamma _1}{9 \gamma _0 \left(\beta _0+4 \gamma _0\right)}-\frac{4 \beta _1 \log (w) \left(3 \beta _0 \gamma _1-3 \beta _1 \gamma _0+\beta _0^2 \left(9 \gamma _0 kl0_0+1\right)\right) w^{-\frac{4 \gamma _0}{\beta _0}-1}}{9 \beta _0^4}+\frac{w^{-\frac{4 \gamma _0}{\beta _0}-1}}{9 \beta _0^4\left(\beta _0+4 \gamma _0\right)} \bigg\lbrace6 \beta _1 \beta _0 \gamma _0 \nonumber\\&\times\left(\beta _1-8 \gamma _1\right)-2 \beta _0^2 \left(\beta _1 \left(9 \gamma _1+8 \gamma _0 \left(9 \gamma _0 kl0_0+1\right)\right)-3 \left(\beta _2 \gamma _0+4 \gamma _1^2\right)\right)+\beta _0^4 \left(36 \gamma _1 kl0_0-3 \gamma^{ii}_2\right)+24 \beta _1^2 \gamma _0^2\nonumber\\&+2 \beta _0^3 \left(3 \gamma _2-2 \left(\beta _1-4 \gamma _1\right) \left(9 \gamma _0 kl0_0+1\right)\right)\bigg\rbrace-\frac{4 w^{-\frac{4 \gamma _0}{\beta _0}-1}}{9 \beta _0^4}\beta _1 \log (w) \left(3 \beta _0 \gamma _1-3 \beta _1 \gamma _0+\beta _0^2 \left(9 \gamma _0 kl0_0+1\right)\right)\nonumber\\&+\frac{2 w^{1-\frac{4 \gamma _0}{\beta _0}}}{3 \beta _0^4 \left(\beta _0-4 \gamma _0\right)}\left(\beta _0^3 \gamma _2-\beta _0^2 \left(\beta _2 \gamma _0+\gamma _1 \left(\beta _1+4 \gamma _1\right)\right)+\beta _1 \beta _0 \gamma _0 \left(\beta _1+8 \gamma _1\right)-4 \beta _1^2 \gamma _0^2\right)+\frac{w^{-\frac{4 \gamma _0}{\beta _0}}}{9 \beta _0^4 \gamma _0}\bigg\lbrace24 \beta _1 \beta _0 \gamma _0 \gamma _1\nonumber\\&-12 \beta _1^2 \gamma _0^2-4 \beta _0^3 \gamma _1 \left(9 \gamma _0 kl0_0+1\right)+\beta _0^2 \left(\beta _1 \left(3 \gamma _1+4 \gamma _0 \left(9 \gamma _0 kl0_0+1\right)\right)-3 \left(\beta _2 \gamma _0+4 \gamma _1^2\right)\right)\bigg\rbrace\\\nonumber\\
			\text{where }w&=(1-\beta_0 x L)\nonumber\,.
		\end{aligned}$}
	\end{equation}

		\section{Contributions to the Adler Function }\label{app:adlercoef}
		\subsection{Dimension-zero contributions}
		In massless case, Adler function is known to $\mathcal{O}(\alpha^4_s)$\cite{Appelquist:1973uz,Zee:1973sr,Chetyrkin:1979bj,Dine:1979qh,Gorishnii:1990vf,Surguladze:1990tg,Chetyrkin:1996ez,Baikov:2008jh,Baikov:2010je,Herzog:2017dtz} and contribution from longitudinal part is zero ($D_{i,j}^{L,(0)}=0$) while $D^{L+T,0}_{i,j}$ is given by:
\begin{equation}
	\resizebox{\textwidth}{!}{$
		\begin{aligned}
			\mathcal{D}^{L+T,0}_{i,j}=&1+x+x^2 \left(\frac{299}{24}-9 \zeta (3)\right)+x^3 \left(-\frac{779 \zeta (3)}{4}+\frac{75 \zeta (5)}{2}+\frac{58057}{288}\right)+x^4 \Big(\frac{4185 \zeta (3)^2}{8}\nonumber\\&+\frac{729 \pi ^2 \zeta (3)}{16}-\frac{1704247 \zeta (3)}{432}+\frac{34165 \zeta (5)}{96}-\frac{1995 \zeta (7)}{16}-\frac{13365 \pi ^2}{256}+\frac{78631453}{20736}\Big)
	\end{aligned}$}
\end{equation}
		\subsection{The Dimension-2 Corrections}\label{app:dim2adler_ms_vus}
		The dimension-2 correction to the Adler function for the quark flavor $i$ and $j$ is known to $\mathcal{O}\left(\alpha^3_s\right)$\cite{Baikov:2004ku,Baikov:2004tk,Chetyrkin:1993hi,Gorishnii:1986pz,Generalis:1989hf,Bernreuther:1981sp} and the analytic expression reads:
\begin{equation}
	\resizebox{\textwidth}{!}{$
		\begin{aligned}
			\mathcal{D}_{2,ij}^{L+T,V/A}(s)&=\frac{3}{4 \pi ^2 s} \Bigg\lbrace \left(m_u^2+m_d^2+m_s^2\right)\bigg(x^2 \left(\frac{8 \zeta (3)}{3}-\frac{32}{9}\right)+x^3 \big(4 \zeta (3)^2+\frac{1592 \zeta (3)}{27}\nonumber\\&-\frac{80 \zeta (5)}{27}-\frac{2222}{27}\big)\bigg)+\left(m_i^2+m_j^2\right)\bigg(1+\frac{13 x}{3}+x^2 \left(\frac{179 \zeta (3)}{54}-\frac{520 \zeta (5)}{27}+\frac{23077}{432}\right)\nonumber\\&+x^3 \bigg(\frac{53 \zeta (3)^2}{2}-\frac{1541 \zeta (3)}{648}+\frac{79835 \zeta (7)}{648}-\frac{54265 \zeta (5)}{108}-\frac{\pi ^4}{36}+\frac{3909929}{5184}\bigg)\bigg)\nonumber\\& \pm  m_i m_j \Bigg(\frac{2 x}{3}+x^2 \left(-\frac{55 \zeta (3)}{27}-\frac{5 \zeta (5)}{27}+\frac{769}{54}\right)+x^3 \Big(-\frac{11677 \zeta (3)^2}{108}+\frac{70427 \zeta (3)}{324}\nonumber\\&\bs\bs+\frac{82765 \zeta (5)}{54}-\frac{555233 \zeta (7)}{864}+\frac{\pi^4}{9}-\frac{7429573}{3888}\Big)\Bigg)\Bigg\rbrace\,,
	\end{aligned}$}
\end{equation}
		where upper and lower signs correspond to vector and axial-vector components, respectively, and this convention is used for the Alder functions in this section. It should be noted that the $\mathcal{O}(\alpha^4_s)$ correction to $m_i m_j$ term has been obtained from eq(15) of Ref.~\cite{Baikov:2004tk}.
		The longitudinal component of the dimension-2 operator is known to $\order{\alpha_s^4}$\cite{Becchi:1980vz,Broadhurst:1981jk,Chetyrkin:1996sr,Baikov:2005rw,Gorishnii:1990zu,Gorishnii:1991zr} and has the form:
		\begin{align}
			\mathcal{D}_{2,ij}^{L,V/A}=&\frac{-3}{8\pi^2}\frac{\left(m_i\mp m_j\right)^2}{M^2_{\tau}}\Bigg\lbrace1+\frac{17 x}{3}+x^2 \left(\frac{9631}{144}-\frac{35 \zeta (3)}{2}\right)+x^3 \Big(-\frac{91519 \zeta (3)}{216}+\frac{715 \zeta (5)}{12}\nonumber\\&-\frac{\pi ^4}{36}+\frac{4748953}{5184}\Big)+x^4 \Big(\frac{192155 \zeta (3)^2}{216}-\frac{46217501 \zeta (3)}{5184}+\frac{455725 \zeta (5)}{432}-\frac{125 \pi ^6}{9072}\nonumber\\&-\frac{52255 \zeta (7)}{256}-\frac{3491 \pi ^4}{10368}+\frac{7055935615}{497664}\Big)\Bigg\rbrace\,.
		\end{align}
		From the above dimension-2 Alder functions, the important piece relevant for Cabibbo suppressed strange quark mass determination\cite{Pich:1999hc} using Eq.~\eqref{eq:delRkl} are:
		\begin{align}
			\delta& \mathcal{D}^{L+T,V+A}_{2}(s)=(\mathcal{D}^{L+T,V+A}_{2,ud}(s)-\mathcal{D}^{L+T,V+A}_{2,us}(s))\nonumber\\&=\frac{-3 m^2_s}{2\pi^2 s}(1-\epsilon^2_d) \Bigg(1+\frac{13 x}{3}+x^2 \left(\frac{179 \zeta (3)}{54}-\frac{520 \zeta (5)}{27}+\frac{23077}{432}\right)\nonumber\\&\bs+x^3 \left(\frac{53 \zeta (3)^2}{2}-\frac{1541 \zeta (3)}{648}+\frac{79835 \zeta (7)}{648}-\frac{54265 \zeta (5)}{108}-\frac{\pi ^4}{36}+\frac{3909929}{5184}\right)\Bigg)\\&\equiv\frac{-3 m^2_s(-\xi^2s)}{2\pi^2 s}(1-\epsilon^2_d)\sum_{i=0}\tilde{d}^{L+T}_{i,0}(\xi^2)x(-\xi^2s)^i
			\label{eq:su3adlerLT}
			\\&=\frac{-3 m^2_s(\xi^2\mtsq)}{2\pi^2 s}(1-\epsilon^2_d)\sum_{i=0}^{4}\sum_{j=0}^{i}\tilde{d}^{L+T}_{i,j}x(\xi^2\mtsq)^i \log^j\left(\frac{\xi^2\mtsq}{-s}\right)\,,
			\label{eq:su3adlerLT1}
		\end{align}
		and the corresponding contribution from the longitudinal component is:
		\begin{align}
			\delta & \mathcal{D}^{L,V+A}_{2}(s)=\mathcal{D}^{L,V+A}_{2,ud}-\mathcal{D}^{L,V+A}_{2,us}\nonumber\\&=\frac{3 m^2_s}{4\pi^2 M^2_{\tau}}(1-\epsilon^2_d) \Biggl\lbrace1+\frac{17 x}{3}+x^2 \left(\frac{9631}{144}-\frac{35 \zeta (3)}{2}\right)+x^3 \Big(-\frac{91519 \zeta (3)}{216}+\frac{715 \zeta (5)}{12}-\frac{\pi ^4}{36}\nonumber\\&\bs+\frac{4748953}{5184}\Big)+x^4 \Bigg(\frac{192155 \zeta (3)^2}{216}-\frac{46217501 \zeta (3)}{5184}+\frac{455725 \zeta (5)}{432}-\frac{52255 \zeta (7)}{256}\nonumber\\&\bs\bs\bs-\frac{125 \pi ^6}{9072}-\frac{3491 \pi ^4}{10368}+\frac{7055935615}{497664}\Bigg)\Biggr\rbrace\\&\equiv\frac{3 m^2_s(-\xi^2s)}{4\pi^2 M^2_{\tau}}(1-\epsilon^2_d)\sum_{i=0}\tilde{d}^{L}_{i,0}(\xi^2)x(-\xi^2s)^i\label{eq:su3adlerL}\\&=\frac{3 m^2_s(\xi^2\mtsq)}{4\pi^2 M^2_{\tau}}(1-\epsilon^2_d)\sum_{i=0}^{4}\sum_{j=0}^{i}\tilde{d}^{L}_{i,j}x(\xi^2\mtsq)^i \log^j\left(\frac{\xi^2\mtsq}{-s}\right)\,.
			\label{eq:su3adlerL1}
		\end{align}
		The RGSPT coefficients for the dimension-2 operators can be written in the following form:
		\begin{equation}
		\resizebox{\textwidth}{!}{$
		\begin{aligned}
			\delta& \mathcal{D}^{J,2}_{V+A}(s)=\text{norm}\times\frac{ 3 m^2_s}{2\pi^2 }(1-\epsilon^2_d)\Bigg\lbrace\frac{1}{w^{8/9}}+x \left(\frac{\tilde{d}^{J}_{1,0}}{w^{17/9}}-\frac{1.79012}{w^{8/9}}+\frac{1.79012}{w^{17/9}}-\frac{1.58025 \log (w)}{w^{17/9}}\right)\nonumber\\&+x^2 \Big(\frac{1.79012 \tilde{d}^{J}_{1,0}}{w^{26/9}}+\frac{\tilde{d}^{J}_{2,0}}{w^{26/9}}-\frac{1.79012 \tilde{d}^{J}_{1,0}}{w^{17/9}}+\frac{\left(-3.35802 \tilde{d}^{J}_{1,0}-8.82061\right) \log (w)}{w^{26/9}}-\frac{0.339459}{w^{8/9}}\nonumber\\&\bs-\frac{4.36949}{w^{17/9}}+\frac{4.70895}{w^{26/9}}+\frac{2.65325 \log ^2(w)}{w^{26/9}}+\frac{2.82884 \log (w)}{w^{17/9}}\Big)\nonumber\\&+x^3 \Big(\frac{6.01952 \tilde{d}^{J}_{1,0}}{w^{35/9}}+\frac{1.79012 \tilde{d}^{J}_{2,0}}{w^{35/9}}+\frac{\tilde{d}^{J}_{3,0}}{w^{35/9}}-\frac{0.339459 \tilde{d}^{J}_{1,0}}{w^{17/9}}-\frac{5.68006 \tilde{d}^{J}_{1,0}}{w^{26/9}}-\frac{1.79012 \tilde{d}^{J}_{2,0}}{w^{26/9}}\nonumber\\&\bs+\frac{\left(8.62308 \tilde{d}^{J}_{1,0}+27.3673\right) \log ^2(w)}{w^{35/9}}+\frac{\left(6.01128 \tilde{d}^{J}_{1,0}+19.7019\right) \log (w)}{w^{26/9}}+\frac{0.593473}{w^{8/9}}\nonumber\\&\bs-\frac{3.28306}{w^{17/9}}+\frac{\left(-15.1635 \tilde{d}^{J}_{1,0}-5.1358 \tilde{d}^{J}_{2,0}-39.8653\right) \log (w)}{w^{35/9}}-\frac{14.9321}{w^{26/9}}+\frac{17.6217}{w^{35/9}}\nonumber\\&\bs-\frac{4.5422 \log ^3(w)}{w^{35/9}}-\frac{4.74965 \log ^2(w)}{w^{26/9}}+\frac{0.53643 \log (w)}{w^{17/9}}\Big)
			\nonumber\\&+x^4 \Bigg(\frac{0.593473 \tilde{d}^{J}_{1,0}}{w^{17/9}}+\frac{27.6536 \tilde{d}^{J}_{1,0}}{w^{44/9}}+\frac{7.3301 \tilde{d}^{J}_{2,0}}{w^{44/9}}+\frac{1.79012 \tilde{d}^{J}_{3,0}}{w^{44/9}}+\frac{\tilde{d}^{J}_{4,0}}{w^{44/9}}-\frac{6.29286 \tilde{d}^{J}_{1,0}}{w^{26/9}}\nonumber\\&\bs-\frac{0.339459 \tilde{d}^{J}_{2,0}}{w^{26/9}}-\frac{21.9542 \tilde{d}^{J}_{1,0}}{w^{35/9}}-\frac{6.99064 \tilde{d}^{J}_{2,0}}{w^{35/9}}-\frac{1.79012 \tilde{d}^{J}_{3,0}}{w^{35/9}}-\frac{12.673}{w^{8/9}}+\frac{11.6487}{w^{17/9}}\nonumber\\&\bs+\frac{\left(1.13991 \tilde{d}^{J}_{1,0}+11.9782\right) \log (w)}{w^{26/9}}+\frac{\left(-19.8721 \tilde{d}^{J}_{1,0}-71.1438\right) \log ^3(w)}{w^{44/9}}-\frac{15.09}{w^{26/9}}\nonumber\\&\bs+\frac{\left(-15.4364 \tilde{d}^{J}_{1,0}-59.0364\right) \log ^2(w)}{w^{35/9}}-\frac{60.9336}{w^{35/9}}+\frac{77.0479}{w^{44/9}}+\frac{8.13109 \log ^3(w)}{w^{35/9}}\nonumber\\&\bs-\frac{\left(68.5739 \tilde{d}^{J}_{1,0}+21.5065 \tilde{d}^{J}_{2,0}+6.91358 \tilde{d}^{J}_{3,0}+192.701\right) \log (w)}{w^{44/9}}-\frac{0.937834 \log (w)}{w^{17/9}}\nonumber\\&\bs+\frac{\left(39.8584 \tilde{d}^{J}_{1,0}+9.19372 \tilde{d}^{J}_{2,0}+111.714\right) \log (w)}{w^{35/9}}+\frac{7.85071 \log ^4(w)}{w^{44/9}}-\frac{0.900672 \log ^2(w)}{w^{26/9}}\nonumber\\&\bs+\frac{\left(67.7471 \tilde{d}^{J}_{1,0}+17.7534 \tilde{d}^{J}_{2,0}+186.459\right) \log ^2(w)}{w^{44/9}}\Bigg)\Bigg\rbrace\,.
			\end{aligned}$}
			\label{eq:summed2_1}
	\end{equation}
		A more compact form for the above equation is:
		\begin{align}
			\delta \mathcal{D}^{J,2}_{V+A}(s)	\equiv\text{norm}\times\frac{ 3 m^2_s}{2\pi^2 }(1-\epsilon^2_d)\sum _{i=0}^4 \sum _{k=0}^i \sum _{j=0}^k x^i \hs\tilde{T}^{J}_{i,j,k} \frac{\log^j(w)}{w^{2\gamma_0/\beta_0+k}}\,,
			\label{eq:summed_D2}
		\end{align}
		where
		\begin{align}
			\text{norm}=
			\begin{cases}
				\frac{-1}{s},\bs \text{if } &J= 0+1\\
				\frac{1}{2\mtsq},\bs &J=0\,,
			\end{cases}
		\end{align}
		and $\tilde{d}_i^{J}$ can be obtained from Eq.~\eqref{eq:su3adlerLT},\eqref{eq:su3adlerL}.
		\section{Details of \texorpdfstring{$m_s$}{} and \texorpdfstring{$\vert V_{us}\vert$}{} determinations in different schemes and the details of sources of uncertainty} \label{app:determination_details}
		\subsection{Strange quark mass  determinations using pQCD inputs}\label{app:pQCD_mass}
		In this section, the strange quark mass is calculated using the longitudinal component calculated using OPE as described in sec.~\eqref{sec:dim_2_behaviour}. As mentioned before, these contributions are poorly convergent, and the strange quark mass determinations will suffer from the large truncation uncertainties in addition to significant dependence on the moments used. The dependence on the moment can be slightly reduced by using prescription I at the cost of enhanced truncation uncertainty. This behavior in the different perturbative schemes and the details of various sources of uncertainties are discussed in the later subsections.
		\subsubsection{\bf{Strange quark mass determination using CIPT scheme}}
		Determination of $m_s(\mtsq)$ using CIPT is based on dimension-2 contributions described in section[\ref{sec:CIPT_intro}]. Using prescription-I, can see that dimension-2 contributions are truncated at $\order{\alpha_s^3},\order{\alpha_s^3},\order{\alpha_s^2}$ and $\order{\alpha_s}$ for $k=0,1,2,3$ and 4 respectively. This truncation results in the enhancement in the total uncertainty in the $m_s(\mtsq)$ determination and can be seen in the Tables~\eqref{tab:msCI},\eqref{tab:msCI1}. However, the main advantage of using prescription-I is that the masses from various moments, using different experimental inputs, agree within the uncertainty, which is not the case in using prescription-2. For this reason, we are presenting $m_s(\mtsq)$ from both prescriptions for different schemes in other schemes too.
		\begin{center}
			\begin{table}[H]
				\centering
				   \begin{adjustbox}{max width=\textwidth}
				\begin{tabular}{|c|c|c|c|c|c|c|c|c|c|c|c|}
					\hline
					Parameter&\multicolumn{5}{c|}{Moments ALEPH\cite{Chen:2001qf}}&\multicolumn{5}{c|}{Moment OPAL\cite{OPAL:2004icu}} \\ \cline{2-6}\cline{7-11}
					\text{} &(0,0)&(1,0)&(2,0)&(3,0)&(4,0)&(0,0)&(1,0)&(2,0)&(3,0)&(4,0)\\ \hline
					$m_s(\mtsq)$&$135_{-37}^{+34}$&$122_{-24}^{+28}$&$120_{-24}^{+32}$&$105_{-23}^{+31}$&$108_{-24}^{+38}$&$124_{-31}^{+30}$&$106_{-26}^{+27}$&$105_{-25}^{+30}$&$94_{-22}^{+29}$&$100_{-24}^{+36}$\\\hline
					$\delta R^{kl}_{\tau}(\text{Exp.})$ &$\text{}^{+27.4}_{-34.8}$&$\text{}^{+15.4}_{-17.7}$&$\text{}^{+11.7}_{-13.0}$&$\text{}^{+9.4}_{-10.4}$&$\text{}^{+9.0}_{-9.9}$&$\text{}^{+23.2}_{-28.7}$&$\text{}^{+17.8}_{-21.6}$&$\text{}^{+14.4}_{-16.8}$&$\text{}^{+11.2}_{-12.8}$&$\text{}^{+10.6}_{-12.0}$\\\hline
					$\xi\in\left[.75,2.0\right]$&$\text{}^{+14.3}_{-6.2}$&$\text{}^{+17.3}_{-8.0}$&$\text{}^{+20.8}_{-10.4}$&$\text{}^{+21.1}_{-10.9}$&$\text{}^{+23.8}_{-13.4}$&$\text{}^{+13.4}_{-5.8}$&$\text{}^{+15.1}_{-7.1}$&$\text{}^{+18.5}_{-9.2}$&$\text{}^{+19.0}_{-9.9}$&$\text{}^{+22.1}_{-12.5}$\\\hline
					Truncation uncertainty&$\text{}^{-8.7}_{+10.7}$&$\text{}^{-9.4}_{+12.2}$&$\text{}^{-12.1}_{+17.3}$&$\text{}^{-11.4}_{+16.8}$&$\text{}^{-15.6}_{+26.7}$&$\text{}^{-8.0}_{+10.0}$&$\text{}^{-8.2}_{+10.6}$&$\text{}^{-10.7}_{+15.3}$&$\text{}^{-10.3}_{+15.2}$&$\text{}^{-14.5}_{+25.1}$\\\hline
					$s_0\in\left[3,\mtsq\right](\GeV^2)$&8.3&11.1&12.7&12.5&6.4&7.7&9.6&11.2&11.2&5.9\\\hline
				\end{tabular}
			\end{adjustbox}
				\caption{Strange quark mass using CIPT in prescription I. Other sources of uncertainties are not shown in the table but are added in the quadrature for $m_s(\mtsq)$ in the second row.}
				\label{tab:msCI}
			\end{table}
			\begin{table}[H]
				\centering
				   \begin{adjustbox}{max width=\textwidth}
				\begin{tabular}{|c|c|c|c|c|c|c|c|c|c|c|c|}
					\hline
					Parameter&\multicolumn{5}{c|}{Moments ALEPH\cite{Chen:2001qf}}&\multicolumn{5}{c|}{Moment OPAL\cite{OPAL:2004icu}} \\ \cline{2-6}\cline{7-11}
					\text{}&(0,0)&(1,0)&(2,0)&(3,0)&(4,0)&(0,0)&(1,0)&(2,0)&(3,0)&(4,0)\\ \hline
					$m_s(\mtsq)$&$126_{-35}^{+31}$&$112_{-22}^{+26}$&$96_{-19}^{+25}$&$82_{-18}^{+25}$&$69_{-17}^{+24}$&$116_{-29}^{+27}$&$97_{-24}^{+25}$&$84_{-20}^{+24}$&$73_{-18}^{+23}$&$64_{-16}^{+23}$\\\hline
					$\delta R^{kl}_{\tau}(\text{Exp.})$ &$\text{}^{+25.7}_{-32.6}$&$\text{}^{+14.1}_{-16.3}$&$\text{}^{+9.5}_{-10.5}$&$\text{}^{+7.4}_{-8.1}$&$\text{}^{+5.9}_{-6.4}$&$\text{}^{+21.7}_{-26.9}$&$\text{}^{+16.4}_{-19.8}$&$\text{}^{+11.6}_{-13.6}$&$\text{}^{+8.8}_{-10.0}$&$\text{}^{+6.9}_{-7.8}$\\\hline
					$\xi\in\left[.75,2.0\right]$&$\text{}^{+13.2}_{-6.4}$&$\text{}^{+16.3}_{-8.3}$&$\text{}^{+18.0}_{-9.4}$&$\text{}^{+18.7}_{-10.0}$&$\text{}^{+18.6}_{-10.1}$&$\text{}^{+12.3}_{-6.0}$&$\text{}^{+14.3}_{-7.2}$&$\text{}^{+15.9}_{-8.3}$&$\text{}^{+16.8}_{-9.0}$&$\text{}^{+17.1}_{-9.3}$\\\hline
					Truncation uncertainty&$\text{}^{-7.3}_{+8.8}$&$\text{}^{-8.0}_{+10.1}$&$\text{}^{-8.0}_{+10.6}$&$\text{}^{-7.7}_{+10.6}$&$\text{}^{-7.2}_{+10.4}$&$\text{}^{-6.7}_{+8.1}$&$\text{}^{-6.9}_{+8.8}$&$\text{}^{-7.0}_{+9.3}$&$\text{}^{-6.9}_{+9.6}$&$\text{}^{-6.6}_{+9.6}$\\\hline
					$s_0\in\left[3,\mtsq\right](\GeV^2)$&7.8&10.2&10.2&9.7&9.1&7.2&8.8&9.0&8.7&8.5\\\hline
				\end{tabular}
			\end{adjustbox}
				\caption{ CIPT determination of strange quark mass using in prescription II. Only significant sources of uncertainty are shown in the table separately, while the rest are added in the quadrature and appear in $m_s(\mtsq)$.}
				\label{tab:msCI1}
			\end{table}
		\end{center}

		\subsubsection{\bf{Strange quark mass determination using FOPT scheme}}
		The effect of different prescriptions used is very significant in the FOPT where perturbative contributions from dimension-2 Alder function are truncated at $\order{\alpha_s^4}$ and $\order{\alpha_s^2}$ for the moment $k=0\text{ and }1$ while rest of them are truncated at $\order{\alpha_s}$. The final results for $m_s(\mtsq)$ determination using FOPT in prescriptions I and II are presented in Tables~\eqref{tab:msFOPT} and \eqref{tab:msFOPT2}, respectively.
		\begin{center}
			\begin{table}[H]
				\centering
				   \begin{adjustbox}{max width=\textwidth}
				\begin{tabular}{|c|c|c|c|c|c|c|c|c|c|c|c|}
					\hline
					Parameter&\multicolumn{5}{c|}{Moments ALEPH\cite{Chen:2001qf}}&\multicolumn{5}{c|}{Moment OPAL\cite{OPAL:2004icu}} \\ \cline{2-6}\cline{7-11}
					\text{} &(0,0)&(1,0)&(2,0)&(3,0)&(4,0)&(0,0)&(1,0)&(2,0)&(3,0)&(4,0)\\ \hline
					$m_s(\mtsq)$&$114_{-34}^{+33}$&$134^{+38}_{-30}$&$147^{+48}_{-34}$&$137^{+45}_{-31}$&$127_{-29}^{+29}$&$106^{+29}_{-29}$&$116^{+35}_{-31}$&$130^{+45}_{-33}$&$124^{+43}_{-30}$&$118^{+40}_{-29}$\\\hline
					$\delta R^{kl}_{\tau}(\text{Exp.})$  &$\text{}^{+23.2}_{-29.3}$&$\text{}^{+16.5}_{-19.0}$&$\text{}^{+13.7}_{-15.4}$&$\text{}^{+11.5}_{-12.8}$&$\text{}^{+9.9}_{-11.0}$&$\text{}^{+19.6}_{-24.2}$&$\text{}^{+19.1}_{-23.2}$&$\text{}^{+17.0}_{-19.9}$&$\text{}^{+13.9}_{-16.0}$&$\text{}^{+11.8}_{-13.5}$\\\hline
					$\xi\in\left[.75,2.0\right]$&$\text{}^{+19.0}_{-11.0}$&$\text{}^{+22.6}_{-12.8}$&$\text{}^{+23.7}_{-13.5}$&$\text{}^{+21.7}_{-12.3}$&$\text{}^{+19.5}_{-11.1}$&$\text{}^{+17.6}_{-10.2}$&$\text{}^{+19.8}_{-11.2}$&$\text{}^{+21.2}_{-12.0}$&$\text{}^{+19.7}_{-11.1}$&$\text{}^{+18.0}_{-10.2}$\\\hline
					Truncation uncertainty&$\text{}^{-8.0}_{+10.1}$&$\text{}^{-14.7}_{+21.7}$&$\text{}^{-21.5}_{+36.7}$&$\text{}^{-20.2}_{+34.5}$&$\text{}^{-18.9}_{+32.2}$&$\text{}^{-7.4}_{+9.3}$&$\text{}^{-12.8}_{+19.0}$&$\text{}^{-19.2}_{+33.3}$&$\text{}^{-18.5}_{+32.1}$&$\text{}^{-17.8}_{+30.7}$\\\hline
					$s_0\in\left[3,\mtsq\right](\GeV^2)$&8.7&13.2&15.7&15.3&15.4&8.0&11.5&13.9&13.8&14.3\\\hline
				\end{tabular}
			\end{adjustbox}
				\caption{Strange quark mass using FOPT in prescription I. Other sources of uncertainties are not shown in the table but are added in the quadrature and appear for $m_s(\mtsq)$ in the second row.}
				\label{tab:msFOPT}
			\end{table}
			\begin{table}[H]
				\centering
				   \begin{adjustbox}{max width=\textwidth}
				\begin{tabular}{|c|c|c|c|c|c|c|c|c|c|c|c|}
					\hline
					Parameter&\multicolumn{5}{c|}{Moments ALEPH\cite{Chen:2001qf}}&\multicolumn{5}{c|}{Moment OPAL\cite{OPAL:2004icu}} \\ \cline{2-6}\cline{7-11}
					\text{}&(0,0)&(1,0)&(2,0)&(3,0)&(4,0)&(0,0)&(1,0)&(2,0)&(3,0)&(4,0)\\ \hline
					$m_s(\mtsq)$&$114_{-34}^{+33}$&$107^{+30}_{-24}$&$97^{+28}_{-20}$&$87^{+27}_{-20}$&$78_{-19}^{+25}$&$106^{+29}_{-29}$&$93^{+28}_{-25}$&$85^{+26}_{-21}$&$78^{+25}_{-19}$&$72^{+24}_{-18}$\\\hline
					$\delta R^{kl}_{\tau}(\text{Exp.})$  &$\text{}^{+23.2}_{-29.3}$&$\text{}^{+13.3}_{-15.2}$&$\text{}^{+9.3}_{-10.4}$&$\text{}^{+7.6}_{-8.4}$&$\text{}^{+6.5}_{-7.1}$&$\text{}^{+19.6}_{-24.2}$&$\text{}^{+15.3}_{-18.5}$&$\text{}^{+11.4}_{-13.3}$&$\text{}^{+9.1}_{-10.4}$&$\text{}^{+7.6}_{-8.6}$\\\hline
					$\xi\in\left[.75,2.0\right]$&$\text{}^{+19.0}_{-11.0}$&$\text{}^{+20.8}_{-12.2}$&$\text{}^{+20.8}_{-12.2}$&$\text{}^{+19.8}_{-11.7}$&$\text{}^{+18.4}_{-10.8}$&$\text{}^{+17.6}_{-10.2}$&$\text{}^{+18.1}_{-10.6}$&$\text{}^{+18.2}_{-10.7}$&$\text{}^{+17.7}_{-10.4}$&$\text{}^{+16.9}_{-9.9}$\\\hline
					Truncation uncertainty&$\text{}^{-8.0}_{+10.1}$&$\text{}^{-9.0}_{+12.0}$&$\text{}^{-9.1}_{+12.7}$&$\text{}^{-8.9}_{+12.8}$&$\text{}^{-8.5}_{+12.5}$&$\text{}^{-7.4}_{+9.3}$&$\text{}^{-7.8}_{+10.5}$&$\text{}^{-8.1}_{+11.2}$&$\text{}^{-8.0}_{+11.5}$&$\text{}^{-7.8}_{+11.6}$\\\hline
					$s_0\in\left[3,\mtsq\right](\GeV^2)$&8.7&10.5&10.6&10.3&10.1&8.0&9.1&9.3&9.3&9.4\\\hline
				\end{tabular}
			\end{adjustbox}
				\caption{ Strange quark mass using FOPT in prescription II. Other sources of uncertainties are not shown in the table but are added in the quadrature and appear for $m_s(\mtsq)$ in the second row}
				\label{tab:msFOPT2}
			\end{table}
		\end{center}
		\subsubsection{\bf{Strange quark mass determination using the RGSPT scheme}}
		The RGSPT determination of strange quark mass is presented in Tables~\eqref{tab:msRGSPT} and \eqref{tab:msRGSPT2}, and the most crucial feature of this scheme is that it provides minimum scale uncertainty compared to the CIPT and FOPT. Another important advantage we can infer from prescription II is that it gives the lowest uncertainty among other perturbative schemes.
		\begin{center}
			\begin{table}[H]
				\centering
				   \begin{adjustbox}{max width=\textwidth}
				\begin{tabular}{|c|c|c|c|c|c|c|c|c|c|c|c|c|}
					\hline
					Parameter&\multicolumn{5}{c|}{Moments ALEPH\cite{Chen:2001qf}}&\multicolumn{5}{c|}{Moment OPAL\cite{OPAL:2004icu}} \\ \cline{2-6}\cline{7-11}
					\text{} &(0,0)&(1,0)&(2,0)&(3,0)&(4,0)&(0,0)&(1,0)&(2,0)&(3,0)&(4,0)\\ \hline
					$m_s(\mtsq)$&$123_{-34}^{+28}$&$121^{+23}_{-23}$&$120^{+26}_{-23}$&$125^{+37}_{-27}$&$113_{-25}^{+35}$&$114^{+24}_{-28}$&$104^{+23}_{-25}$&$105^{+25}_{-24}$&$113^{+35}_{-27}$&$104^{+33}_{-25}$\\\hline
					$\delta R^{kl}_{\tau}$(\text{Exp.})&$\text{}^{+25.2}_{-32.0}$&$\text{}^{+15.2}_{-17.4}$&$\text{}^{+11.6}_{-13.0}$&$\text{}^{+10.9}_{-12.0}$&$\text{}^{+9.1}_{-10.1}$&$\text{}^{+21.3}_{-26.3}$&$\text{}^{+17.5}_{-21.2}$&$\text{}^{+14.3}_{-16.7}$&$\text{}^{+13.1}_{-15.0}$&$\text{}^{+10.8}_{-12.3}$\\\hline
					$\xi\in\left[.75,2\right]$&$\text{}^{+3.3}_{-2.1}$&$\text{}^{+4.3}_{-2.8}$&$\text{}^{+5.1}_{-3.4}$&$\text{}^{+6.2}_{-4.2}$&$\text{}^{+5.9}_{-4.0}$&$\text{}^{+3.0}_{-2.0}$&$\text{}^{+3.7}_{-2.4}$&$\text{}^{+4.6}_{-3.0}$&$\text{}^{+5.7}_{-3.8}$&$\text{}^{+5.5}_{-3.7}$\\\hline
					Truncation uncertainty&$\text{}^{-7.2}_{+8.7}$&$\text{}^{-9.7}_{+12.8}$&$\text{}^{-12.7}_{+18.5}$&$\text{}^{-18.2}_{+31.3}$&$\text{}^{-16.9}_{+29.6}$&$\text{}^{-6.7}_{+8.1}$&$\text{}^{-8.4}_{+11.1}$&$\text{}^{-11.2}_{+16.4}$&$\text{}^{-16.5}_{+28.8}$&$\text{}^{-15.8}_{+27.9}$\\\hline
					$s_0\in\left[3,\mtsq\right](\GeV^2)$&8.1&11.4&13.1&15.1&14.6&7.5&9.9&11.5&13.6&13.5\\\hline
				\end{tabular}
			\end{adjustbox}
				\caption{ Strange quark mass using RGSPT in prescription I. Other sources of uncertainties are not shown in the table but are added in the quadrature and appear for $m_s(\mtsq)$ in the second row.}
				\label{tab:msRGSPT}
			\end{table}
			\begin{table}[H]
				\centering
				   \begin{adjustbox}{max width=\textwidth}
				\begin{tabular}{|c|c|c|c|c|c|c|c|c|c|c|c|}
					\hline
					Parameter&\multicolumn{5}{c|}{Moments ALEPH\cite{Chen:2001qf}}&\multicolumn{5}{c|}{Moment OPAL\cite{OPAL:2004icu}} \\ \cline{2-6}\cline{7-11}
					\text{}&(0,0)&(1,0)&(2,0)&(3,0)&(4,0)&(0,0)&(1,0)&(2,0)&(3,0)&(4,0)\\ \hline
					$m_s(\mtsq)$&$123_{-34}^{+28}$&$110^{+21}_{-21}$&$95^{+18}_{-17}$&$82^{+17}_{-16}$&$70_{-14}^{+16}$&$114^{+24}_{-28}$&$95^{+21}_{-23}$&$84^{+18}_{-18}$&$74^{+17}_{-16}$&$65^{+16}_{-14}$\\\hline
					$\delta R^{kl}_{\tau}(\text{Exp.})$  &$\text{}^{+25.2}_{-32.0}$&$\text{}^{+13.9}_{-16.0}$&$\text{}^{+9.4}_{-10.4}$&$\text{}^{+7.3}_{-8.1}$&$\text{}^{+5.9}_{-6.5}$&$\text{}^{+21.3}_{-26.3}$&$\text{}^{+16.1}_{-19.4}$&$\text{}^{+11.5}_{-13.4}$&$\text{}^{+8.8}_{-10.0}$&$\text{}^{+7.0}_{-7.9}$\\\hline
					$\xi\in\left[.75,2.0\right]$&$\text{}^{+3.3}_{-2.1}$&$\text{}^{+4.0}_{-2.6}$&$\text{}^{+4.3}_{-2.8}$&$\text{}^{+4.4}_{-2.9}$&$\text{}^{+4.4}_{-2.9}$&$\text{}^{+3.0}_{-2.0}$&$\text{}^{+3.4}_{-2.3}$&$\text{}^{+3.8}_{-2.5}$&$\text{}^{+4.0}_{-2.6}$&$\text{}^{+4.1}_{-2.7}$\\\hline
					Truncation uncertainty&$\text{}^{-7.2}_{+8.7}$&$\text{}^{-8.1}_{+10.3}$&$\text{}^{-8.2}_{+10.9}$&$\text{}^{-7.9}_{+11.1}$&$\text{}^{-7.5}_{+10.9}$&$\text{}^{-6.7}_{+8.1}$&$\text{}^{-7.0}_{+8.9}$&$\text{}^{-7.2}_{+9.6}$&$\text{}^{-7.1}_{+10.0}$&$\text{}^{-6.9}_{+10.1}$\\\hline
					$s_0\in\left[3,\mtsq\right](\GeV^2)$&8.1&10.4&10.4&9.9&9.3&7.5&9.0&9.1&8.9&8.6\\\hline
				\end{tabular}
			\end{adjustbox}
				\caption{RGSPT determination of strange quark mass using in prescription II. Only the main sources of uncertainty are shown separately, while the rest are added directly in the quadrature of $m_s(\mtsq)$.}
				\label{tab:msRGSPT2}
			\end{table}
		\end{center}

		\subsection{The strange quark mass determinations using phenomenological inputs}\label{app:pheno_mass}
		In this section, the strange quark mass is calculated using the phenomenological parametrization along with the perturbative $``L+T"-$contributions described in section~\eqref{sec:rev_pheno} and section~\eqref{sec:OPE_contributions}. The quark mass is calculated in the two prescriptions for various moments, and the details of sources of uncertainties are presented in the tables.
		\subsubsection{ \bf{Strange quark mass determination using CIPT scheme}}
		The CIPT determination of $m_s(\mtsq)$ in this section makes use of the full dimension-2 results of $\order{\alpha_s^3}$ as the series presented in Eq.~\eqref{eq:dim2_CI} is convergent for all moments, and prescription I and prescription II yield the same determinations. The results are shown in the Table~\eqref{tab:msphenoCI}.

		\begin{table}[H]
			\centering
			   \begin{adjustbox}{max width=\textwidth}
			\begin{tabular}{|c|c|c|c|c|c|c|c|c|c|c|c|c|}
				\hline
				Parameter&\multicolumn{6}{c|}{Moments ALEPH\cite{Chen:2001qf}}&\multicolumn{6}{c|}{Moment OPAL\cite{OPAL:2004icu}} \\ \cline{2-7}\cline{8-13}
				&(0,0)&(1,0)&(2,0)&(3,0)&(4,0)&$(4,0)\footnotemark[\value{footnote}]$&(0,0)&(1,1)&(2,0)&(3,0)&(4,0)&$(4,0)\footnotemark[\value{footnote}]$\\ \hline
				$m_s(\mtsq)$&$187_{-81}^{+63}$&$162_{-34}^{+31}$&$136_{-24}^{+25}$&$115_{-20}^{+25}$&$98_{-15}^{+22}$&$98_{-19}^{+25}$&$166_{-69}^{+54}$&$134_{-41}^{+34}$&$116_{-27}^{+26}$&$102_{-21}^{+24}$&$91_{-16}^{+21}$&$91_{-19}^{+24}$\\\hline
				$\delta R^{kl}_{\tau}(\text{Exp.})$  &$\text{}^{+57.2}_{-79.0}$&$\text{}^{+25.5}_{-30.1}$&$\text{}^{+15.5}_{-17.5}$&$\text{}^{+11.4}_{-12.7}$&$\text{}^{+8.8}_{-9.7}$&$\text{}^{+8.8}_{-9.7}$&$\text{}^{+49.0}_{-67.8}$&$\text{}^{+30.2}_{-39.0}$&$\text{}^{+19.4}_{-23.3}$&$\text{}^{+13.8}_{-16.0}$&$\text{}^{+10.5}_{-11.9}$&$\text{}^{+10.5}_{-11.9}$\\\hline
				$\xi\in \left[.75,2.0\right]$&$\text{}^{-9.5}_{+22.9}$&$\text{}^{-1.0}_{+9.1}$&$\text{}^{+10.8}_{-0.1}$&$\text{}^{+14.4}_{-1.7}$&$\text{}^{+16.3}_{-3.8}$&$\text{}^{+16.3}_{-3.8}$&$\text{}^{-7.9}_{+19.5}$&$\text{}_{+7.1}^{-0.7}$&$\text{}^{+9.5}_{-0.2}$&$\text{}^{+12.8}_{-1.6}$&$\text{}^{+14.8}_{-3.5}$&$\text{}^{+14.8}_{-3.5}$\\\hline
				Truncation uncertainty&$\text{}^{+3.3}_{-3.1}$&$\text{}_{+4.9}^{-4.5}$&$\text{}_{+8.4}^{-7.1}$&$\text{}_{+10.0}^{-7.9}$&$\text{}_{+10.8}^{-8.1}$&$\text{}_{-2.7}^{+2.8}$&$\text{}_{-2.7}^{+2.8}$&$\text{}^{-3.7}_{+4.0}$&$\text{}^{-6.0}_{+7.1}$&$\text{}^{-7.0}_{+8.9}$&$\text{}^{-7.5}_{+9.9}$&$\text{}^{-7.5}_{+9.9}$\\\hline
				$s_0\in\left[3,\mtsq\right](\GeV^2)$&$11.6$&$14.7$&$14.5$&$13.7$&$5.8$&$13.0$&$10.2$&$12.9$&$12.8$&$12.2$&$5.2$&$11.7$\\\hline
			\end{tabular}
		\end{adjustbox}
			\caption{ Strange quark mass using CIPT using phenomenological inputs for the longitudinal component. Only the main sources of uncertainty are shown separately, while the rest are already added to the quadrature and appear in the total uncertainty in $m_s(\mtsq)$.}
			\label{tab:msphenoCI}
		\end{table}

		\subsubsection{\bf{Strange quark mass determination using FOPT scheme}}
		The FOPT determination of $m_s(\mtsq)$ in this section involves determination in both prescription (I-II) as the perturbation series is not well convergent for different moments, as shown in Eq.~\eqref{eq:dim2_FO}. The results are presented in Table~\eqref{tab:msphenofixed} and Table~\eqref{tab:msphenofixed2}.
		\begin{center}
			\begin{table}[H]
				\centering
				   \begin{adjustbox}{max width=\textwidth}
				\begin{tabular}{|c|c|c|c|c|c|c|c|c|c|c|}
					\hline
					Parameter&\multicolumn{5}{c|}{Moments ALEPH\cite{Chen:2001qf}}&\multicolumn{5}{c|}{Moment OPAL\cite{OPAL:2004icu}} \\ \cline{2-6}\cline{7-11} &(0,0)&(1,0)&(2,0)&(3,0)&(4,0)&(0,0)&(1,0)&(2,0)&(3,0)&(4,0)\\ \hline
					$m_s(\mtsq)$&$141_{-60}^{+46}$&$133_{-31}^{+35}$&$135_{-29}^{+38}$&$145_{-33}^{+50}$&$135_{-31}^{+46}$&$125_{-51}^{+40}$&$111_{-35}^{+34}$&$116_{-30}^{+36}$&$130_{-33}^{+46}$&$125_{-30}^{+44}$\\\hline
					$\delta R^{kl}_{\tau}\left(\mtsq\right)(\text{Exp.})$ & $\text{}^{+41.2}_{-57.9}$&$\text{}^{+20.5}_{-24.2}$&$\text{}^{+15.0}_{-16.9}$&$\text{}^{+13.8}_{-15.4}$&$\text{}^{+11.4}_{-12.7}$&$\text{}^{+35.6}_{-49.8}$&$\text{}^{+24.3}_{-31.2}$&$\text{}^{+18.7}_{-22.5}$&$\text{}^{+16.7}_{-19.5}$&$\text{}^{+13.6}_{-15.6}$\\\hline
					$\xi\in\left[.75,2\right]$&$\text{}_{-6.4}^{+16.5}$&$\text{}_{-10.5}^{+21.7}$&$\text{}^{+24.5}_{-12.0}$&$\text{}^{+25.5}_{-12.6}$&$\text{}^{+23.0}_{-11.4}$&$\text{}^{+14.8}_{-5.8}$&$\text{}^{+18.1}_{-8.8}$&$\text{}^{+21.0}_{-10.3}$&$\text{}^{+22.7}_{-11.2}$&$\text{}^{+21.1}_{-10.4}$\\\hline
					Truncation uncertainty&$\text{}^{-5.8}_{+6.7}$&$\text{}^{-9.7}_{+12.4}$&$\text{}^{-14.4}_{+21.1}$&$\text{}^{-21.2}_{+36.8}$&$\text{}^{-19.9}_{+34.4}$&$\text{}^{-5.1}_{+5.9}$&$\text{}^{-8.0}_{+10.2}$&$\text{}^{-12.4}_{+18.2}$&$\text{}^{-19.0}_{+33.4}$&$\text{}^{-18.6}_{+32.4}$\\\hline
					$s_0\in\left[3,\mtsq\right](\GeV^2)$&$10.7$&$13.1$&$14.4$&$16.2$&$16.3$&$9.5$&$11.6$&$12.8$&$14.6$&$14.9$\\\hline
				\end{tabular}
			\end{adjustbox}
				\caption{FOPT determination of $m_s(\mtsq)$ using prescription I. Only major sources of uncertainty are shown separately, while the rest are added to the quadrature and appear in the total uncertainty in $m_s(\mtsq)$.}
				\label{tab:msphenofixed}
			\end{table}

			\begin{table}[H]
				\centering
				   \begin{adjustbox}{max width=\textwidth}
				\begin{tabular}{|c|c|c|c|c|c|c|c|c|c|c|c|}
					\hline
					Parameter&\multicolumn{5}{c|}{Moments ALEPH\cite{Chen:2001qf}}&\multicolumn{5}{c|}{Moment OPAL\cite{OPAL:2004icu}} \\ \cline{2-6}\cline{7-11} &(0,0)&(1,0)&(2,0)&(3,0)&(4,0)&(0,0)&(1,0)&(2,0)&(3,0)&(4,0)\\ \hline
					$m_s(\mtsq)$&$141_{-60}^{+46}$&$133_{-31}^{+35}$&$121_{-33}^{+26}$&$109_{-23}^{+32}$&$99_{-22}^{+30}$&$125_{-51}^{+40}$&$111_{-35}^{+34}$&$104_{-27}^{+31}$&$97_{-23}^{+29}$&$91_{-22}^{+28}$\\\hline
					$\delta R^{kl}_{\tau}\left(\mtsq\right)(\text{Exp.})$ & $\text{}^{+41.2}_{-57.9}$&$\text{}^{+20.5}_{-24.2}$&$\text{}^{+15.0}_{-16.9}$&$\text{}^{+13.8}_{-15.4}$&$\text{}^{+11.4}_{-12.7}$&$\text{}^{+35.6}_{-49.8}$&$\text{}^{+24.3}_{-31.2}$&$\text{}^{+18.7}_{-22.5}$&$\text{}^{+16.7}_{-19.5}$&$\text{}^{+13.6}_{-15.6}$\\\hline
					$\xi\in\left[.75,2\right]$&$\text8{}_{-6.4}^{+16.5}$&$\text{}_{-10.5}^{+21.7}$&$\text{}^{+24.5}_{-12.0}$&$\text{}^{+25.5}_{-12.6}$&$\text{}^{+23.0}_{-11.4}$&$\text{}^{+14.8}_{-5.8}$&$\text{}^{+18.1}_{-8.8}$&$\text{}^{+21.0}_{-10.3}$&$\text{}^{+22.7}_{-11.2}$&$\text{}^{+21.1}_{-10.4}$\\\hline
					Truncation uncertainty&$\text{}^{-5.8}_{+6.7}$&$\text{}^{-9.7}_{+12.4}$&$\text{}^{-10.4}_{+14.5}$&$\text{}^{-10.7}_{+15.1}$&$\text{}^{-10.4}_{+15.2}$&$\text{}^{-5.1}_{+5.9}$&$\text{}^{-8.0}_{+10.2}$&$\text{}^{-9.2}_{+12.5}$&$\text{}^{-9.6}_{+13.5}$&$\text{}^{-9.7}_{+14.1}$\\\hline
					$s_0\in\left[3,\mtsq\right](\GeV^2)$&$10.7$&$13.1$&$13.3$&$13.0$&$12.8$&$9.5$&$11.6$&$11.7$&$11.6$&$11.7$\\\hline
				\end{tabular}
			\end{adjustbox}
				\caption{ The FOPT determination of $m_s(\mtsq)$ in prescription II. Only major sources of uncertainty are shown separately, while the rest are added to the quadrature and appear in the total uncertainty in $m_s(\mtsq)$.}
				\label{tab:msphenofixed2}
			\end{table}

		\end{center}
		\subsubsection{\bf{Strange quark mass determination using RGSPT scheme}}
		The RGSPT determination in prescriptions I-II is shown in the Table~\eqref{tab:msphenosummed} as the $(4,0)$-moment is not term by term convergent till $\order{\alpha_s^3}$.
		\begin{center}

			\begin{table}[H]
				\centering
				   \begin{adjustbox}{max width=\textwidth}
				\begin{tabular}{|c|c|c|c|c|c|c|c|c|c|c|c|c|c|}
					\hline
					Parameter&\multicolumn{6}{c|}{Moments ALEPH\cite{Chen:2001qf}}&\multicolumn{6}{c|}{Moment OPAL\cite{OPAL:2004icu}} \\ \cline{2-7}\cline{8-13}\hline &(0,0)&(1,0)&(2,0)&(3,0)&(4,0)&$(4,0)\footnotemark[\value{footnote}]$&(0,0)&(1,0)&(2,0)&(3,0)&(4,0)&$(4,0)\footnotemark[\value{footnote}]$\\ \hline
					$m_s(\mtsq)(\text{in}\MeV)$&$178_{-77}^{+57}$&$154_{-33}^{+29}$&$130_{-23}^{+23}$&$111_{-20}^{+21}$&$108_{-21}^{+25}$&$96_{-18}^{+20}$&$157_{-66}^{+49}$&$127_{-39}^{+32}$&$112_{-26}^{+24}$&$99_{-21}^{+21}$&$100_{-21}^{+24}$&$89_{-18}^{+19}$\\\hline
					$\delta R^{kl}_{\tau}(\text{Exp.})$  &$\text{}_{-75.5}^{+55.8}$&$\text{}_{-28.7}^{+24.5}$&$\text{}_{-16.8}^{+14.9}$&$\text{}_{-12.2}^{+11.0}$&$\text{}_{-10.6}^{+9.6}$&$\text{}_{-9.4}^{+8.6}$&$\text{}_{-64.6}^{+47.4}$&$\text{}_{-37.0}^{+28.8}$&$\text{}_{-22.3}^{+18.5}$&$\text{}_{-15.4}^{+13.3}$&$\text{}_{-12.9}^{+11.4}$&$\text{}_{-11.5}^{+10.1}$\\\hline
					$\xi\in\left[.75,2\right]$&$\text{}_{+2.6}^{-2.7}$&$\text{}_{-0.2}^{+1.1}$&$\text{}_{-1.5}^{+2.9}$&$\text{}_{-2.1}^{+3.7}$&$\text{}_{-2.7}^{+4.6}$&$\text{}_{-2.3}^{+4.0}$&$\text{}_{+2.1}^{-2.2}$&$\text{}_{-0.3}^{+1.1}$&$\text{}_{-1.4}^{+2.5}$&$\text{}_{-1.9}^{+3.3}$&$\text{}_{-2.5}^{+4.3}$&$\text{}_{-2.2}^{+3.7}$\\\hline
					Truncation uncertainty&$\text{}_{-2.7}^{+2.9}$&$\text{}_{+6.1}^{-5.4}$&$\text{}^{-7.9}_{+9.7}$&$\text{}^{-8.7}_{+11.3}$&$\text{}^{-11.6}_{+16.9}$&$\text{}^{-8.8}_{+12.1}$&$\text{}_{-2.4}^{+2.5}$&$\text{}_{+4.9}^{-4.4}$&$\text{}^{-6.8}_{+8.3}$&$\text{}^{-7.7}_{+10.1}$&$\text{}^{-10.7}_{+15.6}$&$\text{}^{-8.1}_{+11.1}$\\\hline
					$s_0\in\left[3,\mtsq\right](\GeV^2)$&$11.7$&$14.5$&$14.2$&$13.4$&$12.8$&$10.3$&$12.8$&$12.5$&$12.5$&$12.0$&$12.8$&$11.6$\\\hline
				\end{tabular}
			\end{adjustbox}
				\caption{ Strange quark mass using RGSPT. Other sources of uncertainties are not shown in the table but are added in the quadrature and appear for $m_s(\mtsq)$ in the second row.}
				\label{tab:msphenosummed}
			\end{table}
			
		\end{center}
		
		\footnotetext[\value{footnote}]{prescription II is used for these moments.}
		
		\subsection{Details of the \texorpdfstring{$\vert V_{us}\vert$}{} determinations from OPAL data}\label{app:vus}
		The CKM matrix element $\vert V_{us}\vert$ is calculated using Eq.~\eqref{eq:vus_calc} from the available data on moments for strange and non-strange components. Details of extraction from these moments and associated uncertainties from the inputs parameters are presented in this section. Purely pQCD inputs for the longitudinal component is used to extract $\vert V_{us}\vert$ in the Table~\eqref{tab:VusAllpert1_Opal} and Table~\eqref{tab:VusAll2pert_Opal}. Large theoretical uncertainties in prescription II come from the truncation of perturbative series. Determination of $\vert V_{us}\vert$ from the phenomenological inputs for longitudinal contribution is presented in the Table~\eqref{tab:VusAll1pheno_Opal} and Table~\eqref{tab:VusAll2pheno_Opal} for prescriptions I and II, respectively.\par
		\begin{table}[H]
			\centering
			   \begin{adjustbox}{max width=\textwidth}
				\begin{tabular}{|c|c|c|c|c|c|c|c|c|c|c|c|c|c|c|c|}
					\hline
					Parameters&\multicolumn{5}{c|}{$\vert V_{us}\vert_{CIPT}$}&\multicolumn{5}{c|}{$\vert V_{us}\vert_{FOPT}$}&\multicolumn{5}{c|}{$\vert V_{us}\vert_{RGSPT}$}\\
					\cline{2-6}\cline{7-11}\cline{12-16}
					\text{}&(0,0)&(1,0)&(2,0)&(3,0)&(4,0)&(0,0)&(1,0)&(2,0)&(3,0)&(4,0)&(0,0)&(1,0)&(2,0)&(3,0)&(4,0)\\
					\hline
					$\vert V_{us}\vert$(central)&0.2217&0.2224&0.2219&0.2241&0.2220&0.2227&0.2208&0.2183&0.2181&0.2182&0.2221&0.2229&0.2223&0.2204&0.2217\\ \hline
					$m_s$&$\text{}^{+0.0014}_{-0.0012}$&$\text{}^{+0.0021}_{-0.0018}$&$\text{}^{+0.0025}_{-0.0022}$&$\text{}^{+0.0038}_{-0.0032}$&$\text{}^{+0.0037}_{-0.0031}$&$\text{}^{+0.0017}_{-0.0015}$&$\text{}^{+0.0017}_{-0.0014}$&$\text{}^{+0.0016}_{-0.0014}$&$\text{}^{+0.0021}_{-0.0018}$&$\text{}^{+0.0027}_{-0.0022}$&$\text{}^{+0.0015}_{-0.0013}$&$\text{}^{+0.0022}_{-0.0019}$&$\text{}^{+0.0027}_{-0.0023}$&$\text{}^{+0.0027}_{-0.0023}$&$\text{}^{+0.0037}_{-0.0031}$\\ \hline
					Experimental &$\text{}^{+0.0033}_{-0.0034}$&$\text{}^{+0.0037}_{-0.0037}$&$\text{}^{+0.0036}_{-0.0037}$&$\text{}^{+0.0037}_{-0.0038}$&$\text{}^{+0.0038}_{-0.0038}$&$\text{}^{+0.0033}_{-0.0034}$&$\text{}^{+0.0036}_{-0.0037}$&$\text{}^{+0.0036}_{-0.0036}$&$\text{}^{+0.0036}_{-0.0037}$&$\text{}^{+0.0037}_{-0.0038}$&$\text{}^{+0.0033}_{-0.0034}$&$\text{}^{+0.0037}_{-0.0037}$&$\text{}^{+0.0037}_{-0.0037}$&$\text{}^{+0.0037}_{-0.0037}$&$\text{}^{+0.0038}_{-0.0038}$\\ \hline
					Total theory&$\text{}^{+0.0017}_{-0.0017}$&$\text{}^{+0.0028}_{-0.0029}$&$\text{}^{+0.0040}_{-0.0042}$&$\text{}^{+0.0065}_{-0.0065}$&$\text{}^{+0.0078}_{-0.0074}$&$\text{}^{+0.0024}_{-0.0024}$&$\text{}^{+0.0027}_{-0.0028}$&$\text{}^{+0.0032}_{-0.0032}$&$\text{}^{+0.0042}_{-0.0041}$&$\text{}^{+0.0053}_{-0.0051}$&$\text{}^{+0.0017}_{-0.0015}$&$\text{}^{+0.0028}_{-0.0026}$&$\text{}^{+0.0039}_{-0.0036}$&$\text{}^{+0.0050}_{-0.0047}$&$\text{}^{+0.0070}_{-0.0064}$\\ \hline                    $s_0\in\left[2.50,\mtsq\right]$&0.0032&0.0070&0.0117&0.0188&0.0229&0.0042&0.0089&0.0122&0.0146&0.0200&0.0034&0.0074&0.0126&0.0204&0.0272\\\hline
					total&$\text{}^{+0.0049}_{-0.0050}$&$\text{}^{+0.0084}_{-0.0085}$&$\text{}^{+0.0129}_{-0.0129}$&$\text{}^{+0.0203}_{-0.0203}$&$\text{}^{+0.0245}_{-0.0244}$&$\text{}^{+0.0058}_{-0.0059}$&$\text{}^{+0.0100}_{-0.0100}$&$\text{}^{+0.0131}_{-0.0131}$&$\text{}^{+0.0156}_{-0.0156}$&$\text{}^{+0.0210}_{-0.0210}$&$\text{}^{+0.0051}_{-0.0050}$&$\text{}^{+0.0088}_{-0.0087}$&$\text{}^{+0.0137}_{-0.0136}$&$\text{}^{+0.0213}_{-0.0213}$&$\text{}^{+0.0283}_{-0.0282}$\\ \hline
			\end{tabular}
		\end{adjustbox}
			\caption{Determination of $\vert V_{us}\vert$ in various schemes from the OPAL data using prescription I. The longitudinal component is calculated using the pQCD Adler function.}
			\label{tab:VusAllpert1_Opal}
		\end{table}
		\begin{table}[H]
			\centering
			   \begin{adjustbox}{max width=\textwidth}
				\begin{tabular}{|c|c|c|c|c|c|c|c|c|c|c|c|c|c|c|c|c|c|}
					\hline
					Parameters&\multicolumn{5}{c|}{$\vert V_{us}\vert_{CIPT}$}&\multicolumn{5}{c|}{$\vert V_{us}\vert_{FOPT}$}&\multicolumn{5}{c|}{$\vert V_{us}\vert_{RGSPT}$}\\
					\cline{2-6}\cline{7-11}\cline{12-16}
					&(0,0)&(1,0)&(2,0)&(3,0)&(4,0)&(0,0)&(1,0)&(2,0)&(3,0)&(4,0)&(0,0)&(1,0)&(2,0)&(3,0)&(4,0)\\
					\hline
					$\vert V_{us}\vert$(central)&0.2217&0.2239&0.2275&0.2341&0.2452&0.2227&0.2246&0.2271&0.2310&0.2365&0.2221&0.2246&0.2286&0.2358&0.2477\\ \hline
					$m_s$&$\text{}^{+0.0014}_{-0.0012}$&$\text{}^{+0.0025}_{-0.0021}$&$\text{}^{+0.0043}_{-0.0036}$&$\text{}^{+0.0072}_{-0.0059}$&$\text{}^{+0.0126}_{-0.0097}$&$\text{}^{+0.0017}_{-0.0015}$&$\text{}^{+0.0028}_{-0.0024}$&$\text{}^{+0.0042}_{-0.0035}$&$\text{}^{+0.0062}_{-0.0051}$&$\text{}^{+0.0090}_{-0.0072}$&$\text{}^{+0.0015}_{-0.0013}$&$\text{}^{+0.0027}_{-0.0023}$&$\text{}^{+0.0047}_{-0.0039}$&$\text{}^{+0.0079}_{-0.0064}$&$\text{}^{+0.0138}_{-0.0105}$\\ \hline
					Experimental &$\text{}^{+0.0033}_{-0.0034}$&$\text{}^{+0.0037}_{-0.0038}$&$\text{}^{+0.0037}_{-0.0038}$&$\text{}^{+0.0039}_{-0.0040}$&$\text{}^{+0.0042}_{-0.0043}$&$\text{}^{+0.0033}_{-0.0034}$&$\text{}^{+0.0037}_{-0.0038}$&$\text{}^{+0.0037}_{-0.0038}$&$\text{}^{+0.0039}_{-0.0039}$&$\text{}^{+0.0041}_{-0.0041}$&$\text{}^{+0.0033}_{-0.0034}$&$\text{}^{+0.0037}_{-0.0038}$&$\text{}^{+0.0038}_{-0.0038}$&$\text{}^{+0.0039}_{-0.0040}$&$\text{}^{+0.0043}_{-0.0043}$\\ \hline
					Total theory&$\text{}^{+0.0017}_{-0.0017}$&$\text{}^{+0.0033}_{-0.0035}$&$\text{}^{+0.0063}_{-0.0065}$&$\text{}^{+0.0121}_{-0.0116}$&$\text{}^{+0.0237}_{-0.0206}$&$\text{}^{+0.0024}_{-0.0024}$&$\text{}^{+0.0043}_{-0.0043}$&$\text{}^{+0.0071}_{-0.0068}$&$\text{}^{+0.0110}_{-0.0102}$&$\text{}^{+0.0165}_{-0.0147}$&$\text{}^{+0.0017}_{-0.0015}$&$\text{}^{+0.0033}_{-0.0030}$&$\text{}^{+0.0061}_{-0.0055}$&$\text{}^{+0.0109}_{-0.0096}$&$\text{}^{+0.0200}_{-0.0168}$\\ \hline
					$s_0\in\left[2.50,\mtsq\right]$&0.0030&0.0070&0.0117&0.0188&0.0313&0.0042&0.0089&0.0143&0.0216&0.0325&0.0032&0.0074&0.0126&0.0204&0.0338\\\hline
					total&$\text{}^{+0.0048}_{-0.0048}$&$\text{}^{+0.0086}_{-0.0087}$&$\text{}^{+0.0138}_{-0.0139}$&$\text{}^{+0.0227}_{-0.0225}$&$\text{}^{+0.0395}_{-0.0377}$&$\text{}^{+0.0058}_{-0.0059}$&$\text{}^{+0.0106}_{-0.0162}$&$\text{}^{+0.0164}_{-0.0163}$&$\text{}^{+0.0245}_{-0.0242}$&$\text{}^{+0.0368}_{-0.0359}$&$\text{}^{+0.0049}_{-0.0049}$&$\text{}^{+0.0089}_{-0.0089}$&$\text{}^{+0.0145}_{-0.0145}$&$\text{}^{+0.0235}_{-0.0229}$&$\text{}^{+0.0395}_{-0.0380}$\\ \hline
			\end{tabular}
		\end{adjustbox}
			\caption{Determination of $\vert V_{us}\vert$ in different schemes from the OPAL data using prescription II. The longitudinal component is calculated using the pQCD Adler function.}
			\label{tab:VusAll2pert_Opal}
		\end{table}
		\begin{table}[H]
			\centering
			   \begin{adjustbox}{max width=\textwidth}
			\begin{tabular}{|c|c|c|c|c|c|c|c|c|c|c|c|c|c|c|c|}
					\hline
					Parameters&\multicolumn{5}{c|}{$\vert V_{us}\vert_{CIPT}$}&\multicolumn{5}{c|}{$\vert V_{us}\vert_{FOPT}$}&\multicolumn{5}{c|}{$\vert V_{us}\vert_{RGSPT}$}\\
					\cline{2-6}\cline{7-11}\cline{12-16}
					\text{}&(0,0)&(1,0)&(2,0)&(3,0)&(4,0)&(0,0)&(1,0)&(2,0)&(3,0)&(4,0)&(0,0)&(1,0)&(2,0)&(3,0)&(4,0)\\
					\hline
					$\vert V_{us}\vert$(central)&0.2211&0.2210&0.2212&0.2228&0.2226&0.2222&0.2224&0.2225&0.2208&0.2177&0.2213&0.2213&0.2219&0.2238&0.2233\\ \hline
					$m_s$&$\text{}^{+0.0005}_{-0.0004}$&$\text{}^{+0.0010}_{-0.0009}$&$\text{}^{+0.0018}_{-0.0015}$&$\text{}^{+0.0029}_{-0.0025}$&$\text{}^{+0.0036}_{-0.0030}$&$\text{}^{+0.0008}_{-0.0007}$&$\text{}^{+0.0014}_{-0.0012}$&$\text{}^{+0.0022}_{-0.0019}$&$\text{}^{+0.0024}_{-0.0021}$&$\text{}^{+0.0022}_{-0.0019}$&$\text{}^{+0.0005}_{-0.0004}$&$\text{}^{+0.0011}_{-0.0009}$&$\text{}^{+0.0019}_{-0.0017}$&$\text{}^{+0.0032}_{-0.0028}$&$\text{}^{+0.0038}_{-0.0032}$\\ \hline
					Experimental &$\text{}^{+0.0033}_{-0.0034}$&$\text{}^{+0.0036}_{-0.0037}$&$\text{}^{+0.0036}_{-0.0037}$&$\text{}^{+0.0037}_{-0.0038}$&$\text{}^{+0.0038}_{-0.0038}$&$\text{}^{+0.0033}_{-0.0034}$&$\text{}^{+0.0037}_{-0.0037}$&$\text{}^{+0.0037}_{-0.0037}$&$\text{}^{+0.0037}_{-0.0037}$&$\text{}^{+0.0037}_{-0.0038}$&$\text{}^{+0.0033}_{-0.0034}$&$\text{}^{+0.0036}_{-0.0037}$&$\text{}^{+0.0036}_{-0.0037}$&$\text{}^{+0.0037}_{-0.0038}$&$\text{}^{+0.0038}_{-0.0039}$\\ \hline
					Total theory&$\text{}^{+0.0005}_{-0.0005}$&$\text{}^{+0.0010}_{-0.0010}$&$\text{}^{+0.0020}_{-0.0020}$&$\text{}^{+0.0036}_{-0.0037}$&$\text{}^{+0.0053}_{-0.0054}$&$\text{}^{+0.0009}_{-0.0009}$&$\text{}^{+0.0019}_{-0.0019}$&$\text{}^{+0.0033}_{-0.0033}$&$\text{}^{+0.0039}_{-0.0039}$&$\text{}^{+0.0043}_{-0.0041}$&$\text{}^{+0.0005}_{-0.0004}$&$\text{}^{+0.0011}_{-0.0010}$&$\text{}^{+0.0022}_{-0.0020}$&$\text{}^{+0.0040}_{-0.0036}$&$\text{}^{+0.0055}_{-0.0050}$\\ \hline
					$s_0\in\left[2.5,\mtsq\right]$&0.0032 & 0.0070&0.0117&0.0188&0.0229&0.0042&0.0089&0.0122&0.0146&0.0200&0.0034&0.0074&0.0126&0.0204&0.0272\\
					\hline
					total&$\text{}^{+0.0047}_{-0.0047}$&$\text{}^{+0.0080}_{-0.0080}$&$\text{}^{+0.0124}_{-0.0124}$&$\text{}^{+0.0195}_{-0.0195}$&$\text{}^{+0.0238}_{-0.0239}$&$\text{}^{+0.0054}_{-0.0054}$&$\text{}^{+0.0098}_{-0.0099}$&$\text{}^{+0.0131}_{-0.0132}$&$\text{}^{+0.0155}_{-0.0155}$&$\text{}^{+0.0208}_{-0.0207}$&$\text{}^{+0.0048}_{-0.0048}$&$\text{}^{+0.0084}_{-0.0084}$&$\text{}^{+0.0133}_{-0.0133}$&$\text{}^{+0.0211}_{-0.0211}$&$\text{}^{+0.0280}_{-0.0279}$\\ \hline
			\end{tabular}
		\end{adjustbox}
			\caption{Determination of $\vert V_{us}\vert$ in different schemes from the OPAL data using prescription I. The phenomenological contributions are used for longitudinal components.}
			\label{tab:VusAll1pheno_Opal}
		\end{table}
		\begin{table}[H]
			\centering
			   \begin{adjustbox}{max width=\textwidth}
			   	\begin{tabular}{|c|c|c|c|c|c|c|c|c|c|c|c|c|c|c|c|}
					\hline
					Parameters&\multicolumn{5}{c|}{$\vert V_{us}\vert_{CIPT}$}&\multicolumn{5}{c|}{$\vert V_{us}\vert_{FOPT}$}&\multicolumn{5}{c|}{$\vert V_{us}\vert_{RGSPT}$}\\
					\cline{2-6}\cline{7-11}\cline{12-16}
					\text{ }&(0,0)&(1,0)&(2,0)&(3,0)&(4,0)&(0,0)&(1,0)&(2,0)&(3,0)&(4,0)&(0,0)&(1,0)&(2,0)&(3,0)&(4,0)\\
					\hline
					$\vert V_{us}\vert$(central)&0.2211&0.2210&0.2212&0.2228&0.2260&0.2222&0.2224&0.2225&0.2236&0.2253&0.2213&0.2213&0.2219&0.2238&0.2274\\ \hline
					$m_s$&$\text{}^{+0.0005}_{-0.0004}$&$\text{}^{+0.0010}_{-0.0009}$&$\text{}^{+0.0018}_{-0.0015}$&$\text{}^{+0.0029}_{-0.0025}$&$\text{}^{+0.0039}_{-0.0046}$&$\text{}^{+0.0008}_{-0.0007}$&$\text{}^{+0.0014}_{-0.0012}$&$\text{}^{+0.0022}_{-0.0019}$&$\text{}^{+0.0032}_{-0.0027}$&$\text{}^{+0.0045}_{-0.0038}$&$\text{}^{+0.0005}_{-0.0004}$&$\text{}^{+0.0011}_{-0.0009}$&$\text{}^{+0.0019}_{-0.0017}$&$\text{}^{+0.0032}_{-0.0028}$&$\text{}^{+0.0051}_{-0.0043}$\\ \hline
					Experimental &$\text{}^{+0.0033}_{-0.0034}$&$\text{}^{+0.0036}_{-0.0037}$&$\text{}^{+0.0036}_{-0.0037}$&$\text{}^{+0.0037}_{-0.0038}$&$\text{}^{+0.0038}_{-0.0039}$ &$\text{}^{+0.0033}_{-0.0034}$&$\text{}^{+0.0037}_{-0.0037}$&$\text{}^{+0.0037}_{-0.0037}$&$\text{}^{+0.0037}_{-0.0038}$&$\text{}^{+0.0038}_{-0.0039}$&$\text{}^{+0.0033}_{-0.0034}$&$\text{}^{+0.0036}_{-0.0037}$&$\text{}^{+0.0036}_{-0.0037}$&$\text{}^{+0.0037}_{-0.0038}$&$\text{}^{+0.0039}_{-0.0039}$\\ \hline
					Total theory&$\text{}^{+0.0005}_{-0.0005}$&$\text{}^{+0.0010}_{-0.0010}$&$\text{}^{+0.0020}_{-0.0020}$&$\text{}^{+0.0036}_{-0.0037}$&$\text{}^{+0.0062}_{-0.0065}$&$\text{}^{+0.0009}_{-0.0009}$&$\text{}^{+0.0019}_{-0.0019}$&$\text{}^{+0.0033}_{-0.0033}$&$\text{}^{+0.0051}_{-0.0050}$&$\text{}^{+0.0074}_{-0.0071}$&$\text{}^{+0.0005}_{-0.0004}$&$\text{}^{+0.0011}_{-0.0010}$&$\text{}^{+0.0022}_{-0.0020}$&$\text{}^{+0.0040}_{-0.0036}$&$\text{}^{+0.0068}_{-0.0061}$\\ \hline
					$s_0\in\left[2.5,\mtsq\right]$&0.0030 &0 .0070&0.0117&0.0188&0.0313&0.0042&0.0089&0.0143&0.0216&0.0325&0.0032&0.0074&0.0126&0.0204&0.0338\\
					\hline
					total&$\text{}^{+0.0045}_{-0.0045}$&$\text{}^{+0.0080}_{-0.0080}$&$\text{}^{+0.0124}_{-0.0124}$&$\text{}^{+0.0195}_{-0.0196}$&$\text{}^{+0.0321}_{-0.0322}$&$\text{}^{+0.0054}_{-0.0054}$&$\text{}^{+0.0098}_{-0.0099}$&$\text{}^{+0.0151}_{-0.0151}$&$\text{}^{+0.0225}_{-0.0225}$&$\text{}^{+0.0336}_{-0.0335}$&$\text{}^{+0.0046}_{-0.0047}$&$\text{}^{+0.0084}_{-0.0084}$&$\text{}^{+0.0133}_{-0.0133}$&$\text{}^{+0.0211}_{-0.0211}$&$\text{}^{+0.0347}_{-0.0346}$\\ \hline
			\end{tabular}
			\end{adjustbox}
			\caption{Determination of $\vert V_{us}\vert$ from OPAL data using prescription II. The phenomenological contributions are used for longitudinal components.}
			\label{tab:VusAll2pheno_Opal}
		\end{table}

\end{subappendices}

%% file: Chapters/Chap5.tex
\chapter{Renormalization group summation and analytic continuation from spacelike to timeline regions}
\label{Chapter5}

\lhead{Chapter 5. \emph{RG summation and analytic continuation from spacelike to timeline regions}}

\section{Motivation}
Analytic continuation of the perturbative series from spacelike to timelike regions is performed using renormalization group summed perturbation theory~(RGSPT). This method provides an all-order summation of kinematic ``$\pi^2$-terms'' accessible from a given order of a perturbative series. The impact of the summation of these terms is studied for Higgs boson decay and electromagnetic R-ratio in the perturbative QCD. Results obtained using RGSPT have improved convergence behavior in addition to significantly reduced renormalization scale dependence compared to fixed-order perturbation theory (FOPT). The higher-order behavior using the Pad\'e approximant is also studied for processes considered.

\section{Introduction}
QCD is a theory of strong interaction describing interaction among quarks and gluons. It has asymptotic freedom and is perturbative for large momentum transfers. The strong coupling constant ($\as$) measures the strength of the interaction and has a relatively large value ($\as(2\GeV)\sim 0.3$) compared to the electromagnetic coupling constant ($\alpha=1/137$) for a few $\GeV$s. This large value of $\as$ also introduces certain issues to the perturbation series in the fixed-order perturbation theory (FOPT) formalism. These key issues are mainly convergence, scheme dependence, and renormalization scale dependence. As discussed in the previous chapters, the uncertainties from these sources might constitute a large portion of the total uncertainty in the theoretical predictions. Various perturbative schemes are devised and used in the literature that address these issues. They have been applied in various precision determinations of the parameters of the strong and electroweak interactions of the Standard Model (SM). The perturbative renormalization group (RG) plays a key role in improving the fixed-order perturbation series, and various alternative schemes are also used in the literature. Some of these commonly used schemes are the principle of maximal conformality (PMC)~\cite{Brodsky:2013vpa}, optimized perturbation theory (OPT)~\cite{Stevenson:1981vj}, method of effective charges (MEC)~\cite{Grunberg:1982fw}, complete renormalization group improvement (CORGI)~\cite{Maxwell:1999dv,Maxwell:2000mm}, RGSPT~\cite{Ahmady:2002fd}, RG optimized perturbation theory (RGOPT)~\cite{Kneur:2010ss}, the principle of observable effective matching~\cite{Chishtie:2020cen} etc. \par
The perturbative QCD (pQCD) calculations to higher orders have been performed in the literature for many physical processes. These calculations mainly use the operator product expansion (OPE) formalism in deep Euclidean spacelike regions where perturbative treatment to QCD is applicable. However, the experimental information is obtained in the timelike regions. The physical quantities, such as R-ratios, are related to the discontinuities of the polarization function of a current correlator across the physical cut. Therefore, the analytic continuation from spacelike to timelike region acts as a bridge in relating the experimental observations with the theoretical predictions. For pQCD, the analytic continuation and associated issues have been pointed out in the early days of QCD, especially in Refs.~\cite{Moorhouse:1976qq,Pennington:1983rz,Pennington:1981cw,Bottino:1980rj,Krasnikov:1982fx,Radyushkin:1982kg}. One of the key issues emerged from these studies is large kinematical $\pi^2-$corrections arising from the imaginary part of the logarithms when analytic continuation is performed. These corrections are significant ($\sim (\pi \beta_0 )^n$) at higher orders of the perturbation theory and found to be dominating the genuine perturbative corrections in Refs.~\cite{Kataev:1995vh,Kataev:2008ym,Kataev:2008ntk,Herzog:2017dtz,Baikov:2005rw,Baikov:2008jh}. There have been numerous attempts to sum these kinematical terms in the literature using the RG in Ref.~\cite{Pivovarov:1991bi}, renormalon motivated naive non-abelianization (NNA) in Ref.~\cite{Broadhurst:2000yc}, using contour improved version of pQCD schemes such as CORGI in Ref.~\cite{Maxwell:2001he}, analytic perturbation theory in Refs.~\cite{Bakulev:2006ex,Bakulev:2010gm} and its variants such as fractional analytic perturbation theory in recent Refs.~\cite{Kotikov:2023meh,Kotikov:2023nvz}. This issue has also been addressed in Refs.~\cite{Ahrens:2008qu,Ahrens:2009cxz} for the Higgs production, for Sudakov logarithms in Ref.~\cite{Magnea:1990zb}, for the pion and nucleon electromagnetic form factors in Ref.~\cite{Bakulev:2000uh}, for deep inelastic scattering and Drell-Yan processes in Ref.~\cite{Parisi:1979xd}, for electromagnetic R-ratio in Refs.~\cite{Nesterenko:2020rbj,Nesterenko:2019rag,Nesterenko:2017wpb} and other in Refs.~\cite{Mueller:2012sma,Ralston:1982pa,Pire:1982iv,Gousset:1994yh}  . For hadronic $\tau$ decays, the  contour improved perturbation theory (CIPT) scheme is used and found to be inconsistent with the OPE expansion. For recent developments, we refer to Refs.~\cite{Hoang:2021nlz,Hoang:2021unk,Benitez-Rathgeb:2021gvw,Benitez-Rathgeb:2022yqb,Benitez-Rathgeb:2022hfj,Gracia:2023qdy,Golterman:2023oml}.\par
This chapter discusses an alternative analytical approach to sum the kinematical $\pi^2-$terms by performing the analytic continuation using RGSPT. In the RGSPT scheme, the running logarithms accessible from a given order are summed using the RG equation (RGE) in a closed form. These running logarithms are the key ingredients in the analytic continuation in the complex energy plane. This scheme has already been used in the study of $\tau$-decays in Refs.~\cite{Abbas:2012py,Abbas:2013usa,Abbas:2012fi,Caprini:2017ikn,Ananthanarayan:2016kll,Ananthanarayan:2022ufx} and other processes as well in Refs.~\cite{Ahmady:1999xg,Ahmady:2002fd,Ahmady:2002pa,Ananthanarayan:2020umo,Ahmed:2015sna}. In these studies, RGSPT provides better stability with renormalization scale variations. In some cases, it can also improve the convergence of the perturbative series at higher orders compared to the results from the FOPT scheme. We have discussed these issues involving the polarization functions related to the scalar, vector, electromagnetic current, and Higgs self-energies. These correlators are used in various QCD sum rule studies, and improvements provided by RGSPT can be crucial in precisely determining various SM parameters related to QCD and weak interaction physics.\par 
 This chapter is organized as follows: In section~\eqref{sec:def_RD}, the analytic continuation for the polarization or Adler functions in the FOPT and RGSPT schemes are described. In section~\eqref{sec:applications}, the application of analytic continuation of the perturbative series for various processes in the RGSPT and FOPT are discussed. These processes include the Higgs boson decaying to bottom quark pairs as well as gluon pairs, total hadronic decay width of the Higgs boson, electromagnetic R-ratio, and continuum contribution from the light quarks to muon $g-2$. The summary and conclusion are provided in section~\eqref{sec:summary_ancont}. The supplementary material needed for various sections can be found in the appendices \eqref{app:Rem}, \eqref{app:D_scalar}, \eqref{app:Wilson}, \eqref{app:D_higgs}.
 \section{Analytic continuation from spacelike to timelike regions}\label{sec:def_RD}
 The polarization functions ($\Pi(q^2)$) of current correlators for a physical process are calculated using the OPE formalism in the spacelike region where momentum transfer is large and perturbative treatment is applicable. These contributions are calculated by evaluating the Feynman diagrams appearing in a given order of the perturbation theory. Generally, the higher-order calculations are performed in a very special kinematical limit \emph{i.e.} either a small mass, a large mass, or in the expansion in the ratios of the masses of particles depending upon the scales present in theory. The OPE expansion also factorizes the short-distance perturbative part with the long-distance non-perturbative contributions. The long-distance contributions are encoded in terms of the quark and gluon condensates which are determined using the lattice QCD, chiral perturbation theory, optimized perturbation theory, etc. Once the relevant diagrams are evaluated, we get a fixed-order perturbation theory (FOPT) series given in Eq~\eqref{eq:Pseris} as an expansion in $\as$. \par
 In general, $\Pi(q^2)$ is not an RG invariant quantity and does not obey a homogeneous RGE. The Adler functions ($D(Q^2)$) in QCD are the RG invariant quantities derived from the $\Pi(q^2)$ as:
 \begin{equation}
     D(Q^2)\equiv -Q^2 \frac{d}{d\hs Q^2}\Pi(Q^2)\,,
     \label{eq:adler}
 \end{equation}
where, $\Pi(q^2)$ is calculated at spacelike regions ($Q^2=q^2<0$) and has a cut for the timelike regions ($Q^2=q^2>0$) due to the presence of the logarithms~$\log(\frac{\mu^2}{-q^2})$. The discontinuity across this cut is related to observables that are measured in the experiments in the timelike regions. A systematic study of various processes in the experiments thus requires the theoretical calculations to be valid in the energy regions of interest. It should also have a well-behaved behavior such that a precise determination of the various theoretical parameters can be obtained. For this purpose, proper analytic continuation plays a very important role. In this section, we give a short introduction to quantities that are needed for the analytic continuation and how it is performed using FOPT and RGSPT. These relations include theoretical quantities, such as polarization and Adler functions, and their relation to the experimental quantities, such as R-ratios ($R(s)$) for $e^+e^-$, which are used in the other sections of this chapter.\par   
 Polarization function $\Pi(Q^2)$ is related to the $R(s)$ by the following dispersion relation:
 \begin{equation}
     \Pi(Q^2)=\int_0^{\infty}\frac{R(s)}{(s+Q^2-i\hs\epsilon)} d\hs s\,,
 \end{equation}
 and the $D(Q^2)$ is obtained from the $\Pi(Q^2)$ as:
 \begin{equation}
    D(Q^2)=-Q^2\frac{d}{dQ^2}\Pi(Q^2)=Q^2\int_{0}^{\infty}\frac{R(s)}{(s+Q^2-i\hs\epsilon)^2}ds\,.
\end{equation}
Theoretical value of the $R(s)$ is obtained from the imaginary part of the $\Pi(Q^2)$~\cite{Poggio:1975af} as:
\begin{equation}
	\begin{aligned}
		R(s)&\equiv\frac{1}{2\pi i }\lim_{\epsilon\rightarrow 0}\left[ \Pi(-s-i \epsilon)-\Pi(-s+i \epsilon)\right]\\&=\frac{1}{2\pi i}\int_{-s+i \epsilon}^{-s-i \epsilon} d\hs q^2 \frac{d}{d\hs q^2}\Pi(q^2)\\&=\frac{-1}{2\pi i} \int_{-s+i \epsilon}^{-s-i \epsilon} \frac{dq^2}{q^2} D(q^2)\\ &=\frac{-1}{2\pi i}\oint_{|x_c|=1} \frac{d x_c}{x_c} D(-x_c s)  
		\label{eq:Rs_ancont}
	\end{aligned}
 \end{equation}
	where the contour of the integration does not cross the cut for $q^2>0$. In the intermediate steps of Eq.~\eqref{eq:Rs_ancont}, the path of integration is taken from point $-s-i\hs \epsilon$ on a circle of radius $s$ in the anticlockwise direction to point $-s+i\hs \epsilon$. The RGE plays a key role in the analytic continuation of $\Pi(Q^2)$ or $D(Q^2)$. The running logarithms ($\log(\frac{\mu^2}{-q^2})$) present in them are then analytically continued resulting in the large kinematical ``$\pi^2$" corrections. Once the analytic continuation is achieved, the running logarithms are resumed by setting $\mu^2=s$, leaving behind the large $\pi^2$-terms amplified by $\beta_i$ or $\gamma_i$ coefficients. Such kinematical terms, in some cases, dominate the genuine perturbative corrections at higher orders starting from N$^3$LO. In some cases, the convergence of the fixed order series is also spoiled by them at lower energies. Such examples for the hadronic Higgs decay width and electromagnetic R-ratio can be found in Ref.~\cite{Herzog:2017dtz}. \par
	The timelike perturbative series $\tilde{\mathcal{S}}(s)$ is calculated from a spacelike series $\mathcal{S}(Q^2)$ from Eq.~\eqref{eq:Pseris} as:
	\begin{align}
		\tilde{\mathcal{S}}(s)&\equiv \lim_{\epsilon\rightarrow0}\frac{1}{2 i}(\mathcal{S}(-s-i \epsilon)-\mathcal{S}(-s+i \epsilon))\nonumber\\
  &=\frac{1}{2\pi i}\int_{-s+i \epsilon}^{-s-i \epsilon}dq^2\frac{d}{dq^2}\mathcal{S}(q^2)\nonumber\\
  &=\frac{-1}{2\pi i}\sum_{i=0} j\hs x^{i}T_{i,j}\int_{-s+i \epsilon}^{-s-i \epsilon} \frac{dq^2}{q^2}\log^{j-1}\left(\frac{\mu^2}{-q^2}\right)\label{eq:ancont_1}\\		&=\frac{-1}{2\pi i}\sum_{i=0} j x^{i}  T_{i,j}\oint_{|x_c|=1} \frac{dx_c}{x_c}\log^{j-1}\left(\frac{\mu^2}{s\hs x_c}\right)\,,\quad \left(\text{substituting }q^2\rightarrow-s \hs x_c \right)\label{eq:ancont_2}\\
  &=\frac{-1}{2\pi}\sum_{i=0} j \hs x^{i}  T_{i,j}\int_{-\pi}^{\pi} d\phi\left(L_s-i \phi\right)^{j-1}\,,\quad \left(\text{substituting } x_c\rightarrow e^{i \phi}\right)\label{eq:ancont_3}\\
  &=\frac{1}{2\pi i}\sum_{i=0} x^{i} T_{i,j}\left( (L_s+i \pi )^{j}-(L_s-i \pi )^{j}\right)\,,
  \label{eq:ancont_FOPT}
	\end{align}
	where $L_s\equiv \log\left(\frac{\mu^2}{s}\right)$. So, analytic continuation for a FOPT series given in Eq.~\eqref{eq:Pseris}, can be obtained by taking the imaginary part by substituting $L\rightarrow L_s\pm i\hs\pi$. The logarithmic terms are resumed by setting $s=\mu^2$, leaving behind only the large $\pi^2-$type corrections.\par
 Given the form of the most general term obtained using RGSPT in Eq.~\eqref{eq:coef_rgspt} where running logarithms are present in the numerator and the denominator, summation of $\pi^2-$terms to all orders is very natural. The analytic continuation for RGSPT can also be directly obtained by substituting $L\rightarrow L_s\pm i\hs  \pi$ as for FOPT. Another form in terms of trigonometric functions can also be derived.\par
	The most general term in Eq.~\eqref{eq:coef_rgspt} for the RGSPT series can be written as:
	\begin{equation}
		\frac{\log^m(1-u_1 \log\left(\frac{\mu^2}{-q^2}\right))}{\left(1-u_1 \log\left(\frac{\mu^2}{-q^2}\right)\right)^n}=\partial_{\delta}^m\left(1- u_1\log\left(\frac{\mu^2}{-q^2}\right)\right)^{n-\delta}\Bigg\rvert_{\delta\rightarrow0}
	\end{equation}
where $u_1=x\hs \beta_0$ and $m$ is always an integer. Following the steps as in Eqs.~\eqref{eq:ancont_1},\eqref{eq:ancont_2},\eqref{eq:ancont_3}, we get the following result for RGSPT:
\begin{align}
\frac{1}{2\pi i} \oint_{|q^2|=s} \frac{dq^2}{q^2}&\frac{\log^m(1-u_1 \log\left(\frac{\mu^2}{-q^2}\right))}{\left(1-u_1 \log\left(\frac{\mu^2}{-q^2}\right)\right)^n}\nonumber\\&=\lim_{\delta\rightarrow0}\partial_{\delta}^m 
\begin{cases}\frac{\tan ^{-1}\left(\frac{\pi u_1}{1- u_1\hs L_s}\right)}{\pi  u_1}\,, & n=1 \\
\frac{w_s^{-\frac{1}{2} (n-\delta -1)} \sin \left((n-\delta -1) \tan ^{-1}\left(\frac{\pi u_1}{1-u_1\hs  L_s }\right)\right)}{\pi u_1 (n-\delta -1)}\,,& n\neq 1
\end{cases}
\label{eq:master_eq}
\end{align}
where, $w_s=\left(1- u_1 \hs L_s  \right)^2+\pi^2 u_1^2$. We can see that all the kinematical $\pi^2$-terms are summed in $w_s$ and $\tan^{-1}(\frac{\pi u_1}{1-u_1 \hs L_s}) $. The large logarithms are also under control as they are always accompanied by $\as$, which is one of the important features of the RGSPT. Hence, both RG improvement, as well as all-order summation is naturally achieved in the RGSPT. In addition, results can be calculated in an analytic form, unlike the analytic QCD methods or numerically evaluating imaginary parts along the contour using CIPT. It should be noted that the $n=1$ case has already been known in the literature~\cite{Groote:2012jq,Groote:2001im} and higher terms, to the best of our knowledge, are new and is the prediction from RGSPT. This is our main result, and its implications and the improvements achieved for various physical processes are discussed in the rest of this chapter.
\section{Applications} \label{sec:applications}
In this section, we discuss the application of summation of the kinematical $\pi^2-$terms and its application in various processes. These processes involve energies ranging from a few $\GeV$s to several hundreds of $\GeV$s. They are calculated from the $\Pi(q^2)$ for the gluon field strength tensor as well as vector, axial-vector, and scalar currents. These polarization functions have applications in hadronic $\tau$ decays, $e^+e^-$ annihilation, and hadronic Higgs decays. For more details about these processes, we refer to a recent review in Ref.~\cite{Pich:2020gzz} and references therein. In addition, some continuum contributions to the experimental values of the charmonium and bottomonium moments~\cite{Kuhn:2007vp,Dehnadi:2011gc,Dehnadi:2015fra,Boito:2019pqp,Boito:2020lyp} as well as in hadronic contribution to muon $(g-2)_\mu$~\cite{Boito:2022rkw,Boito:2022dry} are also worth mentioning. Most of such examples for FOPT are already discussed in the literature~\cite{Herzog:2017dtz}. In this study, we have provided a systematic comparison of FOPT results with the RGSPT. First, we discuss the high-energy processes ($\sim 100\GeV$) relevant to the hadronic Higgs decays, and then we move to vector current processes where intermediate energies ($\sim$ few $\GeV$s) are involved. 
\subsection{Higgs decay width in pQCD}\label{subsec:Hdecay}
\subsubsection{\texorpdfstring{$H\rightarrow \overline{b}b$}{} decay.}\label{subsec:Hbb}
Higgs decaying to bottom pair is the dominant decay of the Higgs boson and has been of constant interest from both experimental and theoretical points of view. This process is also looked for as one of the signals in the discovery of the Higgs boson. Theoretically, QCD correction to this process is related to the imaginary part of the $\Pi(q^2)$ of the scalar current correlator. It is known to $\ordas{4}$~\cite{Becchi:1980vz,Broadhurst:1981jk,Chetyrkin:1996sr,Baikov:2005rw,Gorishnii:1990zu,Gorishnii:1991zr,Herzog:2017dtz} and the unknown higher-order coefficients are estimated using the d-Log Pad\'e method in Ref.~\cite{Boito:2021scm}. Recently, a FOPT analysis for this process has been used in Ref.~\cite{Aparisi:2021tym} to calculate the mass of the bottom quark at the scale of Higgs boson mass \emph{i.e.} $m_b(m^2_H)$. It has been found that the FOPT is inapplicable at low energy scale $(\mu\sim m_b)$ due to convergence issues mainly arising from the kinematical $\pi^2-$terms. These shortcomings, however, can be resolved using the RGSPT schemes and discussed in the rest of this subsection. It should be noted that other methods, such as a renormalon motivated large-$\beta_0$ procedure in Ref.~\cite{Broadhurst:2000yc} and analytic QCD approach in Ref.~\cite{Bakulev:2006ex} can also be used to resum the kinematical $\pi^2$-terms. Here, we also show that a significant reduction in the scale uncertainty can also be achieved using RGSPT.\par
The decay of the Higgs to bottom quark pair using pQCD is given by:
\begin{align}
	 \Gamma\left(H\rightarrow \overline{b}b\right) &=\frac{3 G_{F} m_{H}}{4 \sqrt{2} \pi} m^2_{b}(\mu^2) \tilde{\mathcal{S}}(\mu^2)+\text{Other corrections}\,,\\
  &=\Gamma_0 \tilde{\mathcal{S}}(\mu^2)
  \label{eq:hbb}
\end{align}
where $\tilde{\mathcal{S}}(\mu)$ is an analytically continued perturbative series as discussed in Eq.~\eqref{eq:ancont_FOPT}. Other corrections, including electroweak and mixed corrections, are irrelevant to our discussion and therefore ignored in this study.\par
Using the $\ordas{4}$ inputs from Refs.~\cite{Baikov:2005rw,Herzog:2017dtz} and D-log Pad\'e predictions to  $\ordas{8}$ from Ref.~\cite{Boito:2021scm}, the FOPT series has following contributions:
	\begin{equation}
	\resizebox{\textwidth}{!}{$
		\begin{aligned}
	\tilde{\mathcal{S}}(m_H^2)=1+ 0.2030+0.0374+ 0.0019 -0.0014-0.0004+6\times10^{-6}&+3\times10^{-5}+5\times10^{-7}\,.
	 	\label{eq:hbbmu_mh}
	 \end{aligned}$}
\end{equation}
where $\mu=m_H$ with $\alpha_s(m_H^2)=0.1125$ is taken as inputs using RunDec~\cite{Herren:2017osy}.
However, when the above series is calculated at $\mu=\overline{m}_b=4.18\GeV$ and $\alpha_s(\overline{m}_b^2)=0.2245$, numerical contributions from different terms are found to be:
 \begin{align}
 	\tilde{\mathcal{S}}(\overline{m}_b^2)=&1-0.5659+0.0585+ 0.1469-0.1267+0.0297+ 0.0381 -0.0438+ 0.0114\,,
 	\label{eq:hbbmu_mb}
 \end{align}
and we can see that the effect of the running logarithms and the large kinematical corrections spoil the perturbative nature of the series at the scale $\sim \hs m_b$. Due to the poor convergence behavior of the series in Eq.~\eqref{eq:hbbmu_mb}, only results obtained in Eq.~\eqref{eq:hbbmu_mh} are used in Ref.~\cite{Aparisi:2021tym} to extract the $m_b(m^2_H)$.\par
These shortcomings can be cured using the RGSPT scheme, and it can also improve the convergence compared to FOPT results in Eq.~\eqref{eq:hbbmu_mh} and Eq.~\eqref{eq:hbbmu_mb}. At scale $\mu=m_H$, the RGSPT series has the following contributions:
	\begin{equation}
	\resizebox{\textwidth}{!}{$
		\begin{aligned}
		\tilde{\mathcal{S}}^{\small{\Sigma}}(m_H^2)=0.9839+ 0.1894+ 0.0469+ 0.0130+0.0042+ 0.0016+ 0.0007+ 0.0003+0.0002\,.
		\label{eq:hbbmu_mb_sum}
 \end{aligned}$}
\end{equation}
 The perturbative series obtained using  RGSPT is monotonically decreasing, unlike in Eq.~\eqref{eq:hbbmu_mb_sum}), and the truncation uncertainty can also be calculated reliably. When we choose renormalization scale $\mu=m_b$, the perturbation series for RGSPT has the following contributions from different orders:
	\begin{equation}
	\resizebox{\textwidth}{!}{$
		\begin{aligned}
	\tilde{\mathcal{S}}^{\tiny{\Sigma}}(m_b^2)=&0.4949+ 0.0427+ 0.0097+0.0033+0.0011+0.0005+ 0.0002+0.0001+ 5\times10^{-5}\,.
\end{aligned}$}
\end{equation}
Again, the resummation of kinematical $\pi^2-$terms has significantly improved the convergence behavior of the series and it can also be used at such low scales. The RGSPT series to $\order{\alpha_s^4}$ is  monotonically convergent for $\alpha_s\le0.373$, which is around charm mass scale $\mu\sim \overline{m}_c$. While the FOPT series is convergent only for $\alpha_s<0.160$, for $\mu\sim4\overline{m}_b$ for $n_f=5$ active quark flavors. Theoretical uncertainties using FOPT and RGSPT series at scale $\mu=m_H$, when scale is varied in the range $\mu\in\left[m_H/4,2\hs m_H\right]$, have the following numerical values:
\begin{align}
		m^2_b(m_H^2)\tilde{\mathcal{S}}(m_H^2)=9.5655\pm0.0105_{\text{trunc.}}\pm0.0113_{\as}\pm0.0103_{\mu}=9.5655\pm0.0185\\
		m^2_b(m_H^2)\tilde{\mathcal{S}}^{\tiny{\Sigma}}(m_H^2)=9.5384\pm0.0324_{\text{trunc.}}\pm0.0104_{\as}\pm0.0036_{\mu}=9.5384\pm0.0342
\end{align}
where uncertainties are ordered as due to truncation, uncertainties present in the PDG average in $\alpha_s$, and the last one is from the scale variations. We can also rewrite it in the following form:
\begin{align}
    \Gamma^{\text{FOPT}}(H\rightarrow \overline{b}b)&=\Gamma_0 (1.241\pm0.002)\,,\\
    \Gamma^{\text{RGSPT}}(H\rightarrow \overline{b}b)&=\Gamma_0 (1.237\pm0.004)
\end{align}
It should be noted that the truncation uncertainty quoted above FOPT is much smaller than the RGSPT results due to the cancellation between the genuine perturbative contributions and the kinematical $\pi^2-$terms. Hence, the exact nature of these truncation uncertainties directly derived from the FOPT may be misleading and should be dealt with carefully. A safer choice for FOPT would be to estimate such uncertainties from the Adler functions rather than analytically continued series. For the Adler function, we get the following contributions at scale $\mu=m_H$:
\begin{equation}
    S(m_H^2)=1.0000+ 0.2030+ 0.0539+ 0.0162+ 0.0058+\cdots\,,
\end{equation}
and the last term is $\sim 4.3$ times larger than the N$^4$LO term of Eq.~\eqref{eq:hbbmu_mh}.
Other theoretical uncertainties for RGSPT are significantly smaller compared to FOPT.\par
\begin{figure}[H]
\centering
		\includegraphics[width=.49\textwidth]{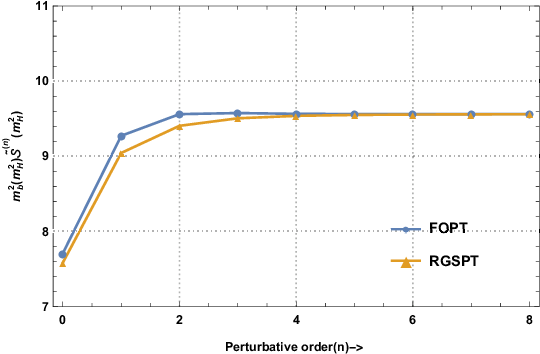}
		\includegraphics[width=.49\textwidth]{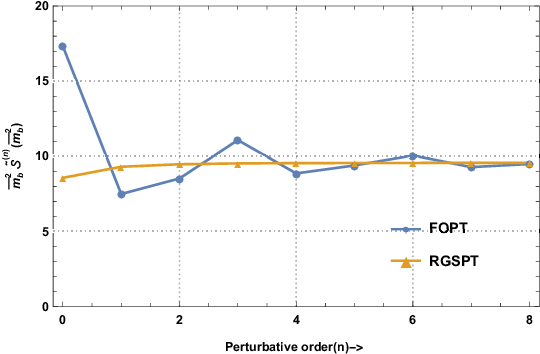}
	\includegraphics[width=.49\textwidth]{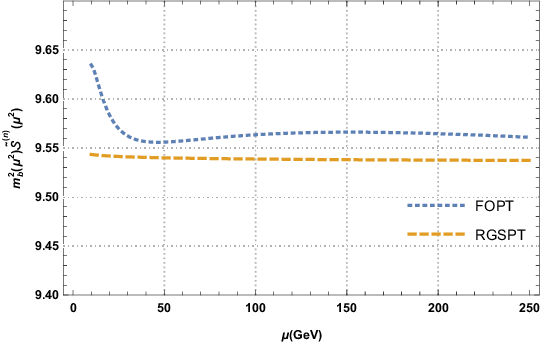}
\caption{\label{fig:Hbb_n} $m_{b}^2(\mu)\delta_{QCD}$ with perturbative order using $\mu=m_H$, $\mu=\overline{m}_b$ and scale dependence for the known $\order{\alpha_s^4}$ terms in the range $\mu\in[10\GeV,2m_H]$.}
\end{figure}

The behavior of $\tilde{\mathcal{S}}_H$ with increasing higher-order contributions for two different scales are presented in Fig.~\eqref{fig:Hbb_n}. The improved convergence over a wider range of renormalization scales using RGSPT allows one to perform the perturbative analysis for a wider energy region compared to FOPT. 
For a relatively large value of the strong coupling at $\mu=m_b$, the convergence of the RGSPT is much better, and it quickly approaches the asymptotic value while the FOPT series oscillates. This can be seen in Fig.~\eqref{fig:Hbb_n}

These improvements are used in the light-quark mass determination using Borel-Laplace sum rules in Ref.~\cite{AlamKhan:2023ili}.

\subsubsection{\texorpdfstring{$H\rightarrow g g$}{}  decay}\label{subsec:hgg}
This decay mode is the second important channel for the Higgs boson decays. It is mediated by a heavy quark loop and the dominant contribution comes from  a top quark loop due to its large value of the Yukawa coupling, and the decay width is calculated in the heavy-top ($M_H\ll M_t$) limit~\cite{Inami:1982xt}. \par
The decay width of this process in QCD is related to the imaginary part of the self-energy of the Higgs boson ($\Pi^{GG}(q^2)$) via optical theorem. It is given by the following relation~\cite{Inami:1982xt}:
\begin{equation}
	\Gamma\left(H\rightarrow gg\right)=\frac{\sqrt{2}G_F}{m_H}|C_1|^2 \im \hs \Pi^{GG}(-m_H^2-i \epsilon)\,,
\end{equation}
where $C_1$ is known to N$^4$LO can be found in Refs.~\cite{Chetyrkin:2016uhw,Herzog:2017dtz}. This Wilson coefficient is obtained using the low-energy theorem and decoupling relations for the $\as$. The decoupling relations for the $\as$ and quark masses can also be obtained using RGSPT and provide further RG improvement to this process by reducing the scale dependence~\cite{Abbas2}.

The perturbative contribution to $\im \hs \Pi^{GG}(q^2)$ are known to $\ordas{4}$~\cite{Kataev:1981gr,Chetyrkin:1997iv,Baikov:2006ch,Schreck:2007um,Herzog:2017dtz} and it can be written as:
\begin{equation}
	\im \hs \Pi^{GG}(q^2)\equiv \frac{2 q^4}{\pi}G(x(q),L=0)=\frac{2 q^4}{\pi}\{1+\sum_{i=1}^{\infty}g_i x^i(q)\}
 \label{eq:G2}
\end{equation}
 where $x=\alpha^{\left(n_l=5\right)}_s(q^2)/\pi$ and $L= \log(\mu^2/q^2)$. Coefficients $g_i$ can be obtained from Ref.~\cite{Herzog:2017dtz}, and their RG evolution of $G(x,L)$ is obtained by the RG invariance of the quantity:
 \begin{equation}
     \mu^2\frac{d}{d\mu^2}\left(\left(\frac{\beta(x)}{x}\right)^2 G(x(\mu), L)\right)=0\,,
     \label{eq:RGE_ancontG2}
 \end{equation}
where $\beta(x)$ is the QCD beta function defined in Eq.~\eqref{anomalous_dim}. We calculate the RGSPT series for this process by rewriting the perturbative series in terms of the RG summed coefficients, $S_i( x(\mu) L)$, as:
\begin{equation}
    G^{\Sigma}(x(\mu),L)=\sum_{i=0}S_i( x(\mu) L) x^i(\mu)\,.
\end{equation}
 When the above series is subjected to Eq.~\eqref{eq:RGE_ancontG2}, we get the following differential equations among various $S_i( x(\mu) L)$ as:
\begin{equation}
	\frac{d S_i(u)}{d u}-\sum _{j=0}^i \frac{\beta _j} {u^{i+j+1}}\frac{d}{d u} \left(u^{i+j+2} S_{i-j}(u)\right)=0\,,
\end{equation}
where $u= x\hs L$. The RGSPT form of the Adler function related to $G(q^2)$ is given by $D_{11}$, and its expression can be found in appendix~\eqref{app:D11_RGSPT}. \par
\begin{figure}[H]
		\includegraphics[width=.49\textwidth]{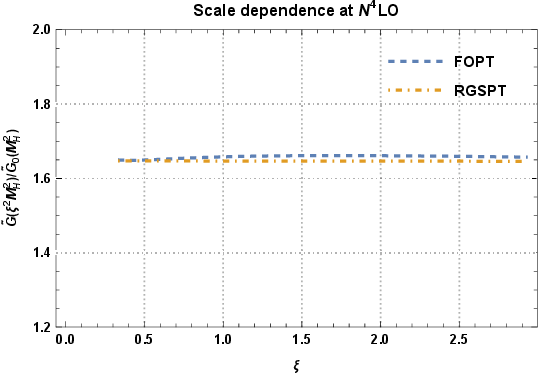}
        \includegraphics[width=.49\textwidth]{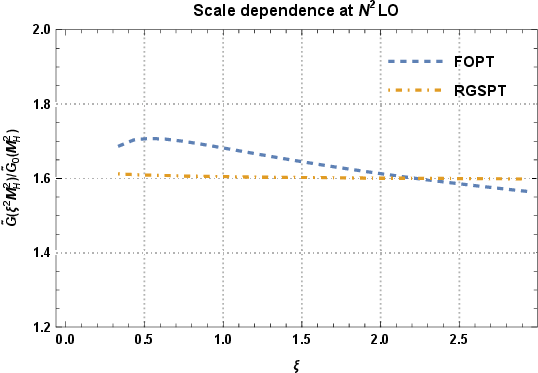}
		\includegraphics[width=.49\textwidth]{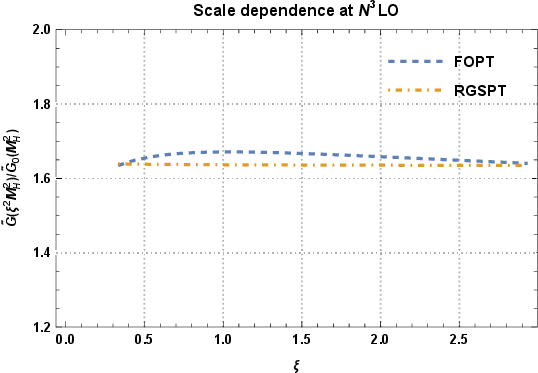}
		\includegraphics[width=.49\textwidth]{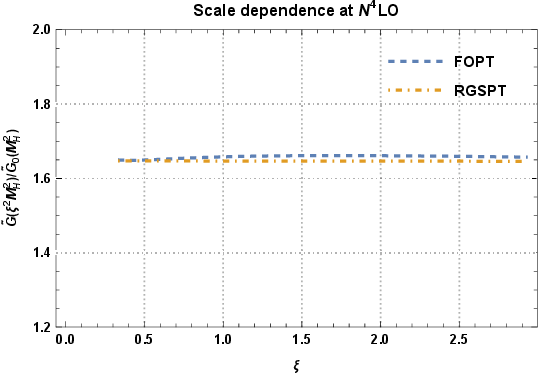}
 \caption{ \label{fig:G2_scdep}The scale variation of $\tilde{G}(\xi^2 m_H^2)/\tilde{G}_0(m_H^2)$ using known N$^4$LO results in the $\msbar$ scheme.}
\end{figure}

Using $n_l=5$ and setting renormalization scale $\mu=q=m_H$, the $G(x(m_H),0)$ has the following numerical form:
\begin{align}
G^{\text{FOPT}}(m^2_H)&=1+3.9523 \alpha _s+6.9555 \alpha _s^2-6.8518 \alpha _s^3-75.2591 \alpha _s^4+\order{\alpha_s^5}\nonumber\\&
=1+ 0.4448+ 0.0881 -0.0098 -0.0121+\cdots\,.
\end{align}
 The N$^4$LO term dominates the N$^3$LO term in the FOPT. In the case of RGSPT, the same quantity is given by:
 \begin{align}
 	G^{\text{RGSPT}}(m^2_H)&=0.9555+3.5357 \alpha_s+8.6098\alpha_s^2+20.4939 \alpha_s^3+56.6920 \alpha_s^4+\order{\alpha_s^5}\nonumber\\&
 	=0.9554+0.3979+0.1090+0.0292+ 0.0091+\cdots\,.
 	\label{eq:summedHgg}
 \end{align}
The convergence of the perturbative series is good for RGSPT compared to FOPT. Summation of the kinematical terms also enhances the range of convergence of the perturbation series, and the N$^4$LO term dominates the N$^3$LO term when $\alpha_s(\mu)>0.3$ for $\mu\sim2\GeV$. In addition, a significant reduction in the scale uncertainty can be seen in Fig.~\eqref{fig:G2_scdep}.

 The higher-order behavior and effects of the summation of kinematical terms can be studied by estimating the unknown terms using the Pad\'e approximants. Since only $\order{\alpha_s^4}$ results are known in the literature, we can choose various $W^{\left(n,m\right)}$ Pad\'e approximants for series $S(q^2)$ (before analytic continuation is performed) defined in Eq.~\eqref{eq:Pseris}. The $n$ and $m$ are the exponents of the polynomial in the numerator and denominator of $W^{\left(n,m\right)}$. We obtain the following Pad\'e approximants:
 \begin{align}
     W^{\left(0,4\right)}&=\frac{1}{1- 12.417 x + 49.268 x^2 - 195.212 x^3 - 466.754 x^4}\,,\\
     W^{\left(1,3\right)}&=\frac{1 - 2.391 x}{1 - 14.808 x + 78.957 x^2 - 313.013 x^3}\,,\\
     W^{\left(3,1\right)}&=\frac{1 + 2.571 x - 17.347 x^2 - 146.841 x^3}{1 - 9.846 x}\,.
 \end{align}
The Pad\'e approximant $W^{\left(2,2\right)}$ results in a negative $\order{\alpha_s^5}$ coefficient therefore discarded. The predictions of these approximations are in agreement, and therefore we take their average:
\begin{align}
  \overline{W}=&1+ 12.417 x + 104.905 x^2 + 886.037 x^3 + 8723.76 x^4 +
  89630 x^5 + 906226 x^6 \nonumber\\&\bs+ 8.9855\times10^6 x^7 + 8.87515\times10^7 x^8 +
  8.80468\times10^8 x^9 + 8.76277\times10^9 x^{10}\,,
  \label{eq:hgg_pred}
\end{align}
that can be used for higher-order behavior. An APAP prediction~\cite{Chishtie:1998rz,Ananthanarayan:2020umo} for $\order{x^5}$ coefficient (without analytic continuation) in the large logarithm limit, we obtain:
 \begin{equation}
     d_{5}^{APAP}=48056\,,
 \end{equation}
 which is nearly half of the predictions from simple Pad\'e approximants in Eq.~\eqref{eq:hgg_pred}. An APAP prediction for the decay width can be found in Ref.~\cite{Abbas:2022wnz}. It will be interesting to compare these predictions with a detailed analysis using the D-log Pad\'e~\cite{Boito:2021scm,Boito:2022rad} or in a large-$\beta_0$ approximation~\cite{Boito:2022fmn}. \par
\begin{figure}[ht]
\centering\includegraphics[width=.7\textwidth]{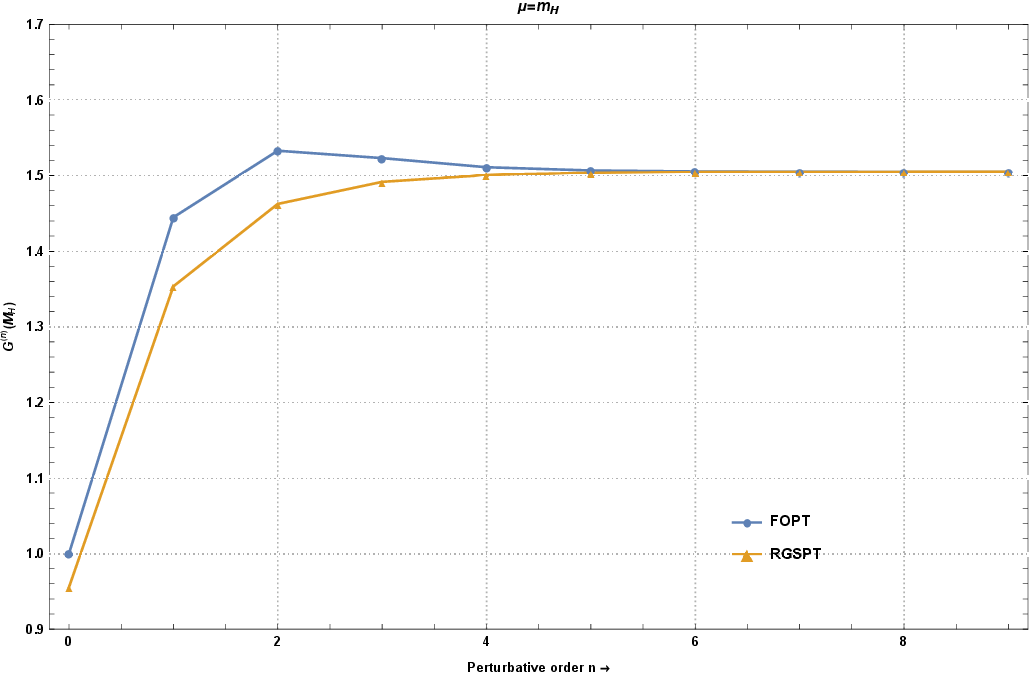}
	\caption{\label{fig:Hgg_n} $G^{(n)}(m_H)$ with perturbative order $n$ in FOPT and RGSPT at scale $\mu=m_H$.}
\end{figure}

 
 Using the above inputs, the higher-order behavior of $H\rightarrow gg$ in FOPT is obtained as:
 \begin{align}
     G^{\text{FOPT}}(m_H^2)=1+0.44477&+ 0.08808 -0.00976 -0.01207 -0.00432 \nonumber\\&-0.00132-0.00051 -0.00015+ 0.00001+\cdots\,,
 \end{align}
 and for RGSPT, the contributions from different order are obtained as:
 \begin{align}
     G^{\text{RGSPT}}(m_H^2)=0.95555&+ 0.39788+ 0.10903+ 0.02921+ 0.00909+ 0.00287\nonumber\\&+ 0.00086+
0.00024+ 0.00006+ 0.00001+\cdots\,,
 \end{align}
 and are shown in Fig.~\eqref{fig:Hgg_n}. It is evident that the 
 RGSPT provides better convergence. The scale dependence from RGSPT and FOPT in different schemes are presented in Fig.~\eqref{fig:G2_scdep}. \par 
  Now, we present our result for this process in different schemes used in the literature. The decay width has already been studied very recently in Refs.~\cite{Herzog:2017dtz,Abbas:2022wnz} for scale-invariant (SI), $\overline{\textrm{MS}}$, and on-shell scheme for the top quark mass present in $C_1$. Using top quark mass for SI and $\overline{\textrm{MS}}$ scheme as $\mu_t=m_t(\mu_t)=165\GeV$, and $m_t=173\GeV$ for on shell scheme, we obtain decay width in different schemes using FOPT as:
\begin{align}
\Gamma^{\text{SI,FOPT}}_{\text{N$^4$LO}}	&=\Gamma_0\left(1.8454\pm0.0116_{\text{trunc.}}\pm0.0346_{\alpha_s}\pm0.0159_{\mu}\right)=\Gamma_0\left(1.8454\pm0.0398\right)\,,\nonumber\\
\Gamma^{\overline{\textrm{MS}},\text{FOPT}}_{\text{N$^4$LO}}&=\Gamma_{0}\left(1.8444\pm0.0115_{\text{trunc.}}\pm0.0346_{\alpha_s}\pm0.0135_{\mu}\right)=\Gamma_{0}\left(1.8444\pm0.0389\right)\,,\nonumber\\ \Gamma^{\text{OS,FOPT}}_{\text{N$^4$LO}}&=\Gamma_{0}\left(1.8452\pm0.0111_{\text{trunc.}}\pm0.0346_{\alpha_s}\pm0.0152_{\mu}\right)=\Gamma_{0}\left(1.8452\pm0.0394\right)\,,
\end{align}
where, $\Gamma_{0}\left( H\rightarrow gg\right)=\frac{G_{F}m_H^3}{36\pi^3\sqrt{2}}\alpha_s^2(m^2_H)$ is leading contribution.
The decay width in the RGSPT scheme read:
\begin{align}
\Gamma^{\text{SI,RGSPT}}_{\text{N$^4$LO}}	&=\Gamma_0\left(1.8273\pm0.0177_{\text{\text{trunc.}}}\pm0.0337_{\alpha_s}\pm0.0037_{\mu}\right)=\Gamma_0\left(1.8273\pm0.0382\right)\,,\nonumber\\
\Gamma^{\overline{\textrm{MS}},\text{RGSPT}}_{\text{N$^4$LO}}&=\Gamma_{0}\left(1.8264\pm0.0177_{\text{trunc.}}\pm0.0346_{\alpha_s}\pm0.0067_{\mu}\right)=\Gamma_{0}\left(1.8264\pm0.0386\right)\,,\nonumber\\	\Gamma^{\text{OS,RGSPT}}_{\text{N$^4$LO}}&=\Gamma_{0}\left(1.8271\pm0.0181_{\text{trunc.}}\pm0.0346_{\alpha_s}\pm0.0042_{\mu}\right)=\Gamma_{0}\left(1.8271\pm0.0385\right)\,.
\end{align}
We can also calculate the decay width in the miniMOM (MM) scheme~\cite{vonSmekal:2009ae,Gracey:2013sca} and top quark mass in the on-shell scheme. The results in FOPT and RGSPT schemes are given by:
\begin{align}
\Gamma^{\text{MM,FOPT}}_{\text{N$^4$LO}}&=\Gamma_{0}\left(1.8444\pm0.0111_{\text{trunc.}}\pm0.0346_{\alpha_s}\pm0.0136_{\mu}\right)=\Gamma_{0}\left(1.8444\pm0.0388\right)\,,\nonumber\\	\Gamma^{\text{MM,RGSPT}}_{\text{N$^4$LO}}&=\Gamma_{0}\left(1.8264\pm0.0181_{\text{trunc.}}\pm0.0337_{\alpha_s}\pm0.0047_{\mu}\right)=\Gamma_{0}\left(1.8264\pm0.0385\right)\,.
\end{align}
It should be noted that the scale dependence is calculated by varying scale in the range $\xi\in\left[m_H/3,3 \hs m_H\right]$. It is clear from these results that the $\pi^2-$terms significantly cancel the genuine perturbative corrections, which led to small truncation uncertainty in the FOPT results. A similar behavior is also observed for the $H \rightarrow bb$ process, but here  $\as$ is a primary source of uncertainty.

\subsubsection{Total Hadronic Higgs Decay Width.}
\label{subsec:H_Hadron}
The branching ratio of Higgs boson decays to hadrons is about 70\%~\cite{Davies:2017xsp}. The hadronic decay width of Higgs boson is calculated by constructing an effective Lagrangian from Yukawa term and strong interaction where heavy top quark is integrated out~\cite{Inami:1982xt,Chetyrkin:1996wr,Chetyrkin:1996ke}. The effective Lagrangian has the form:
\begin{equation}
\mathcal{L}_\text{eff.}=-\frac{H^0}{v^0}\left(C_1 \left[\mathcal{O}'_1\right]+C_2 \left[\mathcal{O}'_2\right]\right)+\mathcal{L}'\,,
\label{eq:L_eff}
\end{equation}
where $H^0$ and $v^0$ are the bare Higgs field and vacuum expectation value. The primed quantities are defined in five-flavored QCD. The coefficients $C_1$ and $C_2$ in Eq.~\eqref{eq:L_eff} are the Wilson coefficients of the operators constructed out of light gluonic and bottom quark degrees of freedom. These Wilson coefficients also carry large logarithms $\sim\log(\mu^2/m_t^2)$, which can be summed using the RGSPT~\cite{Abbas2}. Since these coefficients do not involve analytic continuation, therefore, their FOPT expressions are used in this subsection. Their numerical expression  can be found in the appendix~\eqref{app:Wilson}.
The operators $\mathcal{O}_1^\prime$ and $\mathcal{O}_2^\prime$ are given by:
\begin{align}
    \mathcal{O}_1^\prime&=\left(G_{a,\mu\nu}^{0\prime}\right)^2\,,\\
      \mathcal{O}_2^\prime&=m_b^{0\prime}\overline{b}^{0\prime}b^{0\prime}\,,
\end{align}
where $G^{0\prime}_{a,\mu\nu}$ is the bare gluon field strength, $m_b^{0\prime}$ is the bare bottom quark mass and $b^{0\prime}$ is bare bottom quark field. The current correlators for the operators $\mathcal{O}_1^\prime$ and $\mathcal{O}_2^\prime$ are given by:
\begin{equation}
\Pi_{ij}(q^2)=i \int dx \hs e^{i q x} \langle0\vert \mathcal{T}\left[\mathcal{O}^\prime_i,\mathcal{O}^\prime_j\right]\vert 0\rangle\,.
\label{eq:corr}
\end{equation}
These correlators can be used to define analytically continued quantities:
\begin{align}
    \Delta_{ii}&=K_{ii}\hs \text{Im}\left(\Pi_{ii}(M_H^2)\right)\,,\\
    \Delta_{12}&=K_{12}\hs \text{Im}\left(\Pi_{12}(M_H^2)+\Pi_{21}(M_H^2)\right)\,,
\end{align}
where $\Pi_{12}(M_H^2)=\Pi_{21}(M_H^2)$, $K_{11}=\left(32\pi M_H^4\right)^{-1}$ and $K_{12}=K_{22}=\left(6\pi M_H^2 m_b^2\right)^{-1}$. The $\Delta_{11}$ is proportional to $G(q^2)$ in Eq.~\eqref{eq:G2} and $\Delta_{22}$ is related to the $\tilde{\mathcal{S}}$ in Eq.~\eqref{eq:hbb}.\par
\begin{figure}[H]
		\centering
		\includegraphics[width=.7\textwidth]{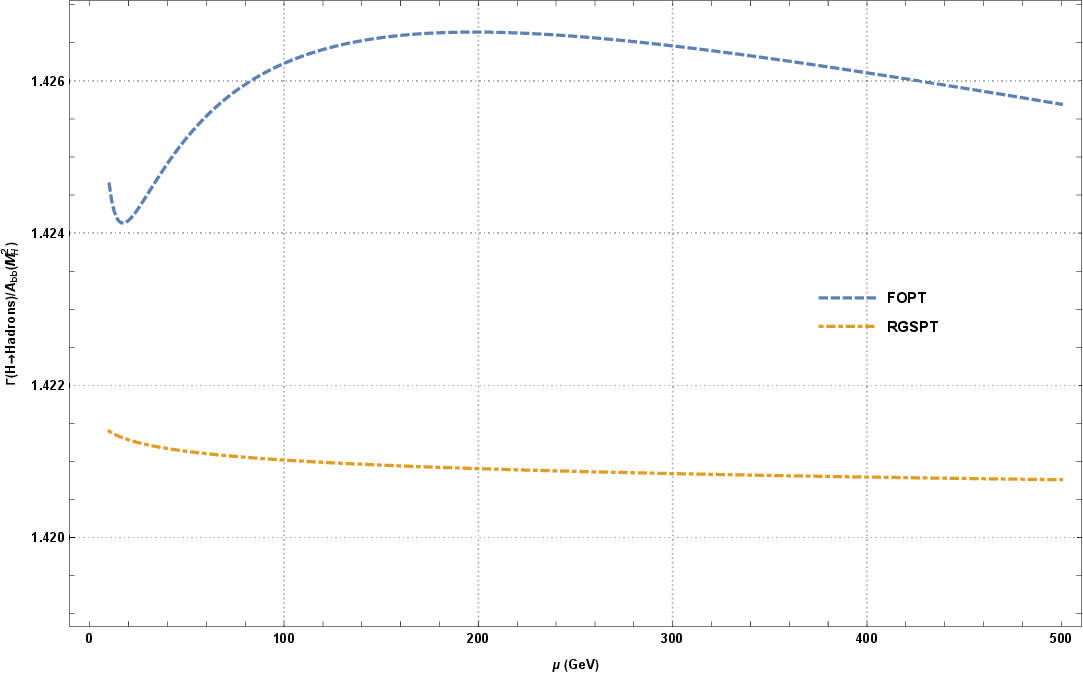}
  \caption{\label{fig:HHad} The scale-dependence of the Hadronic Higgs decay in the $\msbar$ scheme.}
\end{figure}

 
Using the quantities defined above, the total hadronic Higgs decay width is given by:
\begin{align}
    \Gamma\left(\text{H}\rightarrow \text{Hadrons}\right)&=A_{b\overline{b}}\left(C^2_2(1+\Delta_{22})+C_1 C_2 \Delta_{12}\right)+A_{gg}C_1^2 \Delta_{11}\,,
    \label{eq:Hhad}
\end{align}
     where,
     \begin{equation}
         A_{b\overline{b}}\left(\mu^2\right)=\frac{3}{4\pi \sqrt{2}}\gfermi M_H \hs m_b^2\left(\mu^2\right)\,,\quad
         A_{gg}=\frac{4}{\pi \sqrt{2}}\gfermi \pow{\mh}{3}\,.
     \end{equation}
 Recently, the $\ordas{4}$ corrections to hadronic Higgs decay width are presented in Refs.~\cite{Herzog:2017dtz,Davies:2017xsp} with additional bottom quark mass corrections to $H\rightarrow g g$ in Ref.~\cite{Davies:2017rle}. A PMC analysis for $H\rightarrow g g$ and $H\rightarrow b b$ processes can be found in the Ref.~\cite{Wang:2013bla}.  The total Higgs hadronic decay width is presented using these results in Table~\eqref{tab:Hhad1} and scale-dependence in Fig.~\eqref{fig:HHad}. It should be noted that the theoretical uncertainty in RGSPT is dominated by truncation uncertainty from $\Gamma\left(H\rightarrow \overline{b}b\right)$ contributions. The central value obtained from RGSPT is slightly smaller compared to FOPT due to the summation of $\pi^2$-terms but is in agreement within the quoted uncertainty.
 \begin{table}[H]
     \centering
     \begin{tabular}{|c|c|c|c|c|c|c|}
     \hline
          \text{Scheme}&$\Gamma\left(\text{H}\rightarrow \text{Hadrons}\right)(\MeV)$& \multicolumn{5}{c|}{Source of Uncertainty}\\
          \cline{3-7}\text{}&\text{}&\text{Trunc.}& $\delta\as$&$\delta m_b$&$\delta M_H$& $\mu$\\\hline
\text{FOPT}&$2.7082\pm 0.0206$&0.0045&$0.0048$&$0.0183$&$0.0051$&$0.0043$\\\hline
\text{RGSPT}&$2.6978\pm 0.0226$&$0.0115$&$0.0045$&$0.0182$&$0.0051$&$0.0008$\\\hline
     \end{tabular}
     \caption{Total hadronic Higgs decay width in FOPT and RGSPT in the $\msbar-$scheme and the sources of uncertainties. The scale dependence is calculated by varying the renormalization scale in the range $\mu\in\left[10,500\right]\GeV$.}
     \label{tab:Hhad1}
 \end{table} 
 
Now, we move on to the application of analytic continuation of the polarization function involving momentum transfer in the intermediate energy range ($\sim$ few $\GeV$).
\subsection{The electromagnetic R-ratio}
\label{subsec:Rem}
The electromagnetic R-ratio is obtained from the imaginary part of the polarization function for the vector current.  It is an important observable that can be used in the determination of $\as$ using QCD sum rules~\cite{Boito:2018yvl}. The singlet and non-non singlet contributions to process $e^+e^-\rightarrow$ hadrons have been calculated to N$^4$LO in Refs.~\cite{Baikov:2008jh,Baikov:2010je,Baikov:2011pkl,Baikov:2012er,Herzog:2017dtz}. The numerical results presented in Ref.~\cite{Herzog:2017dtz} show that the kinematical $\pi^2$-terms start dominating at N$^3$LO for $n_f=1,2,\cdots6$ flavors of the quark. In this subsection, we study the summation of these terms using RGSPT and their effects, as discussed in the previous sections. The numerical expressions for the Adler functions used in this subsection can be found in the appendix~\eqref{app:Rem}. \par 
The leading-order correction to the Alder function in the massless limit is a constant; therefore, there is no further enhancement arising due to the prefactors from the anomalous dimensions. Hence, the effects of analytic continuation are expected to be milder than the $H\rightarrow gg$ or $H\rightarrow bb $ processes in the hadronic decay width. However, at low energies, these effects can still be enhanced by a relatively large value of the strong coupling. For RGSPT, both the $\as$ and the kinematical terms are present in the denominators, which also results in further reduction in the truncation uncertainties, as observed in previous subsections. \par
For numerical comparison, we are considering the two different cases. In the first case, we consider the case where momentum transfer is below the $J/\psi$ threshold where active quark flavors are $n_f=3$ and $\alpha_s\simeq0.3$. In the second analysis, the momentum transfer is above the mass of the $c\overline{c}$ resonance and below the $\Upsilon$ threshold for which active flavors are $n_f=4$ and $\alpha_s\simeq0.2$. These two cases are already discussed in Ref.~\cite{Herzog:2017dtz} in detail, and we will compare our results with theirs. \par
 \subsubsection{Case-I: Three Active Flavor}\label{subsub:rem3}
\begin{figure}[H]		
  \centering
				\includegraphics[width=.49\textwidth]{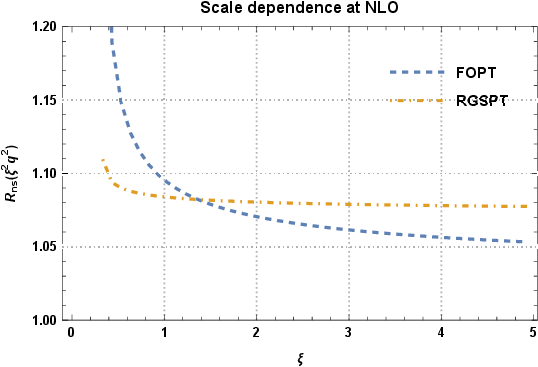}
				\includegraphics[width=.49\textwidth]{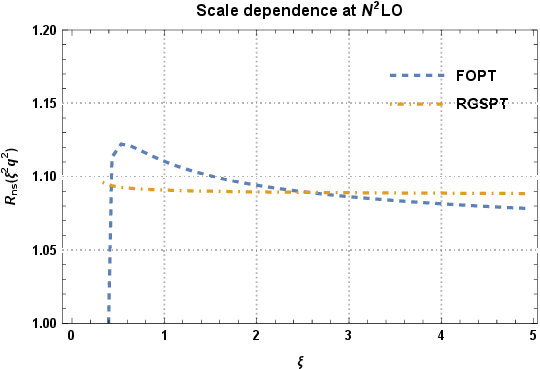}
				\includegraphics[width=.49\textwidth]{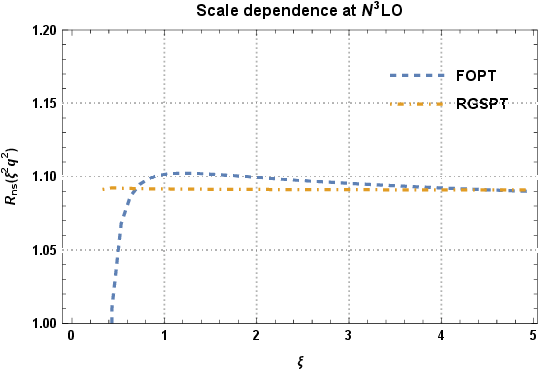}
				\includegraphics[width=.49\textwidth]{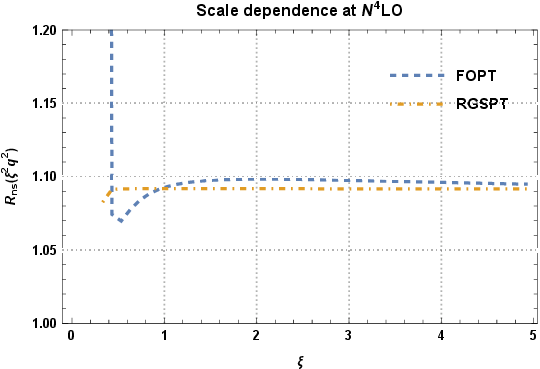}
			\caption{\label{fig:Rem2_scdep} The scale variation of $R_{em}(\xi^2 q^2)$ in the range  $\xi\in\left[1/3,5\right]$ for $n_f=3$  and $\alpha_s(q^2)=0.3$. }
		\end{figure}

 In this case, the leading massless Adler function in Eq.~\eqref{eq:adlerEM} receives contributions only from the non-singlet part of  Eq.~\eqref{eq:adler0}. Using $n_f=3$, scale variation $\xi\epsilon[0.7,5]$  and $\alpha_s(q^2)=0.3$, the FOPT result for $R_{em}(s)$ has following contributions from various orders:
 \begin{align}
 	R_{em}^{\text{(I)}}(s)&=1.0+0.3183 \alpha _s+0.1661 \alpha _s^2-0.3317 \alpha _s^3-1.0972 \alpha _s^4+\order{\alpha_s^5}\nonumber\\&=1.0+ 0.0955+ 0.0150 -0.0090 -0.0089+\cdots\nonumber\\&=1.0926\pm0.0089_{\text{trunc.}}\pm0.0106_{\mu}=1.0926\pm0.0138\,,
 \end{align}
and the corresponding result for RGSPT reads:
\begin{align}
	R_{em}^{\text{(I)},\Sigma}(s)&=1+0.279995 \alpha _s+0.076018 \alpha _s^2+0.0285125 \alpha _s^3+0.022963 \alpha _s^4+\order{\alpha_s^5}\nonumber\\&=1+ 0.0840+ 0.0068+0.0008+ 0.0002+ \cdots \nonumber\\&=1.0918\pm0.0002_{\text{trunc.}}\pm0.00005_{\mu}=1.0918\pm0.0002\,.
\end{align}
	The total uncertainty in the RGSPT is $\sim$ 47 times smaller than the FOPT result because of the summation of the kinematical $\pi^2$-terms. The scale dependence in the two schemes can be seen in Fig.~\eqref{fig:Rem2_scdep}. \par
\begin{figure}[H]
         \centering
            \includegraphics[width=.7\textwidth]{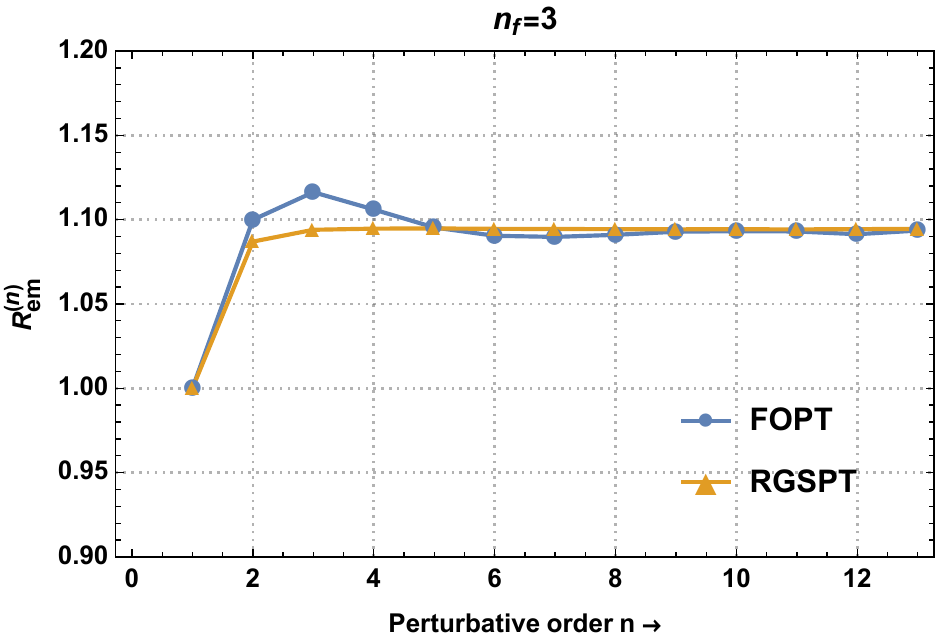}
           \caption{\label{fig:Rem}Stability of the $R_{em}$ for $n_f=3$ in the RGSPT and FOPT schemes.}
    \end{figure}

 
 We can also study the higher-order behavior using the results of Ref.~\cite{Beneke:2008ad} where the predictions for the higher coefficients in the case of hadronic $\tau$ decays using the Borel sum are obtained. These terms are collected in Eq.~\eqref{eq:AdlerTau} in the appendix~\eqref{app:Rem}. The higher-order corrections have the following numerical values:
 \begin{align}
     R_{em}^{\text{(I)}}=&1+0.09549+0.01495-0.00896 -0.00889-0.00394 -0.00050\nonumber\\&\hspace{2mm}+ 0.00062+ 0.00113+ 0.00052+ 0.00025 -0.00103+ 0.00095\,,\nonumber\\
     R_{em}^{\text{(I)},\Sigma}=&1+0.08400+ 0.00684+ 0.00077+ 0.00019 -0.00019-0.00007\nonumber\\&\hspace{2mm} -0.00009-0.00009+ 0.00004-0.00010+ 0.00015-8\times10^{-6}\,.
 \end{align}
 Interestingly, the kinematical terms start dominating the genuine contributions from N$^5$LO for RGSPT. These effects are relatively smaller compared to the FOPT, and reliable predictions from a truncated series using RGSPT can still be obtained. Higher order contributions are under control which can be seen in Fig.~\eqref{fig:Rem}. 

\subsubsection{Case-II: Four Active Flavors}\label{subsub:rem4}
For this case, $n_f=4$, and we use only the non-singlet contributions known to $\ordas{4}$. Using $\alpha_s(q^2)=0.2$, the R-ratio has the following numerical form:
\begin{align}
	R_{em}^{II,\text{FOPT}}(s)=&1+0.3183 \alpha _s+0.1545\alpha _s^2-0.3715\alpha _s^3-0.9536 \alpha _s^4+\order{\alpha_s^5}\nonumber\\=&1.0+ 0.0637+ 0.0062 -0.0030 -0.0015+\cdots\nonumber\\=&1.0653\pm0.0015_{trunc.}\pm0.0010_{\mu}=1.0653\pm0.0018\,,
	\label{eq:Rem_fixed_1}
\end{align}
and the kinematical $\pi^2-$terms starts dominating from $\ordas{3}$. The uncertainty from the scale variations is obtained by varying $\xi$ in the range~$\xi\in\left[1/2,5\right]$.\par
\begin{figure}[H]
 \centering
 \includegraphics[width=.49\textwidth]{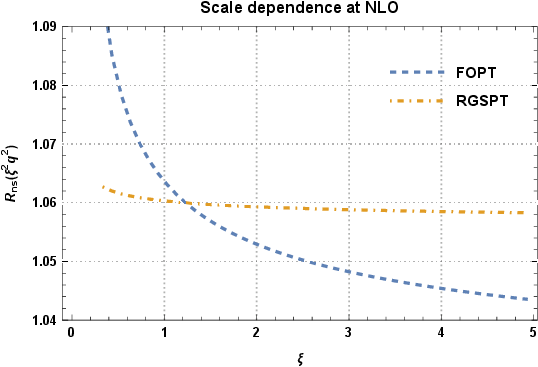}
 \includegraphics[width=.49\textwidth]{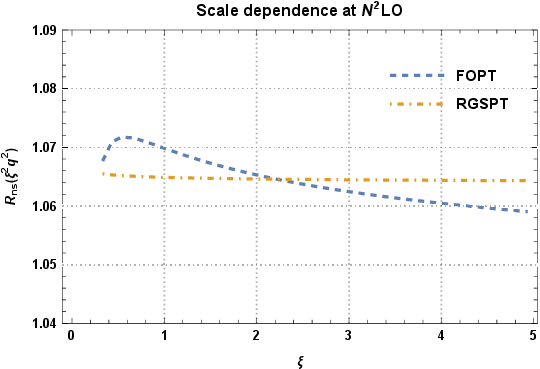}
 \includegraphics[width=.49\textwidth]{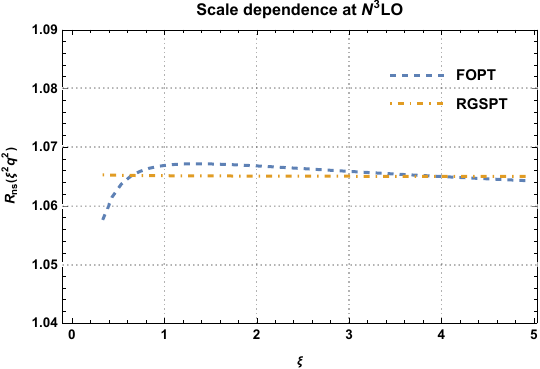}
 \includegraphics[width=.49\textwidth]{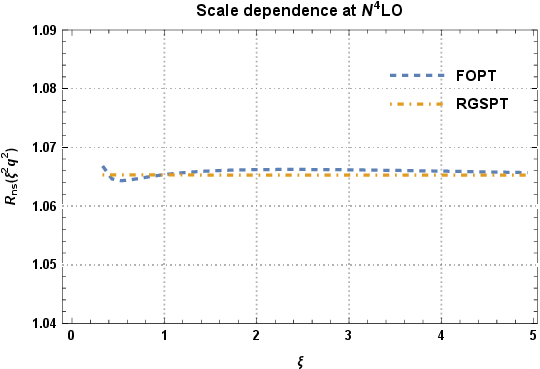}
 \caption{\label{fig:Rem1_scdep} The scale variation of the non-singlet part of the $R_{em}(\xi^2 q^2)$ in the FOPT and RGSPT schemes when scale parameter $\xi$ is varied in the range  $\xi\epsilon\left[1/3,5\right]$ for $n_f=4$ using $\alpha_s(q^2)=0.2$.}
 \end{figure}

Using the same inputs as in Eq.~\eqref{eq:Rem_fixed_1}, the electromagnetic R-ratio in the RGSPT scheme has the following numerical form:
\begin{align}
	R_{em}^{\left(II\right),\text{RGSPT}}(s)&=1+0.3016 \alpha _s+0.1140 \alpha _s^2+0.0293 \alpha _s^3+0.1086 \alpha _s^4+\order{\alpha_s^5}\nonumber\\&=1.0+ 0.0603+ 0.0046+ 0.00023+0.00017+\cdots\nonumber\\&=1.0653\pm0.0002_{\text{trunc.}}\pm0.00005_{\mu}=1.0653\pm0.0002,.
	\label{eq:Rem_summed_1}
\end{align}
 We can see from Eq.~\eqref{eq:Rem_fixed_1} and Eq.~\eqref{eq:Rem_summed_1} that the resummation of the kinematical term improves the convergence as well as the scale dependence for RGSPT compared to the case when FOPT is used. The theoretical uncertainty in the RGSPT results in Eq.~\eqref{eq:Rem_summed_1}) is $\sim$1/10 smaller than that of FOPT result in Eq.~\eqref{eq:Rem_fixed_1}. The scale dependence is minimal after including the N$^4$LO result in FOPT. However, the RGSPT always has better control over the scale dependence than FOPT at each order which can be seen in Figs.~\eqref{fig:Rem1_scdep}.

\begin{figure}[H]
              \centering
            \includegraphics[width=.7\textwidth]{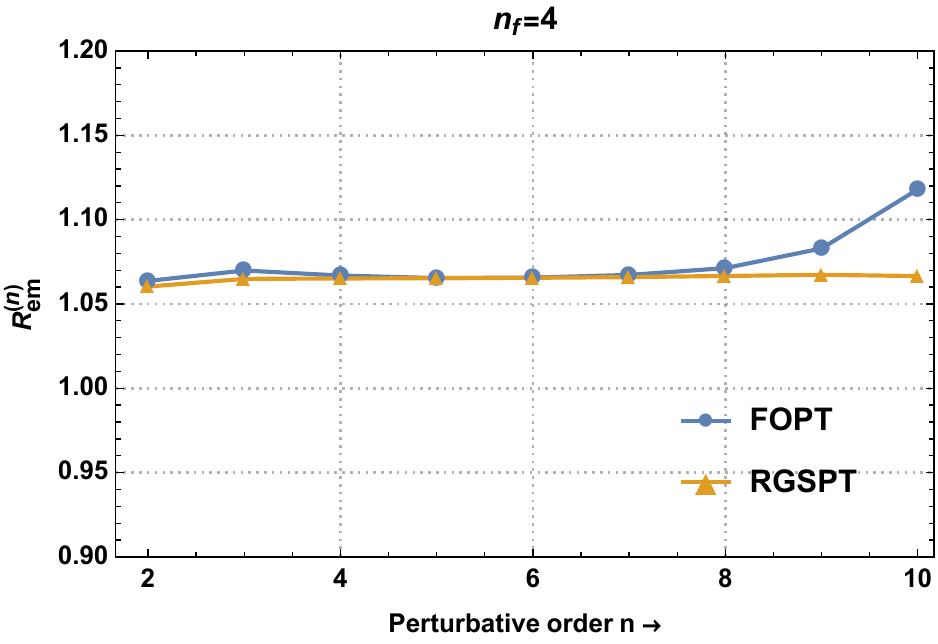}
           \caption{\label{fig:Rem4} Higher order behavior of the non-singlet contribution to $R_{em}$ for $n_f=4$ using $\as=0.2$.}
\end{figure}

To the best of our knowledge, there are no predictions for the higher-order coefficients in this case. We use the Pad\'e approximants as discussed in the subsection~\eqref{subsec:hgg}. Using the $\ordas{4}$ coefficients of non-singlet Adler function from Eq.~\eqref{eq:adler0}, following Pad\'e approximants can be constructed:
\begin{align}
   W^{\left(0,4\right)} =&\frac{1}{1 -  x - 0.524526 x^2 - 0.709564 x^3 - 
 23.121 x^4},\\W^{\left(1,3\right)} =& \frac{1 - 32.5847 x}{1 - 33.5847 x + 
 32.0602 x^2 + 16.382 x^3}\,,\\ W^{\left(2,2\right)} =&\frac{1 - 52.3638 x + 
 26.7566 x^2}{1 - 53.3638 x + 78.5959 x^2}\,,\\ W^{\left(3,1\right)} =&\frac{1 - 
 8.92846 x - 8.40393 x^2 - 12.3776 x^3}{1 - 9.92846 x}\,,
\end{align}
leading to the following average:
\begin{align}
    \overline{W}=&1 +  x + 1.52453 x^2 + 2.75862 x^3 + 27.3888 x^4+ 594.038 x^5 + 
 23309.3 x^6 + 1.05282*10^6 x^7 \nonumber\\&\hspace{2mm} + 5.01986\times10^7 x^8 + 
 2.46615\times10^9 x^9 + 1.23479\times10^{11} x^{10}\,.
\end{align}
 Using these coefficients, the higher-order behavior of $R_{em}$ using $\as=0.2$ for FOPT are obtained as:  
\begin{align}
    R_{em}^{\text{FOPT}}=&1+ x+ 1.52453 x^2-11.5203 x^3 -92.891 x^4+ 308.683 x^5+ 23278.8 x^6\nonumber\\&\hspace{2mm}+ 954842 
    x^7+4.31945\times10^7 x^8+ 2.03702\times10^9 x^9+9.73859\times10^{10} x^{10}+\order{x^{11}}\nonumber\\=&1+ 0.06366+ 0.00618-0.00297 -0.00153+ 0.00032+ 0.00155\nonumber\\&\hspace{2mm}+ 0.00405+0.01165+ 0.03499+ 0.10649+\cdots\,,
\end{align}
and for RGSPT, we obtain the following contributions:
\begin{align}
     R_{em}^{\text{RGSPT}}=&1+0.947499 x+ 1.12508 x^2+ 0.909611 x^3+ 10.5751 x^4+ 221.145 x^5+6241.56 x^6\nonumber\\&\hspace{2mm}+ 157010 x^7+2.61906\times10^6 x^8 -4.44785\times10^7 x^9 -7.62188\times10^9 x^{10}\nonumber\\=&1+0.06032+ 0.00456+0.00023+ 0.00017+ 0.00023+ 0.00042\nonumber\\&\hspace{2mm}+0.00067+0.00071 -0.00076 -0.00833+\cdots\,.
\end{align}
The higher-order behavior using the above results for FOPT and RGSPT are shown in Fig.~\eqref{fig:Rem4}. It is evident that the RGSPT results for $R_{em}$ are more stable than FOPT if we include the higher order predictions using the Pad\'e approximants. In FOPT, these contributions from higher orders are significantly larger compared to RGSPT.  

\subsubsection{Light Quark Hadronic Contribution Muon \texorpdfstring{$g-2$}{}}
Muon anomalous magnetic moment anomaly has been of constant interest in recent years. The tension between the predictions of the standard model and experiments now stands at $4.2\sigma$ ~\cite{Muong-2:2021ojo,Muong-2:2006rrc,Aoyama:2020ynm}. The key issue of the anomaly is contributions coming from the hadron vacuum polarization (HVP). Lattice QCD and dispersive approach based on data-driven methods are used in the literature to quantify these HVP contributions. More details on the subject can be found in the Refs.~\cite{Colangelo:2022jxc,Boito:2022dry,Golterman:2022jvz,Boito:2022njs,Ananthanarayan:2018nyx,Ananthanarayan:2022wsl,Ananthanarayan:2023gzw} and references therein. \par %
 In this subsection, we will focus only on the pQCD contribution to HVP relevant to the summation of the kinematical $\pi^2-$terms. Such contributions enter while evaluating the leading order continuum contributions to HVP using the pQCD input. The electromagnetic R-ratio discussed in subsection~\eqref{subsub:rem3} with some massive corrections are taken as inputs to calculate these contributions using the following relation:
\begin{equation}
	a_\mu^{\text{LO},\text{HVP}}=\frac{1}{3}\frac{\alpha^2_{em}}{\pi^2}\int_{s_0}^{\infty}\frac{G_2(s)}{s}R_{em}(s)\,,
\end{equation}
where kernel $G_2(s)$ is given by ~\cite{Nesterenko:2021byp,Aoyama:2020ynm}:
\begin{equation}
G_2	(s)=\int_{0}^{1}dz\frac{(z^2 (1 - z))}{(z^2 + (1 - z) s/m_\mu^2)}\,,
\end{equation}
and $m_\mu= 105.6583745\MeV$ is the mass of Muon. Recently, three-flavor s-quark connected and disconnected contributions to Muon $g-2$ are calculated in Ref.~\cite{Boito:2022rkw} using the FOPT scheme. These contributions are denoted as $\tilde{a}_{\mu}^{\text{cont.}}$ in this chapter. \par
\begin{figure}[H]
\centering
\includegraphics[width=.49\textwidth]{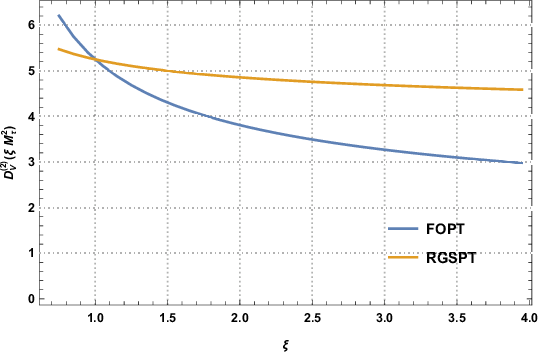}
\includegraphics[width=.49\textwidth]{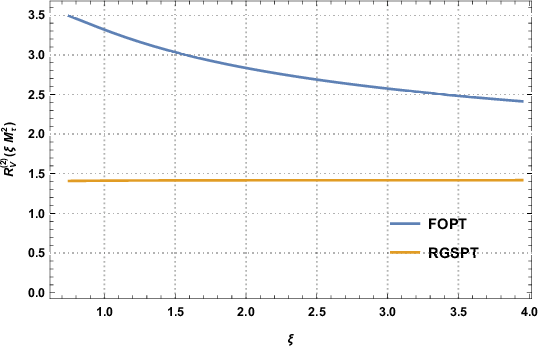}
\includegraphics[width=.49\textwidth]{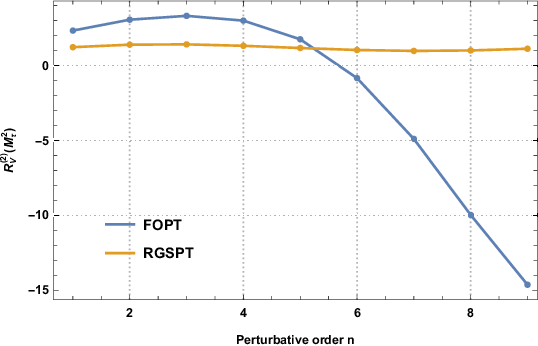}
 \caption{The scale dependence of the $D_V$, and $R_{em}$ using $\ordas{4}$ corrections and the stability of the $R_{em}$ for higher-order as using Pad\'e predictions.
\label{fig:DRV} }
\end{figure}

 The effects of the summation of $\pi^2-$terms to the $R_{em}$ from massless Adler function and higher order behavior are already discussed in the subsection~\eqref{subsub:rem3}. However, there are also some small corrections due to finite strange quark mass ($m_s$). These massive corrections for the Adler function are already known to $\ordas{4}$ 
 from Refs.~\cite{Chetyrkin:1990kr,Chetyrkin:1994ex,Chetyrkin:2000zk,Baikov:2004ku} and can be found in appendix~\eqref{app:Rem_mass_correction}. Higher-order corrections can also be predicted using the Pad\'e approximants. Following the procedure opted in the previous subsection, the Adler function in the massive case is obtained as:
\begin{align}
D_{V,2}^{\text{Pad\'e}}=&12 x + 113.5 x^2 + 1275.89 x^3 + 16496.5 x^4 + 215732 x^5 + 2.87888\times10^6 x^6\nonumber\\&\hspace{.63cm} + 3.91033\times10^7 x^7 + 5.3773\times10^8 x^8 +7.47708\times 10^9 x^9+\cdots\,.
\end{align} 
Now, we can use the above coefficients to study the stability of the $R_{em}^{\left(2\right)}$ with respect to the renormalization scale and with the order of the perturbation theory.  The $R_{em}^{\left(2\right)}$ for FOPT, using $\as(\mtsq)=0.3139$, is obtained as:
\begin{align}
   R_{em}^{\left(2\right)}=& 12 x + 113.5 x^2 + 730.597 x^3 + 2569.98 x^4 - 31969.7 x^5 - 
 1.25\times10^6 x^6 \nonumber\\&- 2.61\times10^7 x^7 - 4.08729\times10^8 x^8 - 
 5.1128\times10^9 x^9 - 4.7013\times10^{10} x^{10}\nonumber\\& - 1.32053\times10^{11} x^{11}+\order{x^{12}}\\
 =&1.199+ 1.1331+ 0.7288+ 0.2562 -0.3184 -1.2469-2.5910 -4.0604\nonumber\\&-5.0750 -4.6626 -1.3086+\cdots\,.
\end{align}
For the RGSPT scheme, 
\begin{align}
R_{em}^{\left(2\right),\Sigma}=&8.2999 x+ 39.6882 x^2+ 167.085 x^3+ 214.167 x^4 -9709.35 x^5
-148249. x^6 \nonumber\\&-1.32872\times10^6 x^7 -6.2473\times10^6 x^8+ 3.26504\times10^7 x^9+ 
1.13115\times10^9 x^{10}\nonumber\\&+ 1.44578\times10^{10} x^{11} +\order{x^{12}}\,,\nonumber\\=& 0.8293+ 0.3962+ 0.1667+ 0.0213 -0.0967 -0.1475 -0.1321\nonumber\\& -0.0621+
0.0324+ 0.1122+ 0.1433+\cdots\,.
\end{align}
From these numerical predictions, it is clear that RGSPT has better convergence than FOPT for the known $\ordas{4}$ results but does not have a convergent behavior when higher order coefficients are used. For FOPT, it is a much more serious issue than RGSPT, as shown in Fig.~\eqref{fig:DRV}.
The scale dependence for massive corrections to the Adler function and $R_{em}(s)$ in the timelike region be found in Fig.~\eqref{fig:DRV}. These quantities are calculated using $\alpha_s(\mtsq)=0.3139$ for three flavors and used in the rest of the subsection. It is clear from these plots that the pQCD analysis for the low energy $s\sim\mtsq$ requires summation in order to make meaningful predictions. The RGSPT series shows that the summation of kinematical $\pi^2-$terms not only improves the scale dependence and radius of convergence for known $\ordas{4}$ terms. 
\par
  \par
\begin{figure}[H]
\centering
			\subfigure[]{\label{fig:muon}\includegraphics[width=.49\textwidth]{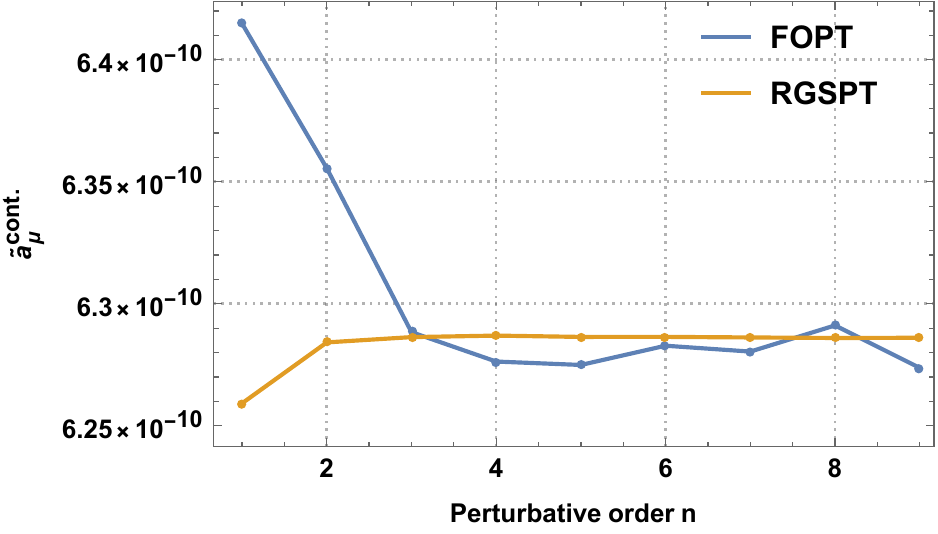}}
				\subfigure[]{\label{fig:muon0_scdep} \includegraphics[width=.49\textwidth]{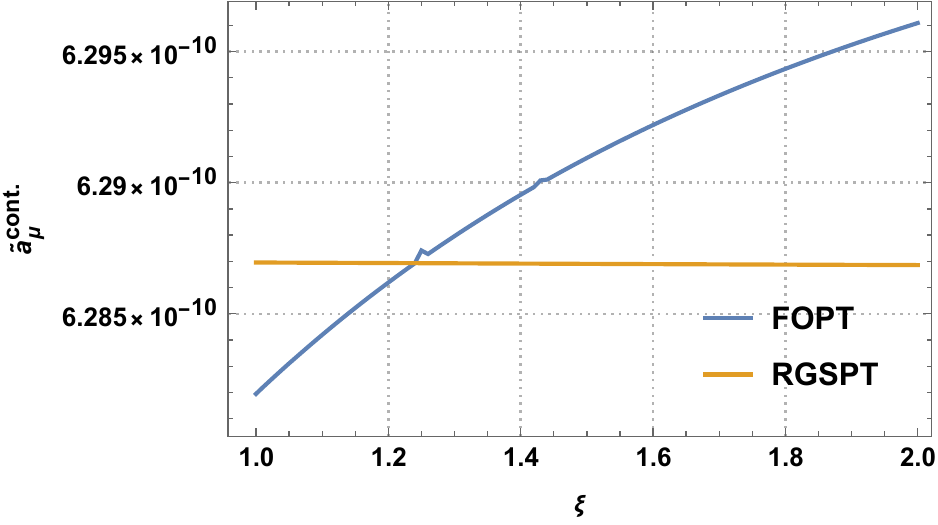}}
				\subfigure[]{\label{fig:muon2_scdep} \includegraphics[width=.49\textwidth]{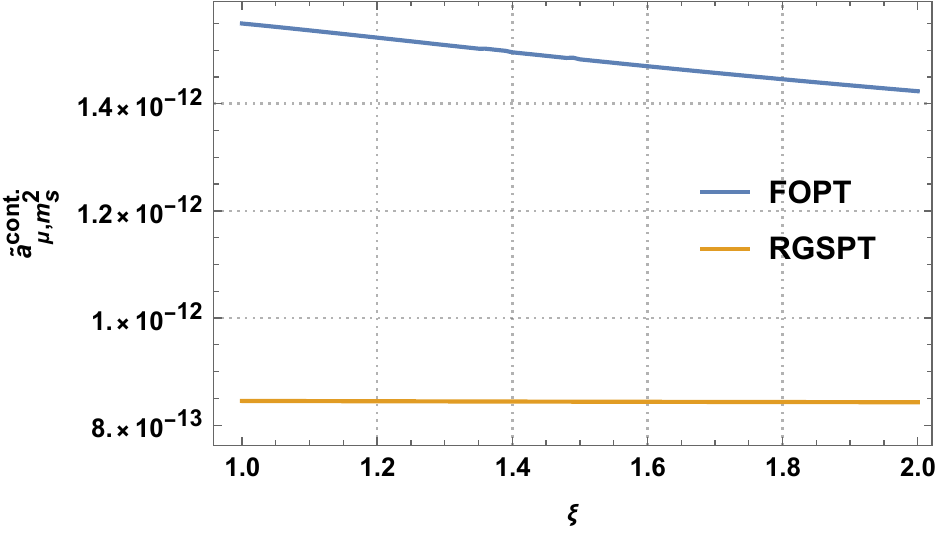}}
   \caption{The behavior of $\tilde{a}_{\mu}^{\text{cont.}}$ with increasing perturbative order in $\ordas{4}$ in Fig.~\eqref{fig:muon} and scale dependence for the exactly known terms in Fig.~\eqref{fig:muon0_scdep}. The scale dependence of the massive corrections to $\tilde{a}_{\mu,m_s^2}^{\text{cont.}}$ in Fig.~\eqref{fig:muon2_scdep}. }
   \end{figure}

 
  The calculation of $\tilde{a}_{\mu}^{\text{cont.}}$ requires both massless as well as massive strange quark mass correction, and relevant expressions are collected in appendix~\eqref{app:Rem}.
  Before moving to $\tilde{a}_{\mu}^{\text{cont.}}$ calculation, it should be noted $\order{\alpha_s^5}$ coefficient to massless correction, $d_5=283$,  provided in Ref.~\cite{Beneke:2008ad} is used to estimate the truncation uncertainties. Using FOPT, we obtain the following contribution to $\tilde{a}_{\mu}^{\text{cont.}}$:
\begin{align}
		\tilde{a}_{\mu}^{\text{cont.}}=\begin{cases}
			&6.277\times10^{-10} \text{ including $\order{\alpha^5_s}$}\\
			&6.282\times 10^{-10} \text{ without $\order{\alpha^5_s}$}
		\end{cases}\,,
\end{align}
 and for RGSPT, we get the following contributions:
 \begin{align}
 	\tilde{a}_{\mu}^{\text{cont.}}=\begin{cases}
 		&6.286\times10^{-10} \text{ including $\order{\alpha^5_s}$}\\
 		&6.287\times10^{-10} \text{ without $\order{\alpha^5_s}$}
 	\end{cases}\,.
 \end{align}
We can see that the truncation error is significantly reduced in the RGSPT. The scale dependence of $\tilde{a}_{\mu}^{\text{cont.}}$ using known $\ordas{4}$ is plotted in Fig.~\eqref{fig:muon0_scdep}. Higher-order terms obtained in Beneke and Jamin~\cite{Beneke:2008ad} can be used to study the behavior of $\tilde{a}_{\mu}^{\text{cont.}}$ in different schemes. The numerical values of these coefficients are presented in the appendix~\eqref{app:Rem}. The numerical stability of the $\tilde{a}_{\mu}^{\text{cont.}}$ using these coefficient is presented in Fig.~\eqref{fig:muon}.  
  
We can also include the massive corrections from the strange quark mass represented as $\tilde{a}_{\mu,m_s^2}^{\text{cont.}}$. The expressions for such corrections in the FOPT and RGSPT schemes are presented in the appendix~\eqref{app:Rem_mass_correction}. Using the strange quark mass as $m_s(2\GeV)=93.4\pm8.6\MeV$ as input from the PDG\cite{Workman:2022ynf}, we obtain the following corrections to $\tilde{a}_{\mu,m_s^2}^{\text{cont.}}$: 
\begin{align}\tilde{a}_{\mu,m_s^2}^{\text{cont.}}&=1.550\times10^{-12}\quad\text{(FOPT)}\,,\\
\tilde{a}_{\mu,m_s^2}^{\text{cont.}}&=8.453\times 10^{-13}\quad\text{(RGSPT)}\,.
\end{align}
The RGSPT contribution is~$\sim54.5\%$ smaller than the FOPT due to the enhanced suppression by the presence of the quark mass anomalous dimension in the denominator. Even though massive corrections to $\tilde{a}_{\mu}^{\text{cont.}}$ are small, analytic continuation effects are numerically significant in this case. The scale dependence of these contributions is presented in Fig.~\eqref{fig:muon2_scdep}.
\section{Summary and Conclusion}\label{sec:summary_ancont}
Effects of kinematical corrections arising from analytic continuation in QCD have a long history. There are various methods used in the literature to control their effects. These effects are related to the RG behavior of the perturbative series. RGSPT uses RGE to sum running logarithms originating from a given order of perturbation series to all orders. We have used this property to sum these kinematical terms to all orders in addition to the reduced dependence on the renormalization scale. These issues are discussed in section~\eqref{sec:intro_RGSPT} and section~\eqref{sec:def_RD} leading to important all order summation formula in RGSPT in Eq.~\eqref{eq:master_eq}.\par
In section~\eqref{sec:applications}, we study the effects of the kinematical terms using FOPT and RGSPT for various processes involving Higgs decay~\eqref{subsec:Hdecay} in pQCD and electromagnetic R-ratio for $e^+e^-$~\eqref{subsec:Rem}.\par 
In section~\eqref{subsec:Hbb}, Higgs decaying to a pair of bottom quarks is studied. The results obtained from RGSPT show enhanced stability with respect to scale variation in addition to good convergence, even if predictions for higher orders are included. For FOPT, cancellation between genuine perturbative correction and $\pi^2$-terms occurs, leading to less truncation uncertainty for the known $\ordas{4}$ results. We argue that the truncation uncertainty for the higher order term should be estimated from the Adler function rather than analytically continued series in FOPT if kinematical $\pi^2$-terms are involved. \par In subsection~\eqref{subsec:hgg}, we study the decay  of Higgs to a pair of gluons. The analytically continued series has better convergent to known orders using RGSPT than FOPT. However, in the total decay width, contributions from the Wilson coefficients result in slightly greater truncation uncertainty for RGSPT. \par
In subsection~\eqref{subsec:H_Hadron}, the results from the previous two subsections as well as some contributions from the mixing of gluonic and fermionic operators, are also included. We get to observe the same patterns as discussed in the previous two subsections.\par 
In subsection~\eqref{subsec:Rem}, electromagnetic R-ratio for $n_f=3$ and $n_f=4$ are studied. Its application in continuum contribution to muon $g-2$ is discussed, including the massive correction from the s-quark. In these examples, in addition to improved scale dependence, a significant reduction in the truncation uncertainties is also achieved. We believe that these results can be used to calculate the experimental values of the low-energy moments of the vector current correlators~\cite{Kuhn:2007vp,Dehnadi:2011gc,Dehnadi:2015fra,Boito:2019pqp,Boito:2020lyp} that are important observables in the determination of the $\as$, charm- and bottom- quark masses. We have used the experimental moments available in the determination of the $\as$, charm- and bottom-quark masses in Ref.~\cite{AlamKhan:2023kgs}.\par
In addition, the results obtained in this study can be used in the QCD sum rules where analytic continuation of the polarization function is required. One such application we have used in the light quark mass determination from the divergences of the axial vector current in chapter~\eqref{Chapter6}.
\begin{subappendices}
\section{Adler function for the vector correlator and electromagnetic current.}
\label{app:Rem}
The Adler function for the vector current correlators is known to  $\order{\alpha_s^4}$ in Refs.~\cite{Appelquist:1973uz,Zee:1973sr,Chetyrkin:1979bj,Dine:1979qh,Gorishnii:1990vf,Surguladze:1990tg,Chetyrkin:1996ez,Baikov:2008jh,Baikov:2010je,Herzog:2017dtz}. The Adler function for $e^+e^-$ also receives additional singlet contributions. The numerical expression for $n_f$ flavor is given by:
\begin{align}
   D_{\text{NS}} =&1 + x +x^2 [1.98571 + L (2.75 - 0.166667  n_f) -   0.115295  n_f] + x^3 [18.2427 - 4.21585  n_f \nonumber\\&+ 0.0862069  n_f^2 +  L^2 (7.5625 - 0.916667  n_f + 0.0277778  n_f^2) +
   L (17.2964 - 2.08769  n_f\nonumber\\&+ 0.0384318  n_f^2)] +x^4 [135.792 -
   34.4402  n_f + 1.87525  n_f^2 - 0.0100928  n_f^3 +
   L (198.14\nonumber\\& - 52.8851  n_f + 3.09572  n_f^2 - 0.0431034  n_f^3) +
   L^2 (88.8789 - 16.1754  n_f + 0.812399  n_f^2 \nonumber\\&- 0.00960795  n_f^3) +
   L^3 (20.7969 - 3.78125  n_f + 0.229167  n_f^2 - 0.00462963  n_f^3))\,,\\
   D_{\text{S}}=&-26.4435 x^3 + (-1521.21 - 218.159 L + (49.0568 + 13.2218 L) n_f]\,,
   \label{eq:adler0}
\end{align}
where $L=\log\left(\frac{\mu^2}{-q^2}\right)$ and $x=\alpha_s(\mu)/\pi$. These contributions are used in the 
 The net contribution to the Adler function is given by:
 \begin{equation}
     D^{em}=\sum_{f}(e_f^2) D_{\text{NS}}+(\sum_{f}e_f)^2D_{\text{S}}
     \label{eq:adlerEM}
 \end{equation}
where $e_f$ is the charge of the $n_f-$flavored quark and the singlet contribution vanishes for the $n_f=3$ case. These expressions are relevant for dimension zero contribution to the $e^+e^-\rightarrow$ hadrons, hadronic Z- and $\tau$ decays, electromagnetic contributions to muon $\left(g-2\right)_{\mu}$, $J/\psi$ and $\Upsilon$ systems. \par
Predictions for the higher-order terms using large$-\beta_0$ approximation and Borel models for hadronic $\tau$ decay width in Ref.~\cite{Beneke:2008ad}:
\begin{align}
    \delta D^{em}\vert_{n_f=3}=&283 x^5 + 3275 x^6 + 18800 x^7 + 388000 x^8 + 919000 x^9 +
    8.37\times10^7 x^{10}\nonumber\\& - 5.19\times10^8 x^{11} + 3.38\times10^{10} x^{12}\,,
    \label{eq:AdlerTau}
\end{align}
Other predictions using D-log Pad\'e approximants for hadronic $\tau$ decay to $\ordas{6}$ can be found in Ref.~\cite{Boito:2018rwt}.
\subsection{Leading mass correction to \texorpdfstring{$\text{R}-$}{}ratio}
\label{app:Rem_mass_correction}
The massless and leading-order massive corrections to $e^+e^-\rightarrow\text{hadrons}$ are now known to $\ordas{4}$ and  can be found in Refs.~\cite{Chetyrkin:1990kr,Chetyrkin:1994ex,Chetyrkin:2000zk,Baikov:2004ku}. The leading mass correction coefficient of the Adler function for the vector current correlator for $n_f=3$ case in FOPT reads:
\begin{align}
	D^{(2)}_{V}=&12 x +x^2 [113.5 + 51 L] +x^3 [1275.89 + 876.75 L + 
	165.75 L^2]+x^4[16496.5 \nonumber\\&\bs+ 13351.6 L + 4233.16 L^2 + 
	483.438 L^3] 
\end{align}
 and the corresponding expression in RGSPT reads:
 \begin{align}
 	D^{(2)}_{V}=&12 x\hs w^{-17/9} +x^2 \left[(134.981 - 40.2963 L_w)w^{-26/9}  - 21.4815w^{-17/9}\right] + x^3 \big[(
 	1551.31 \nonumber\\&- 764.876 L_w + 103.477 L_w^2)w^{-35/9} - (271.34 - 72.135 L_w)w^{-26/9}  - 4.073w^{-17/9}\big]\nonumber\\&+ x^4 \big[w^{-44/9}(
 	19944.3 - 12084.9 L_w + 2827.98 L_w^2 - 238.465 L_w^3) + w^{-35/9} (-3340.89\nonumber\\& + 1521.79 L_w - 185.236 L_w^2) + (-114.043 + 13.679 L_w)w^{-26/9}  + 7.12168w^{-17/9}\big]
 \end{align}

\section{Adler function for the scalar current correlator}\label{app:D_scalar}
The Adler function of the scalar current is known to $\ordas{4}$ \cite{Becchi:1980vz,Broadhurst:1981jk,Chetyrkin:1996sr,Baikov:2005rw,Gorishnii:1990zu,Gorishnii:1991zr,Herzog:2017dtz} relevant for the process $\text{H}\rightarrow\overline{\text{b}}\text{b}$, and leading mass correction for the hadronic $\tau$ decay for the longitudinal contributions. Its numerical value for $n_f$ quark flavor is given by:
\begin{align}
    D_2= &1 + x (5.667 + 2 L) + x^2 (51.57 + 35.33 L +
    4.75 L^2 + (-1.907 - 1.22 L -
    0.167 L^2)  n_f)  \nonumber\\&+ x^3 (648.7 + 509.6 L + 147.29 L^2 +
    11.88 L^3 - (63.74 + 42.12 L + 11.54 L^2 +
    0.94 L^3)  n_f \nonumber\\&+ (0.929 + 0.582 L + 0.204 L^2 +
    0.0185 L^3)  n_f^2) +x^4 (9470.76 + 8286.31 L +
    3047.01 L^2 \nonumber\\&+ 536.76 L^3 +
    30.4297 L^4 - (1454.28 + 1202.9 L + 398.02 L^2 + 68.11 L^3+ 3.91 L^4)  n_f \nonumber\\&+ (54.783 + 45.723 L + 14.688 L^2 +
    2.723 L^3 + 0.166 L^4)  n_f^2 + (-0.454 - 0.453 L \nonumber\\&-
    0.145 L^2 - 0.034 L^3 - 0.0023 L^4)  n_f^3)+\ordas{5}\,,
\end{align}
where $x=\as(\mu^2)/\pi$ and $L=\log\left(\frac{\mu^2}{-q^2}\right)$.

	\section{Wilson coefficients \texorpdfstring{$C_1$}{} and \texorpdfstring{$C_2$}{}}\label{app:Wilson}
   The Wilson coefficients needed for the hadronic Higgs decays, $C_1$ and $C_2$ are known to $\order{\alpha_s^5}$ and are obtained using the four loops decoupling relations~\cite{Liu:2015fxa,Schroder:2005hy,Chetyrkin:2005ia}, and expressions can be found in Refs.~\cite{Herzog:2017dtz,Gerlach:2018hen}. Their value in the different mass schemes are calculated using the quark mass relations \cite{Gray:1990yh,Chetyrkin:1999ys,Chetyrkin:1999qi,Melnikov:2000qh,Marquard:2015qpa,Marquard:2016dcn} and their numerical values in the $\msbar$-, SI-  and OS- schemes are given by:
    \begin{align}
        C_1^{\msbar}&=\frac{-x}{12}\bigg[1 + 2.750 x + x^2[9.642-  0.698 n_l + L_m(1.188 + 0.333 n_l) ]+x^3 [47.370 - 7.69n_l  \nonumber\\&\bs- 0.221n_l^2+   L_m^2 (3.266 + 0.719n_l- 0.056n_l^2) +
        L_m(6.017 + 1.019n_l+ 0.045n_l^2)]  \nonumber\\& \bs+x^4 [311.780 - 62.368n_l+  1.616n_l^2- 0.034n_l^3 +   L_m^2 (21.878 + 1.796n_l\nonumber\\&\hspace{1.8cm}- 0.092n_l^2 - 0.011n_l^3) +   L_m^3 (8.980 + 1.432n_l- 0.273n_l^2 + 0.009n_l^3)\nonumber\\&\hspace{1.8cm} +    L_m(26.504 - 10.211n_l- 2.426n_l^2 + 0.093n_l^3)] \bigg]\\
        C_1^{SI}&=C_1^{\msbar}+x^4 L_m(-2.375 - 0.667 n_l) + x^5 L_m[-21.700 - 4.420n_l+ 0.003n_l^2 \nonumber\\&\bs\bs+
        L_m(-16.526 - 3.649n_l+ 0.278n_l^2)]\\
        C_1^{OS}&=C_1^{\msbar}+x^4[-3.167- 0.889 n_l + L_m(-2.375 - 0.667 n_l) ]\nonumber\\& \hspace{1.3cm}+ x^5[-52.197 -
        10.390n_l+ 0.575n_l^2 -
        L_m(47.825 + 10.170n_l- 0.448n_l^2)\nonumber\\&\hspace{1.3cm}+ L_m^2 (-16.526 - 3.649n_l+ 0.278n_l^2)]\,,
        \end{align}
        \begin{align}
        C_2^{\msbar}&=1 + x^2[0.278 - 0.333 L_m]+ x^3[2.243 - 1.528 L_m- 0.917 L_m^2 + (0.245 + 0.056 L_m^2) n_l] \nonumber\\& \hspace{.6cm}+ x^4 [2.095 - 2.627 L_m- 6.736 L_m^2 - 2.521 L_m^3 + (-0.01 - 0.096 L_m-
        0.009 L_m^3)n_l^2\nonumber\\&\hspace{.6cm}+ (0.31 + 3.142 L_m+ 0.479 L_m^2 +  0.306 L_m^3)n_l]+x^5 [65.144 + 13.373n_l- 3.642n_l^2 \nonumber\\&\hspace{.6cm}+ 0.076n_l^3 + L_m(-121.028 + 27.852n_l- 0.959n_l^2 - 0.001n_l^3)+ L_m^2 (-17.373\nonumber\\&\hspace{.6cm} + 19.162n_l- 1.581n_l^2 + 0.032n_l^3)+ L_m^3 (-25.654 + 3.630 n_l - 0.126n_l^2)  \nonumber\\&\hspace{.6cm}+ L_m^4 (-6.932 + 1.260n_l- 0.076n_l^2 + 0.002n_l^3) ]\\
        C_2^{SI}&= C_2^{\msbar}+x^3 0.667 L_m +x^4 [L_m^2 (4.639 - 0.278 n_l) +
        L_m(5.769 - 0.093 n_l)]  \nonumber\\&\hspace{1.3cm}+ x^5[
        L_m(29.056 - 8.266n_l+ 0.173n_l^2)+L_m^2 (60.290 - 4.793n_l+  0.046n_l^2) \nonumber\\&\hspace{1.3cm}+ L_m^3 (22.261 - 2.673n_l+ 0.080n_l^2)]\\
        C_2^{OS}&=C_2^{\msbar}+x^3[0.889 + 0.667 L_m] + x^4[14.222 - 0.694 n_l + L_m(13.102 - 0.537 n_l) \nonumber\\ &\hspace{1.3cm}+ L_m^2 (4.639 - 0.278 n_l)]+ x^5[206.029 - 30.829n_l+ 0.690n_l^2 \nonumber\\&\hspace{1.3cm}+L_m(203.548 - 27.199n_l+ 0.636n_l^2) +  L_m^2 (100.623 - 9.682n_l+ 0.194n_l^2)\nonumber\\&\hspace{1.3cm}+ L_m^3 (22.261 - 2.673n_l+ 0.080n_l^2) ]\,.
    \end{align}
where $L_m=\log(\mu^2/m_q^2)$ and $m_q$ are the mass of the quark in the mentioned quark mass scheme.\par
 
\section{Adler Functions Relevant for the Hadronic Higgs Decay Width.} \label{app:D_higgs}
The Adler functions relevant to the hadronic Higgs decay to compute the $\Delta_{i,j}$ are presented in this section. The $\Delta_{ij}$ are obtained from the discontinuity of the $D_{ij}$ with appropriate factors.
\subsection{FOPT expressions}
\begin{align}
    D_{11}=&1 + x[12.417 + 3.833 L_H]  + x^2[104.905 + 81.063 L_H + 
    11.0208 L_H^2] \nonumber\\& +x^3 [886.037 + 971.268 L_H + 333.899 L_H^2 + 
    28.1644 L_H^3]\nonumber\\&+ x^4(8723.76 + 10408.8 L_H + 5279.62 L_H^2 + 
    1119.89 L_H^3 + 67.4771 L_H^4)\nonumber\\&+\frac{m_b^2}{M_H^2}\bigg[6 x + x^2[202.046 + 69.5 L_H + 3 L_H^2]  + x^3[4069.51 + 2622.94 L_H  \nonumber\\&\bs\bs+ 
    462.167 L_H^2+ 17.5 L_H^3]\bigg]\,,
    \end{align}
\begin{align}
    D_{12}=&-x[30.67 + 8 L_H] - x^2[524.701 +280.44 L_H + 
    31.33 L_H^2]\nonumber\\&- x^3[7093.07 + 5337.21 L_H +1316.94 L_H^2 + 
    91.39 L_H^3]\,.
\end{align}
The analytic results for $\Delta_{11}$ can be obtained from Refs.~\cite{Davies:2017xsp,Herzog:2017dtz,Davies:2017rle}.
\subsection{RGSPT expressions}
\label{app:D11_RGSPT}
\begin{align}
D_{1,1}^{\Sigma}  =&w^{-2} + x(14.9384w^{-3} - 2.52174 L_w w^{-3} - 2.52174w^{-2})  + x^2(140.526
w^{-4}  \nonumber\\&- 59.6857 L_w w^{-4}+ 4.76938 L_w^2 w^{-4} - 37.441 w^{-3} + 
6.360 L_w w^{-3} + 1.820w^{-2})  \nonumber\\&+ x^3(1233.01 w^{-5} - 783.997 L_w
w^{-5} + 156.525 L_w^2 w^{-5} - 8.018 L_w^3 w^{-5} - 348.927w^{-4}  \nonumber\\&+ 
149.642 L_w w^{-4} - 12.027 L_w^2 w^{-4} + 18.484w^{-3} - 4.589L_w
w^{-3} - 16.533w^{-2}) \nonumber\\& + x^4(11878.4w^{-6} - 8761.86 L_w w^{-6} +
2668.65 L_w^2w^{-6} - 339.04 L_w^3 w^{-6} + 49.6729 w^{-2}  \nonumber\\&+ 12.6372 L_w^4w^{-6} - 
3037.83w^{-5} + 1948.48 L_w w^5 - 392.522 L_w^2w^5 + 
20.2195 L_w^3w^{-5} \nonumber\\&+ 49.8197w^{-4} - 75.7019 L_w w^{-4} + 
8.67931 L_w^2 w^{-4}- 216.349w^{-3} + 41.6915 L_w w^{-3}) \nonumber\\&+\frac{m_b^2}{M_H^2w^{24/23}}\bigg[(0.8166 + 0.8166w^{-2} - 1.633 w^{-1} + x(1.7599 + 30.287w^{-3}\nonumber\\&  - 
3.134 L_w w^{-3}- 34.8133 w^{-2} + 4.20822 L_ww^{-2} + 8.76682 w^{-1} - 
1.07444 L_w w^{-1} )\nonumber\\&  + x^2[1.499 + 638.823 w^{-4} - 158.362 L_ww^{-4} + 
7.988 L_w^2 w^{-4} - 551.269 w^{-3} \nonumber\\&+ 138.9 L_w w^{-3} - 8.074 L_w^2
w^{-3}+ 97.462w^{-2} - 23.943 L_w w^{-2} + 1.384 L_w^2w^{-2}\nonumber\\&  + 
15.531w^{-1} - 2.315 L_w w^{-1}]  + x^3[-12.839 + 10698.6 w^{-6}- 
4262.05 L_w w^{-5} \nonumber\\&+ 513.596 L_w^2 w^{-5}  - 16.93 L_w^3 w^{-5} - 
7966.85w^{-4} + 2985.67 L_w w^{-4} - 364.26 L_w^2 w^{-4}\nonumber\\& + 
13.7218 L_w^3 w^{-4} + 1189.89w^{-3} - 404.192 L_w w^{-3} + 
47.685 L_w^2 w^{-3} - 1.77 L_w^3 w^{-3}\nonumber\\& + 129.34 w^{-2} - 42.9368 L_w
w^{-2} + 2.98299 L_w^2 w^{-2} + 31.3499 w^{-1} - 1.97163 L_w w^{-1}]\bigg]
\end{align}
\begin{align}
    D_{12}^{\Sigma}=w^{-24/23}&\bigg\lbrace (4.17391(1-w^{-1})  + x [-0.409 + (-78.7983 + 10.7543 L_w)w^{-2}  \nonumber\\&+ (48.5405 - 5.49158 L_w)
    w^{-1})] + x^2[-4.37668 + w^{-3}(-1032.38 \nonumber\\&+ 315.943 L_w  - 20.6345 L_w^2)
     + (532.464 - 131.992 L_w + 7.07469 L_w^2)
    w^{-2} \nonumber\\&+ (-20.4045 + 0.53794 L_w)
    w^{-1}]  +x^3 \big[-29.29 + (-12441.7 + 5661.76 L_w  \nonumber\\&- 831.402 L_w^2 + 
    35.067 L_w^3)w^{-4} + (
    5715.82 - 2209.72 L_w + 262.174 L_w^2 \nonumber\\&- 9.04953 L_w^3)
    w^{-3} + (-310.481 + 53.2516 L_w - 0.693018 L_w^2)
    w^{-2} \nonumber\\&+ (-27.3994 + 5.75835 L_w)w^{-1}\big] \bigg\rbrace
\end{align}
\begin{align}
    D_{22}^{\Sigma}=w^{-24/23}& \bigg\lbrace 1 + x[-2.35098 + (8.01764 - 1.31569 L_w)w^{-1}] +x^2 \big[1.144 + (
    59.6174 \nonumber\\&- 22.3168 L_w + 1.69498 L_w^2)
    w^{-2} + (-18.7293 + 3.09316 L_w)w^{-1}\big]\nonumber\\&+ x^3\big[3.74239 + (
    482.958 - 256.916 L_w + 44.9568 L_w^2 - 2.16812 L_w^3)
    w^{-3}\nonumber\\& + (-138.124 + 52.1572 L_w - 3.98485 L_w^2)w^{-2} + (
    4.65204 - 1.50502 L_w)w^{-1}\big]  \nonumber\\&+ x^4\Big[-10.6936+ (
    45.18 - 4.92 L_w)w^{-1} - (8.45 + 13.88 L_w - 1.94 L_w^2)w^{-2}\nonumber\\&+ (
    4598.16 - 2786.2 L_w + 711.6 L_w^2 - 79.135 L_w^3 + 
    2.763 L_w^4)
    w^{-4} \nonumber\\&+ (-1111.99 + 595.803 L_w - 105.099 L_w^2 + 5.09719 L_w^3)
    w^{-3} \Big] \bigg\rbrace
\end{align}
	
\end{subappendices}

%% file: Chapters/Chap6.tex
 \chapter{ Renormalization group improved determination of light quark masses from Borel-Laplace sum rules.}
 
\label{Chapter6}

\lhead{Chapter 6. \emph{RG improved determination of light quark masses from Borel-Laplace sum rules.}}

	\section{Motivation}
		We determine masses of light quarks ($m_u$,$m_d$,$m_s$) using Borel-Laplace sum rules and renormalization group summed perturbation theory (RGSPT) from the divergence of the axial vector current. The RGSPT significantly reduces the scale dependence of the finite order perturbative series for the renormalization group (RG) invariant quantities such as spectral function, the second derivative of the polarization function of the pseudoscalar current correlator, and its Borel transformation. In addition, the convergence of the spectral function is significantly improved by summing all running logarithms and kinematical $\pi^2$-terms. Using RGSPT, we find $m_s(2\GeV)=104.34_{-4.21}^{+4.32}\hs\MeV$, and $m_d(2\GeV)=4.21_{-0.45}^{+0.48}\hs\MeV$ leading to $m_u(2\GeV)=2.00_{-0.38}^{+0.33}\hs\MeV$.
	
\section{Introduction}
The light quark masses are important parameters for quantum chromodynamics (QCD) and electroweak physics. Due to confinement, quarks are not freely observed, and their masses depend on the scheme used. They are taken as input in various quantities related to flavor physics and play a key role in the proton-neutron mass difference and the strong CP violating observable $\epsilon'/\epsilon$, etc. Precise determination of their values has been of constant interest in the past three decades. These masses can be precisely determined in the lattice QCD simulation, and for recent development, we refer to Ref.~\cite{FlavourLatticeAveragingGroupFLAG:2021npn}.\par 
Theoretical tools such as the QCD sum rules~\cite{Shifman:1978bx,Shifman:1978by} have played a key role in their precise determination. These sum rules use both theoretical and experimental input on the spectral function and are based on the assumption of the quark-hadron duality~\cite{Poggio:1975af}. On the hadronic side, the spectral functions for the pseudoscalar channel in the case of the strange and non-strange channels do not have experimental data, and therefore, inputs from chiral perturbation theory (ChPT)~\cite{Weinberg:1978kz,Gasser:1983yg,Gasser:1984gg} become very important. For reviews, we refer to Refs.~\cite{Scherer:2002tk,Scherer:2012zz,Ananthanarayan:2023gzw} and references therein. \par On the theoretical side, operator product expansion (OPE)~\cite{Wilson:1969zs} is used, which has perturbative and non-perturbative contributions. The perturbative corrections are calculated by evaluating the Feynman diagrams, and non-perturbative corrections are the condensates of higher-dimensional operators of quarks and gluons fields. The condensates can be determined from lattice QCD, ChPT, or using QCD sum rules~\cite{Ioffe:2005ym}.\par
Fixed order perturbation theory (FOPT) is the most commonly used prescription used for the finite order perturbation series in the literature. In this prescription, perturbative series is a polynomial in the strong coupling constant ($\as(\mu)$), quark masses ($m_{q}(\mu)$), and the running RG logarithms ($\log(\mu^{2}/Q^{2})$). The RG invariance of an observable ($\mathcal{O}$), known to a finite order in perturbation theory, is obtained using the RG equation (RGE):
\begin{equation}
	\mu^2\frac{d}{d\hs\mu^2}\mathcal{O}=0\,,
	\label{eq:RGE_uds1}
\end{equation} 
that results in a cancellation among the coefficients of different orders in $\as$. Solution to Eq.~\eqref{eq:RGE_uds1} can be used to generate the RG logarithms.\par
RGSPT is a perturbative prescription in which the running RG logarithms arising from a given order of the perturbation theory are summed in a closed form to all orders using RGE. As a result, we get an analytical expression for the perturbative series in which  $\as(\mu)\log(\mu^{2}/Q^{2})\sim\order{1}$. This scheme is useful in reducing the theoretical uncertainties arising from renormalization scale dependence.  The procedure is described in Sec.~\eqref{sec:rgspt} and some of the applications can be found in Refs.~\cite{Abbas:2012py,Ananthanarayan:2016kll,Ananthanarayan:2022ufx,Ahmady:2002fd,Ahmady:2002pa,Ananthanarayan:2020umo,Ahmed:2015sna,Abbas:2022wnz,Chishtie:2018ipg,AlamKhan:2023kgs,AlamKhan:2023dms}.  \par
The Borel-Laplace sum rule is one of the important method widely used in the literature, especially for the determinations of quark mass~\cite{Jamin:1994vr,Dominguez:1997eu,Jamin:2001zr,Jamin:2006tj,Chetyrkin:2005kn,Dominguez:2007my,Narison:2014vka,Yuan:2017foa,Yin:2021cbb}, and in the extraction of hadronic parameters~\cite{Gelhausen:2013wia,Narison:2014ska,Wang:2015mxa,Narison:2015nxh}, etc. However, the dependence of an unphysical Borel parameter ($u$) and free continuum threshold $s_0$ parameter is present in these determinations. In principle, any determination using this sum rule should be independent of the choice of these parameters, but they are tuned to get reliable results. In addition, the determination of the light quark masses from these sum rules is found to be very sensitive to the renormalization scale, and a linear behavior has been reported in Refs.~\cite{Chetyrkin:2005kn,Dominguez:2014vca}. Also, suppression to hadronic spectral function using pinched kernels~\cite{Dominguez:2007my,Dominguez:2008tt}, mainly used in the finite energy sum rules, can not be implemented in this sum rule. \par 
With the limitations in hand, our interest in this sum rule is due to two reasons:
\begin{enumerate}
	\item The formalism developed in the Ref.~\cite{AlamKhan:2023dms} can be used to improve the convergence and reduced renormalization scale dependence for the spectral function by summing kinematical $\pi^2$-terms using RGSPT.
	\item All order summation of the Euler's constant ($\gamma_E$) and Zeta functions arising as a result of the Borel transformation of the RG invariant second derivative of the polarization function using RGSPT.
\end{enumerate}
It should be noted that these improvements are very crucial and can be used in Borel-Laplace sum rule-based studies. On the theoretical side, the leading perturbative $\ordas{4}$ corrections to the pseudoscalar two-point function are now available in Refs.~\cite{Gorishnii:1990zu,Chetyrkin:1996sr,Baikov:2005rw} and other OPE corrections from Refs.~\cite{Chetyrkin:1985kn,Generalis:1990id,Jamin:1992se,Jamin:1994vr,Chetyrkin:2005kn}. For low energy region, there is no experimental information for the pseudoscalar spectral density in the resonance region, but it can be modeled using the experimental values of the resonances~\cite{Bijnens:1994ci,Dominguez:1986aa,Maltman:2001gc}. We have used the results of previous studies on hadronic spectral function from Refs.~\cite{Bijnens:1994ci,Dominguez:1986aa,Maltman:2001gc,Schilcher:2013pvu,Dominguez:1997eu} for the strange and non-strange channel.\par
Hadronic $\tau$ decays are also found to be very useful in the determination of strange quark mass, CKM element $\vert V_{us}\vert$, and strong coupling constant  and more details can be found in Refs.~\cite{Braaten:1991qm,Pivovarov:1991rh,Kambor:2000dj,Maltman:2007ic,Hudspith:2017vew,Boito:2020xli,Ananthanarayan:2022ufx}. These studies use experimental data on the spectral function. Commonly used prescriptions for the perturbative series in these FESR-based studies use FOPT and contour-improved perturbation theory (CIPT). Recently CIPT has been found to be in conflict with the OPE expectations, and for more details, we refer to Refs.~\cite{Hoang:2021nlz,Hoang:2021unk,Benitez-Rathgeb:2021gvw,Benitez-Rathgeb:2022yqb,Benitez-Rathgeb:2022hfj,Gracia:2023qdy,Golterman:2023oml}.  For other light quark mass determinations using sum rules, we refer to Refs.~\cite{Maltman:2001gc,Jamin:2006tj,Dominguez:2014vca,Dominguez:2018azt}.\par 
It should be noted that only the $\msbar$ definition of the $\alpha_s$ and $m_{q}$ are used in this chapter. The value of $\alpha_s(M_z)=0.1179\pm0.0009$ has been taken from PDG~\cite{ParticleDataGroup:2022pth} and evolved to different scales using five-loop $\beta-$function for three flavors using package REvolver~\cite{Hoang:2021fhn} and RunDec~\cite{Chetyrkin:2000yt,Herren:2017osy}. Also, we have used  $x(\mu)\equiv\frac{\as(\mu)}{\pi}$ as expansion parameter in the perturbation series, and if explicit energy scale is not shown, then $x$ and  $m_q$ are assumed to be evaluated at renormalization scale $\mu$.\par
In section~\eqref{sec:formalism}, we briefly introduce the quantities needed for the Borel-Laplace sum rule. In section~\eqref{sec:rgspt}, we have given a brief description of the inputs needed from the RGSPT in our determinations. In section~\eqref{sec:hadronic_info}, hadronic parametrization of the spectral function for the strange and non-strange channel is discussed. In section~\eqref{sec:OPE}, OPE contribution and their results in FOPT and RGSPT prescription are discussed. In section~\eqref{sec:ms_md_determination}, results from the previous sections are used for the light quark mass determinations. In section~\eqref{sec:summary_uds}, we give the summary and conclusion of this chapter, and the supplementary information is provided in appendix~\eqref{app:dim0adler} and other inputs can are presented in appendix~\eqref{app:mass_run}.
\section{Formalism}\label{sec:formalism}
The current correlator for the divergence of the axial currents is defined as:
\begin{equation}
	\Psi_{5}(q^2)\equiv i \int d^4x\hs e^{i q x} \langle 0|\mathcal{T}\{j_{5}(x) j_{5}^{\dagger}(0)\} |0\rangle\,,
	\label{eq:def_corr}
\end{equation}
where $j_{5}$ is given by:
\begin{equation}
	j_{5}=\partial^{\mu}\left(\overline{q}_1\gamma_{\mu} \gamma_5q_2\right)=i \left(m_1+ m_2\right)\left(\overline{q}_1\gamma_5 q_2\right)=i \left(m_1+ m_2\right)j_{0}\,.
	\label{eq:der_J}
\end{equation}
Using Eq.~\eqref{eq:der_J}, the correlation function in Eq.~\eqref{eq:def_corr}, is related to the pseudoscalar correlation function ($\Pi_P(q^2)$), by relation:
\begin{equation}
	\Psi_{5}(q^2)=\left(m_1+ m_2\right)^2\Pi_P(q^2)\,
	\label{eq:Psi2Pi}
\end{equation}
 and the pseudoscalar polarization function is given by:
\begin{equation}
	\Pi_{P}(q^2)= i \int d^4x\hs e^{i q x} \langle 0|\mathcal{T}\{j_{0}(x) j_{0}^{\dagger}(0)\} |0\rangle\,.\nonumber
\end{equation}
Due to the above relation, the sum rule determinations from the correlator in Eq.~\eqref{eq:def_corr} are sometimes known as pseudoscalar determinations.
\par
Using OPE, a theoretical expression for $\Psi_5(q^2)$ is calculated in the deep Euclidean spacelike regions in the limit $m_q^2 \ll q^2$, and the resulting expansion can be arranged as expansion in $1/(q^2)$. At low energies $\sim 1\GeV^2$, instanton effects become relevant, and their contribution is not captured by OPE expansion and therefore are added to it. Further details on the OPE contributions are presented in Sec.~\eqref{sec:OPE}.\newline
The Borel-Laplace sum rules are based on the double-subtracted dispersion relation for the correlation function. Therefore, it involves the double derivative of $\Psi_5(q^2)$ and the dispersion relation is given by:
\begin{equation}
	\Psi''_{5}(q^2)=\frac{d^2}{d (q^2)^2}\Psi_{5}(q^2)=\frac{2}{\pi}\int_{0}^{\infty}ds\frac{\text{Im}\Psi_{5}(s)}{(s-q^2-i\epsilon)^{3}}\,.
	\label{eq:der2Psi}
\end{equation}
The Borel transformation, with parameter ``u",  is obtained using the Borel operator, $\hat{\mathcal{B}_u}$, defined as:
\begin{equation}
	\hat{\mathcal{B}}_u\equiv\lim _{Q^{2}, n \rightarrow \infty \atop Q^{2} / n=u}\frac{(-Q^{2})^n}{\Gamma[n]}\partial_{Q^{2}}^n
	\label{eq:BO}\,,
\end{equation}
where, we have used variables $Q^2=-q^2>0$ for the spacelike and $s=q^2>0$ for timelike regions. It should be noted that we use normalization given in Ref.~\cite{Jamin:1994vr} i.e. $\hat{\mathcal{B}}_{u}[\frac{1}{(x+s)^a}]=\frac{1}{u^a \Gamma[a]}e^{-x/u}$.\par 
The Borel transform of Eq.~\eqref{eq:der2Psi} is obtained as:
\begin{align}
	\Psi''_{5}(u)&\equiv\hat{\mathcal{B}}_u\left[\Psi''_{5}(q^2)\right]=\frac{1}{u^3}\hat{\mathcal{B}}_u\left[\Psi_{5}(q^2)\right](u)\nonumber\\&=\frac{1}{\pi u^3}\int_{0}^{\infty} ds\hs e^{-s/u} \hs \text{Im}\Psi_{5}(s)\nonumber\\&=\frac{1}{u^3}\int_{0}^{\infty}ds \hs\hs e^{-s/u} \rho_{5}(s)\,,
	\label{eq:bs_rule}
\end{align}
where the spectral density is given by:
\begin{equation}
	\rho_{5}(s)=\frac{1}{\pi}\lim_{\epsilon\rightarrow 0}\left[\text{Im}\Psi_{5}(-s-i\hs\epsilon)\right]\,.
	\label{eq:spectral_density}
\end{equation}
It should be noted that the value of the $u\gg\Lambda_{QCD}^2$ in $\Psi_{5}''(u)$ is chosen such that higher order terms of the OPE expansion remain suppressed in the Borel transformed OPE expansion.\par
The Borel-Laplace sum rules in the RHS of Eq.~\eqref{eq:bs_rule} involve an integration ranging from the low energy regime of the strong interactions to the high energy regime. The spectral density is approximated with the quark hadron-duality. For the low-energy regime, the spectral function is parameterized in terms of pion/kaon poles and resonances present in the channel, and for the high-energy region, results from perturbative QCD are used. The spectral density from these two regimes can be written as:
\begin{align}
	\rho_5(s)=\theta(s_0-s)\rho_5^\text{had.}+\theta(s-s_0)\rho_5^{\text{OPE}}\,,
	\label{eq:spectral_decomposition}
\end{align}
where scale $s_0$ separates the two contributions, and its value should be chosen such that the perturbative treatment is justified.\par
Using Eq.~\eqref{eq:spectral_decomposition}, the Borel sum rule in Eq.~\eqref{eq:bs_rule} can be written as:
\begin{equation}
	\Psi''_{5}(u)=\frac{1}{u^3}\int_{0}^{s_0}ds \hs\hs e^{-s/u} \rho^{had}_{5}(s)+\frac{1}{u^3}\int_{s_0}^{\infty}ds \hs\hs e^{-s/u} \rho^{OPE}_{5}(s)\,.
	\label{eq:bs_final}
\end{equation}
and used for the light quark mass determination. \par 
For clarification, various inputs used in Eq.~\eqref{eq:bs_final} are as follows:
\begin{enumerate}
	\item \label{item:1}The $\Psi''_{5}(u)$ is obtained from the Borel transformation of $\Psi''(q^2)$, which involves OPE corrections and addition to the instanton contributions. The instanton contributions are, although small for the choices of the parameters used in this study but relevant, as pointed out in Ref.~\cite{Maltman:2001gc}. These contributions are thus obtained using Eq.~\eqref{eq:psi_Borel}+Eq.~\eqref{eq:borel_instal}.
	\item \label{item:2} The hadronic spectral density $\rho_{5}^{\text{had}}(s)$ is obtained by the parametrization of the experimental information on the hadrons appearing in the strange and non-strange channels. These constitutions are discussed in section~\eqref{sec:hadronic_info} for non-strange and strange channels, and we use Eq.~\eqref{eq:had_NS} or \eqref{eq:had_S}.
	\item \label{item:3}$\rho_5^{\text{OPE}}(s)$ in the RHS of Eq.~\eqref{eq:bs_final} is obtained from the discontinuity of the theoretical expression of the $\Psi_5(q^2)$ which is calculated using the OPE and instanton contributions are also added to it. It has contributions from Eq.~\eqref{eq:rho_exp}. +Eq.~\eqref{eq:rho_insta}.
\end{enumerate}
It should be noted that the main focus of our determination is the RG improvement for the theoretical quantities relevant for point~\eqref{item:1} and point~\eqref{item:3} and its impact in the light quark mass determination. 

	\section{Inputs from RGSPT} \label{sec:rgspt}
In FOPT prescription, a perturbative series $\mathcal{S}(Q^2)$ in pQCD can be written as:
\begin{align}
	\mathcal{S}(Q^2)\equiv \sum_{i=0,j=0} T_{i,j} x^i L^j \,,
	\label{eq:Pseris_uds}
\end{align}
where $x=\as(\mu)/\pi$ and $L=\log(\mu^2/Q^2)$ and corresponding RGSPT series is obtained by rearranging Eq.~\eqref{eq:Pseris_uds} as follows:
\begin{align}
	\mathcal{S}^\Sigma(Q^2)= \sum_{i=0} x^i S_i (x\hs L)\,,
\end{align}
where a closed-form expression for coefficients is obtained for terms:
\begin{equation}
	S_i (z)=\sum_{j=0}^{\infty} T_{i+j,j} z^j\,,
\end{equation}
where $z\equiv x \hs L\hs$. Following the procedure described in chapter~\eqref{Chapter2}, the solution to $S_i (z)$ is found to be the functions of one variable where $z\sim\mathcal{O}(1)$ and the closed-form solution always contains a most general term given by:
\begin{equation}
	\Omega_{n,a}\equiv\frac{\log^n(w)}{w^a}=\frac{\log^n(1-\beta_{0}\hs x(\mu)\log(\mu^2/Q^2))}{(1-\beta_{0}\hs x(\mu)\log(\mu^2/Q^2))^a}\,,
	\label{eq:coef_rgspt_uds}
\end{equation}
where $n$ is a positive integer and $a\propto \gamma_0/\beta_{0}$. The analytic continuation of such terms is already discussed in chapter~\eqref{Chapter5} and another quantity which is the Borel transformation of $\Psi_5''(q^2)$ which is discussed in Section~\eqref{subsec:BorelTransform} of this chapter. One important point to note here is that results from different prescriptions, such as RGSPT and FOPT, are not the same when $\mu^2=Q^2$ is set after the operations like analytic continuation or Borel transformation are performed. The differences arise due to different treatments of the RG logarithms for finite order series for which only a few terms are known.
\section{Hadronic spectral function}\label{sec:hadronic_info}
The hadronic spectral functions are constructed using the contributions from the pion/kaon pole and the data from the experiments on the resonances in a given channel. At low energies, they are dominated by the pion/kaon pole contributions. This section discusses the parametrization of the unknown pseudoscalar spectral function for the non-strange and strange channels. 
\subsection{Non-strange channel}\label{subsec:rho_NS}
\begin{figure}[H]
		\centering
			\includegraphics[width=.49\linewidth]{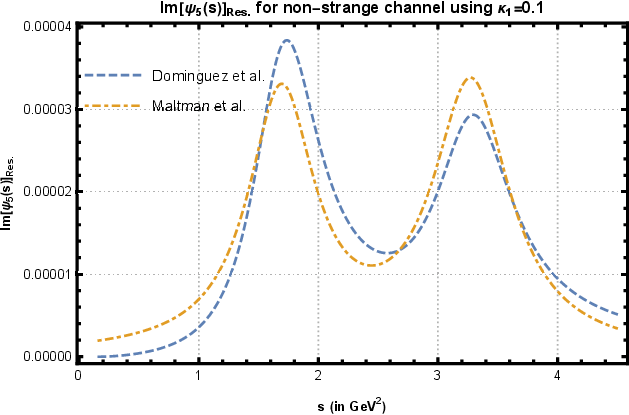}
			\includegraphics[width=.49\linewidth]{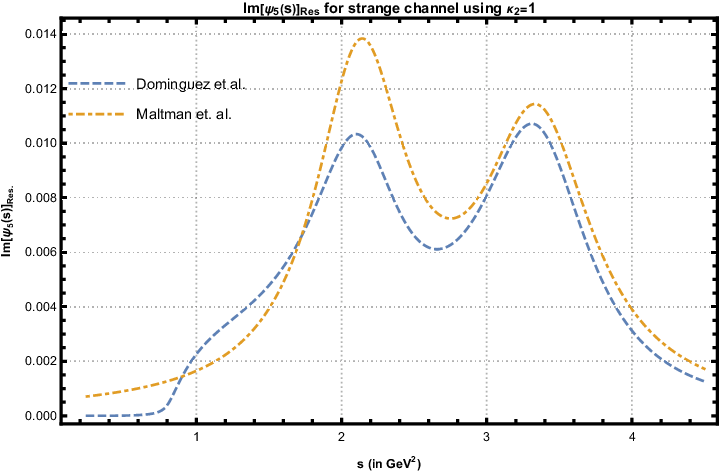}
		\caption{\label{fig:had} Hadronic spectral function in the resonance region for strange and non-strange channels.}
	\end{figure}

For the non-strange pseudoscalar channel, two phenomenological parametrizations are often used in the literature. In Ref.~\cite{Dominguez:1986aa}, Dominguez and Rafael provided a ChPT-based parametrization which is normalized to unity at the threshold. Later some corrections are reported for this parametrization in Ref.~\cite{Bijnens:1994ci}. Another parametrization often used is by Maltman and Kambor~\cite{Maltman:2001gc}, which requires masses and decay constants for the higher resonances. In this study, we have used Dominguez and Rafael's parametrization, which is recently used in Ref.~\cite{Dominguez:2018azt} for the up/down-quark mass determination. We have used their results for the hadronic parametrization, which is given by:
\begin{equation}
	\rho_{\text{NS}}= f_\pi^2 M_\pi^4\hs \delta\left(s-M_\pi^2\right)+\rho_{3\pi} \frac{\text{BW}_1(s)+\kappa_1 \hs \text{BW}_2(s)}{1+\kappa_1}\,,
	\label{eq:had_NS}
\end{equation}
where, $f_\pi=130.2(1.2)\hs\MeV$ and $M_\pi=130.2(1.2)\hs\MeV$ value of $\kappa_1\simeq0.1$ is used in the Ref.~\cite{Dominguez:2018azt} and it controls the relative importance of the resonances. The $3\pi$ resonance contributions are received from the $\pi(1300)$ and $\pi(1800)$ states. Their contributions are encoded in the $\rho_{3\pi}$, which is given by:
\begin{equation}
	\rho_{3\pi}=\frac{1}{\pi}\hs\text{Im}\hs \Psi_{5}(s)\vert_{3\pi}=\frac{1}{9}\frac{M_{\pi}^2}{f_{\pi}^2}\frac{1}{2^7\pi^4}\theta\left(s-9 M_{\pi}^2\right) I_{\pi}(s)\,,
\end{equation}
where, $I_{\pi}(s)$ is the phase space integral given in Eq.~\eqref{eq:I_pi}. In the chiral limit, the phase integral reduces to $I_\pi(s)=3\hs s$ that confirms the prediction for $\rho_{3\pi}$ in Ref.~\cite{Pagels:1972xx}. The $\text{BW}_{1,2}(s)$ is the Breit-Wigner distribution given by:
\begin{equation}
	\text{BW}_i(s)=\frac{\left(M_i^2-s_{\text{th}}\right)^2+M_i^2\Gamma_i^2}{\left(s-M_i^2\right)^2+M_i^2\Gamma_i^2}
\end{equation}
which is normalized to unity at the threshold, i.e., $\text{BW}_{1,2}(s_{th})=1$. For the non-strange spectral function, we use the following data from PDG~\cite{ParticleDataGroup:2022pth} as input:
\begin{align}
	M_\pi&=134.9768(5)\MeV, f_\pi=130.2(1.2)\MeV\nonumber\,,\\
	M_{\pi,1}&=1300(100)\MeV,\quad \Gamma_{\pi,1}=260(36)\MeV\nonumber\,,\\
	M_{\pi,2}&=1810^{+11}_{-9}\MeV,\quad \Gamma_{\pi,2}=215_{-8}^{+7}\MeV\nonumber\,.
\end{align}
Using the above values as input, the non-strange spectral function in the two parameterizations discussed above is plotted in Fig.~\eqref{fig:had}.

\subsection{The Strange Channel}
We use the hadronic parametrization presented in Ref.~\cite{Dominguez:1997eu} for the strange channel. This parametrization is equivalent to the one we have used in the non-strange channel. The hadronic spectral function is given by:
\begin{equation}
	\rho_S(s)=\frac{1}{\pi}\text{Im}\hs \Psi_5(s)\vert_\text{Had.}=f_K^2 M_K^2\delta\left(s-M_K^2\right)+\frac{1}{\pi}\text{Im}\left(\Psi_5(s)\right)\vert_{\text{Res.}}\,.
	\label{eq:had_S}
\end{equation}
where the spectral function for the resonance region is given by 
\begin{equation}
	\frac{1}{\pi}\text{Im}\left(\Psi_5(s)\right)\vert_{\text{Res.}}=
	\rho_{K\pi\pi}(s)\frac{\text{BW}_1(s)+\kappa_2 \hs \text{BW}_2(s)}{1+\kappa_2}
\end{equation}
The Breit-Wigner profile is constructed from $K(1460)$ and $K(1830)$ resonances. The value $\kappa_2\simeq1$ is found to be a reasonable choice in Ref.~\cite{Dominguez:1997eu} to control the contributions from the resonances. In addition, due to its narrow width, there is a significant contribution from the resonant sub-channel $K^*(892)-\pi$. Its contributions are  also included in the $\rho_{K\pi\pi}(s)$ and has the form:
\begin{equation}
	\rho_{K\pi\pi}(s)=\frac{M_K^2}{2f_\pi^2}\frac{3}{2^7\pi^4}\theta(s-M_K^2)\frac{I_K(s)}{s\left(M_K^2-s\right)}\,,
\end{equation}
and the integral $I_K(s)$ is defined in Eq.~\eqref{eq:Is}. For the strange channel, precise values of the resonance masses and the decay width do not exist. We are using the values used in Ref.~\cite{Schilcher:2013pvu} with additional uncertainties of $50 \MeV$ to resonance masses and $10\%$ to the decay widths. For kaons, we use PDG~\cite{ParticleDataGroup:2022pth} values, and the following values for the parameters for the strange channel are used as input:
\begin{align}
	M_K&=497.611(13)\MeV, \quad &f_K=155.7(3)\MeV\nonumber\,,\\
	M_{K,1}&=1460(50)\MeV,\quad &\Gamma_1=260(26)\MeV\nonumber\,,\\
	M_{K,2}&=1830(50)\MeV,\quad &\Gamma_1=250(25)\MeV\nonumber\,,\\
	M_{K^*}&=895.55(2)\MeV,\quad&\Gamma_{K^*}=47.3(5)\MeV\,.
\end{align}
Using the above inputs, the strange spectral function in the resonance region is plotted in Fig.~\eqref{fig:had}.	
	\begin{align}
		I_\pi(s)=&\int_{4 M_\pi^2}^{\left(\sqrt{s}-M_\pi\right)^2}   d\hs u\hs \sqrt{1-\frac{4 M_\pi^2}{u}}\frac{\lambda^{1/2}(s,u,M_\pi^2)}{s}\Bigg\lbrace 5+\frac{1}{\left(s-M_\pi^2\right)}\left[3\left(u-M_\pi^2\right)-s+9M_\pi^2\right]\nonumber\\&+\frac{1}{2\left(s-4 M_\pi^2\right)^2}\bigg[\left(s-3u+3 M_\pi^2\right)^2+3\lambda(s,u,M_\pi^2)\left(1-4\frac{M_\pi^2}{u}\right)+20M_\pi^4 \bigg]\Bigg\rbrace
		\label{eq:I_pi}
	\end{align}
	where, $\lambda(s,u,M_\pi^2)$ is given by:
	\begin{align}
		\lambda(s,u,M_\pi^2)=\left(s-\left(\sqrt{u}-M_\pi\right)^2\right)\left(s-\left(\sqrt{u}+M_\pi\right)^2\right)\,,
	\end{align}
	For the strange spectral function, the integral $I_K$ is given by:
	\begin{equation}
		\begin{split}
		I_K(s)=\int_{M_K^2}^{s}\frac{d\hs u}{u}\left(u-M_K^2\right)\left(s-u\right)&\Bigg \lbrace \left(M_K^2-s\right)\left[u-\frac{(s+M_K^2)}{2}     \right]-\frac{1}{8u}\left(u^2-M_K^4\right)\left(s-u\right) \nonumber\\&\bs+\frac{3}{4} \left(u-M_K^2\right)^2\vert F_{K^*}(u)\vert^2 \Bigg \rbrace\,,
		\end{split}	
			\label{eq:Is}
	\end{equation}
	and
	\begin{equation}
		\vert F_{K^*}(u)\vert^2=\frac{\left(M_{K^*}^2-M_K^2\right)^2+M_{K^*}^2\Gamma_{K^*}^2}{\left(M_{K^*}^2-u\right)^2+M_{K^*}^2\Gamma_{K^*}^2}\,.
	\end{equation}

\section{The OPE corrections}
\label{sec:OPE}
The OPE corrections are calculated in large $Q^2$ limit, and organized as expansion in $1/Q^2$:
\begin{equation}
	\Psi_{5}(Q^2)=Q^2 \hs  \left(m_1+ m_2\right)^2 \sum_{i=0}\frac{\Psi_i(Q^2)}{(Q^2)^i}
	\label{eq:corr_exp}
\end{equation}
and $\Psi_n(Q^2)$ are termed as the contributions from $2n$-dimensional operators. Apart from the quark mass dependence, $\Psi_n(Q^2)$ also receives contributions from the higher dimensional non-perturbative condensate starting from dimension-4. 
$\Psi_0(Q^2)$ and $\Psi_2(Q^2)$ are purely perturbative quantities and have expansion in $m_q$, $\as$ and $\log(\mu^{2}/Q^{2})$ known to $\ordas{4}$
\cite{Gorishnii:1990zu,Chetyrkin:1996sr,Baikov:2005rw}
and $\ordas{1}$~\cite{Chetyrkin:1985kn,Generalis:1990id,Jamin:1992se,Jamin:1994vr}, respectively. Additional correction to $\Psi_2$ proportional to $m_s^4$ is also included from Ref.~\cite{Chetyrkin:2005kn}. The non-logarithmic terms appearing in $\Psi_0(Q^2)$ and $\Psi_2(Q^2)$ are irrelevant for the Sum rule in Eq.~\eqref{eq:bs_final}. This is due to the fact that the contributions calculated from the Borel operator, in Eq.~\eqref{eq:BO}, vanish for any non-negative integer powers of $Q^2$. For the spectral function case, non-logarithmic terms vanish when analytic continuation is performed using Eq.~\eqref{eq:spectral_density}. The relevant expressions for these quantities for FOPT and RGSPT can be found in appendices~\eqref{app:dim0adler} and \eqref{app:dim2adler}. 

The OPE contributions from dimension-4 are known to $\ordas{1}$~\cite{Pascual:1981jr,Jamin:1992se,Jamin:1994vr} and contain condensate terms of quarks and gluon fields. Their RG running should also be taken into account when coupling and masses are evolved with the scale. However, we use the results provided in Refs.~\cite{Spiridonov:1988md,Baikov:2018nzi} to form an RG invariant combination of these condensates. For the quark condensates, this relation is given by:
\begin{equation}
	\langle m_i \overline{q}_j\hs q_j\rangle_{\text{inv.}}=\langle m_i \overline{q}_j \hs q_j \rangle +m_i\hs  m_j^3\left(\frac{3}{7 \pi^2\hs x }-\frac{53}{56\pi^2}\right)\,.
\end{equation}
The RG invariant combinations of the condensates \cite{Spiridonov:1988md} also introduce inverse powers of the $\as$~\cite{Jamin:1994vr,Chetyrkin:1994qu}. For the gluon condensate, we use the following relation:
\begin{equation}
	\frac{\beta(x)}{x^2}\langle\frac{\as}{\pi}G^2\rangle_{\text{inv.}}\equiv\frac{\beta(x)}{x^2}\langle\frac{\as}{\pi}G^2\rangle-4\gamma_m(x)\sum_{k={u,d,s}}\langle m_i \overline{q}_i\hs q_i\rangle-\frac{3}{4\pi^2}\gamma_{\text{vac.}}\sum_{k={u,d,s}} m_k^4\,,
\end{equation}
where, $\gamma_{\text{vac.}}=-1 - (4 x)/3 + x^2 (-223/72 + 2/3 \zeta(3))$ is the vacuum anomalous dimension~\cite{Baikov:2018nzi}. The expression for $\Psi_4$ in RGSPT and FOPT are provided in the appendix~\eqref{app:dim4}.\par
We also consider the dimension-6 contribution to the OPE, given by:
\begin{align}
	\left(I_6\right)_{ij}=-\frac{3}{2}(m_i\langle  \overline{q}_{j}q_{j}\hs G\rangle+m_j\langle \overline{q}_{i}q_{i}\hs G\rangle)-\frac{32}{9}\pi^2 x \left(\langle\overline{q}_{i}q_{i}\rangle^2+\langle\overline{q}_{j}q_{j}\rangle^2-9\langle\overline{q}_{i}q_{i}\rangle\langle\overline{q}_{j}q_{j}\rangle \right)\,,	\label{eq:dim6}
\end{align}
where $i$ and $j$ stand for the quark flavors in the strange and non-strange channels. It should be noted that the structure of the dimension-6 condensate is rather complicated, and in deriving Eq.~\eqref{eq:dim6}, vacuum saturation approximation is used to relate dimension-6 four-quark condensate terms to dimension-4 quark condensates. For more details, we refer to Refs.~\cite{Shifman:1978bx,Jamin:1994vr}. We neglect the contributions to OPE beyond this order. \par 
The numerical values used for the non-perturbative quantities are as follows:
\begin{align}
	\langle \overline{u}u\rangle&=-\frac{f_\pi^2 M_\pi^2}{2\left(m_u+m_d\right)}\text{~\cite{Gell-Mann:1968hlm}}\,,\\
	\langle \overline{s}s\rangle&=\left(0.8\pm0.3\right) \langle \overline{s}s\rangle\text{~\cite{Chetyrkin:2005kn}}\,,\\
	\langle\frac{\as}{\pi}G^2\rangle&=0.037\pm0.015\GeV^4\text{~\cite{Dominguez:2014pga}}\,,\\
	\langle \overline{q}_{i}q_{i}\hs G\rangle&=M_0^2 \langle 
	\overline{q}_{i}q_{i}\rangle\text{\cite{Ioffe:2005ym}}\,,\\
	M_0^2&=0.8\pm0.2\hs \GeV^2 \text{~\cite{Ioffe:2005ym}}\,.
\end{align}
From Eq.~\eqref{eq:corr_exp}, we can obtain $\Psi''_{5}(Q^2)$ which has following form:
\begin{equation}
	\Psi''_{5}(Q^2)=\frac{\left(m_1+ m_2\right)^2} {Q^2}\sum_{i=0}\frac{\tilde{\Psi}''_i(Q^2)}{(Q^2)^i}\,,
	\label{eq:PiD2}
\end{equation}
and the Borel transform as:
\begin{equation}
	\Psi''_{5}(u)=\frac{\left(m_1+ m_2\right)^2} {u}\sum_{i=0}\frac{\tilde{\Psi}''_i(u)}{u^i}\,.
	\label{eq:psi_Borel}
\end{equation}
The spectral function from Eq.~\eqref{eq:corr_exp} is obtained by using Eq.~\eqref{eq:spectral_density}, and it can be organized as:
\begin{equation}
	\rho^{OPE}_{5}(s)= s \hs \mathcal{R}_0(s)+\mathcal{R}_2(s)+\frac{1}{s}\mathcal{R}_4(s)+\frac{1}{s^2}\mathcal{R}_6(s)+\cdots\,,
	\label{eq:rho_exp}
\end{equation}
where $\mathcal{R}_{n}$ are calculated from $\Psi_{n}$ using ~\eqref{eq:master_relation} and analytical expressions for $\mathcal{R}_0$ can be found in Ref.~\cite{Chetyrkin:2005kn}. It should be noted that $\rho^{OPE}_{5}(s)$ and $\Psi''_{5}(Q^2)$ are RG invariant perturbative quantities that enter in Borel-Laplace sum rule in Eq.~\eqref{eq:bs_final}.\par
The  $\Psi_n(q^2)$ for the pseudoscalar current appearing in the OPE expansion of Eq.~\eqref{eq:corr_exp}, are not RG invariant quantity. The Adler function, $\mathcal{D}_{n}(Q^2)$, is obtained from it using the relation:
\begin{equation}
	\mathcal{D}_{n}(Q^2)\equiv -Q^2 \frac{d}{d\hs Q^2}\left[\left(m_1(Q^2)+ m_2(Q^2)\right)^2\hs \Psi_n(Q^2)\right]\,,
	\label{eq:def_Adler}
\end{equation}
which is RG invariant. Both $\Psi_{n}(Q^{2})$ and $\mathcal{D}_{n}(Q^{2})$ have a cut  $Q^2=-q^{2}<0$ due to the term $\log(\frac{\mu^{2}}{-q^{2}})$. The spectral density in Eq.~\eqref{eq:rho_exp} is obtained, by following the procedure decribed in Eq.~\eqref{eq:Rs_ancont}, in the timelike regions ($s=q^2>0$) from the discontinuity of the polarization function:
\begin{align}
	\mathcal{R}_{n}(s)&\equiv\frac{1}{2\pi i }\lim_{\epsilon\rightarrow 0}\left[ \Psi_n(-s-i \epsilon)-\Psi_n(-s+i \epsilon)\right]\nonumber\\&=\frac{1}{2\pi i}\int_{-s+i \epsilon}^{-s-i \epsilon} d\hs q^2 \frac{d}{d\hs q^2}\Psi_n(q^2)\nonumber\\&=\frac{-1}{2\pi i} \int_{-s+i \epsilon}^{-s-i \epsilon} \frac{dq^2}{q^2} \mathcal{D}_{n}(q^2)\nonumber\\ &=\frac{-1}{2\pi i}\oint_{|x_c|=1} \frac{d x_c}{x_c} \mathcal{D}_{n}(-x_c s)\,.
	\label{eq:master_relation}	
\end{align}
The contour integral in the above equation has to be evaluated without crossing the cut for $q ^{2}>0$. It should be noted that for FOPT and RGSPT prescriptions, the imaginary part can be obtained trivially by replacing the $\log(\mu^2/Q^{2})=\log(\mu^2/|Q|^{2})\pm i \hs \pi$ across the cut. For the numerical evaluation methods, such as in the CIPT prescriptions, Eq.~\eqref{eq:master_relation} can be very useful for analytic continuation in the complex plane. To sum $\pi^{2}$-terms in RGSPT, we first perform the RG improvement of the $\Psi_n(q^{2})$ or $\mathcal{D}_n(q^{2})$. The resulting perturbative has the most general term given in Eq.~\eqref{eq:coef_rgspt_uds} for which the imaginary part can be taken by simply setting  $\log(\mu^2/Q^{2})=\log(\mu^2/|Q|^{2})\pm i \hs \pi$. This process results in an analytic expression for which the renormalization scale can be set $\mu^{2}=s$, but the $i\pi$ terms are left behind, which results in improved convergence.  For more details about their effects on the summation of kinematical terms, we refer to Ref.~\cite{AlamKhan:2023dms}. \par
It should be noted that the analytic continuation using the RGSPT expressions for the $\mathcal{R}_n$ are rather lengthy. Therefore, we provide expressions for corresponding Adler functions in appendix~\eqref{app:dim0adler}.
\subsection{Analytic continuation in FOPT and RGSPT }\label{subsec:ancont}
\begin{figure}
		\centering
			\subfigure[]{\label{fig:scdep_R0}	\includegraphics[width=.516\linewidth]{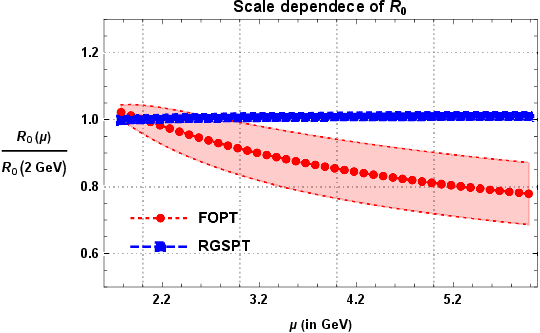}}
		\subfigure[]{ \label{fig:scdep_psi0}	\includegraphics[width=.465\linewidth]{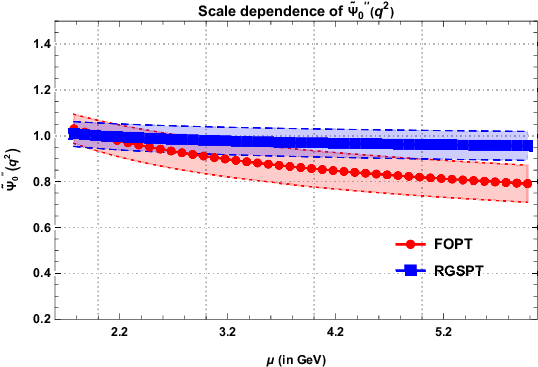}}
		\caption{Renormalization scale dependence of $R_0(s)$ and $\overline{\Psi}''_0(q^2)$ normalized to unity at $2\hs\GeV$ in RGSPT and FOPT. The bands represent the truncation uncertainty.}
		\label{fig:scdep_R0Pd2}
	\end{figure}

The $\mathcal{R}_i(s)$ are obtained from $\Psi_n(q^2)$ by its analytic continuation from spacelike regions to timelike regions (using Eq.~\eqref{eq:master_relation}), which results in the large kinematical $\pi^2$ corrections. These corrections, however, can be summed to all orders using RGSPT, and a good convergence is obtained for the perturbative series. As a demonstration, we define $R_0$ from $\mathcal{R}_0$ as:
\begin{equation}
	R_0\equiv\frac{8\pi^2 }{3 (m_s(2))^2}\mathcal{R}_0\,.
	\label{eq:R0temp}
\end{equation}
Using $\as(2\GeV)=0.2945$ and $m_s(2\GeV)=93.4\MeV$ and setting $m_u=0$, the $R_0$ at different orders of $\as$ has the following contributions:
\begin{align}
	R_0^{\text{FOPT}}&=1.0000+0.6612+ 0.4909+ 0.2912+ 0.1105\,,\\
	R_0^{\text{RGSPT}}&=1.0038+0.4175+0.1760+ 0.0581 -0.0152\,.
\end{align}
We can see that summation of the $\pi^2$-terms enhances the convergence of the perturbation series when RGSPT is employed. The scale dependence of the $R_0$ and truncation uncertainty at different scales are significantly improved, which can be seen in Fig.~\eqref{fig:scdep_R0}. \par
We can also test the RG improvement for the $\tilde{\Psi}_0''(q^2)$ by defining $\overline{\Psi}''_0(q^2)$, analogous to Eq.~\eqref{eq:R0temp}, as:
\begin{equation}
	\overline{\Psi}''_0(q^2)\equiv\frac{8\pi^2 }{3 (m_s(2\GeV))^2}\tilde{\Psi}_0\,.
	\label{eq:psi0temp}
\end{equation}
and using quark masses and $\as$ same as $R_0$ and setting $q=2\GeV$, we get the following contributions to $\overline{\Psi}^{'',\text{FOPT}}_0(q^2)$:
\begin{align}
	\overline{\Psi}^{'',\text{FOPT}}_0&=1.0000+0.4737+ 0.2837+ 0.1917+ 0.1405\,,\\
	\overline{\Psi}^{'',\text{RGSPT}}_0&=1.1508+ 0.5280+ 0.2621+ 0.1670+ 0.1244\,.
\end{align}
In the case of $\overline{\Psi}''_0(q^2)$, we get slightly better convergence than FOPT. The scale dependence of $\overline{\Psi}''_0(q^2)$ normalized to unity at $2\hs\GeV$ is plotted in Fig.~\eqref{fig:scdep_psi0}.

\subsection{Borel Transform in FOPT and RGSPT} \label{subsec:BorelTransform}
The perturbative series in the FOPT prescription is a polynomial form containing $\as$, $m^2_q/Q^2$, and $\log(\frac{\mu^2}{Q^2})$. For Borel transforms in FOPT, only terms $\log(\frac{\mu^2}{Q^2})$ and $Q^2$ are relevant, and we get an analytical expression containing Euler's constant, Zeta functions in addition to terms $\log(\frac{\mu^2}{u})$ and powers of $u$. To obtain the Borel transform, we use the relation from Ref.~\cite{Jamin:1994vr} for the operator in Eq.~\eqref{eq:BO}:
\begin{align}
	\hat{\mathcal{B}}_{u}\Big[\frac{1}{(Q^2)^\alpha}\log^n\left(\frac{\mu^2}{Q^2}\right)\Big]=\frac{1}{(u)^\alpha}\sum_{k=0}^n (-1)^k \hs\hs\text{}^n C_k\log^k\left(\frac{\mu^2}{u}\right)\del_\alpha^{n-k}\left(\frac{1}{\Gamma[\alpha]}\right)\,,
	\label{eq:B_FOPT}
\end{align}
where, $\text{}^n C_k=\frac{n!}{k!(n-k)!}$ is the binomial coefficient. The derivative of the Gamma function results in the appearance of the Euler's constant and Zeta functions that can be seen in Eq.~\eqref{eq:Borel_gammazeta}.\par
For RGSPT, Borel transform is not a trivial task; it involves transcendental function as encountered in Ref.~\cite{deRafael:1981bs} and is evaluated numerically. The most general term in RGSPT, from Eq.~\eqref{eq:coef_rgspt_uds}, can be written as:
\begin{equation}
	\frac{\log^n(w)}{w^\alpha}= \left[\del_\delta^n\hs w^{\delta-\alpha}\right]_{\delta\rightarrow 0}\,,
\end{equation}
where $w=1-\beta_{0} x \log(\mu^2/s)$ and $\alpha$ is some real number depending upon the anomalous dimension of the quantity under consideration. \par
Using Schwinger parametrization, we can write:
\begin{equation}
	\begin{aligned}
		\frac{1}{w^{\alpha}}=&\frac{1}{\Gamma[\alpha]}\int_{0}^{\infty}dt\hs\hs t^{\alpha-1}\hs e^{-t\hs w}\\
		=&\frac{1}{\Gamma[\alpha]}\int_{0}^{\infty}dt\hs\hs t^{\alpha-1}\hs e^{-t\hs \left(1-\beta_0 x \log(\mu^2/s)\right)}\\
		=&\frac{1}{\Gamma[\alpha]}\int_{0}^{\infty}dt\hs\hs t^{\alpha-1}\hs \left(\mu^2/s\right)^{\beta_0 \hs x \hs t}\hs e^{-t}\\
		=&\frac{1}{\Gamma[\alpha]}\sum_{n=0}^{\infty}\frac{(-1)^n}{\Gamma[n+1]}\int_{0}^{\infty}dt\hs\hs t^{\alpha+n-1}\hs \left(\mu^2/s\right)^{\beta_0 \hs x \hs t}
	\end{aligned}
	\label{eq:Sch_par}
\end{equation}
Using the above relation, we can easily perform the Borel operator as follows:
\begin{align}
	\hat{\mathcal{B}}_{u}\left[\frac{1}{s^{z}}\frac{1}{w^{\alpha}}\right]=&\frac{1}{(\mu^2)^z\Gamma[\alpha]}\sum_{n=0}^{\infty}\frac{(-1)^n}{\Gamma[n+1]}\times\int_{0}^{\infty}dt\hs \frac{t^{\alpha+n-1}}{\Gamma[z+\beta_0 x t]}\hs \left(\mu^2/u\right)^{\beta_0 \hs x \hs t+z}\,,
	\label{eq:BO_RGtemp}
\end{align}
where we have used the identity:
\begin{align}
	\hat{\mathcal{B}}_{u}\left[\frac{1}{s^{\alpha}}\right]=\frac{1}{\Gamma[\alpha]u^{\alpha}}\,.
\end{align}
Now, we re-scale integral in Eq.~\eqref{eq:BO_RGtemp} by substituting $\tilde{t}=\beta_0\hs  x\hs t$ and rewrite it as:
\begin{align}
	\hat{\mathcal{B}}_{u}\left[\frac{1}{s^{z}}\frac{1}{w^{\alpha}}\right]=&\frac{1}{(\mu^2)^z\Gamma[\alpha]}\sum_{n=0}^{\infty}\frac{(-1)^n}{\Gamma[n+1] \left(\beta_0 \hs x\right)^{n+\alpha}}\int_{0}^{\infty}d\tilde{t}\hs \frac{\tilde{t}^{\alpha+n-1}}{\Gamma[z+\tilde{t}]}\hs \left(\mu^2/u\right)^{\tilde{t}+z}\,.
	\label{eq:bs1}
\end{align}
We can see that integral in the above relation can not be evaluated analytically~\cite{bateman1981higher}. We use:
\begin{align}
	\tilde{\mu}(z,b,a)\equiv\int_{0}^{\infty}dt\frac{x^{a+t}t^{b}}{\Gamma[b+1]\Gamma[a+t+1]}
	\label{eq:identity1}
\end{align}
to rewrite Eq.~\eqref{eq:bs1} as:
\begin{align}
	\hat{\mathcal{B}}_{u}\left[\frac{1}{s^{z}}\frac{1}{w^{\alpha}}\right]=&\frac{1}{(\mu^2)^z}\sum_{n=0}^{\infty}\frac{(-1)^n}{\left(\beta_0 \hs x\right)^{n+\alpha}} \frac{\Gamma[\alpha+n-1]}{\Gamma[\alpha]\Gamma[n+1]}\times\tilde{\mu}(\mu^2/u,\alpha+n-1,z)\,.
\end{align}
We have to rely on numerical methods beyond this point. However, the identity:
\begin{align}
	\int_{0}^{\infty}e^{-s t} \tilde{\mu}(t,b,a) dt=s^{-\alpha-1}(\log(s))^{-\beta-1}\,,
\end{align}
allows us to recover the original function using Laplace transform.\par
Now, we can demonstrate the impact of the resummation for the Borel transformation. Consider leading mass corrections at different dimension to $\tilde{\Psi}''_j(s)$ from RGSPT, which has the following form:
\begin{equation}
	A_j^{\text{RGSPT}}= \frac{1}{s(1-\beta_0 x L)^{(2j+2)\gamma_0/\beta_0}}\,,
\end{equation}
where $L=\log(\mu^{2}/s)$ is used here for the discussion. Its series expansion to $\ordas{4}$ in FOPT is given by:
\begin{align}
	A_j^{\text{FOPT}}= &\frac{1}{s}\bigg(1+2 \gamma _0 L (j+1) x+\gamma _0 L^2 (j+1) x^2 \left(\beta _0+2 \gamma _0(1+j)\right)\nonumber\\&+\frac{2}{3} \gamma _0 L^3 (j+1) x^3 \left(\beta _0+(1+j)\gamma _0 \right)\times \left(\beta _0+2 \gamma _0(1+j)\right)\nonumber\\&+\frac{1}{6} \gamma _0 L^4 (j+1) x^4\left(\beta _0+\gamma _0(1+ j)\right) \left(\beta _0+2 \gamma _0(1+j) \right) \left(3 \beta _0+2 \gamma _0(1+j)\right)\bigg)+\ordas{5}\,.
	\label{eq:Bexp}
\end{align}
Now, for the Borel transformation of the above series can be obtained by substituting the following values:
\begin{align}
	\label{eq:Borel_gammazeta}
	\hat{\mathcal{B}}_u\left[\frac{1}{s}\right]&= \frac{1}{u}\,,\quad \hat{\mathcal{B}}_u\left[\frac{\log \left(\frac{\mu ^2}{s}\right)}{s}\right]= \frac{\log(\frac{\mu^2}{u})+\gamma_E }{u}\,,\nonumber\\\hat{\mathcal{B}}_u\left[\frac{\log ^2\left(\frac{\mu ^2}{s}\right)}{s}\right]&= \frac{\log^2(\frac{\mu^2}{u})+2 \gamma_E  \log(\frac{\mu^2}{u})+\gamma_E ^2-\zeta(2)}{u}\,,\nonumber\\\hat{\mathcal{B}}_u\left[\frac{\log ^3\left(\frac{\mu ^2}{s}\right)}{s}\right]&=\frac{\left(3 \gamma_E ^2-3\zeta(2)\right) \log(\frac{\mu^2}{u})+3 \gamma_E  \log^2(\frac{\mu^2}{u})}{u}+\frac{\log^3(\frac{\mu^2}{u}) +2 \zeta (3)+\gamma_E ^3-3\gamma_E \zeta(2)}{u}\,,\nonumber\\\hat{\mathcal{B}}_u\left[\frac{\log ^4\left(\frac{\mu ^2}{s}\right)}{s}\right]&=\frac{8 \gamma_E  \zeta (3)+\gamma_E ^4+3/2\zeta(4)-6\gamma_E ^2 \zeta(2)}{u}+\frac{\log(\frac{\mu^2}{u}) \left(8 \zeta (3)+4 \gamma_E ^3-12 \gamma_E \zeta(2)\right)}{u}\nonumber\\&+\frac{\left(6 \gamma_E ^2-6\zeta(2)\right) \log^2(\frac{\mu^2}{u})+ \log^4(\frac{\mu^2}{u})}{u}+\frac{4 \gamma_E  \log^3(\frac{\mu^2}{u})}{u}%
\end{align}
where, $\gamma_E$ is Euler's constant, $\zeta(i)$ are the Zeta functions. These induced terms as a property of Borel-Laplace sum rules are first pointed in Ref.~\cite{Narison:1981ts}. It is interesting to note that all the $\gamma_E$ can be absorbed in the logarithms i.e. $\log(\frac{\mu^2e^{\gamma_E}}{u})$ but not the Zeta functions. A similar case for the Fourier transform of the static potential from momentum to the position space can be found in Ref.~\cite{Jezabek:1998wk}. \par
\begin{figure}[H]
		\centering
		\subfigure[]{\label{fig:RGP2u}\includegraphics[width=.45\linewidth]{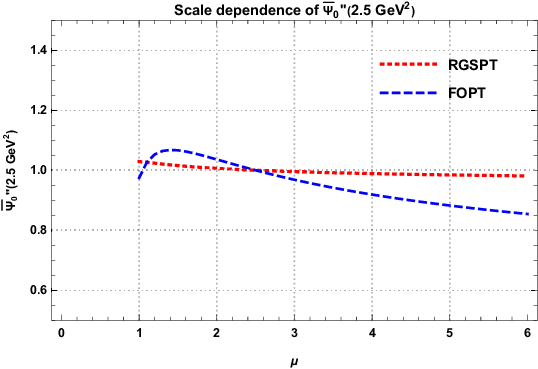}}
		\includegraphics[width=.45\textwidth]{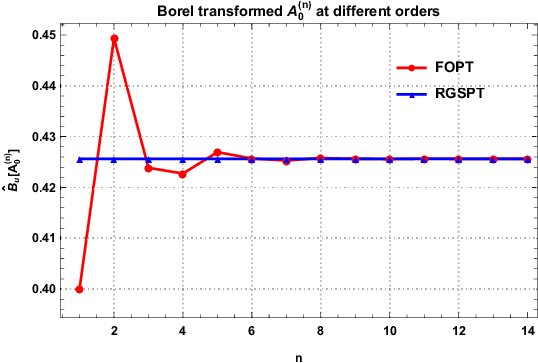}
		\includegraphics[width=.45\textwidth]{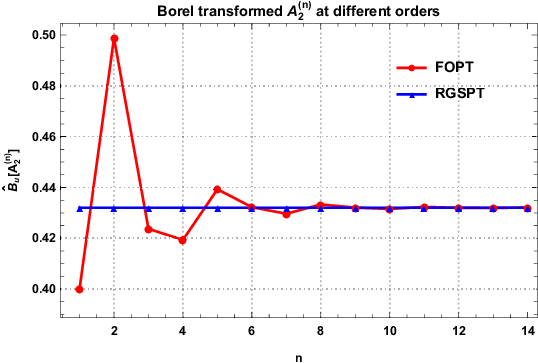}
		\includegraphics[width=.45\textwidth]{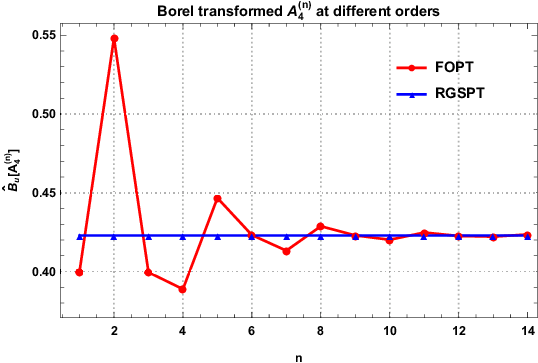}
		\caption{Scale dependence of the $\tilde{\Psi}''_0(u)\vert_{u=2.5\GeV^2}$ in RGSPT and FOPT in Fig~\eqref{fig:RGP2u}. Rest of the figures show convergence of the Taylor expanded $A_j$  calculated at different orders using $u=2.5\GeV^2$ to exact result from RGSPT. }
		\label{fig:A0}
	\end{figure}

Now, we obtain the Borel transform for $A_0$ using Eq.~\eqref{eq:Borel_gammazeta} and by setting $\mu^2=u=2.5\GeV^2$ that resums the logarithms in the case of FOPT. Using $x(\sqrt{2.5})=\as(\sqrt{2.5})/\pi=0.3361/\pi$, the Borel transformation of $A_0$ has the following contributions:
\begin{align}
	\hat{\mathcal{B}}_u [A_0^{\text{RGSPT}}]&=0.4256\,,\nonumber\\
	\hat{\mathcal{B}}_u [A_0^{\text{FOPT}}]&=0.4000+ 0.0494-0.0255 -0.0011+ 0.0042\nonumber\\&=0.4270\,.
\end{align}
It is clear from these numerical contributions that numerical contributions from leading logarithms are oscillatory, and the Borel transformed have poor convergence. The convergence gets worst for higher $j$ values, which can be inferred from Eq.~\eqref{eq:Bexp}, and the first three are plotted in Figs.~\eqref{fig:A0}. The RGSPT value is all order results, but for FOPT, it oscillates and slowly converges to the RGSPT value.\par 
We can use the above results to study the renormalization scale dependence of $\tilde{\Psi}''_0(u)$. To compare FOPT and RGSPT results, we use values of $\tilde{\Psi}''_0(u)\vert_{u=2.5\GeV^2}$ in these prescriptions, normalized to unity at $\mu=2.5\GeV$, and present our results in Fig.~\eqref{fig:RGP2u}. Again, results for RGSPT are very stable for a wide range of renormalization scales.
	\begin{figure}[H]
			\centering
			\includegraphics[width=.49\linewidth]{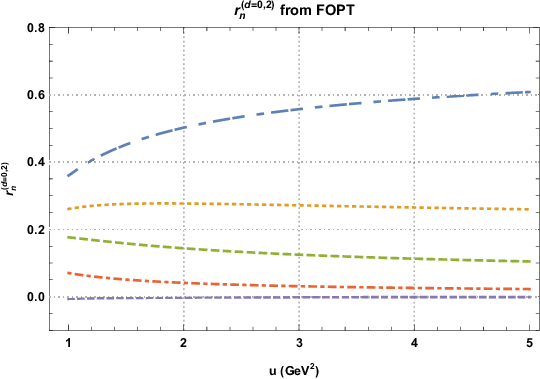}
			\includegraphics[width=.49\linewidth]{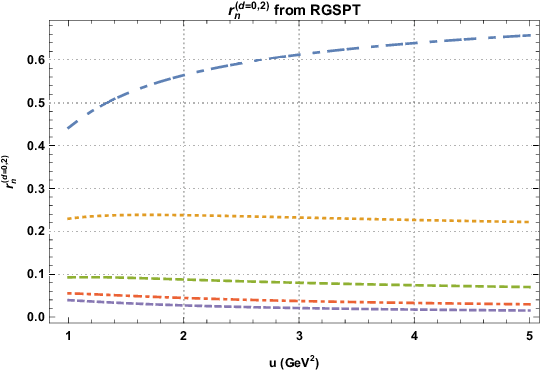}
		
		\caption{$r_n^{d=0,2}$ from FOPT and RGSPT. The lines from up to down correspond to $n=0,1,2,3,4$. }
		\label{fig:rn}
	\end{figure}
\subsection{Convergence of Borel transformed OPE using FOPT and RGSPT.}\label{subsec:conf_borel}
We use the ratio $r_n^{d=0,2}$, defined in Ref.~\cite{Chetyrkin:2005kn}, from Eq.~\eqref{eq:psi_Borel} as:
\begin{equation}
	r_n^{d=0,2}(u)=\frac{\left(\frac{1}{u}\tilde{\Psi}''_0(u)+\frac{1}{u^2}\tilde{\Psi}''_2(u)\right)^{\left(\ordas{n}\right)}}{\Psi''(u)}\,.
\end{equation}
The numerator in the above equation is evaluated using the contributions from $\ordas{n}$ from dimension-0 and dimension-2 corrections to $\Psi''(u)$. Using PDG values for the $m_s(2\GeV)=93.4\MeV$, $m_d(2\GeV)=4.67\MeV$ and setting $u=2.5\GeV^2$, we get the following contributions to $ r_{n}^{d=0,2}$:
\begin{align}
	r_{n}^{d=0,2}\vert_{\text{FOPT}}&=\{53.45\%, 27.46\%, 13.30\%, 3.51\%, -0.22\%\}\nonumber\,,\\
	r_{n}^{d=0,2}\vert_{\text{RGSPT}}&=\{59.25\%, 23.52\%, 8.34\%, 4.04\%, 2.35\%\}\,.\nonumber
\end{align}
From these numerical values, one may suspect that the FOPT has better convergence than RGSPT. This behavior can be attributed to the fact that there are large negative corrections from the Borel transform of the logarithmic  $\log^n\left(\mu^2/Q^2\right)$ terms as depicted in Fig.~\eqref{fig:A0}. The behavior of $r_n$ for different values of the Borel parameter can be found in Fig.~\eqref{fig:rn}.\par
These findings clearly show that RGSPT has the potential to reduce theoretical uncertainty significantly and has been the primary goal of this chapter.
\subsection{Instanton contribution} \label{subsec:insanton}
In addition to the OPE correction, the QCD vacuum structure becomes relevant at low energy, and contributions from the instantons become relevant at energy range$\sim \hs 1\GeV$. Their contributions are estimated using the instanton liquid model (ILM)~\cite{Ilgenfritz:1980vj,Shuryak:1981ff,Shuryak:1982dp} and are added to the pseudoscalar current correlator. These contributions are parameterized in terms of the instanton size $\rho_c$ and number density $n_c$. For the spectral density, we use the results from Refs.~\cite{Elias:1998fs,Maltman:2001gc,Yin:2021cbb}:
\begin{equation}
	\rho^{\text{inst.}}_{i,j}=\frac{1}{2\pi}\text{Im}(\Psi(s)_{\text{inst.}})=\frac{-3\eta_{ij}\left(m_i+m_j\right)^2}{4\pi}J_1\left(\rho_c \sqrt{s}\right) Y_1\left(\rho_c \sqrt{s}\right)\,,
	\label{eq:rho_insta}
\end{equation}
where, $\rho_c=1/0.6$ and $\eta_{ud/us}=1/0.6$~\cite{Shuryak:1982dp}. In addition, we also need the Borel transform of the second derivative of the polarization function for instanton, which is given by~\cite{Narison:2014vka}:
\begin{equation}
	\Psi_{5,ij}''(u)^{\text{inst.}}=\hat{\mathcal{B}}_u[(\Psi_5''(s))_{\text{inst.}}]= \frac{3\eta_{ij} \rho_c^2\left(m_i+m_j\right)^2}{8 \pi ^2} e^{-\frac{1}{2} \rho_c^2\hs u} \times\left[K_0\left(\frac{1}{2}\rho_c^2 u\right)+K_1\left(\frac{1}{2}\rho_c^2 u\right)\right]\,,
	\label{eq:borel_instal}
\end{equation}
where $K_0$ and $K_1$ are the modified Bessel functions. These contributions are numerically relevant for low values of the Borel parameter $u\sim 1\GeV^2$.\par
Now, we have all the theoretical and phenomenological needed as input for the Borel sum rule in Eq.~\eqref{eq:bs_final}. In the next section, light quark mass determination using FOPT and RGSPT is performed. 
\section{Light quark mass determination}\label{sec:ms_md_determination}
In this section, we determine that masses of the strange quark mass using the Borel-Laplace sum rule in Eq.~\eqref{eq:bs_final} from the divergences of the axial vector current. It should be noted that the $m_u$ is determined using the ratio $\epsilon_{ud}\equiv m_u/m_d=0.474_{-0.074}^{+0.056}$~\cite{ParticleDataGroup:2022pth}.\par
Before moving to mass determination from the Sum rule, we need to fix the values for the continuum threshold $s_0$ and Borel parameter $u$. In principle, any determination from the Borel-Laplace sum rule should be independent of the choice of these parameters in the limit $u\gg s_0$. However, in practical cases, there is a dependence on the determinations of light quark masses on these parameters. For practical purposes, these parameters are tuned to get stable results for a given range. The Borel parameter is chosen large enough to suppress the contributions from non-perturbative condensate terms and resonances. However, the continuum threshold $s_0$ is chosen in a region where contributions from the higher resonances are negligible and spectral function can be approximated with the continuum pQCD correction. A proper window for $s_0$ and $u$ is crucial for the stable determination of the Borel-Laplace sum rule, and we have discussed them for FOPT and RGSPT and in this section for the individual as well as simultaneous $m_d$ and $m_s$ determination.\par
\begin{figure}[H]
	\centering
	\includegraphics[width=.48\textwidth]{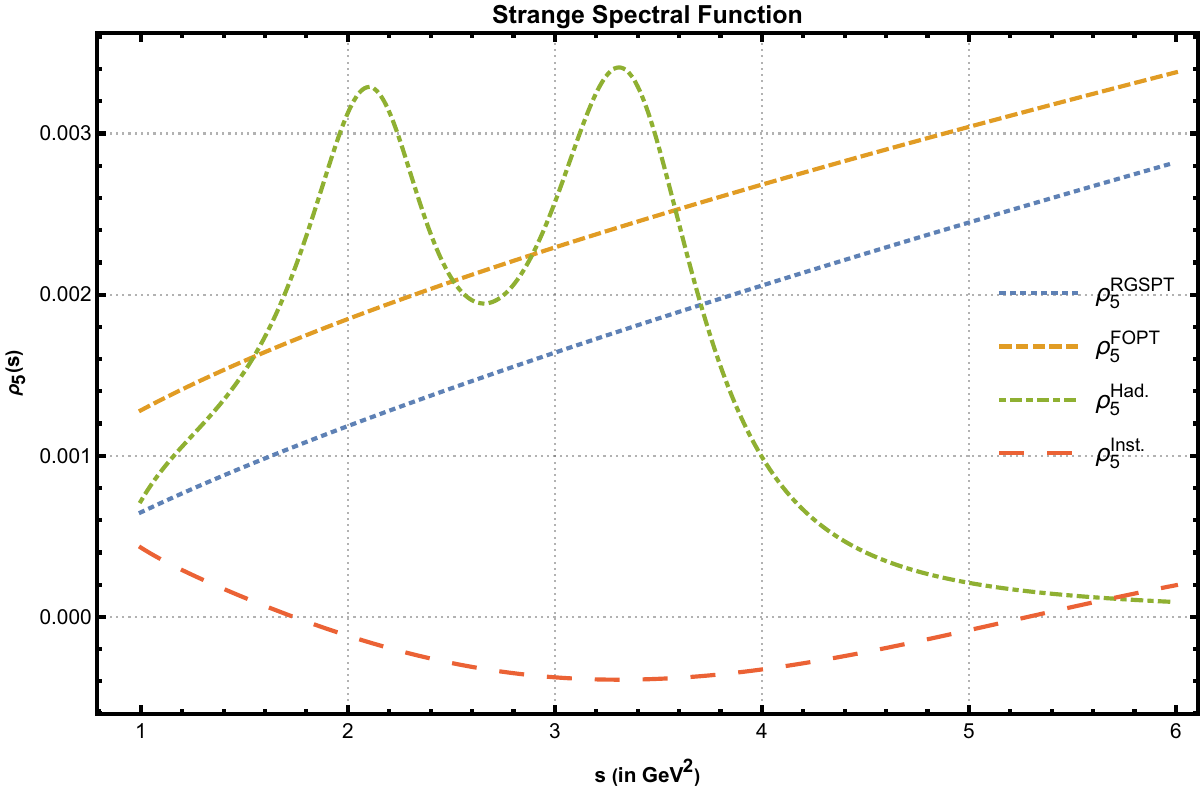}
	\includegraphics[width=.49\textwidth]{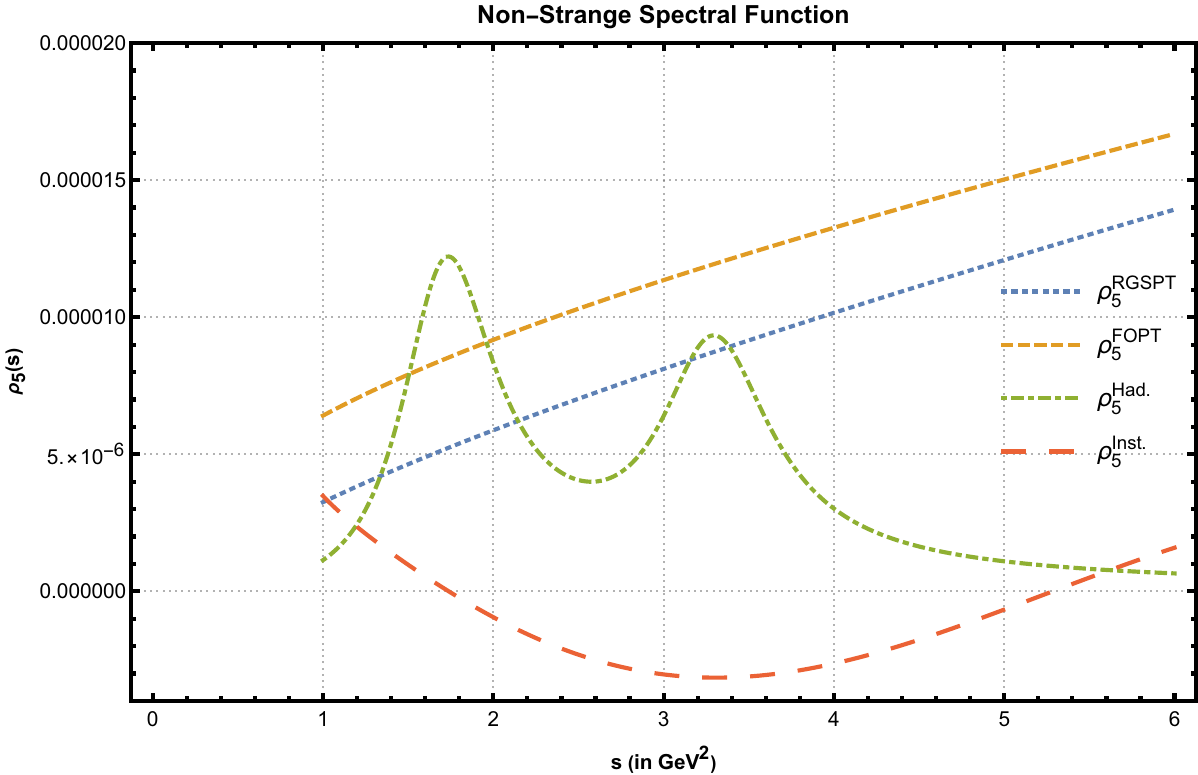}	
	\caption{Hadronic and theoretical spectral functions in the strange and non-strange channel.}
	\label{fig:rho_all}
\end{figure}

We can also perform quark mass determination by choosing the value of $s_0$ for which both hadronic and perturbative spectral functions are in agreement. However, these determinations are going to be very sensitive to the second resonance present in the hadronic spectral function. Another issue is the absence of information about higher resonances which are already neglected in this study. Various contributions to the spectral functions are presented in Figs.~\eqref{fig:rho_all} for Non-strange and strange channels. For these channels, this agreement is found in the range $s_0\in \left[3.38,3.79\right]$ for which we have taken $s_0=3.58\pm 0.20\GeV^2$ in such determinations. However, such determinations are not taken in our final average due to the issues discussed above.

\subsection{\texorpdfstring{$m_s$}{} determination}\label{subsec:ms}
To fix the free $s_0$ and $u$ parameters, we first perform $m_s$ determination at different values. In the FOPT prescription, there is a crossover around $s_0=4.5\hs\GeV^2$ for different values of the Borel parameter that can be seen in Fig.~\eqref{fig:msFOPT_u_s0_1}. Therefore, we choose $s_0=4.5\pm0.5\GeV^2$ to minimize variation in the $m_s$ determination with respect to the Borel parameter from FOPT determinations. However, this is not the case for RGSPT as there is no crossing point in Fig.~\eqref{fig:msRGSPT_u_s0_1}. There is a stability window for $s_0\in\hs\left[3.5,4.2\right]\GeV^2$ region for RGSPT, but we do not use this value as $\sqrt{s_0}$ is close to the mass of the second resonance. This results in slightly more uncertainty from the variation of $u$ in the $m_s$ determination compared to FOPT, which can be seen in Fig.~\eqref{fig:msmdu1}. However, we find that there is a linear increase in the difference of maximum and minimum values of strange quark mass ($\Delta(m_s)$) determination for $s_0\in\left[3,5\right]\GeV^2$ with $u\in\left[2,3\right]$, which can be seen in Fig.~\eqref{fig:delms}. This linear behavior is milder in the case of RGSPT compared to FOPT. \par
Now, we move on to our final determination for which we adopt the choice of parameters used in Ref.~\cite{Chetyrkin:2005kn}. For the Borel parameter, we use $ u=2.5\pm0.5\GeV^2$ and the renormalization scale is varied in the range $u/2\leq\mu^2\leq2 u$. We take the continuum threshold value $s_0=4.5\pm0.5\hs\GeV^2$ and $m_u=2.16_{-0.26}^{+0.49}\MeV$~\cite{ParticleDataGroup:2022pth} as input in our determination. We obtain the following value of $m_s(2\GeV)$, using FOPT:
\begin{equation}
	m_s(2\hs\GeV)=103.64_{-4.61}^{+6.45}\hs\MeV\,,
\end{equation} 
and for RGSPT, we obtain:
\begin{equation}
	m_s(2\hs\GeV)=104.20_{-4.29}^{+4.37}\hs\MeV\,.
\end{equation}
The details of significant sources of uncertainties can be found in Table~\eqref{tab:md_ms_indiv_final}. The pQCD uncertainties containes uncertainties arising from uncertainties present in the quark and gluon condensates, $\as$, renormalization scale variation, and  tuncation uncertainty. The truncation uncertainty is calculated from the contribution of the last terms present in the expansion of $\as$ in the perturbative series.   Uncertainties from other parameters are included in the hadronic uncertainties.\par
It is worth mentioning that the uncertainties coming from scale variation in RGSPT are significantly smaller than in FOPT, leading to small pQCD uncertainties compared to the hadronic uncertainties. It is important to note from Table.~\eqref{tab:md_ms_indiv_final} is that the total theoretical uncertainty from pQCD parameters is smaller than the hadronic uncertainties when RGSPT is used. We present the scale dependence in our determinations in Fig.~\eqref{fig:md_scdep}. Another point to note is that the exclusion of the instanton term for the RGSPT and FOPT series leads to a decrease of strange quark mass about~$1.26\MeV$ and $1.24\MeV$, respectively. \par
Now, we also present our results for theoretical and hadronic spectral functions are in agreement. Using $s_0=3.58\pm 0.20\GeV^2$ and taking the rest of the parameters discussed above, we get the following determinations for FOPT and RGSPT scheme:
\begin{align}
	m_s(2\GeV)&=107.29_{-5.83}^{+7.57}\MeV\,,\quad (\text{FOPT})\\
	m_s(2\GeV)&=106.02_{-4.57}^{+4.36}\MeV\,,\quad (\text{RGSPT})
\end{align}
The dependence of these determinations on the Borel parameter is presented in Fig.~\eqref{fig:msmdQH_u0}.
\subsection{\texorpdfstring{$m_d$}{} determination}\label{subsec:md}
\begin{figure}[H]
		\centering
			\subfigure[]{ \label{fig:md_scdep}	\includegraphics[width=.49\linewidth]{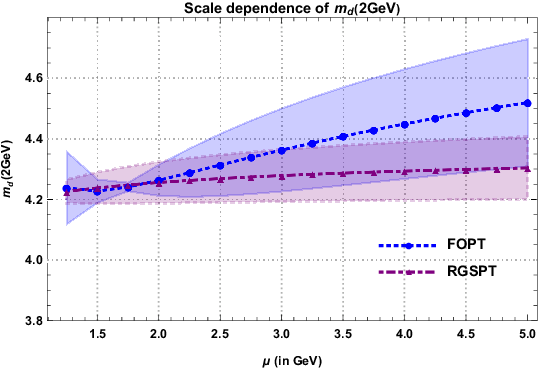}}
			\subfigure[]{ \label{fig:ms_scdep_uds}	\includegraphics[width=.49\linewidth]{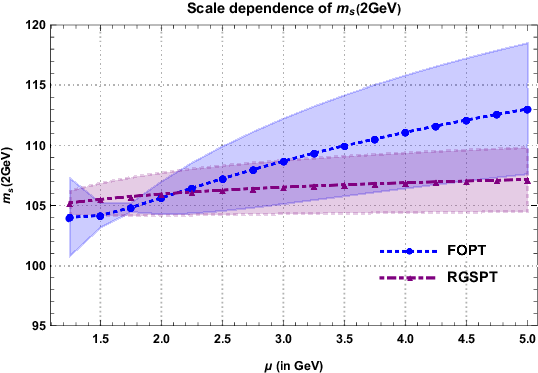}}
		\caption{The scale dependence in the individual $m_d$ and $m_s$ determinations using FOPT and RGSPT. The bands in the plot represent truncation uncertainty at different scales.}
		\label{fig:msmd_scdep}
	\end{figure}

\begin{figure}[H]
		\centering
			\subfigure[]{ \label{fig:msFOPT1}\includegraphics[width=.49\linewidth]{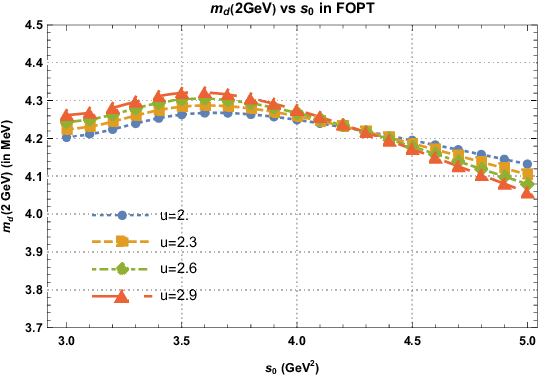}}
			\subfigure[]{  \label{fig:mdRGSPT1}\includegraphics[width=.49\linewidth]{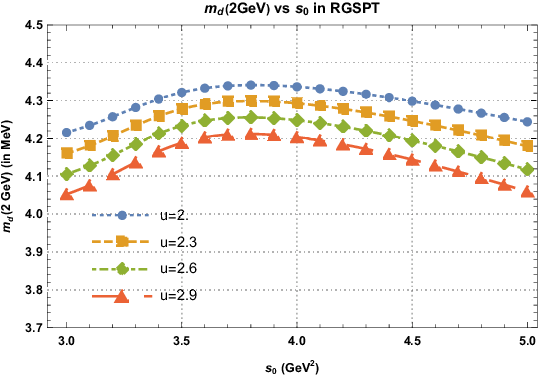}}		
		\caption{$m_d(2\GeV)$ calculated at different values of Borel parameter and $s_0$ using: \eqref{fig:msFOPT1} FOPT and, \eqref{fig:mdRGSPT1} RGSPT prescription.}
		\label{fig:md_s0_u0}
	\end{figure}

Similar to $m_s$ determination, there is a crossover point for $m_d$ in FOPT, but near to $\pi(1800)$ resonance mass as we can see in Fig.~\eqref{fig:md_s0_u0}. Due to this, we choose $s_0=4.5\pm0.5\GeV^2$ as in the previous subsection. Using the same parameters and $\epsilon_{ud}$~\cite{ParticleDataGroup:2022pth}, we obtain the following values:
\begin{align}
	m_d(2\hs\GeV)&=4.18_{-0.44}^{+0.51}\hs\MeV\,,\\
	\implies m_u(2\hs\GeV)&=1.98_{-0.40}^{+0.34}\hs\MeV\,,
\end{align}
and for RGSPT, we obtain the following value:
\begin{align}
	m_d(2\hs\GeV)&=4.21_{-0.39}^{+0.48}\hs\MeV\,.\\
	\implies m_u(2\hs\GeV)&=2.00_{-0.40}^{+0.33}\hs\MeV\,.
\end{align}
We present the scale dependence in our determinations in Fig.~\eqref{fig:md_scdep}. The details of the sources of uncertainties can be found in Table~\eqref{tab:md_ms_indiv_final}. Exclusion of the instanton terms leads to a decrease in the central value of $m_d(2\GeV)$ by~$0.20\MeV$ and~$0.13\MeV$ in determinations using FOPT and RGSPT prescriptions, respectively.\par
\begin{figure}[H]
	\centering
	\centering
	\includegraphics[width=.7\linewidth]{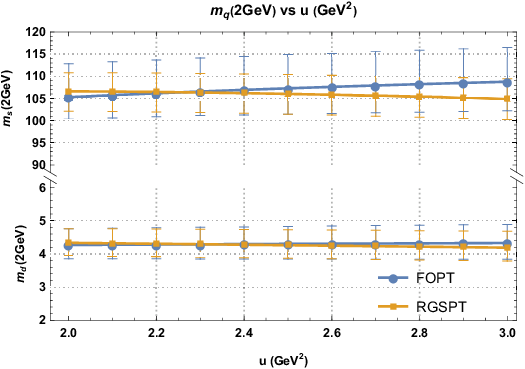}
	\caption{\label{fig:msmdQH_u0} Borel parameter dependence of the individual determinations of $m_d(2\GeV)$ assuming quark hadron duality is obeyed at $s_0=$ from FOPT and RGSPT prescriptions.}
\end{figure}

Now, using $s_0=3.58\pm 0.20\GeV^2$ and taking the rest of the parameters discussed above, we get the following determinations for FOPT and RGSPT scheme:
\begin{align}
	m_d(2\GeV)&=4.30_{-0.46}^{+0.52}\MeV\,,\quad (\text{FOPT})\\
	\implies m_u(2\GeV)&=2.04_{-0.40}^{+0.35}\MeV\,,\\
	m_d(2\GeV)&=4.26_{-0.39}^{+0.46}\MeV\,,\quad (\text{RGSPT})\\
	\implies m_u(2\GeV)&=2.02_{-0.40}^{+0.32}\MeV\,,\,.
\end{align}
The dependence of these determinations on the Borel parameter can be found in Fig.~\eqref{fig:msmdQH_u0}.
\subsection{Simultaneous \texorpdfstring{$m_d$ and $m_s$}{} determination.}
We can also perform the simultaneous determination of the $m_s$ and $m_d$ using the sum rule in Eq.~\eqref{eq:bs_final} for strange and non-strange channels. Using FOPT, we obtain:
\begin{align}  
	m_s(2\hs\GeV)&=102.36_{-4.36}^{+6.40}\hs\MeV\,,\\
	m_d(2\hs\GeV)&=4.09_{-0.43}^{+0.50}\hs\MeV\,,\\
	\implies m_u(2\hs\GeV)&=1.98_{-0.40}^{+0.33}\hs\MeV\,,
\end{align} 
and for RGSPT, we obtain the following values:
\begin{align}
	m_s(2\hs\GeV)&=104.34_{-4.21}^{+4.23}\hs\MeV\,,\\
	m_d(2\hs\GeV)&=4.21_{-0.45}^{+0.48}\hs\MeV\,,\\
	\implies m_u(2\hs\GeV)&=2.00_{-0.40}^{+0.33}\hs\MeV\,.
\end{align}
The details of sources of uncertainties in the determination of quark masses can be found in Table~\eqref{tab:mdms_final}. In this case, uncertainty in $m_s$ determination is smaller than one obtained in subsection~\eqref{subsec:ms}. The values obtained for $m_s$ and $m_d$ are very close to the determination from subsection~\eqref{subsec:ms} and ~\eqref{subsec:md}. \par
Using $s_0=3.58\pm 0.20\GeV^2$ for FOPT, we obtain:
\begin{align}  
	m_s(2\hs\GeV)&=107.39_{-5.08}^{+6.95}\hs\MeV\,,\\
	m_d(2\hs\GeV)&=4.30_{-0.44}^{+0.51}\hs\MeV\,,\\
	\implies m_u(2\hs\GeV)&=2.04_{-0.40}^{+0.34}\hs\MeV\,,
\end{align} 
and for RGSPT, we obtain the following values:
\begin{align}
	m_s(2\hs\GeV)&=106.14_{-4.52}^{+4.32}\hs\MeV\,,\\
	m_d(2\hs\GeV)&=4.26_{-0.43}^{+0.46}\hs\MeV\,,\\
	\implies m_u(2\hs\GeV)&=2.02_{-0.40}^{+0.32}\hs\MeV\,.
\end{align}

	\begin{table}[H]
		\centering
		   \begin{adjustbox}{max width=\textwidth}
		\begin{tabular}{||c|c|c|c|c|c|c|c|c|c|c|c|c|c|c||}\hline\hline
			\text{} &\multicolumn{7}{c|}{\textbf{FOPT}}&\multicolumn{7}{c|}{\textbf{RGSPT}}\\\cline{2-15}
			\text{Quark mass}&Final Value&$\mu$&$\as$&u&$s_0$&\text{pQCD}&\text{Had.}&Final Value&$\mu$&$\as$&u&$s_0$&\text{pQCD}&\text{Had.}
			\\\hline\hline
			\multirow{2}{5em}{$m_d(2\GeV)$}&\multirow{2}{5em}{$4.18^{+0.51}_{-0.44}$}&+0.18&+0.06&+0.07&+0.08&+0.20&+0.47&\multirow{2}{5em}{$4.21^{+0.48}_{-0.39}$}&+0.02&+0.06&+0.10&+0.05&+0.10&+0.47\\\cline{3-8}\cline{10-15}
			\text{}&\text{}&-0.07&-0.06&-0.05&-0.09&-0.11&-0.43&\text{}&-0.01&-0.06&-0.08&-0.07&-0.10&-0.38\\\hline\hline\hline
			\multirow{2}{5em}{$m_s(2\GeV)$}&\multirow{2}{5em}{$103.64^{+6.45}_{-4.61}$}&+4.59&+1.40&+0.76&+2.59&+5.14&+3.91&\multirow{2}{5em}{$104.20^{+4.37}_{-4.29}$}&+0.55&+1.42&+1.40&+1.76&+2.45&+3.61\\\cline{3-8}\cline{10-15}
			\text{}&\text{}&-1.66&-1.40&-0.52&-2.66&-2.85&-3.62&\text{}&-0.52&-1.40&-1.48&-2.13&-2.43&-3.54\\\hline
		\end{tabular}
	\end{adjustbox}
		\caption{\label{tab:md_ms_indiv_final} $m_d$ and $m_s$ determination using FOPT and RGSPT and the sources of uncertainties denoted in the column.}
	\end{table}
	\begin{table}[ht]
		\centering
		   \begin{adjustbox}{max width=\textwidth}
		\begin{tabular}{||c|c|c|c|c|c|c|c|c|c|c|c|c|c|c||}\hline\hline
			\text{} &\multicolumn{7}{c|}{\textbf{FOPT}}&\multicolumn{7}{c|}{\textbf{RGSPT}}\\\cline{2-15}
			\text{Quark mass}&Final Value&$\mu$&$\as$&u&$s_0$&\text{pQCD}&\text{Had.}&Final Value&$\mu$&$\as$&u&$s_0$&\text{pQCD}&\text{Had.}
			\\\hline\hline
			\multirow{2}{5em}{$m_d(2\GeV)$}&\multirow{2}{5em}{$4.18^{+0.50}_{-0.44}$}&+0.18&+0.06&+0.07&+0.08&+0.19&+0.47&\multirow{2}{5em}{$4.21^{+0.48}_{-0.45}$}&+0.02&+0.06&+0.10&+0.05&+0.10&+0.47\\\cline{3-8}\cline{10-15}
			\text{}&\text{}&-0.07&-0.06&-0.05&-0.09&-0.09&-0.43&\text{}&-0.01&-0.06&-0.08&-0.07&-0.10&-0.43\\\hline\hline
			\multirow{2}{5em}{$m_s(2\GeV)$}&\multirow{2}{5em}{$103.80^{+6.14}_{-4.22}$}&+4.54&+1.38&+0.72&+2.56&+4.77&+3.87&\multirow{2}{5em}{$104.34^{+4.32}_{-4.24}$}&+0.55&+1.41&+1.34&+1.74&+2.42&+3.57\\\cline{3-8}\cline{10-15}
			\text{}&\text{}&-1.66&-1.38&-0.49&-2.62&-2.24&-3.58&\text{}&-0.52&-1.38&-1.44&-2.11&-2.40&-3.50\\\hline
		\end{tabular}
	\end{adjustbox}
		\caption{\label{tab:mdms_final} $m_d$ and $m_s$ determination using FOPT and RGSPT and the sources of uncertainties denoted in the column.}
	\end{table}
	
	\begin{figure}[H]
		\centering
				\subfigure[]{\label{fig:msFOPT_u_s0_1} \includegraphics[width=.48\linewidth]{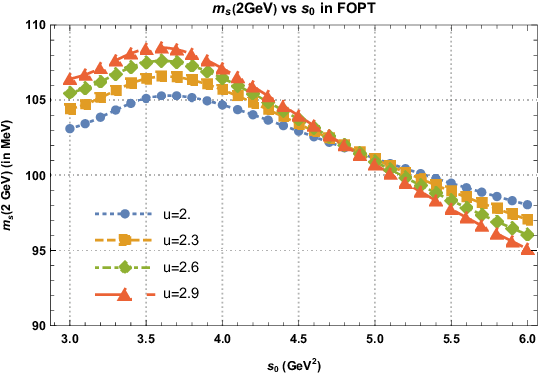}}
				\subfigure[]{ \label{fig:msRGSPT_u_s0_1}\includegraphics[width=.48\linewidth]{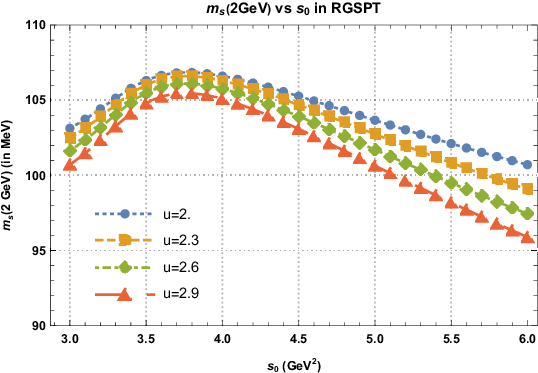}}
				\newline
			\subfigure[]{ \label{fig:msmdu1}\includegraphics[width=.495\linewidth]{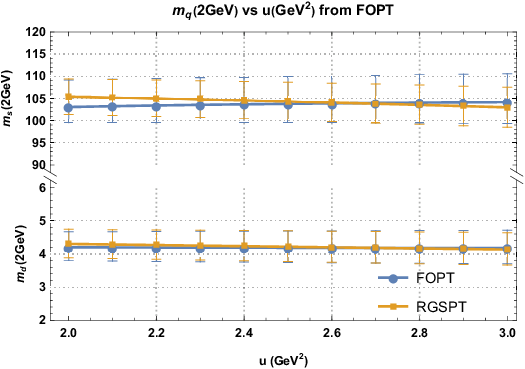}}
				\subfigure[]{ \label{fig:delms}\includegraphics[width=.48\linewidth]{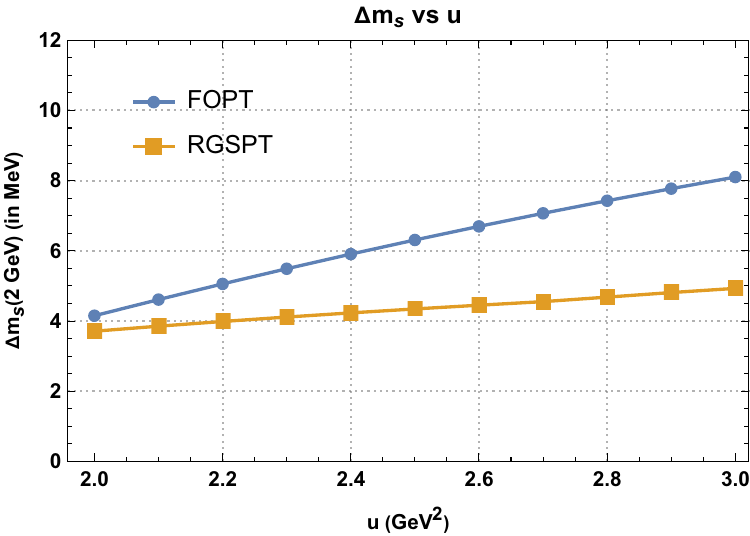}}
		\caption{$m_s(2\GeV)$ calculated at different values of Borel parameter and $s_0$ using: \eqref{fig:msFOPT_u_s0_1} FOPT and, \eqref{fig:msRGSPT_u_s0_1} RGSPT scheme. In Fig.~\eqref{fig:msmdu1}, $m_s(2\GeV)$ and $m_d(2\GeV)$ at different values of the Borel parameter in the range $u\in\left[2,3\right]\GeV^2$. In \eqref{fig:delms}, $\Delta m_s(2\GeV)$ obtained by varying $s_0\in\left[3,5\right]\GeV^2$ at different values of Borel parameter $u$.}
		\label{fig:msFOPT_u_s0}
	\end{figure}

\section{Summary and conclusion}\label{sec:summary_uds}
We have used the Borel-Laplace sum rule to determine the light quark masses from the correlator of the divergence of the axial vector current. The sum rule uses both hadronic as well as perturbative contributions. In section~\eqref{sec:rgspt}, we briefly review the procedure of RG summation in RGSPT and its importance in the RG improvement for the theoretical quantities used in the Borel-Laplace sum rule. \par In section~\eqref{sec:hadronic_info}, we discuss the hadronic pseudoscalar spectral function for which no experimental information is available. However, these contributions can be parameterized in terms of the information available on the masses and decay width of the spectral function. We use commonly used hadronic parametrization from Dominguez and Rafael~\cite{Dominguez:1986aa} in this chapter for the light quark mass determination, and it has a good agreement with another parametrization by Maltman and Kambor~\cite{Maltman:2001gc}, which can be seen in Fig.~\eqref{fig:had}.\par
In section~\eqref{sec:OPE}, the continuum contributions are discussed in detail. The most commonly used FOPT prescription results are already available in the literature. These determinations have large uncertainties from the variation of the renormalization scale. RGSPT can reduce such uncertainties and is inspired by the findings of Ref.~\cite{AlamKhan:2023dms}, we first sum the kinematical $\pi^2$-terms appearing due to analytic continuation of the spectral function in section~\eqref{subsec:ancont}. The analytic continuation using RGSPT also results in better convergence of the perturbation series for the dimension-0 contribution. We also find better convergence and improved scale dependence for the $\Psi''(Q^2)$, which is also an RG invariant quantity. These improvements can be seen in Fig.~\eqref{fig:scdep_R0Pd2}. \par 
In section~\eqref{subsec:BorelTransform}, we first calculate the Borel transformation for the $\Psi''(Q^2)$ in the RGSPT prescription. The Borel transformation for RGSPT can only be performed numerically, and it resums all the Euler's constant and various Zeta functions that arise due to Borel transformation for RGSPT. The FOPT results are found to be poorly convergent and oscillate around the all-order result from the RGSPT. RGSPT also improves the scale dependence of the Borel transformed $\Psi''(Q^2)$ which is used as input in the Borel-Laplace sum rule in Eq.~\eqref{eq:bs_final}. These improvements can be seen in Figs.~\eqref{fig:A0},\eqref{fig:RGP2u}. The result obtained is used in section~\eqref{subsec:conf_borel} to test the convergence of $\Psi''(u)$. FOPT series is found to be slightly more convergent than RGSPT, but it is argued that RGSPT results are more trustworthy as Borel transformation in FOPT has oscillatory behavior for the known results. We also include small instanton contributions using results from ILM in section~\eqref{subsec:insanton}.\par
We determine the light quark masses in section~\eqref{sec:ms_md_determination} using the free parameters $s_0$ and $u$ used in Ref.~\cite{Chetyrkin:2005kn}. This particular choice leads to small $u$ dependence in the FOPT determinations of the $m_s$, which can be seen in Fig.~\eqref{fig:msFOPT_u_s0_1}. For the RGSPT determination, the stability region is closer to the second resonance; therefore, we have used the choices for these free parameters from FOPT. This leads to slightly large uncertainty in the $m_s$ determination from the variations of $u$. In addition to the individual determination of the $m_s$ and $m_d$, we have performed simultaneous determination and found a slightly more precise value for $m_s$. These results are presented in Table~\eqref{tab:md_ms_indiv_final} and Table~\eqref{tab:mdms_final}. In addition to this, we have also presented our determination by choosing $s_0=3.58\pm0.20\GeV^2$ in the resonance region where there is good agreement between theoretical and hadronic spectral function. These values result in higher values of the quark masses which can be seen in Fig.~\eqref{fig:msFOPT_u_s0} and Fig.~\eqref{fig:msFOPT_u_s0}. Since this value choice of $s_0$ is sensitive to the parameters of the second resonance and higher resonances are neglected, we do not consider them in our final determinations. \par
Now, we give our final determination for the light quark masses, which comes from the simultaneous determination of the $m_s$ and $m_d$ and their values at $2\GeV$ are:
\begin{align}
	m_s(2\hs\GeV)&=104.34_{-4.21}^{+4.23}\hs\MeV\,,\\
	m_d(2\hs\GeV)&=4.21_{-0.45}^{+0.48}\hs\MeV\,,\\
	m_u(2\hs\GeV)&=2.00_{-0.40}^{+0.33}\hs\MeV\,.
\end{align}
Corresponding PDG average values~\cite{ParticleDataGroup:2022pth} are:
\begin{align}
	m_s(2\hs\GeV)&=93.4_{-3.4}^{+8.6}\hs\MeV\,,\\
	m_d(2\hs\GeV)&=4.67_{-0.17}^{+0.48}\hs\MeV\,,\\
	m_u(2\hs\GeV)&=2.16_{-0.26}^{+0.49}\hs\MeV\,.
\end{align}
which is in agreement with our findings.

\begin{subappendices}

  \section{Contribution to current correlator } \label{app:dim0adler}
		\subsection{Dimension zero contributions}
  The zero-dimensional contribution to OPE is known to five loops ($\as^4$)~\cite{Gorishnii:1990zu,Chetyrkin:1996sr,Baikov:2005rw}. We are using the following expression for $\Psi_0$:
\begin{align}
    \Psi_0(q^2)&=\frac{3  }{8 \pi ^2}\Bigg\lbrace L+\left(L^2+\frac{17 L}{3}\right) x+x^2 \left(\frac{17 L^3}{12}+\frac{95 L^2}{6}+L \left(\frac{9631}{144}-\frac{35 \zeta (3)}{2}\right)\right)\nonumber\\&+x^3 \Big[L^2 \left(\frac{4781}{18}-\frac{475 \zeta (3)}{8}\right)+\frac{221 L^4}{96}+\frac{229 L^3}{6}+L \Big(-\frac{91519 \zeta (3)}{216}+\frac{715 \zeta (5)}{12}\nonumber\\&-\frac{\pi ^4}{36}+\frac{4748953}{5184}\Big)\Big]+x^4 \Big[L \Big(\frac{192155 \zeta (3)^2}{216}-\frac{46217501 \zeta (3)}{5184}+\frac{455725 \zeta (5)}{432}-\frac{125 \pi ^6}{9072}\nonumber\\&-\frac{52255 \zeta (7)}{256}-\frac{3491 \pi ^4}{10368}+\frac{7055935615}{497664}\Big)+L^2 \Big(-\frac{1166815 \zeta (3)}{576}+\frac{24025 \zeta (5)}{96}\nonumber\\&-\frac{\pi ^4}{36}+\frac{97804997}{20736}\Big)+L^3 \left(\frac{3008729}{3456}-\frac{5595 \zeta (3)}{32}\right)+\frac{51269 L^4}{576}+\frac{1547 L^5}{384}\Big]\Bigg\rbrace\,,
\end{align}
where $x\equiv \as(\mu))/\pi $ and $L=\log(\frac{\mu^2}{-q^2})$. This expression reproduces the results for $\mathcal{R}_0$ and $\tilde{\Psi}_0''(u)$ in Ref.~\cite{Chetyrkin:2005kn}.\par 
For RGSPT, these quantities can be derived using Eq.~\eqref{eq:master_relation} and Eq.~\eqref{eq:def_Adler} from the Adler function. The RGSPT expression for the dimension-0 Adler function is given by:
		\begin{align}
			\mathcal{D}_0(q^2)=&\frac{3 \left(m_{i}+m_{j}\right)^2 }{8 \pi ^2 w^{\frac{8}{9}}}\Bigg\lbrace1+x \left(-1.790+\frac{7.457}{w}-\frac{1.580 \log (w)}{w}\right)+x^2 \Big[-0.339+\frac{60.699}{w^2}\nonumber\\&+\frac{2.653 \log ^2(w)}{w^2}-\frac{14.514}{w}+\left(\frac{2.829}{w}-\frac{27.849}{w^2}\right) \log (w)\Big]+x^3 \Big[-\frac{129.189}{w^2}\nonumber\\&+\frac{599.649}{w^3}-\frac{4.542 \log ^3(w)}{w^3}+\left(\frac{76.231}{w^3}-\frac{4.750}{w^2}\right) \log ^2(w)-\frac{5.207}{w}+0.593\nonumber\\&+\left(\frac{53.766}{w^2}-\frac{361.248}{w^3}+\frac{0.536}{w}\right) \log (w)\Big]+x^4 \Bigg[-12.673+\frac{15.012}{w}-\frac{66.312}{w^2}\nonumber\\&-\frac{5.207}{w}+0.593-\frac{1339.755}{w^3}+\frac{6992.440}{w^4}+\frac{7.851 \log ^4(w)}{w^4}\nonumber\\&+\left(\frac{8.131}{w^3}-\frac{183.752}{w^4}\right) \log ^3(w)+\left(-\frac{146.509}{w^3}+\frac{1384.280}{w^4}-\frac{0.901}{w^2}\right) \log ^2(w)\nonumber\\&+\left(\frac{18.438}{w^2}+\frac{759.073}{w^3}-\frac{4787.937}{w^4}-\frac{0.938}{w}\right) \log (w)\Bigg]\Bigg\rbrace\,,
		\end{align}
	where $w=1-x(\mu)\beta_0 \log(\frac{\mu^2}{-q^2})$. 

		\subsection{Dimension-2 Corrections\label{app:dim2adler}}
  The dimension-2 contributions with full mass dependence are available to $\ordas{1}$ in Refs.~\cite{Chetyrkin:1985kn,Generalis:1990id,Jamin:1992se,Jamin:1994vr}. Additional $\ordas{2}$ correction is taken from Ref.~\cite{Chetyrkin:2005kn}. For FOPT, we use the following expression for dimension-2 contribution to the current correlator: 
	\begin{align}
	    \Psi_2=&\frac{3 }{8 \pi ^2} \Bigg\lbrace\left(m_{i}^2+m_{j}^2\right) \bigg[-2 L+\left(-4 L^2-\frac{32 L}{3}\right) x-2 L+x^2 \bigg(-\frac{25 L^3}{3}-\frac{97 L^2}{2}\nonumber\\&+L \left(\frac{154 \zeta (3)}{3}-\frac{5065}{36}\right)\bigg)\bigg]-m_{i} m_{j} \bigg[x \left(-4 L^2-\frac{56 L}{3}+8 \zeta (3)-\frac{88}{3}\right)-2 L-4\bigg]\Bigg\rbrace\,.
	   \end{align}
 For RGSPT, we derive the spectral function from the dimension-2 Adler function:
 \begin{align}
     \mathcal{D}_2=&\frac{\left(m_{i}+m_{j}\right)^2\left(m_{i}^2+m_{j}^2\right)  }{139968 \pi ^2 w^{34/9}}\Bigg[ 8 x \big(729 \left(20 w^2-251\right) x \zeta (3)+32 \log (w) (8 w (290 x-81)\nonumber\\&+1600 x \log (w)-12607 x)\big)+52488 w^2+(5497360-w (7643 w+1797332)) x^2\nonumber\\&+1296 (361-145 w) w x\Bigg]+\frac{\left(m_{i}+m_{j}\right)^2 m_{i} m_{j} }{216 \pi ^2 w^{25/9}} \left[w (290 x-81)+256 x \log (w)-1046 x\right]
 \end{align}

\subsection{Dimension-4 contributions}
\label{app:dim4}
The dimension-4 contributions can be obtained from Refs.~\cite{Pascual:1981jr,Jamin:1992se,Jamin:1994vr}. For dimension-4 contributions, we use an RG-invariant combination of the condensates given in Refs.~\cite{Spiridonov:1988md,Baikov:2018nzi}. The constant terms at this order are important for the Borel-Laplace operator. We give a summed expression for the current correlator:
	\begin{align}
	\Psi_4=&-\frac{\langle \sum_{i=u,d,s}\overline{q}_i \hs q_i)\rangle_{\text{inv.}}}{162 w^{17/9}} (w (145 x-81)+128 x \log (w)-442 x) +\frac{2 x }{9  w^{17/9}}(\langle\overline{q}_d\hs q_d+\overline{q}_u\hs q_d\rangle_{\text{inv.}}\nonumber\\&+\frac{\langle \overline{q}_i q_j+\overline{q}_j q_j\rangle_{\text{inv.}}}{81  w^{17/9}} (w (145 x-81)+128 x \log (w)-523 x) -\frac{ \langle \frac{\as}{\pi}G^2\rangle_{\text{inv.}} }{1296  w^{17/9}}(2 w (145 x-81)\nonumber\\&+256 x \log (w)-1181 x)+\frac{1}{1512 \pi ^2  w^{11/3} x}\Bigg\lbrace \left(m_{j}^4+ m_{j}^4 \right)(-324 w^2 - 189 w x)\nonumber\\&-81 w x \sum_{k=u,d,s} m_k^4+m_{i}^2 m_{j}^2 \left(x^2 (-6090 (w-1)-5376 \log (w))+1134 w x\right)\nonumber\\&+\left(m_{i}^3 m_{j}+ m_{i}m_{j}^3\right)\bigg[x \left(-3480 w^2+5073 w-3072 w \log (w)\right)+648 w^2\nonumber\\&\bs+x^2 (-8555 w-7552 \log (w)+1877))\bigg]  \Bigg\rbrace\,.
\end{align}
Corresponding FOPT expression is given by:
\begin{align}
	\Psi_4=&\frac{1}{6}((6 L+11) x+3) \langle \sum_{i=u,d,s}\overline{q}_i \hs q_i\rangle_{\text{inv.}}-\frac{1}{3}(2 (3 L+7) x+3)  (\langle \overline{q}_i q_j+\overline{q}_j q_j\rangle_{\text{inv.}})\nonumber\\&+\frac{1}{9}2 x  \langle\overline{q}_d\hs q_d+\overline{q}_u\hs q_d\rangle_{\text{inv.}}+\frac{1}{16}\langle \frac{\as}{\pi}G^2\rangle_{\text{inv.}}((4 L+11) x+2) \nonumber\\&-\frac{1}{56 \pi ^2 x} \Big(3x \sum_{k=u,d,s} m_k^4-m_i^3 m_j ((90 L+59) x+24)-m_i m_j^3 ((90 L+59) x+24)\nonumber\\&\bs\bs\bs-42 x m_i^2 m_j^2+m_i^4 ((45 L+7) x+12)+m_j^4 (45 L x+7 x+12) \Big)
\end{align}
\end{subappendices}

%% file: Chapters/Chap7.tex
 \chapter{Renormalization group improved determination of \texorpdfstring{$\as$}{}, \texorpdfstring{$m_c$}{}, and \texorpdfstring{$m_b$}{} from the low energy moments of heavy quark current correlators}
\label{Chapter7}

\lhead{Chapter 7. \emph{RG improved determination of \texorpdfstring{$\as$}{}, \texorpdfstring{$m_c$}{}, and \texorpdfstring{$m_b$}{} from relativistic sum rules}}
 \onehalfspacing

\section{Motivation}
We determine $\as$, $m_c$, and $m_b$ using the relativistic quarkonium sum rule and the renormalization group summed perturbation theory (RGSPT). Theoretical uncertainties, especially originating from the variation of the renormalization scale, are considerably reduced for the higher moments. Our determinations using RGSPT are also found to be stable with respect to the use of $\msbar$ quark mass for the condensate terms. We obtain $\alpha_s^{\left(n_f=5\right)}(M_z)=0.1171(7)$, $\overline{m}_c=1281.1(3.8)\hs\MeV$, and $\overline{m}_b=4174.3(9.5)\hs\MeV$.

\section{Introduction}
 The strong interaction in the standard model of particle physics describes the interactions of the quarks and gluon. These interactions are very precisely studied under QCD framework which is non-perturbative at low-energy regions and has asymptotic freedom~\cite{Gross:1973id,Politzer:1973fx} at high energies. The QCD scale, $\lqcd$, is a scale parameter that separates these energy regimes. At low energies, where momentum transfer ($q$) is of the order of $\lqcd$, chiral perturbation theory (ChPT)  and lattice QCD are powerful methods to describe strong interactions. ChPT describes the interactions of pion and kaons while the lattice QCD computations are improving over the years, and now predictions even for the bottom quark systems are also available~\cite{FlavourLatticeAveragingGroupFLAG:2021npn}. The perturbative nature of QCD at high energies ($q\gg\lqcd$) allows one to use methods like operator product expansion (OPE) to systematically calculate the various quantities as an expansion of strong coupling constant ($\as$) by evaluating the Feynman diagrams appearing at different orders of $\as$. The OPE also parametrizes the non-perturbative physics in the condensates involving the quarks and gluon fields. These condensates can be calculated using lattice QCD and ChPT~\cite{Ioffe:2005ym}, Optimized perturbation theory (OPT)~\cite{Kneur:2020bph} or using powerful tools such as QCD sum rules~\cite{Shifman:1978bx,Shifman:1978by}. For more details about the applications of the QCD sum rules, we refer to Refs.~\cite{Dominguez:2018zzi,Narison:2022paf}. \par
The effective field theories (EFT) of the strong interactions play a key role in studying systems ranging from a few $\MeV$ to several $\GeV$. For reviews, we refer to Refs.~\cite{Ananthanarayan:2023gzw,Meng:2022ozq,Casalbuoni:1996pg,Brambilla:1999xf,Brambilla:2010cs,Brambilla:2004jw,Mannel:2020ups}. Since EFTs are formulated for a very specific energy range, they are sensitive to fewer parameters than full QCD. These features allow an efficient determination of the parameters of the SM using QCD sum rules with the experimental information taken as inputs.\par 
The low-energy moments ($\mathcal{M}_n^{X}$) of the current correlators, defined in Eq.~\eqref{eq:Moments_V}, are important quantities that can be theoretically calculated. The corresponding quantity is obtained from the experimental data on the resonances, which are only available for the vector channel (V). The moments for the pseudoscalar channel (P) can not be obtained from real experiments but can be obtained using the lattice QCD simulations~\cite{HPQCD:2008kxl,McNeile:2010ji,Maezawa:2016vgv,Nakayama:2016atf,Petreczky:2019ozv,Petreczky:2020tky}. From these simulations, the dimensionless quantities such as $\mathcal{M}_0^{P}$ and the ratios of the higher moments ($\mathcal{R}_n^{P}$), defined in Eq.~\eqref{eq:Def_R}, can be reliably obtained. These results are used in the determination of $\as$, bottom quark mass ($m_b$), charm quark mass ($m_c$), and the non-perturbative quantities such as the gluon condensates~\cite{Kuhn:2001dm,Ahmady:2004er,Chetyrkin:2009fv,Narison:2010cg,Chetyrkin:2010ic,Narison:2011xe,Dehnadi:2011gc,Beneke:2014pta,Dehnadi:2015fra,Erler:2016atg,Boito:2020lyp,Boito:2019pqp,Erler:2022mzd}. Other QCD sum rules-based determinations can be found in Refs.~\cite{Peset:2018ria,Signer:2007dw,Signer:2008da,Bodenstein:2011ma,Bodenstein:2011fv,Bodenstein:2010qx,Dominguez:2014pga,Kiyo:2015ufa,Narison:2019tym} and for recent lattice QCD determinations, we refer to Refs.~\cite{FermilabLattice:2018est,Komijani:2020kst}. \par
The determination of these parameters using traditional fixed-order perturbation theory (FOPT) series from the lower moments is dominated by experimental uncertainties, but higher moments are dominated by theoretical uncertainties. Theoretical uncertainties arise when parameters such as $\as$, quark masses ($m_q$), the gluon condensate ($\langle\frac{\as}{\pi} G^2\rangle$) are taken as input, and the renormalization scale ($\mu$) is varied in a certain range. Higher moments are more sensitive to the renormalization scale dependence and therefore dominated by its uncertainties. Also, the $\msbar$ definition of the quark mass for the vector channel, when used in the non-perturbative gluon condensate terms, gives unreliable determinations for the strong coupling and quark masses. This problem is cured by using the on-shell mass taken as input~\cite{Chetyrkin:2017lif,Dehnadi:2011gc}.\par 
In this chapter, we have addressed these issues by summing the running logarithm using the RGSPT. In this scheme, the running logarithm arising from a given order is summed to all orders in closed form using the renormalization group equation (RGE). This scheme has already been found to be useful in other processes in Refs.~\cite{Abbas:2012py,Ananthanarayan:2016kll,Ananthanarayan:2022ufx,Ahmady:2002fd,Ahmady:2002pa,Ananthanarayan:2020umo,Ahmed:2015sna,Abbas:2022wnz,Chishtie:2018ipg,Abbas:2022wnz,Abbas2}.\par It should be noted that there is already an existing $m_b$ determination using RGSPT by Ahmady et al. in Ref.~\cite{Ahmady:2004er}. Special emphasis was given to the scale reduction in the $\msbar$ and its conversion to the pole mass scheme and the $1S$ scheme. Since then, there has been a significant reduction in the uncertainties in the experimental moments and the value of the strong coupling constant. Also, the first four moments to four-loop ($\as^3$), and the quark mass relations to four-loop ($\as^4$) are now available. This information can be used to further reduce the theoretical uncertainties. With these advantages in hand, we take one step further and extend its application in the $m_c$ and $\as$ determinations. \par 
In section \eqref{sec:formulas}, we briefly discuss various quantities relevant to this chapter. In section \eqref{sec:RGmom}, we discuss the renormalization group (RG) improvement of the moments using RGSPT. Since the pseudoscalar and vector channel moments are available for the charm case, we use these moments in the determination of $m_c$ and $\as$ in sections \eqref{sec:mc_det} and \eqref{sec:as_det}, respectively. In section \eqref{sec:mb_det}, $m_b$ is obtained only from the vector moments. In section \eqref{sec:summary}, we provide our final determination, and the importance of the RGSPT is discussed in detail. The supplementary material needed in this study is presented in the appendices~\eqref{app:pert_coef} and \eqref{app:RGcoefs}.\par 
Before moving to the next section, it should be noted that we use the following numerical inputs in this chapter:
\begin{equation}
\begin{aligned}
    M_c&=1.67\pm0.07\GeV\,,\\
    M_b&=4.78\pm0.06\GeV\,,\\
    \as^{\left(n_f=5\right)}(M_Z)&=0.1179\pm0.0009\,,\\
    m_c(m_c)&=1.27\pm0.02\GeV\,,\\
    m_b(m_b)&=4.18\pm0.04\GeV\,.
\end{aligned}
\label{eq:num_input}
\end{equation}
and decoupling and the running of $\as$ is performed at the $\msbar$ scheme values of the charm and bottom quark masses using REvolver~\cite{Hoang:2021fhn} and RunDec~\cite{Herren:2017osy} packages.

\section{Theoretical inputs}\label{sec:formulas}

The normalized total hadronic cross-section ($R_{\overline{q}q}$), defined as:
\begin{align}
	R_{\overline{q}q}\equiv\frac{3s}{4\pi\alpha^2}\sigma\left(e^+e^-\rightarrow q\overline{q}+X\right)\simeq\frac{\sigma\left(e^+e^-\rightarrow q\overline{q}+X\right)}{\sigma\left(e^+e^-\rightarrow \mu^+\mu^-\right)}\,,
 \label{eq:Rratio}
\end{align}
is one of the most important observable sensitive to the quark mass ($m_q$). The inverse moment for the vector channel ($\mathcal{M}_q^{V,n}$), are derived from $R_{\overline{q}q}$ as:
\begin{align}
	\mathcal{M}_q^{V,n}=\int\frac{ds}{s^{n+1}}R_{q\overline{q}}\,.
 \label{eq:Moments_V}
\end{align}

It is evident from Eq.~\eqref{eq:Moments_V} that for higher moments, significant contributions come from low energy resonances. To quantify these contributions, theoretical inputs from the non-relativistic QCD (NRQCD)~\cite{Caswell:1985ui,Bodwin:1994jh} play a crucial role. Their results can be taken as input in the determination of the $m_c$ and $m_b$ ~\cite{Signer:2007dw,Signer:2008da} and the sum rules are usually referred to as the non-relativistic sum rule. \par

 Using analyticity and unitarity, the moments are related to the coefficients of the Taylor expansion for the quark-heavy correlator evaluated around $s=0$ as:
\begin{align}
	\mathcal{M}_{n}^{V,\text{th}}=\frac{12\pi^2Q^2_q}{n!}\frac{d^n}{ds^n}\Pi^V(s)\Big|_{s=0}
\end{align}
where $Q_q$ is the electric charge, $s=\sqrt{q^2}$ is the $e^+e^-$ center of mass energy, and $\Pi_V(s)$ are the current correlators of two vector currents given by:
\begin{align}
	\left(s\hs g_{\mu\nu}-q_\mu q_\nu\right)\Pi^V(s)=-i \int dx e^{i q x}\langle0\vert T\{j_\mu(x)j_\nu(0)\}\vert0 \rangle\,,\nonumber
\end{align}
where, \begin{align}
	j_\mu=\overline{q}(x)\gamma^\mu q(x)\,.\nonumber
\end{align}
\par 
For the pseudoscalar channel, slightly different definitions are used in Ref.~\cite{Dehnadi:2015fra} and which is also adopted in this study. The pseudoscalar current correlator is defined as
\begin{align}
    \Pi^{P}(s)&\equiv i\int d\hs x e^{i\hs q\hs x}\langle 0\vert T\lbrace j_P(x)\hs j_P(0)\rbrace\vert 0\rangle\,,
    \label{eq:vacP}
\end{align}
where \begin{align}
    j_P=2\hs i\hs m_q\hs \overline{q}(x)\gamma^5 q(x)\,,
\end{align}
and the double subtracted polarization function is obtained from Eq.~\eqref{eq:vacP} as:
\begin{align}
    P(s)=\frac{1}{s^2}\left(\Pi^{P}(s)-\Pi^{P}(0)-s\hs\left[ \frac{d}{d\hs s}\Pi^{P} (s)\right]_{s=0}\right)\,,
\end{align}
from which the moments are obtained as:
\begin{align}
    \mathcal{M}^{P,\text{th}}_n(s)=\frac{12\pi^2 Q_q^2}{n!} \frac{d^n}{d\hs s^n}P(s)\Big|_{s=0}\,.
\end{align}
Theoretical moments are calculated using the OPE and have contributions from purely perturbative ($\mathcal{M}_n^{X,\text{pert}}$) as well as non-perturbative ($\mathcal{M}_n^{X,\text{n.p}}$) origin. Therefore, we can write the theoretical moments as follows:
 \begin{align}
\mathcal{M}_{n}^{X,\text{th}}=\mathcal{M}_n^{X,\text{pert.}}+\mathcal{M}_n^{X,\text{n.p.}}\,.
\label{eq:Def_MX}
 \end{align}
 The fixed order perturbative series for $\mathcal{M}_n^{X,\text{pert.}}$ have the following form:
\begin{align}
    \mathcal{M}_n^{X,\text{pert}}=m_q^{-2n}\sum_{i=0}T_{i,j}^X x^i L^j
    \label{eq:mom_fopt}
\end{align}
where $m_q\equiv m_q(\mu)$, $x\equiv\as(\mu)/\pi$ and $L\equiv \log(\mu^2/q^2)$. \par The $T_{i,0}^{X,Y}$ are RG inaccessible terms calculated using the perturbation theory by evaluating the Feynman diagrams appearing in a given order. Their numerical values are presented in appendix~\eqref{app:pert_coef}. Other $T_{i,j}^{X,Y}$ coefficients can be obtained using the RGE and are known as RG-accessible terms. 
The two-loop correction to $\mathcal{M}_n^{X,\text{pert}}$ are calculated in Ref.~\cite{Kallen:1955fb}, three-loops in Refs.~\cite{Chetyrkin:1995ii,Chetyrkin:1996cf,Boughezal:2006uu,Czakon:2007qi,Maier:2007yn}, the first four moments at four-loop (or $\as^{3}$)  from Refs.~\cite{Hoang:2008qy,Maier:2009fz}. Predictions for higher moments using Pad\'e approximants can be found in Ref.~\cite{Kiyo:2009gb} and using the analytic reconstruction method in Refs.~\cite{Greynat:2010kx,Greynat:2011zp}. A large-$\beta_0$ renormalon-based analysis for the low energy moments of the current correlators can be found in Ref.~\cite{Boito:2021wbj}.\par
 The $\mathcal{M}_n^{X,\text{n.p}}$ include the contributions from the condensate terms and has the following form:
\begin{equation}
\mathcal{M}_n^{X,\text{n.p.}}=\frac{1}{\left(2\hs m_q\right)^{4n+4}}\Big\langle\frac{\as}{\pi}G^2\Big\rangle_{\text{RGI}}\left(T^{X,\text{n.p.}}_{0,0}+x(m_q) T^{X,\text{n.p.}}_{1,0}\right)+\ord{x^2}\,.
    \label{eq:def_cond}
\end{equation}
 where, $T^{X,\text{n.p.}}_{i,0}$ are the perturbative correction as prefactors to the gluon condensate and are known to NLO~\cite{Broadhurst:1994qj} and can be found in the appendix~\eqref{app:pert_coef}. For the RG invariant  gluon condensate, we use the following numerical value~\cite{Ioffe:2005ym}:
 \begin{equation}
     \langle\frac{\as}{\pi}G^2\Big\rangle_{\text{RGI}}=0.006\pm0.012\GeV^4\,.
 \end{equation}
In addition, we also need quark mass relations to convert it from the $\msbar$ scheme to the on-shell scheme. These relations are now known to four-loops~\cite{Tarrach:1980up,Gray:1990yh,Fleischer:1998dw,Chetyrkin:1999qi,Marquard:2015qpa,Marquard:2016dcn}. The one-loop relation relevant for this study is given by:
\begin{equation}
    m_q(\mu)=M_q \left(1-x(\mu)\left(\frac{4}{3}+\log\left(\frac{\mu^2}{M_q^2}\right)\right)\right)+\ord{x^2}\,,
\end{equation}
which will be used in Eq.~\eqref{eq:def_cond} for the quark condensate terms.\par
From theoretical moments, defined in Eq.~\eqref{eq:Def_MX}, the ratio of the moments ($\mathcal{R}^{X}_n$) can be obtained as:
\begin{equation}
\mathcal{R}^{X}_n\equiv\frac{\left(\mathcal{M}^{X}_n\right)^{\frac{1}{n}}}{\left(\mathcal{M}^{X}_{n+1}\right)^{\frac{1}{n+1}}}\,,
\label{eq:Def_R}
\end{equation}
which are more sensitive to the $\as$ and less sensitive to the quark masses. The mass dependence arises only from the running logarithms present in the perturbative expansion. This quantity is very useful in the determination of the $\as$.\par With an introduction to these RG invariant quantities, we are in a position to discuss their RG improvement using RGSPT in the next section.

\section{RG Improvement of Moments} 
\label{sec:RGmom}
The FOPT expression for the $\mathcal{M}_n^{X,\text{pert.}}$, in Eq.~\eqref{eq:mom_fopt}, is a RG invariant perturbative expansion quark mass and $\as$. The evolution of the quark masses and $\as$ is dictated by their RGE. There are also some studies where $m_q(\mu)$ is expanded in the $\msbar$ scheme, and an extra scale ($\mu_m$) is introduced whose effects appear in $\ordas{2}$ in running logarithm. Although this procedure is very general, independent scale variations of $\left(\mu,\mu_m\right)$ give more renormalization scale uncertainty as the RG invariance of the moments $\mathcal{M}_n^{X,\text{pert.}}$ is broken in the case of the finite order results. Since this study is only focused on the RG improvement, we restrict ourselves to the single renormalization scale ($\mu$), which is known as a correlated choice of scale approach.\par  
We use the formalism developed in chapter~\eqref{Chapter2} to obtain a closed form summed expression by rewriting the perturbative series in Eq.~\eqref{eq:mom_fopt} as follows:
    \begin{equation}
        \mathcal{M}_n^{X,\Sigma}= m_q^{-2\hs n}\sum_{i=0}x^i\hspace{.4mm}S_{i}(x \hspace{.4 mm} L)\,,
        \label{eq:ser_summed}
    \end{equation}
    where the $S_{i}\left(x L\right)$ are the RG summed coefficients, defined in Eq.~\eqref{eq:summed_coefs},  are given by:
    \begin{equation}
        S_{i}\left(x\hspace{.4mm}L\right)=\sum_{n=i}^{\infty} T^X_{n,n-i}  (x\hspace{.4mm}L)^{n-i}\,.
        \label{eq:summed_coefs_asmcmb}
    \end{equation}
    Since, $\mathcal{M}_n^{X,\text{pert.}}$ is an observable, Eq.~\eqref{eq:mom_fopt} has a homogeneous RGE given by:
    \begin{equation}
        \mu^2 \frac{d}{d\mu^2}\mathcal{M}_n^X=\left(\beta(x) \partial_x+ \gamma_m(x) \partial_m+\partial_L\right)\mathcal{M}_n^X=0\,,
    \end{equation}
    where $\beta(x)$ and $\gamma_m$ are the QCD beta function~\cite{vanRitbergen:1997va,Gross:1973id,Caswell:1974gg, Jones:1974mm,Tarasov:1980au,Larin:1993tp,Czakon:2004bu,Luthe:2016ima,Baikov:2016tgj,Herzog:2017ohr} and quark mass anomalous dimension~\cite{Tarrach:1980up,Tarasov:1982plg,Larin:1993tq,Vermaseren:1997fq,Chetyrkin:1997dh,Baikov:2014qja,Luthe:2016xec,Baikov:2017ujl} given by:
     \begin{align}
      \beta(x)&\equiv \mu^2\frac{d}{d\mu^2}x(\mu)=-\sum_i \beta_i x^{i+2} \,, \label{eq:beta_function}\\
 \gamma_m&\equiv \mu^2\frac{d}{d\mu^2}m_q(\mu)=-m_q(\mu)\sum_i \gamma_i\hs x^{i+1} \,. \label{eq:mass_anom}
    \end{align}
     Now, we can follow the steps described in Ref.~\cite{Ahmady:2002fd} by collecting coefficients corresponding to summed coefficients defined in Eq.~\eqref{eq:summed_coefs_asmcmb}. This process results in a set of coupled differential equations for $S_{i}( x\hs L)$, which can be written in a compact form as:
    \begin{align}
        \sum _{i=0}^k \bigg[\beta _i &(\delta_{i,0}+w-1)  S_{k-i}'(w)+S_{k-i}(w) \left(-2 n \gamma_i+\beta _i (-i+k)\right)\bigg]=0\,
        \label{eq:summed_de}
    \end{align}
   where, $w\equiv 1-\beta_0\hs x\hs L$. The solutions for the above differential equation are presented in the appendix~\eqref{app:RGcoefs}. From these solutions, we can obtain various $\mathcal{M}_n^{X,\Sigma}$. It should be noted that the corresponding expression in the on-shell scheme is obtained by setting quark mass anomalous dimension $\gamma_i=0$.\par

\begin{figure}[ht]	
\centering
		\includegraphics[width=.48\textwidth]{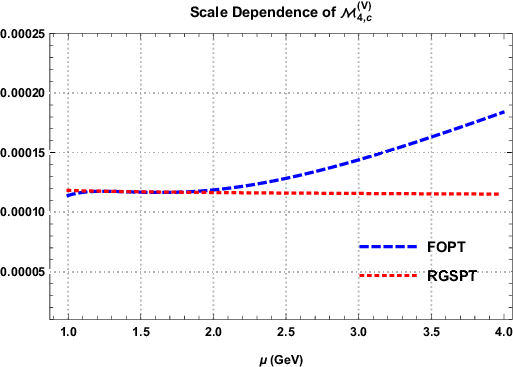}
		\includegraphics[width=.48\textwidth]{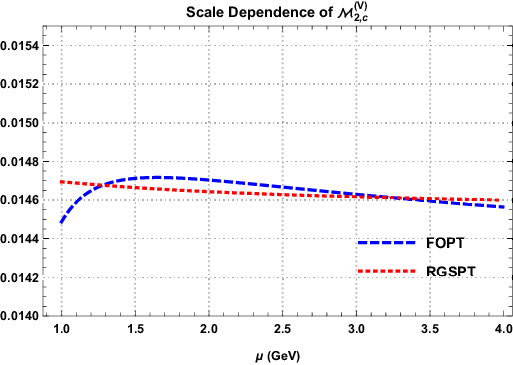}
	\includegraphics[width=.48\linewidth]{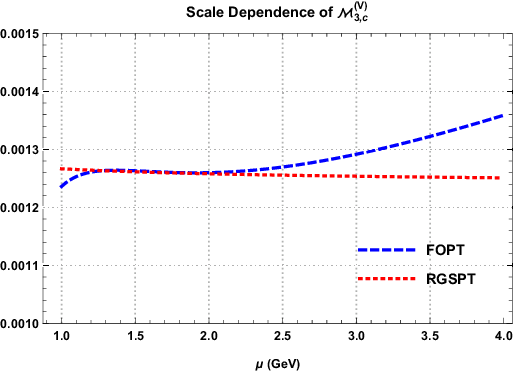}
	\includegraphics[width=.48\linewidth]{Pictures/mc_mb_as//Mc4.eps}
\caption{\label{fig:MomV_c} Renormalization scale dependence of the first four  vector moments for the charm quark.}
\end{figure}

\begin{figure}[ht]\centering
		\includegraphics[width=.48\linewidth]{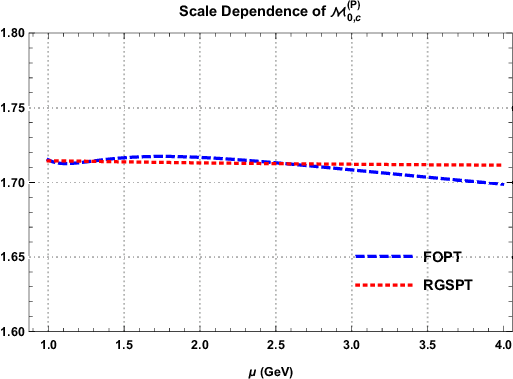}
		\includegraphics[width=.48\linewidth]{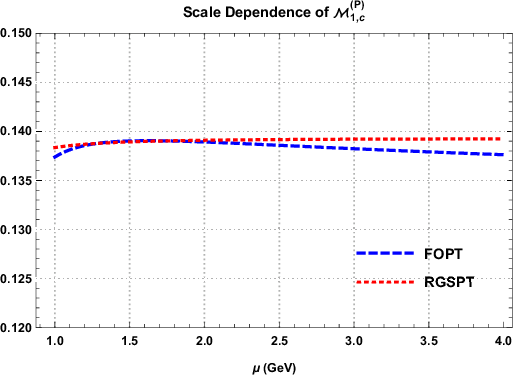}
	\includegraphics[width=.48\linewidth]{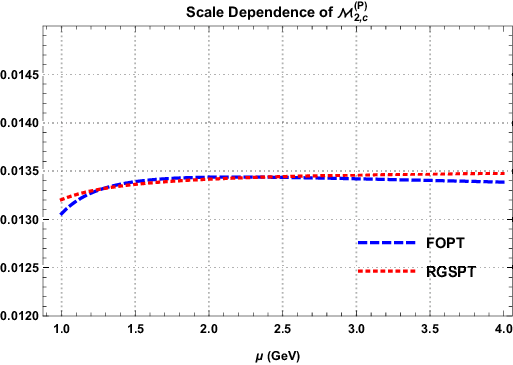}
	\includegraphics[width=.48 \textwidth]{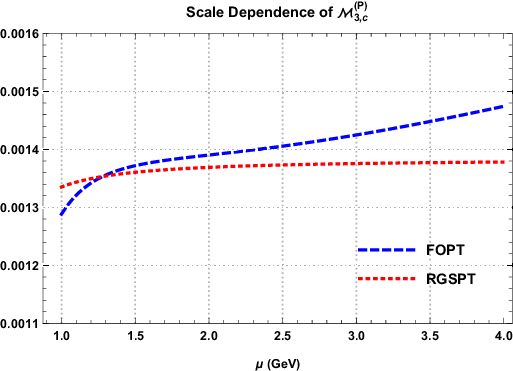}
\caption{\label{fig:MomP_c} Renormalization scale dependence of the first four  pseudoscalar moments for the charm quark.}
\end{figure}

\begin{figure}[ht]
\centering
		\includegraphics[width=.48\linewidth]{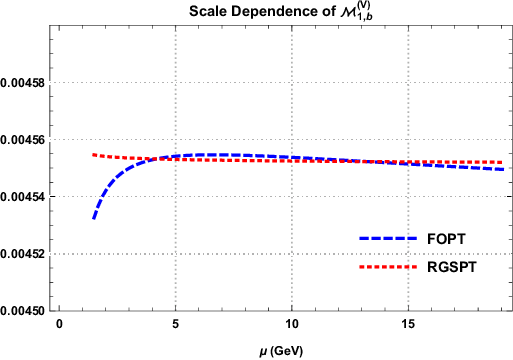}
		\includegraphics[width=.48\linewidth]{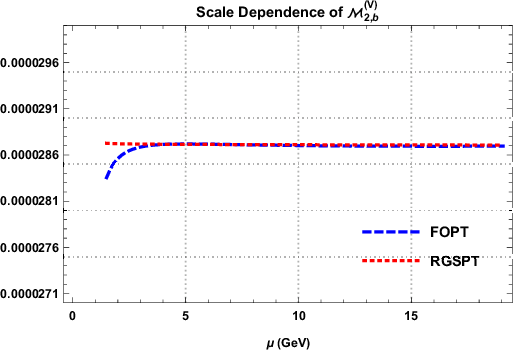}
	\includegraphics[width=.48\linewidth]{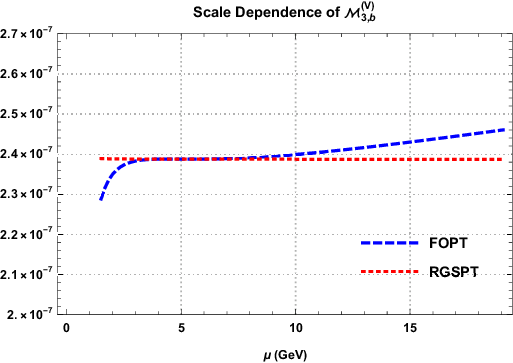}
	\includegraphics[width=.48\linewidth]{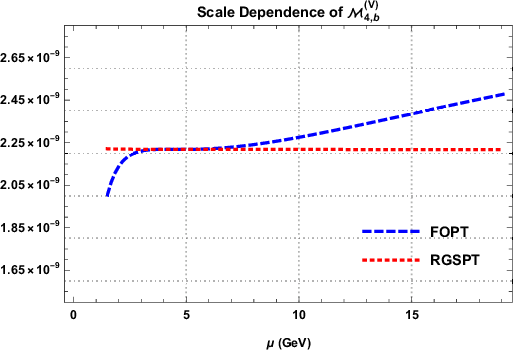}
\caption{\label{fig:Mom_b} Renormalization scale dependence of the first four  vector moments for the bottom quark.}
\end{figure}

Once RG improved perturbative series is obtained for different $\mathcal{M}^{X}_n$, we can study their scale dependence. For the charm moments, we take $\as^{\left(n_f=4\right)}(3\GeV)=0.2230$~and $m_c(3\GeV)=993.9 \MeV$. For the bottom moments, we take $\as^{\left(n_f=5\right)}(10\GeV)=0.1780$ and $m_b(10\GeV)=3619.4\MeV$. These values are obtained from Eq.~\eqref{eq:num_input} using the REvolver package. The scale dependence of the first four moments for the vector and pseudoscalar channel for the charm case can be found in Fig.~\eqref{fig:MomV_c} and Fig.~\eqref{fig:MomP_c}, respectively. For the bottom quark case, we only used vector moments, and the scale dependence can be found in Fig.~\eqref{fig:Mom_b}. It is evident from these figures that the RGSPT has better control of the scale variations compared to the FOPT. For the vector moments, the third and fourth moments in the FOPT scheme are very sensitive to scale variations and contribute to a large theoretical uncertainty even though their experimental values are known more precisely. With these advantages in hand, we have used FOPT and RGSPT in the determinations of the $\as$, $m_c$, and $m_b$ in the next sections. 
\section{Charm mass determination}
Charm quark is very interesting for the low as well as high energy regime of the QCD. It is not heavy enough that heavy quark effective theory can be used for precise prediction nor close enough $\lqcd$ such that formalism like ChPT can be applied. Various technical issues arise when it is used in quarkonium physics~\cite{Ananthanarayan:2023gzw,Meng:2022ozq,Casalbuoni:1996pg,Brambilla:1999xf,Brambilla:2010cs,Brambilla:2004jw,Mannel:2020ups}. However, Lattice QCD methods have significantly developed over the years, and now precise predictions for the charm quarks are now available in the literature. We use experimental inputs as well as the results from the lattice QCD on the moments in the extraction of the $m_c$ in this section.\par In this section, we determine the $m_c$ from the vector moments and pseudoscalar moments using FOPT and RGSPT. As observed in section~\eqref{sec:RGmom}, these determinations from higher moments in the vector channel are very sensitive to scale variations, and the $\msbar$ definition in the condensate terms also causes trouble. The determination from the pseudoscalar channel does not suffer very much from these issues. 
\label{sec:mc_det}
\subsection{\texorpdfstring{$m_c$}{} determination using experimental inputs for the vector channel.}
For the vector channel, we use the experimental moments provided in Refs.~\cite{Chetyrkin:2017lif,Dehnadi:2011gc} in the $m_c$ determination. These moments are presented in Table~\eqref{tab:momentsV_charm}, and our results for the $m_c$ determination in the $\msbar$ scheme are presented in Table~\eqref{tab:mc_V_ms}.  A significant condensate contributions in the $m_c$ determination using FOPT is found for $\mathcal{M}^{V}_3$ and $\mathcal{M}^{V}_4$. This is caused by the use of quark mass in the $\msbar$ scheme in the condensate terms for FOPT, as pointed out in Refs.~\cite{Chetyrkin:2010ic,Dehnadi:2011gc}, and the pole quark mass in the condensate terms is used to avoid such large contributions. However, this is not the case for the RGSPT determinations, which are also stable with respect to the scale variations. It should be noted that this issue also appears in RGSPT if the renormalization scale is chosen such the RG logarithms vanish (see Table~\eqref{tab:as_V_ms} and \eqref{tab:as_P_ms_charm} for $\as$-determination). For this particular choice of renormalization scale, RGSPT reduces to the FOPT scheme. When we use the pole mass in the condensate, the $m_c$ determination using FOPT and RGSPT are presented in Table~\eqref{tab:mc_V_pole}. In this case, the situation is a little bit improved for FOPT, but scale dependence is still the major source of theoretical uncertainties. 
\begin{table}[H]
    \centering
    \begin{tabular}{|c|c|c|}
    \hline
        \textbf{Moments} &  \textbf{Ref.~\cite{Dehnadi:2011gc}}& \textbf{Ref.~\cite{Chetyrkin:2017lif}}\\\hline\hline
        $\mathcal{M}^{V,\text{exp.}}_1$&$2.121\pm0.036$ &$2.154\pm0.023$\\ \hline 
        $\mathcal{M}^{V,\text{exp.}}_2$&$1.478\pm0.028$ &$1.490\pm0.017$\\\hline
        $\mathcal{M}^{V,\text{exp.}}_3$&$1.302\pm0.027$ &$1.308\pm0.016$\\ \hline
        $\mathcal{M}^{V,\text{exp.}}_4$&$1.243\pm0.028$ &$1.248\pm0.016$\\ \hline
    \end{tabular}
    \caption{Moments for the vector channel for the charm case. These moments are in the units of $10^{-n}\GeV^{-2n}$.}
    \label{tab:momentsV_charm}
\end{table}
\begin{table}[H]
    \centering
       \begin{adjustbox}{max width=\textwidth}
    \begin{tabular}{||c|c|c|c|c|c|c|c|c|c|c|c|c|c||}\hline\hline
    \text{}&\text{} &\multicolumn{6}{c|}{\textbf{FOPT}}&\multicolumn{6}{c|}{\textbf{RGSPT}}\\\cline{3-14}
    \textbf{Sources}&\textbf{Moments}&\text{}&\multicolumn{4}{c|}{Theo. Unc.}& \text{}&\text{}&\multicolumn{4}{c|}{Theo. Unc.}& \text{} \\
\cline{4-7}\cline{10-13}
\text{}&\text{}&$m_c(3\GeV)$&$\as$&$\mu$&n.p.&total&Exp. Unc.&$m_c(3\GeV)$&$\as$&$\mu$&n.p.&total&Exp. Unc.
\\\hline\hline
        \multirow{3}{4em}{ Ref.~\cite{Dehnadi:2011gc}}
               &$\mathcal{M}^{V}_1$&$1005.4( 13.9)$&3.2&7.6&0.2&8.3&11.2&$1000.2( 12.3)$&3.7&1.9&2.3&4.8&11.3\\\cline{2-14}
        \text{}&$\mathcal{M}^{V}_2$&$997.2( 19.8)$&4.7&11.3&14.4&18.9&6.1&$988.5( 9.2)$&5.4&1.6&3.5&6.6&6.3\\ \cline{2-14}
         \text{}&$\mathcal{M}^{V}_3$&$1022.1( 127.8)$&3.4&41.8&120.6&127.7&4.0&$983.4( 9.2)$&6.7&1.4&3.9&7.8&4.9\\ \cline{2-14}
         \text{}&$\mathcal{M}^{V}_4$&$1077.3( 113.6)$&1.0&100.5&52.9&113.6&2.8&$980.5( 8.9)$&7.7&0.9&1.8&8.0&3.9\\\hline\hline
       \multirow{3}{4em}{Ref.~\cite{Chetyrkin:2017lif}}
                &$\mathcal{M}^{V}_1$&$995.4( 10.8)$&3.3&7.6&0.0&8.3&6.9&$990.1( 8.5)$&3.6&2.0&2.3&4.8&7.0\\ \cline{2-14}
         \text{}&$\mathcal{M}^{V}_2$&$994.6( 19.5)$&4.7&11.3&14.7&19.1&3.7&$985.8( 7.7)$&5.4&1.6&3.5&6.6&3.8\\ \cline{2-14}
         \text{}&$\mathcal{M}^{V}_3$&$1021.3( 126.5)$&3.4&41.9&126.5&133.3&2.3&$982.3( 8.3)$&6.7&1.4&3.9&7.8&2.8\\ \cline{2-14}
         \text{}&$\mathcal{M}^{V}_4$&$1076.8( 113.8)$&1.0&100.7&52.9&113.8&1.6&$979.8( 8.3)$&7.7&0.9&1.8&8.0&2.2\\\hline\hline
    \end{tabular}
\end{adjustbox}
    \caption{$m_c$ determinations using FOPT and RGSPT in $\msbar$ scheme using experimental inputs from Table~\eqref{tab:momentsV_charm}. Results are in the units of $\MeV$ and the scale dependence is calculated for the energy range $\mu\in[1,4]\hs\GeV$.}
    \label{tab:mc_V_ms}
\end{table}
%
\begin{table}[H]
    \centering
       \begin{adjustbox}{max width=\textwidth}
    \begin{tabular}{||c|c|c|c|c|c|c|c|c|c|c|c|c|c||}\hline\hline
    \text{}&\text{} &\multicolumn{6}{c|}{\textbf{FOPT}}&\multicolumn{6}{c|}{\textbf{RGSPT}}\\\cline{3-14}
    \textbf{Sources}&\textbf{Moments}&\text{}&\multicolumn{4}{c|}{Theo. Unc.}& \text{}&\text{}&\multicolumn{4}{c|}{Theo. Unc.}& \text{} \\
\cline{4-7}\cline{10-13}
\text{}&\text{}&$m_c(3\GeV)$&$\as$&$\mu$&n.p.&total&Exp. Unc.&$m_c(3\GeV)$&$\as$&$\mu$&n.p.&total&Exp. Unc.
\\\hline\hline
        \multirow{3}{4em}{ Ref.~\cite{Dehnadi:2011gc}}
               &$\mathcal{M}^{V}_1$&$1004.8(13.7)$&3.3&7.1&1.4&7.9&11.2&$1000.9( 12.1)$&3.7&1.9&0.9&4.3&11.3\\\cline{2-14}
        \text{}&$\mathcal{M}^{V}_2$&$989.4( 9.2)$&5.4&3.5&2.0&6.7&6.3&$989.6 (8.6)$&5.5&1.8&1.0&5.8&6.3\\ \cline{2-14}
         \text{}&$\mathcal{M}^{V}_3$&$990.9( 13.5)$&5.4&10.9&2.5&12.6&4.7&$984.8( 8.7)$&6.8&2.4&1.0&7.3&4.8\\ \cline{2-14}
         \text{}&$\mathcal{M}^{V}_4$&$1014.5( 37.7)$&3.8&37.2&2.9&37.5&3.4&$980.9( 9.8)$&8.0&3.9&0.9&9.0&3.9\\\hline\hline
       \multirow{3}{4em}{Ref.~\cite{Chetyrkin:2017lif}}
                &$\mathcal{M}^{V}_1$&$995.1( 10.3)$&3.3&6.8&0.7&7.6&7.0&$990.8( 8.2)$&3.8&2.0&0.9&4.3&7.0\\ \cline{2-14}
         \text{}&$\mathcal{M}^{V}_2$&$987.3( 7.4)$&5.4&3.2&0.8&6.3&3.8&$986.9 (7.0)$&5.5&1.8&1.0&5.9&3.8\\ \cline{2-14}
         \text{}&$\mathcal{M}^{V}_3$&$990.7( 12.7)$&5.8&10.9&2.7&12.4&2.7&$983.7( 7.8)$&6.8&2.4&1.1&7.3&2.8\\ \cline{2-14}
         \text{}&$\mathcal{M}^{V}_4$&$1014.9( 37.0)$&3.8&36.7&0.8&36.9&2.0&$980.2( 9.3)$&8.1&3.9&1.0&9.0&2.2\\\hline\hline
    \end{tabular}
\end{adjustbox}
    \caption{$m_c$ determinations using FOPT and RGSPT  using experimental inputs from Table~\eqref{tab:momentsV_charm}. The pole mass of the charm quark is used as input in the non-perturbative condensate terms. Results are in the units of $\MeV$ and the scale dependence is calculated for the energy range $\mu\in[1,4]\hs\GeV$.}
    \label{tab:mc_V_pole}
\end{table}
\subsection{\texorpdfstring{$m_c$}{} determination using the lattice QCD inputs.}
The moments for the vector currents are obtained using the experimental data on hadrons from the $e^+ e^-$ collision. However, this is not the case for the pseudoscalar channel, which is not realized in nature but can be computed using the lattice QCD simulations. We use the results for the reduced moments ($R_n$), a dimensionless quantity is reliably calculable from the lattice QCD in the Refs.~\cite{HPQCD:2008kxl,McNeile:2010ji,Maezawa:2016vgv,Petreczky:2019ozv,Petreczky:2020tky}. The regular moments calculated in the perturbative QCD are related to these reduced moments by the following relations~\cite{Boito:2020lyp,Dehnadi:2015fra}:
\begin{equation}
    \mathcal{M}_n^P=T^P_{n,0}\left(\frac{R_{2n+4}}{m_{\eta_c}}\right)^{2n}\,,
    \label{eq:reduced_mom}
\end{equation}
and the results are collected in Table~\eqref{tab:momP_charm}. It should be noted that the reduced moments provided in the Refs.~\cite{HPQCD:2008kxl,McNeile:2010ji} are converted to regular moments using the current PDG~\cite{ParticleDataGroup:2022pth} value $m_{\eta_c}=2.9839\pm0.0004$ for the $\eta_c$-meson.\par
The $m_c$ determination using FOPT and RGSPT from the lattice QCD moments are presented in Table~\eqref{tab:mc_P_ms}. These determinations do not suffer issues from the condensate terms; the determinations from the first two moments are precise and close. The RGSPT determinations are even better for all three moments. Since the results from the pseudoscalar channel in the $\msbar$ scheme are good enough, we do not find it necessary to give our determinations using on-shell mass as input for the condensate terms. 
\begin{table}[ht]
    \centering
       \begin{adjustbox}{max width=\textwidth}
    \begin{tabular}{|c|c|c|c|c|c|}\hline
         \textbf{Moments}& \textbf{Ref.~\cite{HPQCD:2008kxl}} &\textbf{Ref.~ \cite{McNeile:2010ji}} &\textbf{Ref.~ \cite{Maezawa:2016vgv}} & \textbf{Ref.~\cite{Petreczky:2019ozv}}& \textbf{Ref.~\cite{Petreczky:2020tky}} \\\hline\hline
         $\mathcal{M}^{P}_1$&$1.404\pm0.019$&$1.395\pm0.005$&$1.385\pm0.007$&$1.386\pm0.005$&$1.387\pm0.004$\\\hline
         $\mathcal{M}^{P}_2$&$1.359\pm0.041$&$1.365\pm0.012$&$1.345\pm0.032$&$1.349\pm0.012$&$1.344\pm0.010$\\\hline
         $\mathcal{M}^{P}_3$&$1.425\pm0.059$&$1.415\pm0.010$&$1.406\pm0.048$&$1.461\pm0.050$&$1.395\pm0.022$\\\hline
    \end{tabular}
\end{adjustbox}
    \caption{Pseudoscalar moment calculated from lattice QCD for the charm case. These $\mathcal{M}^{P}_n$ are in the units of $10^{-n}\GeV^{-2n}$.}
    \label{tab:momP_charm}
\end{table}
\begin{table}[ht]
    \centering
       \begin{adjustbox}{max width=\textwidth}
    \begin{tabular}{||c|c|c|c|c|c|c|c|c|c|c|c|c|c||}\hline\hline
    \text{}&\text{} &\multicolumn{6}{c|}{\textbf{FOPT}}&\multicolumn{6}{c|}{\textbf{RGSPT}}\\\cline{3-14}
    \textbf{Sources}&\textbf{Moments}&\text{}&\multicolumn{4}{c|}{Theo. Unc.}& \text{}&\text{}&\multicolumn{4}{c|}{Theo. Unc.}& \text{} \\
\cline{4-7}\cline{10-13}
\text{}&\text{}&$m_c(3\GeV)$&$\as$&$\mu$&n.p.&total&Exp. Unc.&$m_c(3\GeV)$&$\as$&$\mu$&n.p.&total&Exp. Unc.
\\\hline\hline
          \multirow{3}{4em}{ Ref.~\cite{HPQCD:2008kxl}}
               &$\mathcal{M}^{P}_1$&$983.6( 10.0)$&1.1&5.0&2.4&5.7&8.2&$989.3( 9.0)$&1.4&3.5&0.7&3.8&8.1\\\cline{2-14}
        \text{}&$\mathcal{M}^{P}_2$&$988.3 (12.5)$&1.7&6.8&3.6&12.5&9.8&$990.5( 11.4)$&1.5&5.8&0.9&6.0&9.7\\ \cline{2-14}
         \text{}&$\mathcal{M}^{P}_3$&$998.9( 29.9)$&2.2&26.6&10.5&28.6&8.5&$985.4( 11.6)$&3.2&6.4&2.0&7.4&9.0\\\hline\hline
          \multirow{3}{4em}{ Ref.~\cite{McNeile:2010ji}}
               &$\mathcal{M}^{P}_1$&$987.1( 6.1)$&1.1&5.0&2.3&5.5&2.4&$992.8( 4.5)$&1.4&3.5&3.5&0.7&2.3\\\cline{2-14}
        \text{}&$\mathcal{M}^{P}_2$&$986.9( 8.3)$&1.7&6.7&3.6&7.8&2.7&$989.1( 6.6)$&1.5&5.8&0.9&6.0&2.7\\ \cline{2-14}
         \text{}&$\mathcal{M}^{P}_3$&$1000.2( 28.6)$&2.2&26.5&10.4&28.6&1.4&$986.9( 7.6)$&3.2&6.4&2.0&7.4&1.5\\\hline\hline
          \multirow{3}{4em}{ Ref.~\cite{Maezawa:2016vgv} }
               &$\mathcal{M}^{P}_1$&$991.7( 6.4)$&1.1&4.9&2.2&5.5&3.2&$997.3( 5.0)$&1.4&3.5&0.7&3.8&3.2\\\cline{2-14}
        \text{}&$\mathcal{M}^{P}_2$&$991.5 (10.9)$&1.6&6.8&3.5&7.9&7.5&$993.6 (9.6)$&1.5&5.8&0.9&6.1&7.5\\ \cline{2-14}
         \text{}&$\mathcal{M}^{P}_3$&$1001.5( 29.4)$&2.2&26.5&10.3&28.5&7.1&$988.2( 10.5)$&3.2&6.4&2.0&7.4&7.5\\\hline\hline
        \multirow{3}{4em}{ Ref.~\cite{Petreczky:2019ozv}}
               &$\mathcal{M}^{P}_1$&$991.2( 6.0)$&1.1&4.9&2.3&5.5&2.4&$996.8( 4.5)$&1.4&3.5&0.7&3.8&2.3\\\cline{2-14}
        \text{}&$\mathcal{M}^{P}_2$&$990.5( 8.3)$&1.6&6.8&3.5&7.9&2.8&$992.7(6.6)$&1.5&5.8&0.9&6.0&2.7\\ \cline{2-14}
         \text{}&$\mathcal{M}^{P}_3$&$993.8( 29.7)$&2.2&26.7&10.9&28.9&7.0&$980.0( 10.5)$&3.3&6.3&2.0&7.4&7.4\\\hline\hline
       \multirow{3}{4em}{Ref.~\cite{Petreczky:2020tky}}
                &$\mathcal{M}^{P}_1$&$990.6( 5.9)$&1.1&4.9&2.3&5.5&1.9&$996.2( 4.2)$&1.4&3.5&0.7&3.8&1.9\\ \cline{2-14}
         \text{}&$\mathcal{M}^{P}_2$&$991.6( 8.2)$&1.6&3.5&7.9&2.3&9.8&$993.7(6.5)$&1.5&5.8&0.9&6.1&2.3\\ \cline{2-14}
         \text{}&$\mathcal{M}^{P}_3$&$1003.2( 28.6)$&2.2&26.5&10.2&28.6&3.2&$989.9( 8.2)$&3.2&6.4&2.0&7.4&3.4\\\hline\hline
    \end{tabular}
\end{adjustbox}
    \caption{$m_c$ determinations using FOPT and RGSPT  using experimental inputs from Table~\eqref{tab:momP_charm}. The pole mass of the charm quark is used as input in the non-perturbative condensate terms. Results are in the units of $\MeV$, and the scale dependence is calculated for the energy range $\mu\in[1,4]\hs\GeV$.}
    \label{tab:mc_P_ms}
\end{table}

\section{Bottom quark mass determination} \label{sec:mb_det}
The bottom quark is heavy, and to the best of our knowledge, there are no direct lattice data on the moments in the case of the bottom quark. We use the experimental information on the moments provided in Refs.~\cite{Kuhn:2007vp,Chetyrkin:2009fv,Dehnadi:2015fra} for the vector channel in the bottom quark mass determinations. These moments from different sources are tabulated in Table~\eqref{tab:momV_bottom}. The $m_b$ determination using FOPT and RGSPT in the $\msbar$ scheme are presented in Table~\eqref{tab:mb_ms_cond_ms}. Similar to the charm case, FOPT again suffers from large non-perturbative contributions and scale dependence. The uncertainty from the condensate term is alone $~70\%$ of the central value for the third and fourth moment. When bottom quark mass in the on-shell scheme is used as input, this issue is resolved, and our $m_b$ determinations are presented in Table~\eqref{tab:mb_ms_cond_pole}.

\begin{table}[ht]
    \centering
    \begin{tabular}{||c|c|c|c|c||}\hline\hline
         \textbf{Moments} & \textbf{Ref.~\cite{Dehnadi:2015fra}} &\textbf{Ref.~ \cite{Chetyrkin:2009fv}} &\textbf{Ref.~ \cite{Kuhn:2007vp}}\\\hline\hline
         $\mathcal{M}^{V,\text{exp.}}_1$&$4.526\pm0.112$&$4.592\pm0.031$&$4.601\pm0.043$\\\hline
         $\mathcal{M}^{V,\text{exp.}}_2$&$2.834\pm0.052$&$2.872\pm0.028$&$2.881\pm0.037$\\\hline
         $\mathcal{M}^{V,\text{exp.}}_3$&$2.338\pm0.036$&$2.362\pm0.026$&$2.370\pm0.034$\\\hline
         $\mathcal{M}^{V,\text{exp.}}_4$&$2.154\pm0.030$&$2.170\pm0.026$&$2.178\pm0.032$\\\hline
    \end{tabular}
    \caption{Vector moments from different sources used as input for the bottom quark mass determination. These $\mathcal{M}^{V}_n$ are in the units of $10^{-\left(2n+1\right)}\GeV^{-2n}$.}
    \label{tab:momV_bottom}
\end{table}

\begin{table}[ht]
    \centering
       \begin{adjustbox}{max width=\textwidth}
    \begin{tabular}{||c|c|c|c|c|c|c|c|c|c|c|c|c|c||}\hline\hline
    \text{}&\text{} &\multicolumn{6}{c|}{\textbf{FOPT}}&\multicolumn{6}{c|}{\textbf{RGSPT}}\\\cline{3-14}
    \textbf{Sources}&\textbf{Moments}&\text{}&\multicolumn{4}{c|}{Theo. Unc.}& \text{}&\text{}&\multicolumn{4}{c|}{Theo. Unc.}& \text{} \\
\cline{4-7}\cline{10-13}
\text{}&\text{}&$m_b(10\GeV)$&$\as$&$\mu$&n.p.&total&Exp. Unc.&$m_b(10\hs\GeV)$&$\as$&$\mu$&n.p.&total&Exp. Unc.
\\\hline\hline
         \multirow{4}{4em}{ Ref.~\cite{Dehnadi:2015fra}}
          &$\mathcal{M}^{V}_1$&$3632.2( 53.4)$&3.2&5.4&0.0&6.3&53.0&$3631.6(53.1)$&3.2&0.7&0.0&3.3&53.0\\\cline{2-14}
        \text{}&$\mathcal{M}^{V}_2$&$3632.8( 21.0)$&5.2&5.3&0.0&7.4&19.7&$3633.1( 20.4)$&5.2&0.4&0.1&5.2&19.7\\ \cline{2-14}
         \text{}&$\mathcal{M}^{V}_3$&$3637.8( 2746.7)$&6.2&20.2&2746.6&2746.7&11.0&$3634.1( 12.8)$&6.5&0.3&0.2&6.5&11.1\\\cline{2-14}
         \text{}&$\mathcal{M}^{V}_4$&$3648.8( 2491.7)$&6.5&49.9&2491.2&2491.7&7.4&$3635.0( 10.5)$&7.4&0.4&0.2&7.4&7.5\\
         \hline\hline
        \multirow{4}{4em}{ Ref.~\cite{Chetyrkin:2009fv}}
               &$\mathcal{M}^{V}_1$&$3601.9(14.1)$&3.2&5.3&0.0&6.2&14.1&$3601.3 (14.5)$&3.2&0.7&0.0&3.3&14.1\\\cline{2-14}
        \text{}&$\mathcal{M}^{V}_2$&$3618.7( 12.7)$&5.2&5.2&0.0&7.4&10.4&$3619.0( 11.6)$&5.2&0.4&0.1&5.2&10.4\\ \cline{2-14}
         \text{}&$\mathcal{M}^{V}_3$&$3630.5( 2739.5)$&6.2&20.2&2739.4&2739.5&7.8&$3626.8( 10.2)$&6.5&0.3&0.2&6.5&7.9\\\cline{2-14}
         \text{}&$\mathcal{M}^{V}_4$&$3644.9( 2487.8)$&6.5&49.9&2487.3&2487.8&6.4&$3631.0( 9.8)$&7.4&0.4&0.2&7.4&6.5\\
         \hline\hline
       \multirow{4}{4em}{Ref.~\cite{Kuhn:2007vp}}
                &$\mathcal{M}^{V}_1$&$3597.8( 20.6)$&3.2&5.3&0.0&6.2&19.6&$3597.2( 19.9)$&3.2&0.7&0.0&3.3&19.6\\ \cline{2-14}
         \text{}&$\mathcal{M}^{V}_2$&$3615.4( 15.5)$&5.2&5.2&0.0&7.4&13.7&$3615.7( 14.6)$&5.2&0.4&0.1&5.2&13.7\\ \cline{2-14}
         \text{}&$\mathcal{M}^{V}_3$&$3628.1( 2737.1)$&6.2&20.3&2737.0&2737.1&10.2&$3624.4( 12.1)$&6.5&0.3&0.2&6.5&10.3\\ \cline{2-14}
         \text{}&$\mathcal{M}^{V}_4$&$3642.9( 2485.9)$&6.5&50.0&2485.3&2485.3&7.8&$3629.0(10.8)$&7.4&0.4&0.2&7.4&7.9\\\hline\hline
    \end{tabular}
    \end{adjustbox}
    \caption{$m_b$ determinations using FOPT and RGSPT in the $\msbar$ scheme using experimental inputs from Table~\eqref{tab:momV_bottom}. Results are in the units of $\MeV$ and the scale dependence is calculated for the energy range $\mu\in[2,20]\hs\GeV$.}
    \label{tab:mb_ms_cond_ms}
\end{table}

\begin{table}[H]
    \centering
     \begin{adjustbox}{max width=\textwidth}
    \begin{tabular}{||c|c|c|c|c|c|c|c|c|c|c|c|c|c||}\hline\hline
    \text{}&\text{} &\multicolumn{6}{c|}{\textbf{FOPT}}&\multicolumn{6}{c|}{\textbf{RGSPT}}\\\cline{3-14}
    \textbf{Sources}&\textbf{Moments}&\text{}&\multicolumn{4}{c|}{Theo. Unc.}& \text{}&\text{}&\multicolumn{4}{c|}{Theo. Unc.}& \text{} \\
\cline{4-7}\cline{10-13}
\text{}&\text{}&$m_b(10\GeV)$&$\as$&$\mu$&n.p.&total&Exp. Unc.&$m_b(10\hs\GeV)$&$\as$&$\mu$&n.p.&total&Exp. Unc.
\\\hline\hline
         \multirow{4}{4em}{ Ref.~\cite{Dehnadi:2015fra}}
          &$\mathcal{M}^{V}_1$&$3632.2( 53.4)$&3.2&5.4&0.0&6.3&53.0&$3631.6( 53.0)$&3.2&0.7&0.0&3.3&53.0\\\cline{2-14}
        \text{}&$\mathcal{M}^{V}_2$&$3632.8( 21.0)$&5.2&5.3&0.0&7.4&19.7&$3633.2( 20.4)$&5.2&0.4&0.0&5.2&19.7\\ \cline{2-14}
         \text{}&$\mathcal{M}^{V}_3$&$3637.6( 23.2)$&6.2&19.5&0.0&20.5&11.0&$3634.2( 12.8)$&6.5&0.3&0.0&6.5&11.1\\\cline{2-14}
         \text{}&$\mathcal{M}^{V}_4$&$3648.3( 48.8)$&6.5&47.8&0.1&48.3&7.4&$3635.1( 10.5)$&7.4&0.4&0.0&7.4&7.5\\
         \hline\hline
        \multirow{4}{4em}{ Ref.~\cite{Chetyrkin:2009fv}}
               &$\mathcal{M}^{V}_1$&$3601.9( 15.5)$&3.2&5.3&0.0&6.2&14.1&$3601.3( 14.5)$&3.2&0.7&0.0&3.3&14.1\\\cline{2-14}
        \text{}&$\mathcal{M}^{V}_2$&$3618.7( 12.7)$&5.2&5.2&0.0&7.4&10.4&$3619.1( 11.6)$&5.2&0.4&0.0&5.2&10.4\\ \cline{2-14}
         \text{}&$\mathcal{M}^{V}_3$&$3630.4( 22.0)$&6.2&19.6&0.0&20.5&7.8&$3626.9( 10.2)$&6.5&0.3&0.0&6.5&7.9\\\cline{2-14}
         \text{}&$\mathcal{M}^{V}_4$&$3644.3( 48.7)$&6.5&47.9&0.1&48.3&6.4&$3631.1( 9.8)$&7.4&0.4&0.0&7.4&6.5\\
         \hline\hline
       \multirow{4}{4em}{Ref.~\cite{Kuhn:2007vp}}
                &$\mathcal{M}^{V}_1$&$3597.8( 20.6)$&3.2&5.3&0.0&6.2&19.6&$ 3597.3( 19.9)$&3.2&0.7&0.0&3.3&19.6\\ \cline{2-14}
         \text{}&$\mathcal{M}^{V}_2$&$3615.4( 15.5)$&5.2&5.2&0.0&7.3&13.7&$3615.8( 14.6)$&5.2&0.4&0.0&5.2&13.7\\ \cline{2-14}
         \text{}&$\mathcal{M}^{V}_3$&$3628.0( 22.9)$&6.2&19.6&0.0&20.5&10.2&$3624.5( 12.1)$&6.5&0.3&0.0&6.5&10.3\\ \cline{2-14}
         \text{}&$\mathcal{M}^{V}_4$&$3642.4( 48.9)$&6.5&47.9&0.1&48.3&7.8&$3629.1( 10.8)$&7.4&0.4&0.0&7.4&7.9\\\hline\hline
    \end{tabular}
\end{adjustbox}
    \caption{$m_b$ determinations using FOPT and RGSPT  using experimental inputs from Table~\eqref{tab:momV_bottom}. The pole mass of the bottom quark is used as input in the non-perturbative condensate terms. Results are in the units of $\MeV$ and the scale dependence is calculated for the energy range $\mu\in[2,20]\hs\GeV$.}
    \label{tab:mb_ms_cond_pole}
\end{table}
\section{\texorpdfstring{$\as$}{} determination} \label{sec:as_det}
For the $\as$ determination, instead of the $\mathcal{M}^{X}_n$, we use the dimensionless quantities such as $\mathcal{M}^{P}_0$ and the ratio of moments $\mathcal{R}^{X}_n$ defined Eq.~\eqref{eq:Def_R}. Theoretical expressions are sensitive to the $\as$, and the quark mass dependence appears at NNLO via running logarithms. These quantities are very important observables for $\as$ determination. These ratios can also be calculated from the lattice QCD for the charm case in the pseudoscalar channel. We do not get any reliable determinations of the $\as$ for the bottom moments in the vector channel. Therefore, our determinations are only based on the charmonium sum rules.\par 
We use the ratios of the moments for the vector channel provided in Refs.~\cite{Boito:2019pqp,Boito:2020lyp} in the $\as$ determinations. For the pseudoscalar channel, we use results on the moments and from Refs.~\cite{HPQCD:2008kxl,McNeile:2010ji,Maezawa:2016vgv,Nakayama:2016atf,Petreczky:2019ozv,Petreczky:2020tky}. \par 
It should be noted that the $\as$ determination in this section is first performed at charm quark mass scale $m_c(m_c)=1.27\pm0.2\hs\GeV$ and then evolved to boson mass scale ($M_Z=91.18\hs\GeV$) by performing the matching and decoupling at the bottom quark mass scale using $m_b(m_b)=4.18\hs\GeV$ in the $\msbar$ scheme~\cite{ParticleDataGroup:2022pth}. We have used the REvolver package to perform the running and decoupling once $\as(\overline{m}_c)$ is obtained.\par Another technical point is the evolution of quark mass when uncertainties coming from scale variation are calculated. For this, we have taken our $x(\overline{m}_c)=\as(\overline{m}_c)/\pi$ determination as input and numerically solved for $x(q)$ at different scale, $q$, using relation:
\begin{equation}
    m_c(q)=\overline{m}_c \int_{x(\overline{m}_c)}^{x(q)} \hs d\hs x\hs e^{\left(\frac{\gamma(x)}{\beta(x)}\right)}\,,
    \label{eq:mq_evolve}
\end{equation}
where $\gamma(x)$ and $\beta(x)$ are the five-loop quark mass anomalous dimension and QCD beta function. It should be noted that Eq.~\eqref{eq:mq_evolve} is used to evolve the masses at different scales using the extracted central value of $\as(\overline{m}_c)$ at $\mu=\overline{m}_c$ from the moments.
%
\subsection{\texorpdfstring{$\as$}{} from the vector channel.}
For the vector channel moments, one needs the experimental information about the resonances, and additional continuum contributions are modeled using the theoretical expression for the hadronic $R-$ratio for $e^+e^-$ in Eq.~\eqref{eq:Rratio}. We do not calculate these moments in this section, instead using very recent results provided in Refs.~\cite{Boito:2019pqp,Boito:2020lyp}. We have collected experimental inputs in Table~\eqref{tab:RV_charm}. We have tabulated our determinations in the $\msbar$ scheme in Table~\eqref{tab:as_V_ms}. Contrary to the previous sections, the effects of the non-perturbative terms in the $\msbar$ scheme are the same for $\as$ determinations from FOPT and RGSPT. Even though RGSPT has significant control over the renormalization scale uncertainties, uncertainties arising from the non-perturbative contributions dominate in the higher moments. \par 
We also perform $\as$ determinations to control these uncertainties using the numerical value of the on-shell mass for the charm quark in the condensate terms. The results obtained are presented in Table~\eqref{tab:as_V_pole}. We have found that using the on-shell mass in the non-perturbative term significantly improves our $\as$ determination. It is remarkable to note that the theoretical uncertainties are of similar size to the experimental ones for RGSPT.\par
\begin{figure}[ht]
\centering
		\includegraphics[width=.49\textwidth]{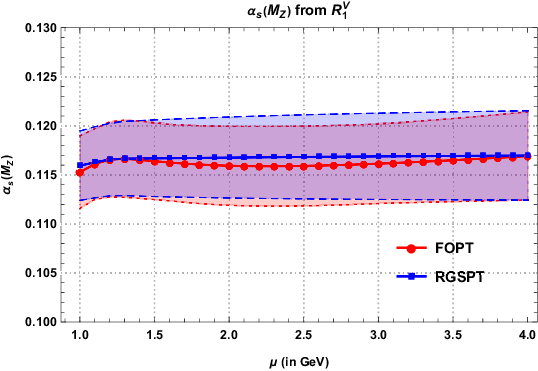}
		\includegraphics[width=.49\textwidth]{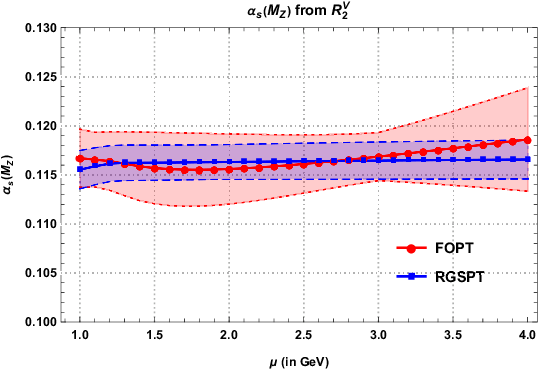}
	\includegraphics[width=.49\textwidth]{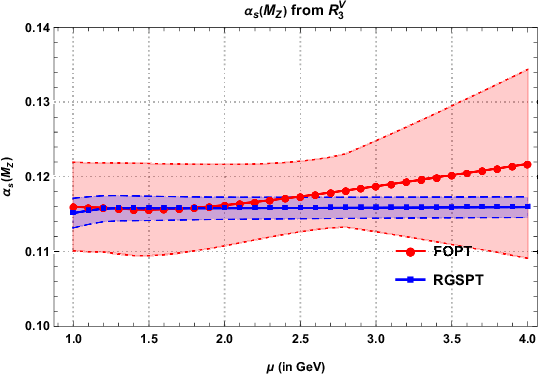}
\caption{\label{fig:asmz_V} $\as$ determination from the $R_n^{V}$ using $\msbar$ value of the charm quark mass as input in the condensate term. The bands represent the total uncertainty in the determinations.}
\end{figure}
\begin{figure}[ht]
\centering
		\includegraphics[width=.49\linewidth]{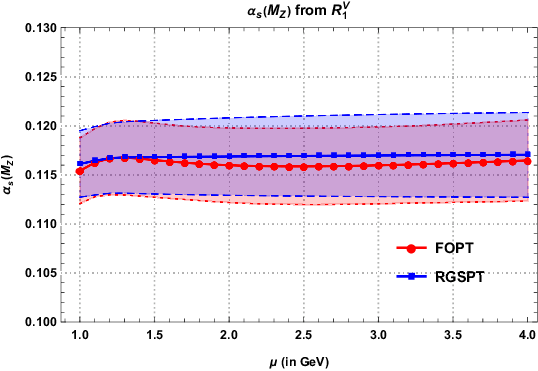}
		\includegraphics[width=.49\linewidth]{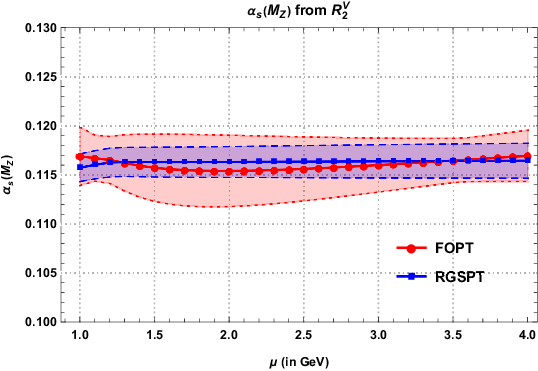}
	\includegraphics[width=.49\linewidth]{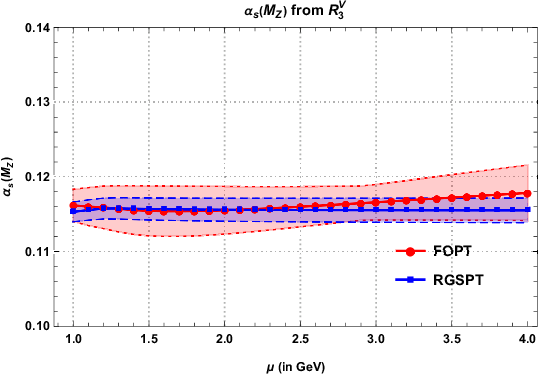}
\caption{\label{fig:asmz_V_pole} $\as$ determination from the $R_n^{V}$ using the on-shell value of the charm quark mass as input in the condensate term. The bands represent the total uncertainty in the determinations.}
\end{figure}

We can also notice that our determination in Tables~\eqref{tab:as_V_ms},\eqref{tab:as_V_pole} have the same central value for both RGSPT and FOPT. Due to the choice of scale $\mu=\overline{m}_c$, FOPT and RGSPT have the same expressions for the moments. Different scale choices result in different central values and uncertainties. This behavior can be seen in Fig.~\eqref{fig:asmz_V} and Fig.~\eqref{fig:asmz_V_pole} for the two scenarios considered above for the treatment of the non-perturbative terms. These plots also show remarkable stability in the $\as$ determinations from the RGSPT.  
\begin{table}[H]
    \centering
    \begin{tabular}{||c|c||}\hline\hline
         $n$& $\mathcal{R}^{V}_n$ \\\hline\hline
         1&$1.770\pm0.017$\\\hline
         2&$1.1173\pm0.0023$\\\hline
         3&$1.03536\pm0.00084$\\\hline\hline
    \end{tabular}
    \caption{Ratio of the experimental moments for the vector current obtained using the current PDG~\cite{ParticleDataGroup:2022pth} value of the $\as^{\left(5\right)}(M_Z)=0.1179\pm 0.0009$.}
    \label{tab:RV_charm}
\end{table}

\begin{table}[ht]
    \centering
     \begin{adjustbox}{max width=\textwidth}
    \begin{tabular}{||c|c|c|c|c|c|c|c|c|c|c|c|c||}
    \hline\hline
    \text{}&\multicolumn{6}{c|}{\textbf{FOPT}}&\multicolumn{6}{c|}{\textbf{RGSPT}}\\\cline{2-13}
    \textbf{Moment}&\text{}&\multicolumn{4}{c|}{Theo. Unc.}&\text{}&\text{}&\multicolumn{4}{c|}{Theo. Unc.}& \text{} \\
		\cline{3-6}\cline{9-12}
\text{}&$\as(M_Z)$&$m_c$&$\mu$&n.p.&total& Exp. Unc.&$\as(M_Z)$&$m_c$&$\mu$&n.p.&total& Exp. Unc. \\ \hline\hline
  $\mathcal{R}^{V}_1$&$0.1167(39)$&3&13&8&16&36&$0.1167(38)$&3&7&8&11&36\\\hline
  $\mathcal{R}^{V}_2$&$0.1163(31)$&4&27&12&29&11&$0.1163(18)$&3&8&12&15&11\\\hline
  $\mathcal{R}^{V}_3$&$0.1159(60)$&4&58&14&60&5&$0.1159(17)$&3&6&14&16&5\\\hline\hline
    \end{tabular}
\end{adjustbox}
    \caption{$\as$ determination using FOPT and RGSPT for the vector channel and sources of uncertainties from different sources. The scale dependence is calculated for the energy range $\mu\in[1,4]\hs\GeV$. The $\msbar$ scheme value for the charm quark is used in condensate terms.}
    \label{tab:as_V_ms}
\end{table}

\begin{table}[ht]
    \centering
     \begin{adjustbox}{max width=\textwidth}
    \begin{tabular}{||c|c|c|c|c|c|c|c|c|c|c|c|c||}
    \hline\hline
    \text{}&\multicolumn{6}{c|}{\textbf{FOPT}}&\multicolumn{6}{c|}{\textbf{RGSPT}}\\\cline{2-13}
    \textbf{Moment}&\text{}&\multicolumn{4}{c|}{Theo. Unc.}&\text{}&\text{}&\multicolumn{4}{c|}{Theo. Unc.}& \text{} \\
		\cline{3-6}\cline{9-12}
\text{}&$\as(M_Z)$&$m_c$&$\mu$&n.p.&total& Exp. Unc.&$\as(M_Z)$&$m_c$&$\mu$&n.p.&total& Exp. Unc. \\ \hline\hline
 $\mathcal{R}^{V}_1$&$0.1169(38)$&3&13&5&15&35&$0.1169(36)$&2&6&5&8&35\\\hline
  $\mathcal{R}^{V}_2$&$0.1164(28)$&4&24&9&26&10&$0.1164(15)$&2&5&6&8&10\\\hline
  $\mathcal{R}^{V}_3$&$0.1159(30)$&3&27&13&30&5&$0.1159(14)$&2&2&6&6&5\\\hline\hline
    \end{tabular}
\end{adjustbox}
    \caption{$\as$ determination using FOPT and RGSPT for the vector channel and sources of uncertainties from different sources. The scale dependence is calculated for the energy range $\mu\in[1,4]\hs\GeV$. The on-shell mass of the charm quark is used in condensate terms.}
    \label{tab:as_V_pole}
\end{table}
\begin{table}[ht]
    \centering
     \begin{adjustbox}{max width=\textwidth}
    \begin{tabular}{|c|c|c|c|c|c|c|}\hline
        \textbf{ Moments} & \textbf{Ref.~\cite{HPQCD:2008kxl}}& \textbf{Ref.~\cite{McNeile:2010ji}} &\textbf{Ref.~ \cite{Maezawa:2016vgv} }&\textbf{Ref.~\cite{Nakayama:2016atf}} & \textbf{Ref.~\cite{Petreczky:2019ozv}}& \textbf{Ref.~\cite{Petreczky:2020tky}} \\\hline\hline
         $\mathcal{M}^{P}_0$&$1.708\pm0.007$    &$1.708\pm0.005$    &$-$                &$1.699\pm0.008$    &$1.705\pm0.005$    &$1.7037\pm0.0027$\\\hline
         $\mathcal{R}^{P}_1$&$1.197\pm0.004$    &$-$                &$1.188\pm0.004$    &$1.199\pm0.004$    &$1.1886\pm0.013$   &$1.1881\pm0.0007$\\\hline
         $\mathcal{R}^{P}_2$&$1.033\pm0.004$    &$-$                &$1.0341\pm0.0018$  &$1.0344\pm0.0013$  &$1.0324\pm0.0016$  &$-$\\\hline
    \end{tabular}
\end{adjustbox}
    \caption{Pseudoscalar moment calculated from lattice QCD for the charm case. These $\mathcal{M}^{P}_n$ are in the units of $10^{-n}\GeV^{-2n}$.}
    \label{tab:momP_as_charm}
\end{table}

\begin{table}[ht]
    \centering
    \begin{adjustbox}{max width=\textwidth}
    \begin{tabular}{|c|c|c|c|c|c|c|c|c|c|c|c|c|c|}\hline
    \text{}&\text{} &\multicolumn{6}{c|}{\textbf{FOPT}}&\multicolumn{6}{c|}{\textbf{RGSPT}}\\\cline{3-14}
    \textbf{Sources}&\textbf{Moments}&\text{}&\multicolumn{4}{c|}{Theo. Unc.}& \text{}&\text{}&\multicolumn{4}{c|}{Theo. Unc.}& \text{} \\
\cline{4-7}\cline{10-13}
\text{}&\text{}&$\as(M_Z)$&$m_c$&$\mu$&n.p.&total&Exp. Unc.&$\as(M_Z)$&$m_c$&$\mu$&n.p.&total&Exp. Unc.
\\\hline\hline
         \multirow{3}{4em}{ Ref.~\cite{HPQCD:2008kxl}}&$\mathcal{M}^{P}_0$&0.1172(20)&3&19&3&19&6&0.1172(8)&3&3&3&5&6\\
         \cline{2-14}
         \text{}&$\mathcal{R}^{P}_1$&0.1182(43)&4&42&5&43&6&0.1181(15)&3&12&5&13&6\\
         \cline{2-14}\text{}&$\mathcal{R}^{P}_2$&0.1150(53)&4&50&9&51&15&0.1149(18)&3&7&9&11&15\\\hline\hline
         \multirow{1}{4em}{  Ref.~\cite{McNeile:2010ji} }&$\mathcal{M}^{P}_0$&0.1172(20)&3&19&3&19&5&0.1172(7)&3&3&3&5&5\\\hline\hline
         \multirow{2}{4em}{ Ref.~\cite{Maezawa:2016vgv}}&$\mathcal{M}^{P}_0$&0.1168(48)&3&8&6&47&7&0.1168(13)&3&9&6&11&7\\
         \cline{2-14}
         \text{}&$\mathcal{R}^{P}_1$&0.1152(51)&4&50&8&50&6&0.1152(13)&3&7&8&11&6\\\hline\hline
        \multirow{3}{4em}{ Ref.~\cite{Nakayama:2016atf}}&$\mathcal{M}^{P}_0$&0.1164(20)&3&18&4&19&7&0.1164(9)&3&3&4&5&7\\
          \cline{2-14}\text{}&$\mathcal{R}^{P}_1$&0.1182(43)&4&42&5&43&6&0.1184(15)&3&13&5&14&6\\
         \cline{2-14}
         \text{}&$\mathcal{R}^{P}_2$&0.1153(50)&4&49&8&50&5&0.1153(12)&3&7&8&7&5\\\hline\hline
       \multirow{3}{4em}{Ref.~\cite{Petreczky:2019ozv}}&$\mathcal{M}^{P}_0$&0.1169(20)&3&19&3&19&5&0.1169(7)&3&13&3&5&5\\
         \cline{2-14} \text{}&$\mathcal{R}^{P}_1$&0.1169(47)&4&47&6&47&2&0.1169(12)&3&10&6&12&2\\
         \cline{2-14}
         \text{}&$\mathcal{R}^{P}_2$&0.1146(53)&4&52&9&53&6&0.1146(13)&3&6&9&11&6\\\hline\hline
       \multirow{2}{4em}{ Ref.~\cite{Petreczky:2020tky}}&$\mathcal{M}^{P}_0$&0.1168(19)&3&19&3&19&2&0.1168(13)&3&9&6&11&7\\
         \cline{2-14}
         \text{}&$\mathcal{R}^{P}_1$&0.1168(47)&4&47&6&47&1&0.1168(12)&3&1&6&11&1\\\hline
    \end{tabular}
    
     \end{adjustbox}
     
    \caption{$\as$ determination from the pseudoscalar channel in the $\msbar$ scheme from various sources as input from Table~\eqref{tab:momP_as_charm}.}
    \label{tab:as_P_ms_charm}

\end{table}


\subsection{\texorpdfstring{$\as$}{} determination using the lattice QCD data}
The lattice QCD is the only source where the simulations can calculate pseudoscalar moments, and then the results are extrapolated to the continuum limits. Their results have their limits as the dimensionless moments and their ratios $\mathcal{R}^P_{n}$ for the higher moments suffer from the lattice artifacts. Despite these limitations, lower moments can still be used in the $\as$ determination. The lattice QCD simulations calculate the reduced moments $R_{2n+4}$, which will be transformed into the regular moments from the perturbative QCD. In addition to Eq.~\eqref{eq:reduced_mom}, we also need zeroth moment and dimensionless $\mathcal{R}_n^P$ using Eq.~\eqref{eq:Def_R} are obtained as~\cite{Boito:2020lyp,Dehnadi:2015fra}:
\begin{equation}
\mathcal{M}_0^P=T^P_{0,0} R_4\,,\quad  \mathcal{R}_n^P=\frac{\left(T^P_{n,0}\right)^{1/n}}{\left(T^P_{n+1,0}\right)^{1/\left(n+1\right)}} \left(\frac{R_{2n+4}}{R_{2n+6}}\right)^2\,.
\label{eq:def_Rn}
\end{equation}
Using above relations, the results of Refs.~\cite{HPQCD:2008kxl,McNeile:2010ji,Maezawa:2016vgv,Petreczky:2019ozv,Petreczky:2020tky} are tabulated in Table~\eqref{tab:momP_as_charm}. We use these results in the $\as$ determination from the pseudoscalar moments, and the result in the $\msbar$ scheme are presented in Table~\eqref{tab:as_P_ms_charm}. The non-perturbative effects are under control since only the first few moments are used as input. They do not cause any issues in our determination, so there is no need to reiterate the exercise using the on-shell mass for the charm quark in condensate terms. Again, our determinations from the FOPT are dominated by theoretical uncertainties, especially the scale variation, but RGSPT gives very stable results. It is interesting to note that the determinations of $\as$ from $\mathcal{R}^{P}_2$ give smaller central values than the other moments considered. 
\section{Summary and Conclusion}\label{sec:summary}
In section~\eqref{sec:formulas}, the perturbative quantities related to the low energy moments of the current correlators are RG improved using RGSPT in section~\eqref{sec:RGmom}. The scale dependence of these RG invariant quantities using FOPT and RGSPT are plotted in Figs.~\eqref{fig:MomV_c}, \eqref{fig:MomP_c}, \eqref{fig:Mom_b}. It is evident from these plots that a more precise determination of the $m_c$, $m_b$, and $\as$ can be obtained.\par In section~\eqref{sec:mc_det}, the determination of the $m_c$ is performed using experimental vector and lattice pseudoscalar moments. The $m_c$ determinations using FOPT from the vector moments suffer from large uncertainties originating from the non-perturbative terms. This problem is not encountered with RGSPT; results are presented in Table~\eqref{tab:mc_V_ms}. To improve FOPT determination, we take quark mass in the on-shell scheme as input for the non-perturbative terms. This choice leads to improved determinations and results are presented in Table.~\eqref{tab:mc_V_pole}. For the pseudoscalar moments, the non-perturbative effects are not as problematic as in the vector case. Slightly more precise values are obtained from the lattice moments than vector moments. These results are presented in Table~\eqref{tab:mc_P_ms}.\par
In section~\eqref{sec:mb_det}, the $m_b$ determination only using the vector moments is performed using FOPT and RGSPT. The condensate terms are more troubling in this case than in the $m_c$ determination. For the third and fourth moments, the determinations using FOPT suffer largely from the uncertainty from non-perturbative terms, which are nearly $70\%$ of the central value. These results are presented in Table~\eqref{tab:mb_ms_cond_ms}. When the numerical value of the bottom quark mass in the on-shell scheme is taken as input in the non-perturbative term, this problem for FOPT determination disappears. These results are presented in Table~\eqref{tab:mb_ms_cond_pole}.\par
In section~\eqref{sec:as_det}, the $\as$ determination is performed using dimensionless moments and ratios of the moments for the charm vector and pseudoscalar moments. The values obtained using the vector moments have similar issues from the non-perturbative terms as in the case of the $m_c$ and $m_b$ determination. These are again solved using the on-shell charm quark mass. These results are presented in Tables~\eqref{tab:as_V_ms} and \eqref{tab:as_V_pole}. Determinations using the pseudoscalar moments are presented in Table~\eqref{tab:as_P_ms_charm}.\par
In addition, it is worth mentioning that the RGSPT can also be used to calculate the continuum contributions to experimental moments where electromagnetic R-ratio is taken as input. In Ref.~\cite{AlamKhan:2023dms}, a significant reduction in the theoretical uncertainties originating from renormalization scale variation and truncation of the perturbation series is obtained for R-ratio. As an application of the method developed, light quark masses determined in Ref.~\cite{AlamKhan:2023ili} are more precise compared to the FOPT. These results can also be used for the method used in Refs.~\cite{Erler:2016atg,Erler:2022mzd} for the continuum contributions.\par
Now, we turn to the final values for the $m_c$, $m_b$, and $\as$ determination. We take the most precise values obtained in this study. Interestingly, all of them are obtained using RGSPT and lattice inputs except for the bottom quark mass, for which no lattice moments are available. For the charm mass, we give our final determination that is obtained from Table~\eqref{tab:mc_P_ms} using Ref.~\cite{Petreczky:2020tky} as:
\begin{align}
    m_c(3\GeV)&=0.9962(42)\hs\GeV\,,\\
    \implies m_c(m_c)&=1.2811(38)\hs\GeV\,.
\end{align}
\par
For the bottom quark mass, we take the most precise value obtained in Table~\eqref{tab:mb_ms_cond_pole} from Ref.~\cite{Chetyrkin:2009fv} as:
\begin{align}
    m_b(10\GeV)&=3.6311(98)\hs\GeV\,,\\
    \implies m_b(m_b)&=4.1743(95)\hs\GeV\,.
\end{align}
For strong coupling constant, we have two most precise determinations in Table~\eqref{tab:as_P_ms_charm} from Refs.~\cite{McNeile:2010ji,Petreczky:2019ozv}. We average out these values and obtain the final determination:
\begin{align}
    \as(M_Z)&=\{0.1172(7),0.1169(7)\}\,,\\
    \implies  \as(M_Z)&=0.1171(7)\,.
\end{align}
These values are in full agreement with the current PDG~\cite{ParticleDataGroup:2022pth} values which read:
\begin{align}
    \as(M_Z)&=0.1179(9)\,,\\
     m_c(m_c)&=1.27\pm0.02\hs\GeV\,,\\
     m_b(m_b)&=4.18\pm0.03\hs\GeV\,.
\end{align}

\begin{subappendices}

\section{Perturbative coefficients of the moments} \label{app:pert_coef}
 The perturbative coefficients of the vector and pseudoscalar moments for the charm case are presented in Table~\eqref{tab:coef_V_charm} and Table~\eqref{tab:coef_P_charm}, respectively. For the bottom case, only vector moments are used and their coefficients are presented in Table~\eqref{tab:coef_V_bottom}. 
\begin{table}[H]
    \centering
    \begin{tabular}{||c||c|c|c|c|c|c||}
    \hline\hline
        Moments& $T^V_{0,0}$& $T^V_{1,0}$& $T^V_{2,0}$& $T^V_{3,0}$& $T^{V,\text{n.p.}}_{0,0}$& $T^{V,\text{n.p.}}_{1,0}$\\\hline 
        $\mathcal{M}^V_1$ &1.067& 2.555& 2.497& -5.640&-0.251& -0.235\\\hline 
        $\mathcal{M}^V_2$ &0.457&1.110& 2.777& -3.494 & -0.104 & 0.051\\\hline
        $\mathcal{M}^V_3$ &0.271 & 0.519 & 1.639 & -2.840 & -0.038 & 0.077 \\\hline
        $\mathcal{M}^V_4$ &0.185 & 0.203 & 0.796 & -3.348 & -0.013 & 0.047\\\hline\hline
    \end{tabular}
    \caption{The perturbative coefficients of the vector moments for the charm case in the $\msbar$ scheme.}
    \label{tab:coef_V_charm}
\end{table}
\begin{table}[H]
    \centering
    \begin{tabular}{||c||c|c|c|c|c|c||}
    \hline\hline
        Moments& $T^P_{0,0}$& $T^P_{1,0}$& $T^P_{2,0}$& $T^P_{3,0}$&$T^{P,\text{n.p.}}_{0,0}$& $T^{P,\text{n.p.}}_{1,0}$\\\hline 
         $\mathcal{M}^P_0$ & 0.333& 0.778 & 0.183 & -1.806 & 0.877 & -0.045\\\hline 
        $\mathcal{M}^P_1$ & 0.133 & 0.516 & 1.881 & 1.515 & 0.125 & -0.391\\\hline 
       $\mathcal{M}^P_2$ & 0.076 & 0.303 & 1.567 & 3.711 & 0.0 & -0.142\\\hline 
       $\mathcal{M}^P_3$ & 0.051 & 0.178 & 1.138 & 3.583 & -0.009 & -0.022\\\hline\hline
    \end{tabular}
    \caption{The perturbative coefficients of the pseudoscalar moments for the charm case in the $\msbar$ scheme.}
    \label{tab:coef_P_charm}
\end{table}

\begin{table}[H]
    \centering
    \begin{tabular}{||c||c|c|c|c|c|c||}
    \hline\hline
        Moments& $T^V_{0,0}$& $T^V_{1,0}$& $T^V_{2,0}$& $T^V_{3,0}$&$T^{V,\text{n.p.}}_{0,0}$& $T^{V,\text{n.p.}}_{1,0}$\\\hline 
       $\mathcal{M}^V_1$ & 0.267 & 0.639 & 0.790 & -1.941 & -0.110 & -0.271\\\hline 
        $\mathcal{M}^V_2$ & 0.114 & 0.277 & 0.808 & -0.661 & -0.063 & -0.076 \\\hline 
      $\mathcal{M}^V_3$ & 0.068 & 0.130 & 0.517 & -0.294 & -0.026 & 0.005\\\hline 
     $\mathcal{M}^V_4$ &  0.046 &  0.051 & 0.305 & -0.346 & -0.009 & 0.017\\\hline \hline 
    \end{tabular}
    \caption{The perturbative coefficients of the vector moments for the bottom case in the $\msbar$ scheme.}
    \label{tab:coef_V_bottom}
\end{table}

\section{Solution to summed coefficients}
\label{app:RGcoefs}
The solution to the differential equation in Eq.~\eqref{eq:summed_de} are obtained as:
\begin{align}
S_0(w)=& w^{\frac{2 n \gamma _0}{\beta _0}}\,,\\
S_1(w)=& \left(T_{1,0}^X+\frac{2 n L_w \gamma _0 \left(\beta _1-2 \beta _0 \gamma _0\right)}{\beta _0^2}+\frac{2 n \left(\beta _1 \gamma _0-\beta _0 \gamma _1\right)}{\beta _0^2}\right) w^{\frac{2 n \gamma _0}{\beta _0}-1}\nonumber\\&+\frac{2 n \left(\beta _0 \gamma _1-\beta _1 \gamma _0\right) w^{\frac{2 n \gamma _0}{\beta _0}}}{\beta _0^2}\,,\\
S_2(w)=&\frac{n w^{\frac{2 n \gamma _0}{\beta _0}}}{\beta _0^4} \Bigg[\gamma _2 \beta _0^3-\left(\beta _2 \gamma _0+\gamma _1 \left(\beta _1-2 n \gamma _1\right)\right) \beta _0^2+\beta _1 \gamma _0 \left(\beta _1-4 n \gamma _1\right) \beta _0+2 n \beta _1^2 \gamma _0^2\Bigg] \nonumber\\&+ w^{\frac{2 n \gamma _0}{\beta _0}-2}\Bigg[T_{2,0}^X+\frac{L_w \left(\beta _1-2 \beta _0 \gamma _0\right)}{\beta _0^4} \Big[-\beta _0^2 \left(\beta _0-2 n \gamma _0\right) T_{1,0}^X+4 n \left(n \beta _1-\beta _0^2\right) \gamma _0^2\nonumber\\&\bs+2 n \beta _0 \left(\beta _0-2 n \gamma _0\right) \gamma _1\Big]+\frac{n}{\beta _0^4} \Big[\left(-2 \gamma _1 T_{1,0}^X+4 \gamma _0 \gamma _1-\gamma _2\right) \beta _0^3+2 n \beta _1^2 \gamma _0^2\nonumber\\&+\left(2 n \gamma _1^2-\beta _2 \gamma _0+\beta _1 \left(\gamma _1+2 \gamma _0 \left(T_{1,0}^X-2 \gamma _0\right)\right)\right) \beta _0^2+\beta _1 \gamma _0 \left(\beta _1-4 n \gamma _1\right) \beta _0\Big]\nonumber\\&-\frac{n L_w^2 \gamma _0 \left(\beta _0-2 n \gamma _0\right) \left(\beta _1-2 \beta _0 \gamma _0\right){}^2}{\beta _0^4}\Bigg] +w^{\frac{2 n \gamma _0}{\beta _0}-1}\Bigg[\frac{2 n}{\beta _0^4} \Big[\gamma _1 \left(T_{1,0}^X-2 \gamma _0\right) \beta _0^3\nonumber\\&+\left(\gamma _0 \left(\beta _2+\beta _1 \left(2 \gamma _0-T_{1,0}^X\right)\right)-2 n \gamma _1^2\right) \beta _0^2-\beta _1 \gamma _0 \left(\beta _1-4 n \gamma _1\right) \beta _0-2 n \beta _1^2 \gamma _0^2\Big]\nonumber\\&-\frac{4 n^2 L_w \gamma _0 \left(\beta _1-2 \beta _0 \gamma _0\right) \left(\beta _1 \gamma _0-\beta _0 \gamma _1\right)}{\beta _0^4}\Bigg]\,.
\end{align}
\begin{align}
S_3(w)= &\frac{2 n w^{\frac{2 n \gamma _0}{\beta _0}} }{3 \beta _0^6}S_{3,0}(w)+w^{\frac{2 n \gamma _0}{\beta _0}-1}S_{3,1}(w) +w^{\frac{2 n \gamma _0}{\beta _0}-2} S_{3,2}(w) +w^{\frac{2 n \gamma _0}{\beta _0}-3} S_{3,3} \,,
\end{align}
where,
\begin{align}
    S_{3,0}(w)=&\Bigg[\gamma _3 \beta _0^5-\left(\beta _3 \gamma _0+\beta _2 \gamma _1+\left(\beta _1-3 n \gamma _1\right) \gamma _2\right) \beta _0^4+3 n \beta _1^2 \gamma _0^2 \left(2 n \gamma _1-\beta _1\right) \beta _0\nonumber\\&+\left(\gamma _1 \beta _1^2+\left(2 \beta _2 \gamma _0-3 n \left(\gamma _1^2+\gamma _0 \gamma _2\right)\right) \beta _1+n \gamma _1 \left(2 n \gamma _1^2-3 \beta _2 \gamma _0\right)\right) \beta _0^3\nonumber\\&-\beta _1 \gamma _0 \left(\beta _1^2-6 n \gamma _1 \beta _1+6 n^2 \gamma _1^2-3 n \beta _2 \gamma _0\right) \beta _0^2-2 n^2 \beta _1^3 \gamma _0^3\Bigg]\,,
    \end{align}
    \begin{align}
    S_{3,1}(w)&=\Bigg[\frac{2 n^2 L_w \gamma _0 \left(\beta _1-2 \beta _0 \gamma _0\right)}{\beta _0^6} \Big[\gamma _2 \beta _0^3-\left(\beta _2 \gamma _0+\gamma _1 \left(\beta _1-2 n \gamma _1\right)\right) \beta _0^2+\beta _1 \gamma _0 \left(\beta _1-4 n \gamma _1\right) \beta _0\nonumber\\&\bs+2 n \beta _1^2 \gamma _0^2\Big]+\frac{n}{\beta _0^6} \bigg[\gamma _2 \left(T_{1,0}^X-2 \gamma _0\right) \beta _0^5+\beta _1 \gamma _0 \Big(\beta _1^2+2 n \left(\gamma _0 T_{1,0}^X-4 \left(\gamma _0^2+\gamma _1\right)\right) \beta _1\nonumber\\&+6 n \left(2 n \gamma _1^2-\beta _2 \gamma _0\right)\Big) \beta _0^2+\big(2 \beta _1 \left(n \left(8 \gamma _1 \gamma _0^2+\left(\gamma _2-2 \gamma _1 T_{1,0}^X\right) \gamma _0+\gamma _1^2\right)-\beta _2 \gamma _0\right)\big) \beta _0^3\nonumber\\&-4 n^2 \gamma _1^3+6 n \beta _2 \gamma _0 \gamma _1+\beta _1^2 \gamma _0 \left(T_{1,0}^X-2 \gamma _0\right)+\Big(-\left(\left(\beta _2 \gamma _0+\gamma _1 \left(\beta _1-2 n \gamma _1\right)\right) T_{1,0}^X\right)+\beta _3 \gamma _0\nonumber\\&+2 \gamma _0 \left(\beta _2 \gamma _0+\gamma _1 \left(\beta _1-4 n \gamma _1\right)\right)-2 n \gamma _1 \gamma _2\Big) \beta _0^4+6 n \beta _1^2 \gamma _0^2 \left(\beta _1-2 n \gamma _1\right) \beta _0+4 n^2 \beta _1^3 \gamma _0^3\bigg]\Bigg]\,,
    \end{align}
    \begin{align}
    S_{3,2}(w)&=\Bigg[-\frac{2 n^2 L_w^2 \gamma _0}{\beta _0^6}\Bigg[ \left(\beta _0-2 n \gamma _0\right) \left(\beta _0 \gamma _1-\beta _1 \gamma _0\right) \left(\beta _1-2 \beta _0 \gamma _0\right)^2\Bigg]-\frac{2 n L_w \left(\beta _1-2 \beta _0 \gamma _0\right)}{\beta _0^6} \nonumber\\&\bs\bigg[4 n^2 \beta _1^2 \gamma _0^3+\left(\beta _2 \gamma _0+\beta _1 \left(2 \gamma _0-T_{1,0}^X\right) \gamma _0-2 n \gamma _1 \left(\gamma _1+\gamma _0 \left(T_{1,0}^X-4 \gamma _0\right)\right)\right) \beta _0^3\nonumber\\&\bs-\gamma _0 \left(\beta _1^2+2 n \left(-\gamma _0 T_{1,0}^X+4 \gamma _0^2-\gamma _1\right) \beta _1+2 n \left(\beta _2 \gamma _0-2 n \gamma _1^2\right)\right) \beta _0^2\nonumber\\&\bs+\gamma _1 \left(T_{1,0}^X-2 \gamma _0\right) \beta _0^4+2 n \beta _1 \gamma _0^2 \left(\beta _1-4 n \gamma _1\right) \beta _0\bigg]+\frac{1}{\beta _0^6}\bigg[2 \beta _0^6 \gamma _1 T_{1,0}^X-4 n^3 \beta _1^3 \gamma _0^3\nonumber\\&\bs+6 n^2 \beta _0 \beta _1^2 \gamma _0^2 \left(2 n \gamma _1-\beta _1\right)+2 n^2 \beta _0^2 \beta _1 \gamma _0 \big(2 \beta _1 \left(-\gamma _0 T_{1,0}^X+4 \gamma _0^2+\gamma _1\right)\nonumber\\&\bs-6 n \gamma _1^2+3 \beta _2 \gamma _0\big)+2 n \beta _0^3 \big[\left(-\gamma _0 T_{1,0}^X+2 \gamma _0^2-\gamma _1\right) \beta _1^2+n \gamma _1 \left(2 n \gamma _1^2-3 \beta _2 \gamma _0\right)\nonumber\\&\bs+n \big(-16 \gamma _1 \gamma _0^2+\left(4 \gamma _1 T_{1,0}^X+\gamma _2\right) \gamma _0+\gamma _1^2\big) \beta _1\big]\nonumber\\&\bs+\beta _0^5 \left(-\beta _2 T_{1,0}^X-2 \gamma _0 \left(\beta _1+2 n \gamma _1\right) T_{1,0}^X+8 n \gamma _0^2 \gamma _1+2 n \gamma _1 \left(T_{2,0}^X-2 \gamma _1\right)\right)\nonumber\\&\bs+\beta _0^4 \big[2 n^2 \gamma _1 \left(-2 \gamma _1 T_{1,0}^X+8 \gamma _0 \gamma _1-\gamma _2\right)-2 n \beta _1 \gamma _0 \left(-2 \gamma _0 T_{1,0}^X+T_{2,0}^X+4 \gamma _0^2-2 \gamma _1\right)\nonumber\\&\bs+2 n \beta _2 \left(\gamma _1+\gamma _0 \left(T_{1,0}^X-2 \gamma _0\right)\right)+\beta _1^2 T_{1,0}^X\big]\bigg]\Bigg] \,,
    \end{align}
    \begin{align}
     S_{3,3}(w)&=\Bigg[\frac{2 n L_w^3 \gamma _0}{3 \beta _0^6} \left(\beta _0-n \gamma _0\right) \left(\beta _0-2 n \gamma _0\right) \left(\beta _1-2 \beta _0 \gamma _0\right)^3-\frac{L_w \left(\beta _1-2 \beta _0 \gamma _0\right) }{\beta _0^6}\Bigg[-4 n^3 \beta _1^2 \gamma _0^3\nonumber\\&+\Big[\beta _1 T_{1,0}^X-2 n \left(-2 \gamma _0^2 T_{1,0}^X+2 \gamma _1 T_{1,0}^X+4 \gamma _0^3+\gamma _2+\gamma _0 \left(T_{2,0}^X-6 \gamma _1\right)\right)\Big] \beta _0^4\nonumber\\&+2 \left(T_{2,0}^X-\gamma _0 T_{1,0}^X\right) \beta _0^5+2 n^2 \beta _1 \gamma _0^2 \left(4 n \gamma _1-\beta _1\right) \beta _0+2 n \big[\gamma _0 \left(\beta _1+2 n \gamma _1\right) T_{1,0}^X\nonumber\\&-2 \beta _1 \gamma _0^2+2 n \gamma _1^2-\beta _2 \gamma _0-8 n \gamma _0^2 \gamma _1+n \gamma _0 \gamma _2\big] \beta _0^3+2 n \gamma _0 \big[\beta _1^2+n \left(\beta _2 \gamma _0-2 n \gamma _1^2\right)\nonumber\\&+n \left(-2 \gamma _0 T_{1,0}^X+8 \gamma _0^2-3 \gamma _1\right) \beta _1\big] \beta _0^2\Bigg]+\frac{L_w^2 \left(\beta _1-2 \beta _0 \gamma _0\right)^2}{\beta _0^6} \Big[\beta _0^2 \left(\beta _0-2 n \gamma _0\right) \left(\beta _0-n \gamma _0\right) T_{1,0}^X\nonumber\\&-n \big(2 \left(\gamma _1-3 \gamma _0^2\right) \beta _0^3+\gamma _0 \left(8 n \gamma _0^2+\beta _1-6 n \gamma _1\right) \beta _0^2+2 n \gamma _0^2 \left(\beta _1+2 n \gamma _1\right) \beta _0-4 n^2 \beta _1 \gamma _0^3\big)\Big]\nonumber\\&+\frac{1}{3 \beta _0^6}\big[3 \left(T_{3,0}^X-2 \gamma _1 T_{1,0}^X\right) \beta _0^6+\big(3 \left(\beta _2-n \gamma _2\right) T_{1,0}^X-6 n \gamma _1 T_{2,0}^X+12 n \gamma _1^2-24 n \gamma _0^2 \gamma _1-2 n \gamma _3\nonumber\\&+6 \gamma _0 \left(\left(\beta _1+2 n \gamma _1\right) T_{1,0}^X+n \gamma _2\right)\big) \beta _0^5+\beta _0^4\big(-3 \beta _1^2 T_{1,0}^X-12 n \beta _1 \gamma _0^2 T_{1,0}^X+6 n^2 \gamma _1^2 T_{1,0}^X\nonumber\\&+3 n \beta _1 \gamma _1 T_{1,0}^X+6 n \beta _1 \gamma _0 T_{2,0}^X+24 n \beta _1 \gamma _0^3-24 n^2 \gamma _0 \gamma _1^2-n \beta _3 \gamma _0-18 n \beta _1 \gamma _0 \gamma _1+2 n \beta _1 \gamma _2\nonumber\\&+6 n^2 \gamma _1 \gamma _2+n \beta _2 \left(-3 \gamma _0 T_{1,0}^X+6 \gamma _0^2-4 \gamma _1\right)\big) +n \beta _0^3\big[-4 n^2 \gamma _1^3+6 n \beta _2 \gamma _0 \gamma _1\nonumber\\&+\beta _1^2 \left(4 \gamma _1+3 \gamma _0 \left(T_{1,0}^X-2 \gamma _0\right)\right)+2 \beta _1 \left(\beta _2 \gamma _0+3 n \left(8 \gamma _1 \gamma _0^2-\left(2 \gamma _1 T_{1,0}^X+\gamma _2\right) \gamma _0-\gamma _1^2\right)\right)\big] \nonumber\\&+n \beta _1 \gamma _0\beta _0^2 \big(-\beta _1^2+6 n \gamma _0 \left(T_{1,0}^X-4 \gamma _0\right) \beta _1+6 n \left(2 n \gamma _1^2-\beta _2 \gamma _0\right)\big) \nonumber\\&+6 n^2 \beta _1^2 \gamma _0^2 \left(\beta _1-2 n \gamma _1\right) \beta _0+4 n^3 \beta _1^3 \gamma _0^3\big]\Bigg]
\end{align}
where we have taken the $T_{0,0}^X=1$ in Eq.~\eqref{eq:mom_fopt}.

\end{subappendices}

%% file: Chapters/Chap8.tex
\chapter{Outlook}
\label{Chapter8}

\lhead{Chapter 8. \emph{Outlook}}
The standard model is an effective theory that works well for the energy range accessible with current technology. Its prediction depends on the free parameters present in it, which are fixed by various experimental inputs. Therefore, its prediction for any process can only be reliable if the uncertainties in the parameters are as small as possible. Perturbation theory is often used in theoretical calculations, and it is very important that the resulting series has a good convergence behavior. Moreover, it should also be less sensitive to the choice of the renormalization scheme and scale variations. To achieve these improvements, the renormalization group plays a key role. In addition to this, the renormalization group also allows for the comparison of a parameter calculated from different processes at different energies. In this thesis, we have explored the improvements that can be achieved using one of the renormalization group-improved prescription RGSPT and its application in the determination of some of the standard model parameters.\par
In chapter~\eqref{Chapter2}, we discussed the formalism of RGSPT and its application in summation of RG logarithms for strong coupling constant, and quark masses in the the $\msbar$-scheme as well as quark mass relations in the $\msbar$ and pole mass scheme. These relations are important if one wants an analytical expression that can be used for other processes in QCD. One obvious extension of the RGSPT is extension to the cases where apart from the renormalization scale, the factorization scale is also involved in the case of effective field theories such as NRQCD. We have considered this possibility in Ref.~\cite{Alam:matching}. Another extension is in the summation of large logarithms that arise when heavy quark decouples and these relations are also useful for the matching coefficients in the hadronic Higgs decays discussed in chapter~\eqref{Chapter5}. This work is also under progress~\cite{Abbas2}.\par
In chapter~\eqref{Chapter3}, we used asymptotic Pad\'e approximants to estimate the unknown perturbative contribution to the static energy to $\ordas{5}$. We have also discussed the Fourier transform of momentum space potential to position space, in which lattice results are often presented, which leads to pathological corrections from low-energy modes and is subtracted using a restricted version of the Fourier transform. We have provided the analytical expressions for the restricted and unrestricted Fourier-transformed version of static potential and static energy in the fixed-order perturbation theory (FOPT). As an application, we determine $\Lambda^{\msbar}_{QCD}$ by fitting RG improved perturbation series with the Cornell-type potential from lattice simulations~\cite{Karbstein:2018mzo} for two flavors QCD. \par
For a heavy quark system, the total energy is the sum of the masses of the quarks and the static energy between them. It can be used to calculate the spectrum of the bottomonium and charmonium systems. Since the coefficients of the anomalous dimension for the quark masses and the relation between the pole and $\msbar$ are already known to five and four loops, respectively. Our work can provide a further extension to related studies in one more order.  Also, a restricted Fourier transform for the RGSPT scheme can be implemented numerically and can be used for the $\as$ determination by comparing it with 3-flavor lattice QCD data. RGSPT can be used in the determination of the heavy quark masses from $\Upsilon$ $\left(1S\right)$ system~\cite{Ayala:2014yxa} by reducing renormalization scale uncertainties.
\par 
In chapter~\eqref{Chapter4}, strange quark mass and CKM matrix element $|V_{us}|$ are determined from the moments of the Cabibbo suppressed spectral moments of hadronic $\tau$ decays using finite energy sum rule. We used the publicly available information on the spectral moment in our determination of these parameters. We found that our results using RGSPT are significantly more stable with respect to the scale variations compared to the commonly used schemes such as FOPT and CIPT. It will be worth extending this study to non-spectral weight functions where more stability with respect $s_0$ can be obtained. It is also important to study the renormalon aspects of RGSPT in determining the $\as$ as it has been recently found out that the CIPT is incompatible with the OPE expectations in Ref.~\cite{Hoang:2020mkw}.\par

In chapter~\eqref{Chapter5}, the application of RGSPT in the analytic continuation and summation of large kinematical $\pi^2-$corrections to all orders is discussed for various processes. These corrections dominate the genuine perturbative corrections at higher orders~\cite{Herzog:2017dtz} and also result in the poor convergence of the perturbative series at low energies~\cite{Aparisi:2021tym} when FOPT prescription is used. We have studied processes such as the decay of Higgs boson to bottom quark pair ($H\rightarrow b\hs b$), decay of Higgs boson to gluon pair ($H\rightarrow g\hs g$), hadronic Higgs decays, electromagnetic R-ratio ($R_{em}$). The $R_{em}$ is also used in the data-driven methods to calculate the continuum contribution to the hadronic vacuum polarization contribution to $\left(g-2\right)_\mu$. Three-flavor quark-disconnected contributions are evaluated and compared with the results obtained in Ref.~\cite{Boito:2022rkw}. The formalism developed in the paper has a significant role to play in studies related to the Borel-Laplace sum rule determinations using pQCD results. One of such application in the light quark mass determination is also discussed in chapter~\eqref{Chapter6}. \par
In chapter~\eqref{Chapter6}, we determined the masses of the light quarks ($m_u$, $m_d$ and $m_s$) using the Borel-Laplace sum rule from the strange and non-strange pseudoscalar current correlator. Theoretical inputs used in such analysis are the Borel transform of the polarization function of the pseudoscalar current correlator and its spectral density of the correlator, which is related to the discontinuity across the cut. The spectral density gets extra kinematical $\pi^2-$terms which are already resummed using RGSPT in chapter~\eqref{Chapter5}. A significant improvement in reducing the uncertainties coming from the scale variation and due to truncation of the perturbation series is obtained compared to FOPT studies. Determinations from the Borel-Laplace rules using the FOPT series inherently suffer from large renormalization scale uncertainties that are in fact linear in nature~\cite {Chetyrkin:2005kn}. Using RGSPT, we have significantly reduced such uncertainties in comparison to FOPT in our determinations. One of the drawbacks of using RGSPT in this method is that the Borel transform has to be calculated numerically, as it involves transcendental functions. \par
Borel-Laplace rules are widely used in the literature to extract the various QCD parameters such as quark masses, decay constants of the heavy and heavy-light quark systems, etc. It will be worth exploring these ideas in the determination of the strong coupling constant, quark masses, and condensate terms using the ratio of moments of $e^+e^-$~\cite{Narison:2023srj}. \par

In the final chapter~\eqref{Chapter7}, $\as$ and $\overline{m}_c$ and $\overline{m}_b$ are determined from the moments of the vector and pseudoscalar heavy quark currents. The first four moments are already known to four loops ($\ordas{3}$) and have a significant renormalization scale dependence leading the large-scale uncertainties in determinations. In addition to these, determinations using $\msbar$ quark mass definition in the higher dimensional condensate results in unstable determinations from higher moments. On the experimental side, higher moments are relatively precisely known for which theoretical predictions from FOPT are unstable~\cite{Kuhn:2007vp}. We have used RGSPT to get improved determinations of these parameters which do not suffer from the use of the quark mass definition in the condensate terms. \par
The experimental moments are calculated using the hadronic inputs, such as the width and masses of the resonances. The continuum contributions to it use theoretical inputs from electromagnetic R-ratio, for which results obtained in chapter~\eqref{Chapter5} for kinematical $\pi^2$ term summation can be very useful in controlling the uncertainties related to the truncation error and scale variations.

%% file: Appendices/AppendixA.tex
\chapter{Appendix 1} 

\label{app:mass_run} 

\lhead{Appendix A. \emph{Coupling}} 

		\section{QCD beta function coefficients and quark mass anomalous dimension.}
		The running of strong coupling and the quark masses are computed by solving the following differential equations:
		\begin{align}
			\mu^2\frac{d}{d\mu^2}x=\beta(x)=-\sum_i x^{i+2} \beta_i\,, \quad
			\mu^2 \frac{d}{d\mu^2}m\equiv&\hspace{2mm}m \hspace{.4mm}\gamma_m =	-m\sum_{i}\gamma_i \hspace{.4mm}x^{i+1} \label{anomalous_dim}\,.
		\end{align}
		where $x\equiv x(\mu)$, $L=\log(\mu^2/q^2)$, the $\beta_i$ are the QCD beta function coefficients and $\gamma_i$ are the quark mass anomalous dimension.\par 
		The QCD beta function coefficients are known to five-loop \cite{vanRitbergen:1997va,Gross:1973id,Caswell:1974gg, Jones:1974mm,Tarasov:1980au,Larin:1993tp,Czakon:2004bu,Baikov:2016tgj,Herzog:2017ohr}  and their analytic expression for $n_f$-active flavor are: 
		\begin{align}
			&\beta_0 = \frac{11}{4}-\frac{1}{6}n_f\, \quad 	\beta_1= \frac{51}{8} - \frac{19}{24}n_f\,,\quad
			\beta_2 = \frac{2857}{128} - \frac{5033}{1152} n_f + \frac{325}{3456}n_f^2\,, \nonumber\\
			&\beta_3 = \frac{149753}{1536} - \frac{1078361}{41472} n_f + \frac{50065}{41472} n_f^2 + \frac{1093}{186624} n_f^3 + \frac{891}{64} \zeta(3) - \frac{1627}{1728} n_f \zeta(3) + \frac{809}{2592} n_f^2 \zeta(3)\nonumber\,,
		\end{align}
		\begin{align}
			\beta_{4} =&\frac{8157455}{16384}+\frac{621885 \zeta (3)}{2048}-\frac{9801 \pi ^4}{20480} -\frac{144045 \zeta (5)}{512}+ n_{f}\Big[-\frac{336460813}{1990656}-\frac{1202791 \zeta (3)}{20736}\nonumber\\&+ \frac{6787 \pi ^4}{110592}+\frac{1358995 \zeta (5)}{27648} \Big] +n_{f}^{2}\Big[\frac{25960913}{1990656}+\frac{698531 \zeta (3)}{82944}-\frac{5263 \pi ^4}{414720}-\frac{5965 \zeta (5)}{1296}\Big]\nonumber\\&+n_{f}^{3}\Big[-\frac{630559}{5971968}-\frac{24361 \zeta (3)}{124416}+\frac{809 \pi ^4}{1244160}+\frac{115 \zeta (5)}{2304}\Big]+n_{f}^{4}\Big[\frac{1205}{2985984}-\frac{19 \zeta (3)}{10368}\Big]\, .
		\end{align}			
		
		The known five-loop quark mass anomalous dimension coefficients~\cite{Tarrach:1980up,Tarasov:1982plg,Larin:1993tq,Vermaseren:1997fq,Chetyrkin:1997dh,Baikov:2014qja,Luthe:2016ima,Luthe:2016xec} are:
		\begin{align}
			\gamma_m^{(0)}=&1 \,, \quad \gamma_m^{(1)}=\frac{1}{4^2} \left(\frac{202}{3} +\frac{-20}{9} n_f\right)\,,\nonumber\\ 
			\gamma_m^{(2)}=&\frac{1}{4^3}\Big[1249+n_f \left(-\frac{160 \zeta (3)}{3}-\frac{2216}{27}\right)-\frac{140 n_f^2}{81}\Big]\,,\nonumber\\
			\gamma_m^{(3)}=&\frac{1}{4^4} \Bigg[\frac{135680 \zeta (3)}{27}-8800 \zeta (5)+\frac{4603055}{162}+n_f^2 \left(\frac{800 \zeta (3)}{9}-\frac{160 \zeta (4)}{3}+\frac{5242}{243}\right)\nonumber\\&\bs+n_f \left(-\frac{34192 \zeta (3)}{9}+880 \zeta (4)-\frac{18400 \zeta (5)}{9}-\frac{91723}{27}\right)+n_f^3 \left(\frac{64 \zeta (3)}{27}-\frac{332}{243}\right)\Bigg]\,,\nonumber\\
			\gamma_m^{(4)} = & \frac{1}{4^5}\Bigg\lbrace\frac{99512327}{162}+\frac{46402466 \zeta (3)}{243}+96800 \zeta (3)^2-\frac{698126 \zeta (4)}{9}-\frac{231757160 \zeta (5)}{243}\nonumber\\&\bs+242000 \zeta (6)+412720 \zeta (7)+n_f \Big[-\frac{150736283}{1458}-\frac{12538016 \zeta (3)}{81}\nonumber\\&\bs-\frac{75680 \zeta (3)^2}{9}+\frac{2038742 \zeta (4)}{27}+\frac{49876180 \zeta (5)}{243}-\frac{638000 \zeta (6)}{9}-\frac{1820000 \zeta (7)}{27}\Big]\nonumber\\&\bs+n_f^2 \Big[\frac{1320742}{729}+\frac{2010824 \zeta (3)}{243}+\frac{46400 \zeta (3)^2}{27}-\frac{166300 \zeta (4)}{27}-\frac{264040 \zeta (5)}{81}\nonumber\\&\bs+\frac{92000 \zeta (6)}{27}\Big]+n_f^3 \Big[\frac{91865}{1458}+\frac{12848 \zeta (3)}{81}+\frac{448 \zeta (4)}{9}-\frac{5120 \zeta (5)}{27}\Big]\nonumber\\&\bs+n_f^4 \Big[-\frac{260}{243}-\frac{320 \zeta (3)}{243}+\frac{64 \zeta (4)}{27}\Big]\Bigg\rbrace\,.
		\end{align}
		
		\section{Running of The Perturbative QCD Coupling Constant\label{app:alpha_run}}
		The running of the strong coupling constant in terms of known $\beta$ functions and the strong coupling at renormalization scale $\mu$ \cite{Jezabek:1998wk}, is given by:
		\begin{equation}
		\resizebox{\textwidth}{!}{$
			\begin{aligned}
			x(p) =& x \Bigg(1+ x \beta _0 L+x^2 \left(\beta _1 L+\beta _0^2 L^2\right)+x^3 \left(\beta _2 L+\frac{5}{2} \beta _1 \beta _0 L^2+\beta _0^3 L^3\right)\nonumber\\&+x^4 \big(\beta _3 L+\big(\frac{3 \beta _1^2}{2}+3 \beta _0 \beta _2\big) L^2+\frac{13}{3} \beta _1 \beta _0^2 L^3+\beta _0^4 L^4\big)+x^5\Big( \beta _4 L+\big(\frac{7 \beta _1 \beta _2}{2}+\frac{7 \beta _0 \beta _3}{2}\big) L^2  \nonumber\\&+\big(6 \beta _2 \beta _0^2+\frac{35}{6} \beta _1^2 \beta _0\big) L^3+\frac{77}{12} \beta _1 \beta _0^3 L^4+\beta _0^5 L^5\Big) \Bigg)+ \order{x^6}
			\label{alphasmu}
	\end{aligned}$}
\end{equation}
		where $L=\log(\mu^2/p^2)$. \par 
		The coupling used to extract the $\Lambda^{\overline{\textrm{MS}}}_{\textrm{QCD}}$ at scale $\mu$ is given by:
		\begin{equation}
		\resizebox{\textwidth}{!}{$
			\begin{aligned}
			x(\mu)=&\frac{ y }{b_1}\Bigg(1-\ell y+y^2 \left(\frac{b_2}{b_1^2}+\ell^2-\ell-1\right)-y^3 \left(-\left(2-\frac{3 b_2}{b_1^2}\right) \ell+\frac{1}{2} \left(1-\frac{b_3}{b_1^3}\right)+\ell^3-\frac{5 \ell^2}{2}\right)\nonumber\\&+y^4 \bigg[-\left(\frac{3}{2}-\frac{6 b_2}{b_1^2}\right) \ell^2+\left(-\frac{3 b_2}{b_1^2}-\frac{2 b_3}{b_1^3}+4\right) \ell+\frac{b_4}{3 b_1^4}-\frac{b_2 \left(3-\frac{5 b_2}{3 b_1^2}\right)}{b_1^2}-\frac{b_3}{6 b_1^3}\nonumber\\&\bs\bs+\ell^4-\frac{13 \ell^3}{3}+\frac{7}{6}\bigg]\Bigg)+\order{y^6}
		\end{aligned}$}
		\label{as_lam}
\end{equation}
		
		where $b_i\equiv \frac{\beta_i}{\beta_0}$, $\ell\equiv\log(\log(\mu^2/(\Lambda^{\overline{\textrm{MS}}}_{\textrm{QCD}})^2))$ and $y\equiv\frac{b_1}{\beta_0 \log(\mu^2/(\Lambda^{\overline{\textrm{MS}}}_{\textrm{QCD}})^2)}$.
		
		\section{Running of quark masses}
			\begin{align}
			m(q^2)=&m(\mu^2)\Bigg\lbrace1+x \gamma _0 L+x^2 \left(\gamma _1 L+\frac{1}{2} \gamma _0 L^2 \left(\beta _0+\gamma _0\right)\right)\nonumber\\&+x^3 \Big[\frac{1}{6} \gamma _0 L^3 \left(\beta _0+\gamma _0\right) \left(2 \beta _0+\gamma _0\right)+\gamma _2 L+L^2 \left(\frac{\beta _1 \gamma _0}{2}+\gamma _1 \left(\beta _0+\gamma _0\right)\right)\Big]\nonumber\\&+x^4 \Bigg[\gamma _3 L+L^2 \Big(\beta _1 \gamma _1+\frac{\beta _2 \gamma _0}{2}+\frac{3 \beta _0 \gamma _2}{2}+\frac{\gamma _1^2}{2}+\gamma _0 \gamma _2\Big)\nonumber\\&\bs+L^3 \Big(\frac{1}{6} \left(\beta _1 \gamma _0 \left(5 \beta _0+3 \gamma _0\right)+3 \gamma _1 \left(\beta _0+\gamma _0\right) \left(2 \beta _0+\gamma _0\right)\right)\Big)\nonumber\\&\bs+\frac{1}{24} \gamma _0 L^4 \left(\beta _0+\gamma _0\right) \left(2 \beta _0+\gamma _0\right) \left(3 \beta _0+\gamma _0\right)\Bigg]\nonumber\\&+x^5\Bigg[\gamma _4 L+\frac{1}{2} L^2 \Big[\beta _3 \gamma _0+3 \beta _1 \gamma _2+2 \gamma _1 \left(\beta _2+\gamma _2\right)+2 \gamma _3 \left(2 \beta _0+\gamma _0\right)\Big]\nonumber\\&+\frac{1}{6} L^3 \left(3 \beta _1^2 \gamma _0+\beta _1 \gamma _1 \left(14 \beta _0+9 \gamma _0\right)+3 \left(2 \beta _0+\gamma _0\right) \left(\beta _2 \gamma _0+\gamma _2 \left(2 \beta _0+\gamma _0\right)+\gamma _1^2\right)\right)\nonumber\\&+\frac{1}{12} L^4 \left(\beta _1 \gamma _0 \left(13 \beta _0 \gamma _0+13 \beta _0^2+3 \gamma _0^2\right)+2 \gamma _1 \left(\beta _0+\gamma _0\right) \left(2 \beta _0+\gamma _0\right) \left(3 \beta _0+\gamma _0\right)\right)\nonumber\\&+\frac{1}{120} \gamma _0 L^5 \left(\beta _0+\gamma _0\right) \left(2 \beta _0+\gamma _0\right) \left(3 \beta _0+\gamma _0\right) \left(4 \beta _0+\gamma _0\right)\Bigg]\Bigg\rbrace+\order{\alpha_s^6}\,,
			\label{eq:massmu}
		\end{align}